\documentclass[a4paper,12pt,twoside]{article}

\title{The Effect of Hole Ice on the Propagation and Detection of Light in IceCube}
\author{Sebastian Fiedlschuster}
\date{September 2018}
\newcommand\shorttitle{Hole Ice}
\newcommand\documentid{FAU-ECAP-DIPL-2018-001}

\usepackage{geometry}
\usepackage{microtype}
\usepackage{parskip}

\usepackage[T1]{fontenc}
\usepackage[utf8]{inputenc}
\usepackage{lipsum}
\usepackage{graphicx}
\usepackage[font=footnotesize]{caption}

\newcommand\image[1]{\includegraphics[width=\textwidth]{img/#1}}
\newcommand\smallerimage[1]{\centering\includegraphics[width=0.7\textwidth]{img/#1}}
\newcommand\halfimage[1]{\includegraphics[width=0.48\textwidth]{img/#1}}
\newcommand\halfcropimage[2]{\includegraphics[width=0.48\textwidth, trim={#2}, clip]{img/#1}}
\newcommand\thirdimage[1]{\includegraphics[width=0.32\textwidth]{img/#1}}

\usepackage{placeins} %

\usepackage{fontawesome}

\usepackage[font=scriptsize, position=b]{subcaption}

\usepackage{pgfplots}

\usepackage{tikz-feynman}
\usepackage{tkz-euclide}
\usetkzobj{angles}

\usetikzlibrary{calc}
\usetikzlibrary{decorations.pathmorphing}
\usetikzlibrary{intersections,through}
\usetikzlibrary{arrows,decorations.markings}
\usetikzlibrary{shapes,decorations.shapes}

\tikzset{
  photon/.style={draw, postaction={decorate}, decoration={markings, mark=at position 1 with {\arrow[scale=2, >=latex]{>}}}},
  holeice/.style={draw, dash pattern=on \pdflinewidth off 2pt}
}
\usetikzlibrary{arrows.meta}

\tikzset{%
  >={Latex[width=2mm,length=2mm]},
            base/.style = {rectangle, rounded corners, draw=black,
                           minimum width=4cm, minimum height=1cm,
                           text centered, font=\sffamily},
           start/.style = {base, fill=blue!30, rounded corners = 15},
            stop/.style = {start},
         process/.style = {base, minimum width=2.5cm, fill=orange!15,
                           font=\ttfamily},
        decision/.style = {base, diamond, inner sep = 0pt, fill = green!30, minimum width = 2cm, minimum height = 2cm, node distance = 2.5cm},
        substeps/.style = {font = \sffamily, align = left},
}

\newcommand\substep{- }

\providecommand{\tightlist}{}

\providecommand{\texorpdfstring}[2]{#1}

\usepackage{color}

\usepackage[backend=biber, style=alphabetic]{biblatex}

\usepackage[nottoc, numbib]{tocbibind}
\usepackage{amsmath}
\usepackage{amssymb}

\renewcommand{\vec}[1]{\overrightarrow{#1}}
\newcommand{\norm}[1]{\left\lVert#1\right\rVert}
\newcommand{\len}[1]{\overline{#1}}

\newcommand{\identical}{\equiv}

\newcommand{\reals}{\mathbb{R}}
\newcommand{\naturals}{\mathbb{N}}

\newcommand{\e}{\mathrm{e}}

\usepackage{xfrac}
\usepackage{pdftexcmds}
\makeatletter
\ifcase\pdf@shellescape
  \usepackage[frozencache]{minted}\or %
  \usepackage[finalizecache]{minted}\or %
  \usepackage[frozencache]{minted}\fi %
\makeatother

\setminted{
  tabsize=2, linenos, breaklines, breakafter=/_, autogobble, breakautoindent, frame=lines, framesep=2mm, fontsize=\footnotesize
}

\newenvironment{python}{\minted{python}}{\endminted}

\newenvironment{ccode}{\minted{c}}{\endminted}

\setminted[bash]{
  linenos=false
}

\newenvironment{bash}{\minted{bash}}{\endminted}

\usepackage[ngerman, english]{babel}

\usepackage{iflang}

\IfLanguageName{ngerman}{
  \newcommand\pageStr{Seite}
}{
  \newcommand\pageStr{Page}
}
\usepackage[explicit]{titlesec}

\newcommand\currentsection{}
\newcommand\currentsubsection{}
\newcommand\currentsubsubsection{}

\newcommand\preparesectionmarks{
  \renewcommand\sectionmark[1]{\renewcommand\currentsection{##1}\renewcommand\currentsubsection{}}
  \renewcommand\subsectionmark[1]{\renewcommand\currentsubsection{##1}}
  \renewcommand\subsubsectionmark[1]{\renewcommand\currentsubsubsection{##1}}
}

\let\oldsection\section
\def\section{\cleardoublepage\oldsection}
\input{lib/header}
\usepackage{xspace}

\newcommand\unit[1]{\text{\,#1}\xspace}
\newcommand\m{\unit{m}}
\newcommand\cm{\unit{cm}}
\newcommand\mm{\unit{mm}}
\newcommand\nm{\unit{nm}}

\newcommand\ns{\unit{ns}}

\newcommand\GeV{\unit{GeV}}
\newcommand\TeV{\unit{TeV}}
\newcommand\PeV{\unit{PeV}}

\usepackage{siunitx}
\usepackage{physics}
\usepackage{hepnames}
\usepackage{tabularx}
\usepackage{multirow}

\newcommand{\oldarraystretch}{}
\newcolumntype{C}{>{\centering\arraybackslash}X}
\newcolumntype{R}{>{\raggedleft\arraybackslash}X}
\newcolumntype{L}{>{\raggedright\arraybackslash}X}

\newenvironment{tabelle}[1]%
{%
\renewcommand{\oldarraystretch}{\arraystretch}
  \renewcommand{\arraystretch}{1.5} %
  \noindent\tabularx{\textwidth}{#1}
}{%
  \endtabularx
  \renewcommand{\arraystretch}{\oldarraystretch}
}

\usepackage{adjustbox}

\usepackage{colortbl}
\usepackage{url}

\usepackage{footmisc}

\usepackage{xcolor}
\definecolor{bluelink}{rgb}{.247,.443,.729}
\definecolor{greenlink}{rgb}{.23,.57,.23}
\definecolor{redlink}{rgb}{.51,.169,.129}

\newcommand\defaulthypersetup{\hypersetup{allcolors=black, citecolor=greenlink, linkcolor=bluelink, urlcolor=black}}

\usepackage[breaklinks, colorlinks]{hyperref}\defaulthypersetup
\usepackage[hyphenbreaks]{breakurl}

\newcommand\docframe[1]{%
  \vspace{1em}\fbox{\minipage{\textwidth}#1\endminipage}\vspace{1em}%
}

\newcommand\iconpar[2]{%
  \docframe{\iconparwithoutframe{#1}{#2}}%
}

\newcommand\iconparwithoutframe[2]{%
  \begin{tabularx}{\textwidth}{lL}%
    \adjustbox{minipage=6mm, center=6mm, valign=t, raise=-2mm}{\center{#1}} & \footnotesize\sffamily #2%
  \end{tabularx}%
}

\newcommand\documentationicon{\faBook}
\newcommand\docpar[1]{\iconpar{\documentationicon}{#1}}
\newcommand\docparwithoutframe[1]{\iconparwithoutframe{\documentationicon}{#1}}

\newcommand\sourcecodeicon{\faFileTextO}
\newcommand\sourcepar[1]{\iconpar{\sourcecodeicon}{#1}}
\newcommand\sourceparwithoutframe[1]{\iconparwithoutframe{\sourcecodeicon}{#1}}

\newcommand\youtubeicon{\faVideoCamera} %
\newcommand\youtubepar[1]{\iconpar{\youtubeicon}{#1}}

\usepackage{csquotes} %

\newcommand\noun[1]{\textsc{#1}}
\usepackage[acronym]{glossaries}

\usepackage{pifont}

\renewcommand\d{\text{d}}
\newcommand\dx{\d x}
\newcommand\dn{\d n}

\renewcommand\abs{_\text{abs}}
\newcommand\sca{_\text{sca}}
\newcommand\esca{_\text{e}}
\newcommand\hi{^\text{H}}
\newcommand\lambdaabs{\lambda\abs}
\newcommand\dom{_\text{DOM}}
\newcommand\domhi{_\text{DOM,HI}}
\newcommand\domdima{_\text{DOM,HI}^\text{Dima}}
\newcommand\llh{\text{LLH}}

\newcommand\imc{_{i,\text{MC}}}
\newcommand\meancostheta{\langle \cos \theta \rangle}

\newcommand\issue[1]{\texttt{issues/#1} on the CD-ROM as well as online at \url{https://github.com/fiedl/hole-ice-study/issues/#1}}
\newcommand\script[1]{\texttt{hole-ice-study/scripts/#1} on the CD-ROM as well as online at \url{https://github.com/fiedl/hole-ice-study/tree/master/scripts/#1}}

\newcommand\followup{}

\newcommand\authorname[1]{\noun{#1}\xspace}
\newcommand\softwarename[1]{\noun{#1}\xspace}
\newcommand\icecube{\noun{IceCube}\xspace}
\newcommand\icesim{\softwarename{IceSim}}
\newcommand\clsim{\softwarename{Clsim}}
\newcommand\ppc{\softwarename{Ppc}}
\newcommand\photonics{\softwarename{Photonics}}
\newcommand\steamshovel{\softwarename{Steamshovel}}
\newcommand\rongen{\authorname{Rongen}}
\newcommand\chirkin{\authorname{Chirkin}}

\newcommand\HH{\mathcal{H}}
\newcommand\HHbest{{\HH_\text{b}}}

\newcommand\nbsp{\nobreakspace{}}

\bibliography{diplomarbeit}
\begin{document}

  \pagestyle{empty}
  \input{config/titlepage}

\makeatletter
\begin{titlepage}
  \begin{center}

    \LARGE \@title

    \vspace{1cm}

    \large

    \@author \medskip

    \@date

    \vspace{1cm}
    \includegraphics[width=5cm, decodearray={0.2 0.5}]{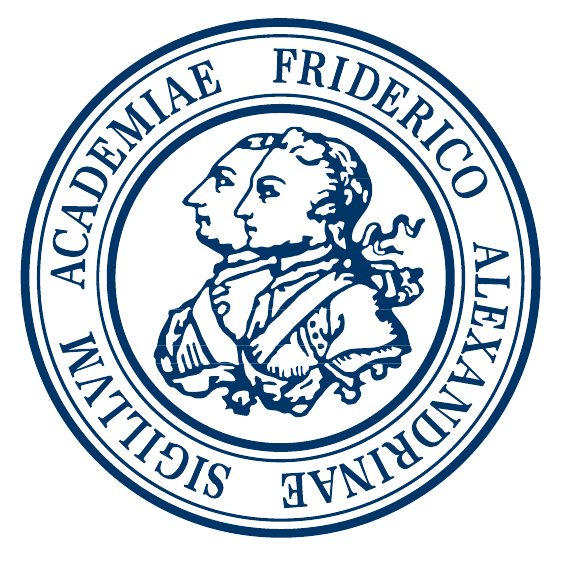}
    \vspace{1cm}

    Erlangen Centre For Astroparticle Physics \medskip

    Friedrich-Alexander-Universität Erlangen-Nürnberg \medskip

    Prof.\,Dr.~Uli Katz \medskip

    Dr.~Thorsten Glüsenkamp

    \vfill
    \includegraphics[width=4.5cm]{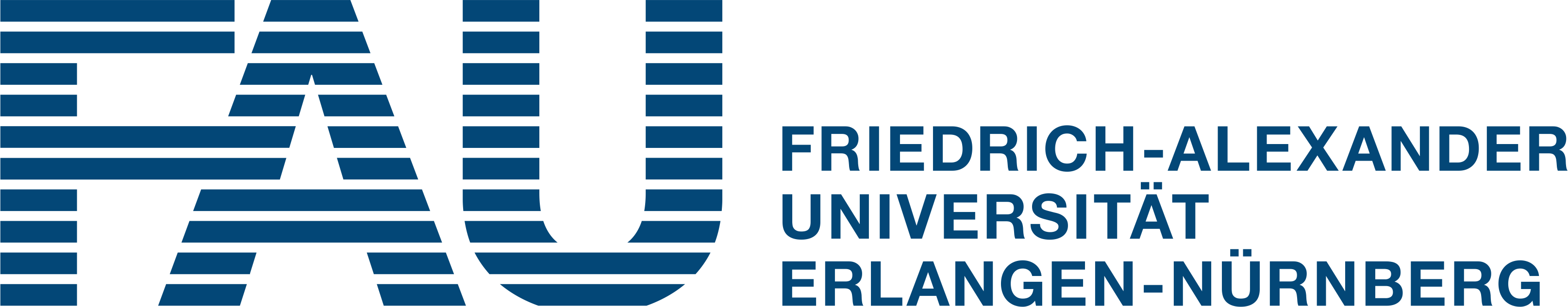}\hfill
    \raisebox{-5mm}{\includegraphics[width=4cm]{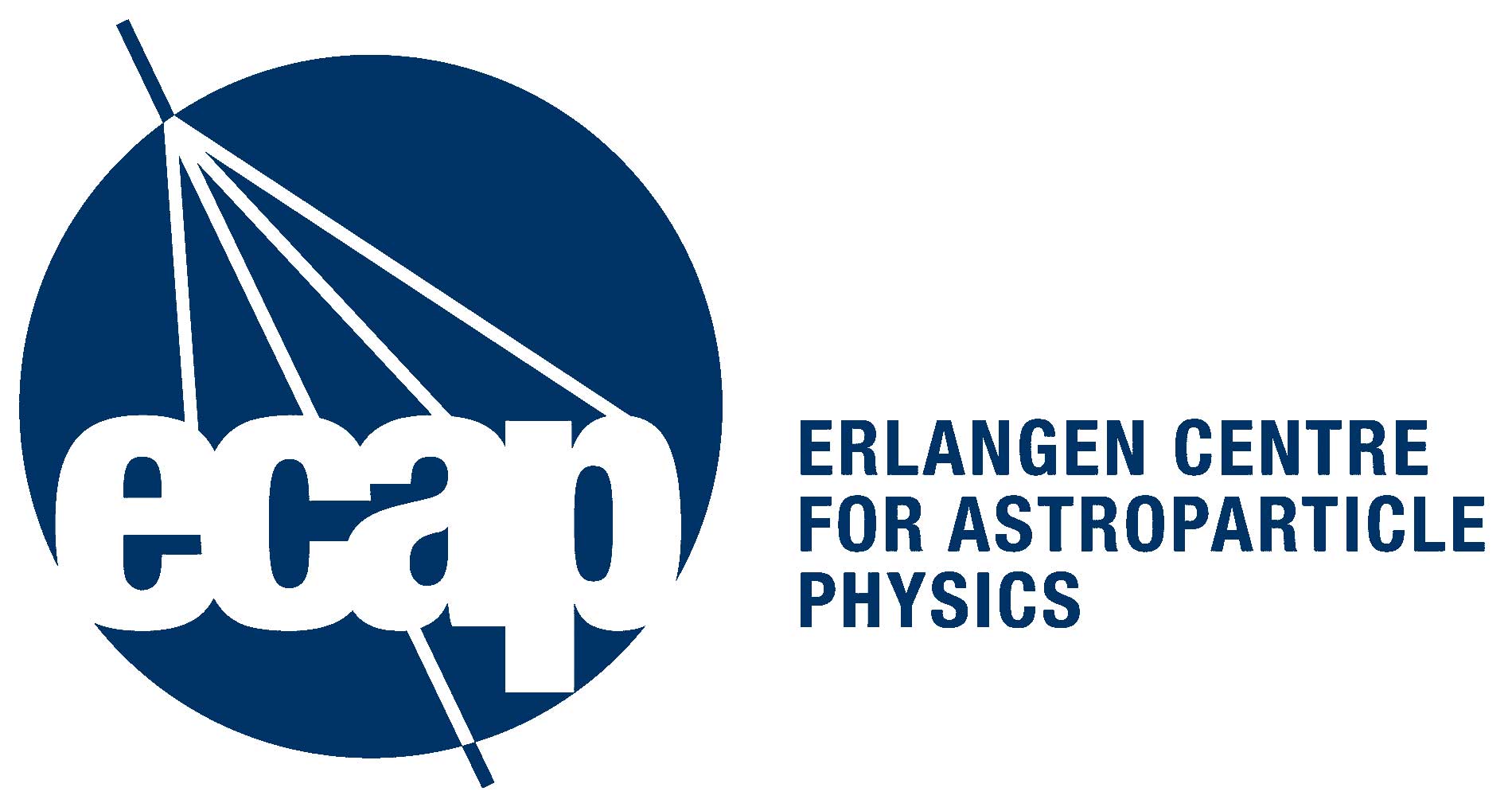}}\hfill
    \raisebox{-3mm}{\includegraphics[width=4.5cm]{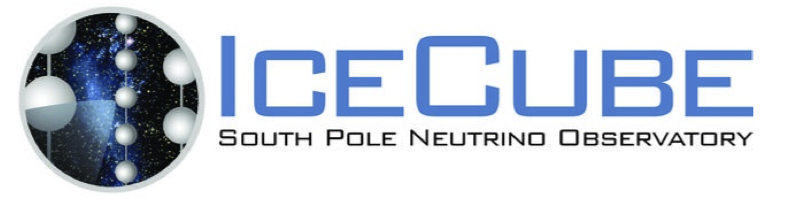}}

  \end{center}
\end{titlepage}
\makeatother

\newpage
\thispagestyle{empty}
\mbox{}\vfill
Copyright 2018, Sebastian Fiedlschuster.

Printed in Erlangen, Germany.

The \LaTeX\xspace version of this document can be found online at \url{https://github.com/fiedl/hole-ice-latex}.

Additional material can be found at \url{https://github.com/fiedl/hole-ice-study}.

  \cleardoublepage
\thispagestyle{empty}

\begin{abstract}

\icecube is a neutrino observatory at Earth's South Pole that uses glacial ice as detector medium. Secondary particles from neutrino interactions produce Cherenkov light, which is detected by an array of photo detectors deployed within the ice. In distinction from the glacial bulk ice, hole ice is the refrozen water in the drill holes around the detector modules, and is expected to have different optical properties than the bulk ice.
Aiming to improve detector precision, this study presents a new method to simulate the propagation of light through the hole ice, introducing several new calibration parameters. The validity of the method is supported by a series of statistical cross checks, and by comparison to measurement and simulation results from other calibration studies.
Evaluating calibration data indicates a strongly asymmetric shielding of the detector modules. A preliminary analysis suggests that this cannot be accounted for by the shadow of cables, but can be explained by hole ice with a suitable scattering length, size, and position relative to the detector modules.
The hole-ice approximation, which is used in the standard simulation chain is found to disagree with all existing direct-propagation methods and should be recalculated with a new direct-simulation run.

\end{abstract}

  \cleardoublepage
\thispagestyle{empty}

\begin{otherlanguage}{ngerman}

\makeatletter
\begin{center}
  \large Der Einfluss von Loch-Eis auf die Ausbreitung und Detektion von Licht im IceCube-Neutrino-Observatorium

  \medskip
  \normalsize \@author

  \@date
\end{center}
\makeatother

\vspace{1cm}

\begin{abstract}

Das \icecube-Neutrino-Observatorium am Südpol verwendet das Eis eines Gletschers als Detektor-Medium, in dem Teilchen aus Neutrino-Reaktionen auf ihrem Weg durch das Eis Licht erzeugen, das von Photo-Detektoren im Eis registriert wird. Loch-Eis (engl. \glqq hole ice\grqq) ist das erneut gefrorene Wasser in den Bohrlöchern, in denen die Detektor-Module in das Eis eingelassen wurden, und hat voraussichtlich andere optische Eigenschaften als das übrige Eis des Gletschers.

Um die Detektor-Kalibrierung und damit die Genauigkeit von \icecube zu verbessern, stellt diese Arbeit neue Algorithmen vor, mit denen die Propagation von Photonen durch Loch-Eis mit verschiedenen Eigenschaften simuliert werden kann. Die Tauglichkeit der Algorithmen wird durch eine Reihe von statistischen Überprüfungen sowie durch einen Vergleich mit Messungen und Simulationen anderer Kalibrierungsstudien untermauert.

Als Anwendungsbeispiele werden in dieser Arbeit die Simulation eines oder mehrerer Loch-Eis-Zylinder mit unterschiedlichen Eigenschaften, etwa der Lage, der Größe, der Absorptions- und der Streulänge des Loch-Eises, die Simulation eines lichtabsorbierenden Kabels sowie die beispielhafte Kalibrierung anhand von Daten des Leuchtdioden-Kalibrierungs-Systems von \icecube durchgeführt.

Die großräumige Ausbreitung von Licht durch das Eis des Gletschers erfolgt nahezu unbeeinflusst von den Eigenschaften des Loch-Eises. Allerdings muss jedes Photon, das von \icecube registriert oder vom Kalibrierungssystem abgegeben wird, zunächst durch das Loch-Eis gelangen, sodass sowohl die Detektor-Einheiten als auch die Kalibrierungs-Leuchtdioden in Abhängigkeit von den Eigenschaften des Loch-Eises effektiv abgeschirmt werden, da ein Teil des Lichtes vom Loch-Eis absorbiert, ein anderer Teil reflektiert werden. Für Detektor-Module, die im Loch-Eis nicht völlig zentrisch zum Liegen gekommen sind, ist dieser Effekt abhängig von der Azimuth-Richtung der Photonen, was sich auch in den Kalibrierungsdaten widerspiegelt.

Aus Kalibrierungsdaten ergibt sich, dass die Detektor-Module richtungsabhängig abgeschirmt werden. Vorläufige Ergebnisse dieser Simulations-Studie zeigen, dass die beobachtete Abschirmung nicht von Kabeln verursacht werden kann, die an der Seite der Detektor-Module verlaufen, sondern die Annahme eines Loch-Eises geeigneter Lage, Größe und optischer Eigenschaften erforderlich ist, um die Kalibrierungsdaten zu erklären.

Die neuen Propagations-Algorithmen werden in Kürze in das Simulations-System von \icecube integriert werden. Studien, die eine geringe Statistik aufweisen, sich also mit Neutrino-Reaktionen befassen, die nur wenig Licht hervorrufen, oder, bei denen das Licht nur von wenigen Detektor-Modulen registriert wird, können mit den in dieser Arbeit vorgestellten Simulationsmethoden systematische Unsicherheiten verringern.

Da die direkte Simulation der Licht-Propagation durch das Loch-Eis jedoch zusätzliche Simulations-Laufzeit erfordert, ist es empfehlenswert, dass Studien mit einer hohen Statistik, die sich also mit Neutrino-Reaktionen befassen, die eine große Menge an Licht hervorrufen, das wiederum von einer Vielzahl von Detektor-Modulen registriert wird, den Einfluss des Loch-Eises durch bereits gebräuchliche Verfahren nur näherungsweise berücksichtigen. Die Parameter dieser Näherungsverfahren sollten jedoch aufgrund der Erkenntnisse aus direkten Simulationen korrigiert werden.

\end{abstract}

\end{otherlanguage}

  \cleardoublepage
  \pagestyle{fancy}
  \hypersetup{allcolors=black}
  \setcounter{tocdepth}{2}
  \tableofcontents
  \defaulthypersetup

  \cleardoublepage
\section*{CD-ROM}
\label{sec:cd_rom}
\addcontentsline{toc}{section}{\protect\numberline{}CD-ROM}

For the contents of the CD-ROM, see appendix \ref{sec:cd_rom_contents}.

  \section{Introduction}
\label{sec:intro}

\icecube is a neutrino observatory located at Earth's South Pole. It
uses a cubic-kilometer of glacial ice as detector medium where secondary
particles from neutrino interactions produce light as they move through
the ice. The light is detected by an array of photo detector modules
that are deployed throughout the ice. \cite{evidence2013}

The primary scientific objective of \icecube is the study of neutrinos
with energies ranging from \(10\TeV\) to \(10\PeV\) produced in
astrophysical processes, and the identification and characterization of
their sources. In collaboration with other neutrino detectors as
\noun{Antares}, with optical, x-ray, gamma-ray, radio, and
gravitational-wave observatories, \icecube participates in efforts for
multi-messenger astronomy. Other objectives include the indirect
detection of dark matter, the search for other exotic particles, and the
study of neutrino-oscillation physics.
\cite{instrumentation, evidence2013}

As \icecube detects neutrinos indirectly through the interaction with
other particles, involving a chain of processes and components, a key
requirement for precise measurements is to minimize uncertainties for
each process and component involved. Some components such as technical
instruments in the detector modules can be tested and calibrated in
isolation in laboratories. Other components involved such as the glacial
ice cannot be extracted and need to be studied where they are.
Uncertainties concerning the properties of the glacial ice can affect
the precision for measurements of the direction of the detected
neutrinos by several percent. \cite{wrede}

All light that is detected by the detector modules needs to travel
through the refrozen water of the drill holes that were needed to deploy
the detector modules within the ice. This so-called \textit{hole ice}
may have properties significantly different from the surrounding bulk
ice regarding the propagation of light through this medium. The
properties of the hole ice are less known than the properties of the
bulk ice and pose the largest systematic uncertainty for study of
neutrino oscillations and a number of other analyses. \cite{icrc17pocam}

This study aims to provide the necessary tools to improve detector
calibration by introducing the means to simulate the propagation of
light through the hole ice. By comparing different simulation scenarios,
involving hole ice of different respective properties, to calibration
data, it is then possible to study the properties of the hole ice and
its effect on the propagation of light, and on the detection of light by
the detector modules, reducing the systematic uncertainties imposed by
the hole ice, and in the long run improving the precision of the
\icecube observatory.

After providing some background information in sections
\ref{sec:theoretical_background} to \ref{sec:simulation_background}, two
algorithms and their integration into the existing \icecube software
framework will be presented in section \ref{sec:methods} that allow to
simulate the direct propagation of photons through hole ice of different
properties. The validity of the algorithms will be supported by a series
of tests and cross checks in section
\ref{sec:unit_tests_and_cross_checks}.

Examples of application such as the simulation of one or several
hole-ice cylinders with different sizes and photon scattering lengths,
the simulation of shadowing cables, and a calibration method using LED
flasher data are given in section\nbsp\ref{sec:applications}.
Section\nbsp\ref{sec:discussion} presents a brief comparison of methods
and preliminary results to other studies. This section also discusses
performance considerations and lists ice properties not considered in
this study.

The material needed to reproduce this study is provided on the
accompanying CD-ROM and can be found online at
\url{https://github.com/fiedl/hole-ice-study}.

  \section{Theoretical Concepts}
\label{sec:theoretical_background}

\subsection{Neutrinos}
\label{sec:neutrinos}

Neutrinos are particles that primarily interact with other particles
through the \textit{weak interaction} and have very small probabilities
of interacting with other particles, allowing them to cross matter
almost unhindered. While this makes them interesting messenger particles
for observing far-distant astronomical sources and phenomena, because
they are long ranging and arrive undeflected and unscattered, it also
makes their detection challenging due to the large amount of detector
medium required. \cite{lexikonderphysik, instrumentation}

\begin{figure}[htbp]
  \smallerimage{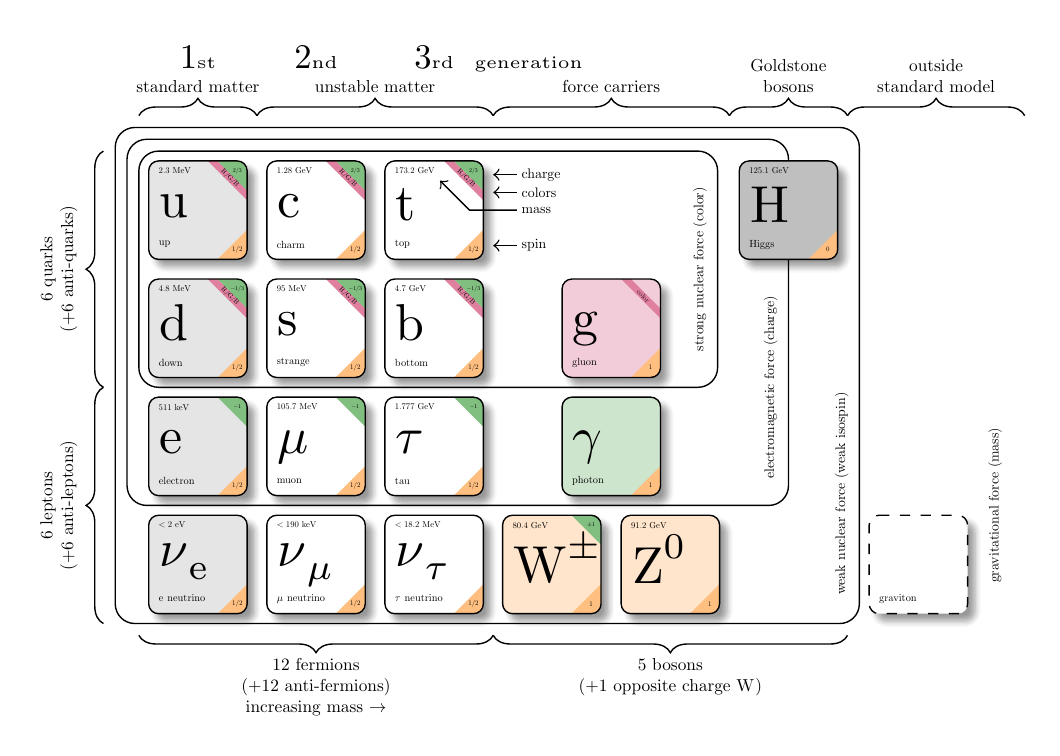}
  \caption{Particles of the standard model of particle physics. Together with the electron, the muon, and the tau, the neutrinos form the group of leptons, which do not participate in \textit{strong interactions}. As neutrinos also do not participate in the \textit{electromagnetic interaction}, their primary interaction channel is through \textit{weak interactions}. Image based on: \cite{standardmodel}}
  \label{fig:Pheith9i}
\end{figure}

Within the \textit{standard model of particle physics} (see figure
\ref{fig:Pheith9i}), the electron neutrino, \(\Pnue\), the muon
neutrino, \(\Pnum\), and the tau neutrino \(\Pnut\), are
\textit{leptons} without electrical charge and couple to the \(\PWpm\)
and \(\PZz\) bosons.

As the \textit{mass eigenstates} \(\Pnu_i\) of neutrinos that describe
neutrinos in the context of their propagation through spacetime, are not
identical to the neutrino \textit{flavor eigenstates} \(\Pnulepton\)
that describe neutrinos participating in weak interactions, neutrinos
are subject to quantum mechanical phenomena as
\textit{neutrino oscillations}. \cite{particledatareview}

\[
  \ket{\Pnulepton} = \sum_{i=1}^N U_{\Plepton i}\,\ket{\Pnu_i}, \ \ \ \Plepton \in \{ \Pe, \Pmu, \Ptau \}
\]

Determining the mixing matrix \(U_{\Plepton i}\) that connects the mass
eigenstates to the flavor eigenstates, is part of \icecube's scientific
objectives.

Neutrinos generated by interactions of cosmic-ray particles in the
Earth's atmosphere, \textit{atmospherical neutrinos}, have energies in
the GeV to TeV scale. \textit{Astrophysical neutrinos}, generated by
high-energy phenomena in the universe, have energies up to the PeV
scale. \cite{instrumentation}

\subsection{Neutrino Interactions Relevant to \icecube}
\label{sec:neutrino_interactions}

\begin{figure}[htbp]
  \centering
  \subcaptionbox{Neutral-current neutrino interaction through $\PZz$ bosons.}{\includegraphics[width=0.3\textwidth]{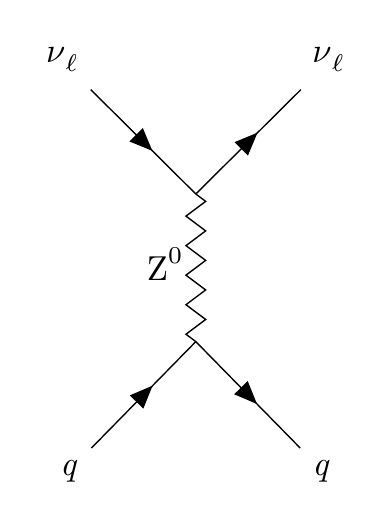}\vspace*{2mm}}\hspace{1cm}
  \subcaptionbox{Charged-current neutrino interactions through $\PWpm$ bosons.}{\includegraphics[width=0.3\textwidth]{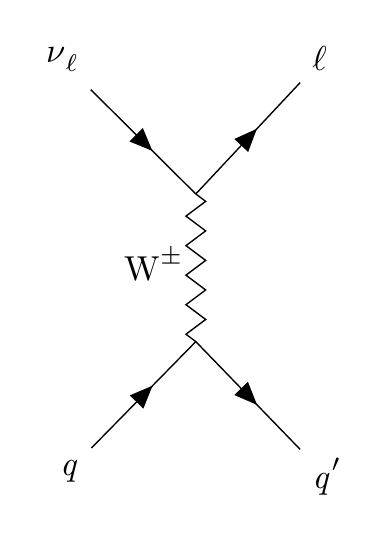}}
  \caption{Feynman diagrams showing neutrinos $\Pnulepton$ interacting with quarks~$\Pquark$ of nuclei of the ice or nearby rock through $\PZz$ and $\PWpm$ bosons, producing leptons~$\Plepton$ (electrons~$\Pe$, muons~$\Pmu$, tau particles~$\Ptau$) and quarks. Time evolves to the right in these diagrams.}
  \label{fig:Phei1oob}
\end{figure}

The primary interaction channel for detecting neutrinos in \icecube is
deep-inelastic scattering of neutrinos with quarks of nuclei in the
detector material or nearby rock (see figure \ref{fig:Phei1oob}).
\cite{energyreco}

\begin{figure}[htbp]
  \centering
  \subcaptionbox{Cascade-like event: The cascade is completely contained within the detector and deposits a total of $1141\TeV$ energy within the detector. Image and data source: \cite{evidence2013,energyreco}}{\halfimage{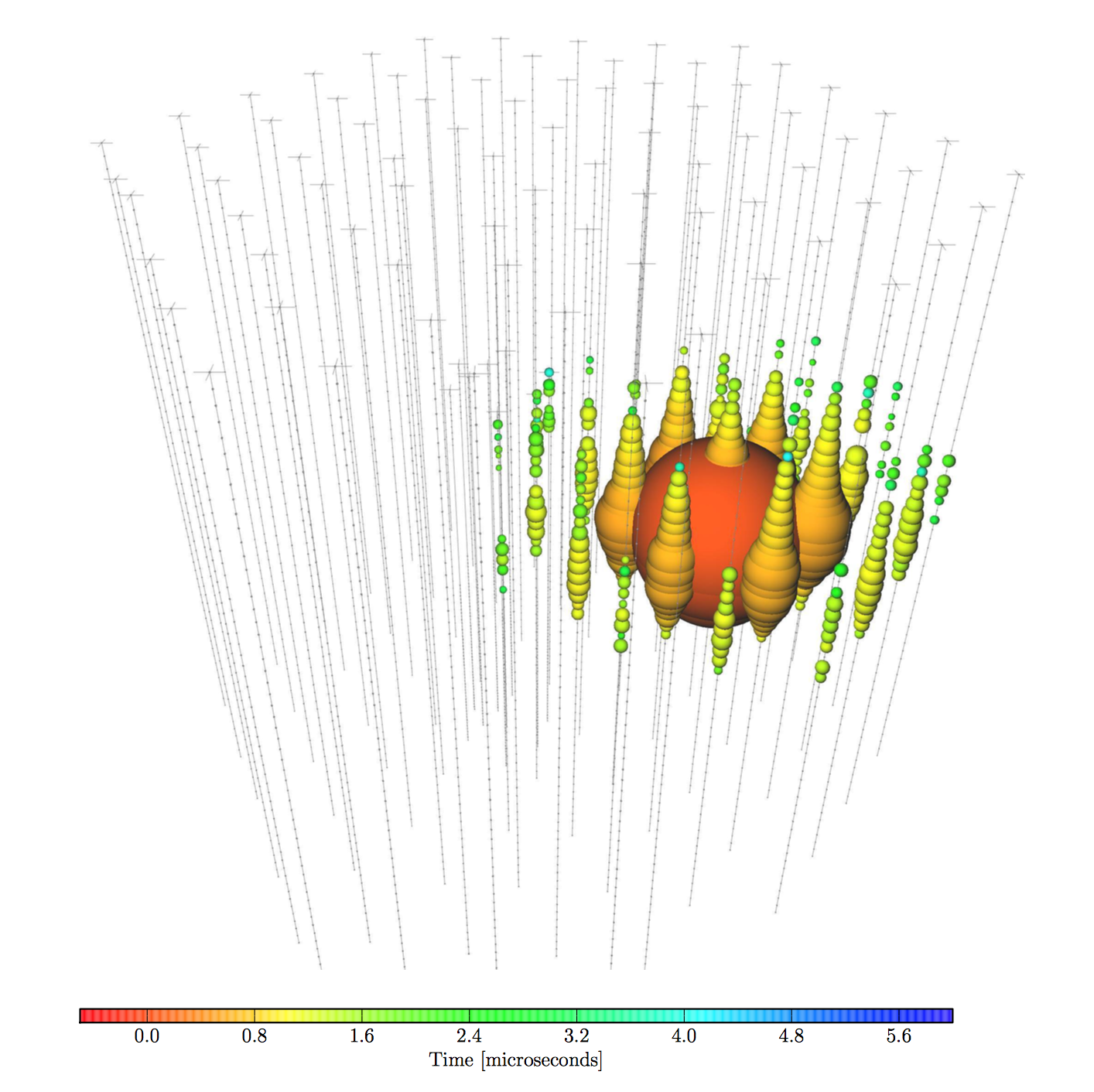}}\hfill
  \subcaptionbox{Track-like event: The muon track starts within the detector and deposits a total of $71\TeV$ of energy within the detector before it leaves the detector volume. Image and data source: \cite{evidence2013,energyreco}}{\halfimage{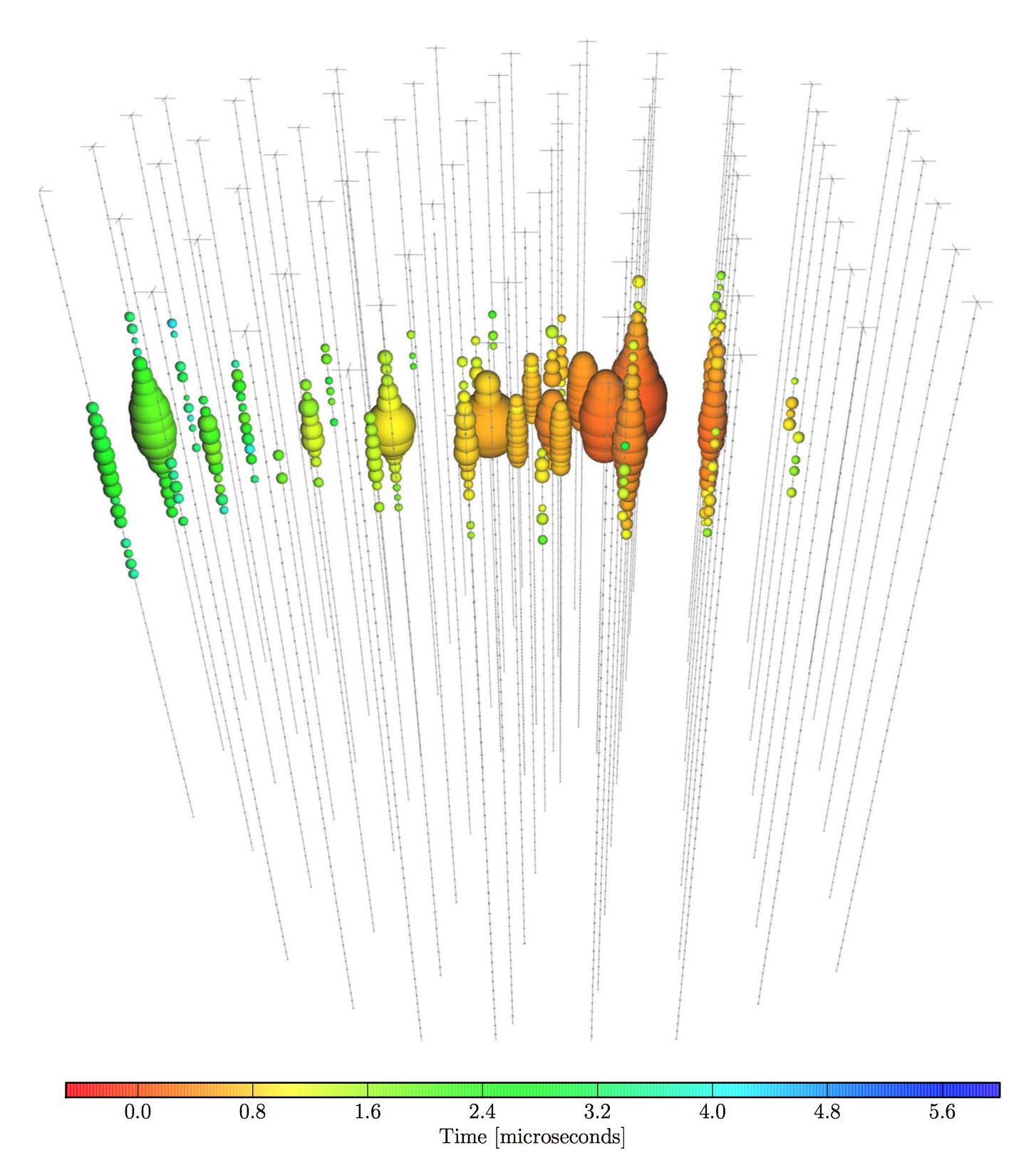}}\hfill
  \subcaptionbox{Simulated double-bang event: The earlier (red) cascade has been created from the primary neutrino interaction vertex. The tau travels from the position of the first cascade for a short distance, creating a track, before decaying into another later (green) cascade. Image source: \cite{nufact2018}}{\halfimage{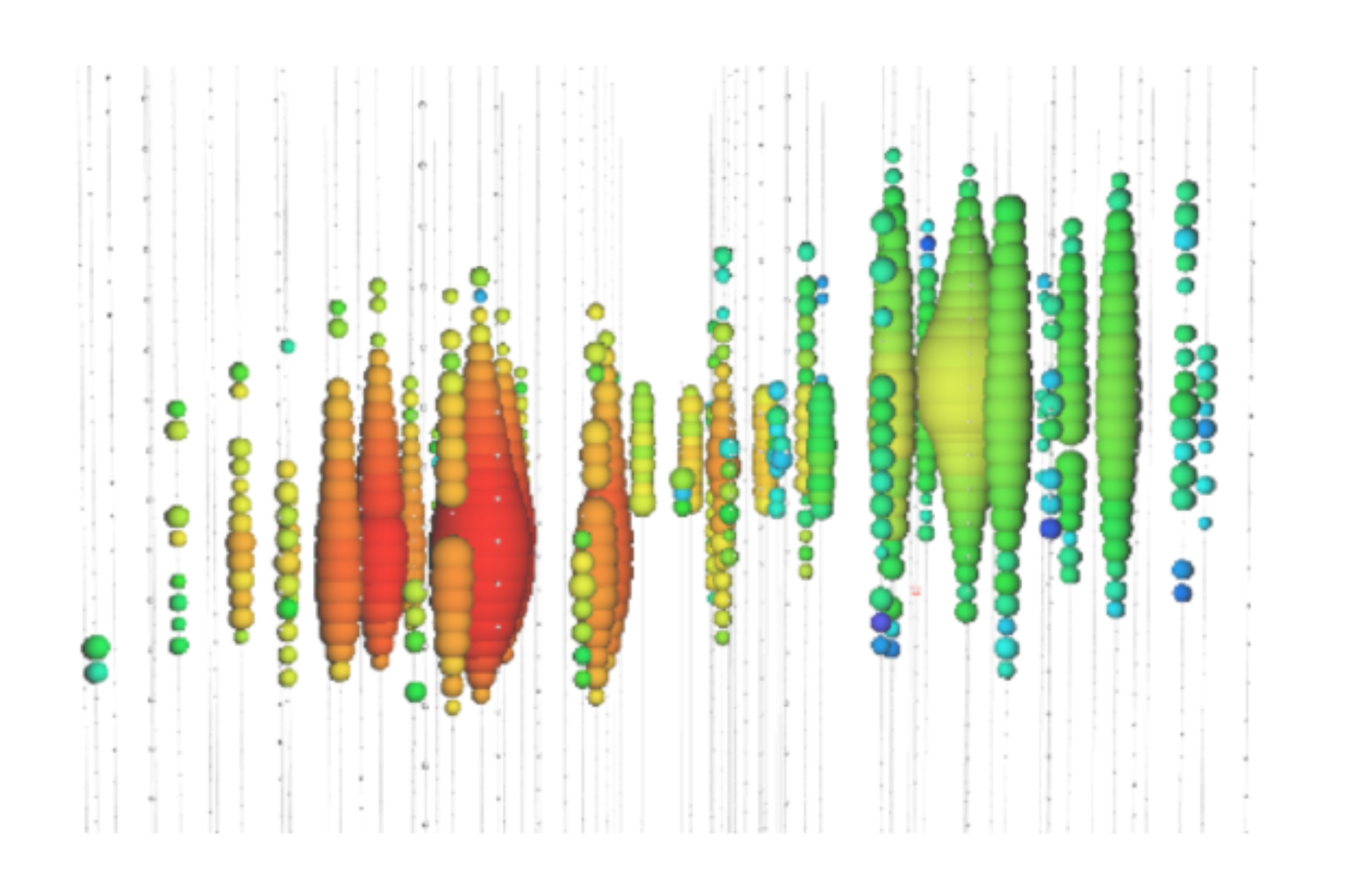}}\hfill
  \caption{Visualization of examples for the different event signatures of neutrino events observed with \icecube. The spheres represent the energy registered by the respective detector module. The color indexes the time information: Red is the beginning of the event, green the middle, and blue the end of the event.}
  \label{fig:eeQuaef6}
\end{figure}

These interactions lead to three different kind of event signatures in
the \icecube detector, visualized in figure \ref{fig:eeQuaef6}.

\paragraph{Shower- or Cascade-Like Events}

In both, interactions through \(\PZz\) bosons
(\textit{neutral-current interactions}) and through \(\PWpm\) bosons
(\textit{charged-current interactions}), hadronic showers are created as
energy is transferred from the incoming neutrino to the outgoing quark.
Hadronic showers are particle cascades originating from hadron
particles. If the outgoing lepton is an electron, this may also create
an accompanying electromagnetic shower, which is a particle cascade
originating from particles primarily interacting through the
electromagnetic interaction. \cite{energyreco}

\begin{align*}
  & \text{neutral current:} & \Pnulepton + \text{nucleon} & \rightarrow \Pnulepton + \text{hadron} \ \ \ \ \ \ \Plepton \in \{ \Pe, \Pmu, \Ptau \}                             & \\
  &                         &                             & \rightarrow \Pnulepton \text{ (escapes) } + \text{hadronic shower}  & \\ \\
  & \text{charged current:} & \Pnue + \text{nucleon}      & \rightarrow \Pe + \text{hadron}                                     & \\
  &                         &                             & \rightarrow \text{electromagnetic shower} + \text{hadronic shower}  & \\
\end{align*}

\paragraph{Track-Like Events}

If the outgoing lepton is a muon, this creates track-like event
signatures as muons travel long distances. TeV muons may travel several
kilometers in the antarctic ice. \cite{skysearch, mmc}

\begin{align*}
  & \text{charged current:} & \Pnum + \text{nucleon}      & \rightarrow \Pmu + \text{hadron}                                    & \\
  &                         &                             & \rightarrow \text{muon track} + \text{hadronic shower}              & \\
\end{align*}

\paragraph{Double-Bang Events}

If the outgoing lepton is a tau, the tau will create a track. But the
track will only be a couple of meters long as the tau then decays into
another hadronic shower. This creates a so-called
\textit{double-bang signature}, where two hadronic showers are joined by
a short track. The shorter the joining track is, the more similar the
event looks to a single-cascade-like event.
\cite{skysearch, energyreco, particledatareview}

\begin{align*}
  & \text{charged current:} & \Pnut + \text{nucleon}      & \rightarrow \Ptau + \text{hadron}                                   & \\
  &                         &                             & \rightarrow \text{tau track} + \text{hadronic shower}               & \\
  &                         &                             & \rightarrow \text{hadronic shower} + \text{hadronic shower}         & \\
\end{align*}

\subsection{Cherenkov-Light Emission}
\label{sec:cherenkov}

When charged particles, both the primary lepton and particles within the
cascades, move through the detector medium faster than the phase
velocity of light within this medium, they emit so-called
\textit{Cherenkov radiation}, which are photons with wavelengths in the
visible and the near ultra violet spectrum.
\cite{energyreco, instrumentation, skysearch}

The light emission is not isotropic: Photons are emitted under an angle
\(\phi\) relative to the direction of propagation of the charged
particle. \cite{physiklexikon}

\[
  \cos \phi = \frac{1}{\beta\,n} = \frac{c'}{v}, \ \ \ \beta = \frac{v}{c}, \ \ \ c' = \frac{c}{n}
\]

\(n\) is the refractive index of the medium, \(c'\) the speed of light
within the medium, \(c\) the speed of light in vacuum.

The virtual photon field of the charged particle moving through the
medium polarizes the atoms in the medium. Each resulting electric dipole
is a source of electromagnetic radiation. But as each dipole arranges
towards the charged particle, integrating over the whole spatial sphere
around the charged particle, the net emitted radiation vanishes if the
charged particle's velocity \(v\) is smaller than the speed of light
\(c'\) in the medium. For higher velocities, the Coulomb field of the
charged particle can only polarize the atoms within a cone with an
opening angle of \(2\phi\), the \textit{Cherenkov cone}, resulting in a
net photon emission perpendicular to the surface of the cone.
\cite{physiklexikon}

\begin{figure}[htbp]
  \centering\includegraphics[width=0.5\textwidth]{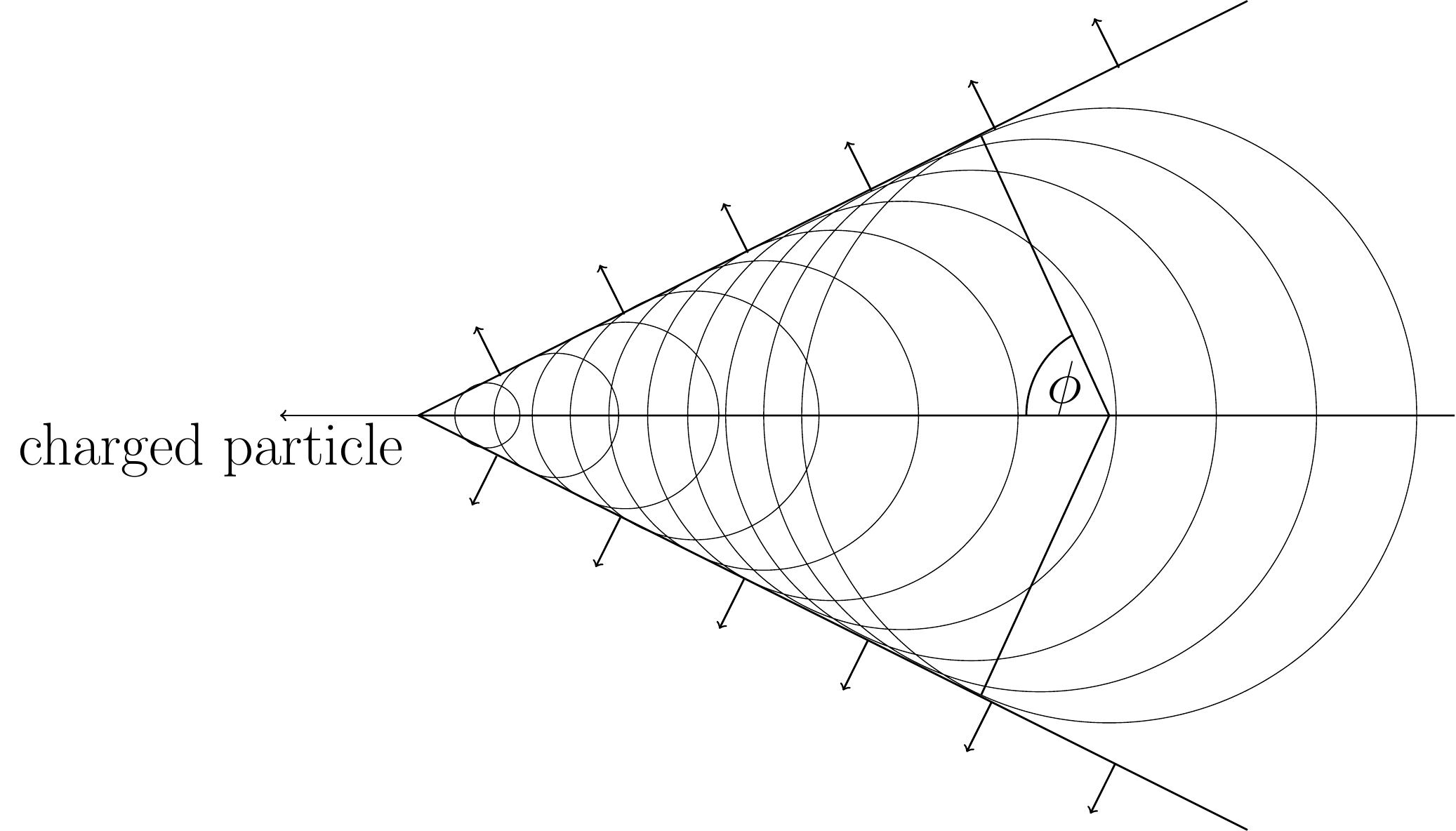}
  \caption{Visualization of the \textit{Cherenkov cone}: A charged particle moves through a medium with a velocity $v>c'$ greater than the phase velocity $c'$ of light within this medium. Due to non-isotropic polarization effects, \textit{Cherenkov photons} are emitted under an angle $\phi$ relative of the direction of motion of the charged particle.}
  \label{fig:ehai2Ahj}
\end{figure}

The spectral distribution of Cherenkov photons is given by equation
\ref{eq:cherenkovspectrum}. A particle with \(\pm z\) elementary charges
that travels a distance \(\d x\) emits \(\d^2 N\) Cherenkov photons with
a wavelength range of \([\nu; \nu + \d\nu]\). \(\alpha\) is the fine
structure constant. \cite{katz2012} The spectrum prefers photons with
shorter wavelengths \(\nu\).

\begin{equation}
  \frac{\d^2 N}{\d x\,\d\nu} = \frac{2\pi\,\alpha\,z^2}{\nu^2}\cdot
  \left( 1 - \frac{1}{\beta^2\,n^2} \right)
  \label{eq:cherenkovspectrum}
\end{equation}

Per GeV of secondary particle shower energy within the detector, an
order of \(10^5\)\nbsp visible Cherenkov photons are created.
\cite{instrumentation}

\subsection{Photon Absorption and Scattering}
\label{sec:scattering}

The propagation of light through a medium depends on the optical
properties of that medium, in particular the velocity of light within
that medium, the scattering probability and the absorption probability.
\cite{lundberg}

The absorption of light for the relevant wavelength range is caused by
electronic and molecular excitation processes \cite{lundberg} and is
quantified by the \textbf{absorption length} \(\lambda\abs\), which is
the mean of the exponentially distributed free path length to absorption
\cite{lundberg}. Therefore, in accordance with the
\noun{Beer-Lambert law}, the absorption length is the path length that
light needs to travel within a medium to have its intensity drop to
\(\sfrac{1}{\e}\) of its original intensity. \cite{lexikonderphysik}
Absorption lengths in the South-Polar ice vary between \(10\m\) in dusty
regions and \(280\m\) in very clear ice layers.
\cite{ackermann, ppcpaper, icepaper}

The scattering of light off microscopic scattering centers, such as
sub-millimeter-sized air bubbles and micron-sized dust grains
\cite{Price1997, ackermann} is the dominant scattering mechanism in
glacial ice \cite{Askebjer1997, lundberg}. This scattering can be
modeled using the more general \noun{Mie scattering} theory, which
describes the scattering of electromagnetic radiation off small,
spherical masses of material with refractive indices differing from the
refractive index of its surroundings.
\cite{Mie1908, ackermann, lundberg}

\noun{Mie scattering} gives the distribution of the scattering angle
\(\theta\) for any wavelength and scattering center size. For ice, this
distribution is approximated using a one-parameter
\noun{Henyey-Greenstein} (HG) phase function
\(p_\text{HG}(\theta; \tau)\), where the one parameter \(\tau\) is the
mean cosine of the scattering angle. \cite{lundberg}

\[ p_\text{HG}(\theta; \tau) = \frac{1 - \tau^2}{2(1 + \tau^2 - 2\tau\, \cos \theta)^\frac{3}{2}}, \ \ \ \tau = \meancostheta \]

The South-Polar ice has shown to be preferentially forward scattering
with a mean cosine of the scattering angle of \(\meancostheta = 0.94\)
with only a weak dependence on the wavelength. \cite{ackermann}

The \textbf{scattering length} \(\lambda\sca\), which is also called
\textbf{geometric} scattering length, is the mean of the exponentially
distributed free path and thereby the average distance between
scatterings. \cite{ackermann} A related and often used quantity is the
\textbf{effective scattering length} \(\lambda\esca\), which is the
distance that light needs to propagate through a turbid medium before
the photon directions are completely randomized. \cite{lundberg}

\begin{equation}
  \lambda\esca = \frac{\lambda\sca}{1 - \meancostheta}
\end{equation}

In a medium with isotropic scattering, the geometric and the effective
scattering lengths are the same. In a preferentially forward scattering
medium like the South-Polar ice, the original direction of a sample of
photons is tendentially retained for several scattering steps until the
photon direction of the sample is isotropized. The projection of the net
velocity vector on the original direction is decreased on average by
\(\meancostheta\) for each scattering. \cite{lundberg}

After \(n\) scatterings, the effectively transported forward distance
along the original direction is
\(\lambda\sca\,\sum_{i=0}^n \meancostheta^i\), such that in the limit of
many scatterings, \(n \rightarrow \infty\), the effectively transported
forward distance becomes the effective scattering length.
\cite{lundberg, ackermann}

\[ \lim_{n \rightarrow \infty} \lambda\sca\,\sum_{i=0}^n \meancostheta^i = \frac{\lambda\sca}{1 - \meancostheta} = \lambda\esca \]

Typical effective scattering lengths within the South-Polar ice are on
the order of \(25\m\) \cite{lundberg} and range from \(5\m\) to \(90\m\)
in the detector volume \cite{icepaper}, corresponding to the geometric
scattering length ranging from \(0.3\m\) to \(5.4\m\).

Light interference effects are ignored during photon propagation as the
average distance between the scattering centers is large compared to the
photon wavelength. \cite{ackermann} Also, despite modeling different ice
regions with abrupt boundaries in the simulations of this study, the
physical boundaries are assumed such that refractive index variations
are continuous. Hence reflection at the medium boundaries is ignored in
simulations. \cite{lundberg}

  \section{Experimental Concepts}
\label{sec:experimental_background}

\subsection{\icecube Detector}

The \icecube neutrino detector is built into a cubic-kilometer of the
glacial ice at Earth's South Pole. 5160 photo detecting optical modules
have been deployed between \(1450\m\) and \(2450\m\) below the surface.
Construction has begun in 2005 with the deployment of the first optical
modules. The detector is fully operational since 2010.
\cite{instrumentation}

The optical modules are anchored on 86 vertical cables called
\textit{strings}, which are positioned on a triangular grid with an
overall hexagonal footprint. The strings are about \(125\m\) apart. Each
string holds 60 optical modules with a vertical spacing of \(17\m\).
\cite{instrumentation}

\begin{figure}[htbp]
  \smallerimage{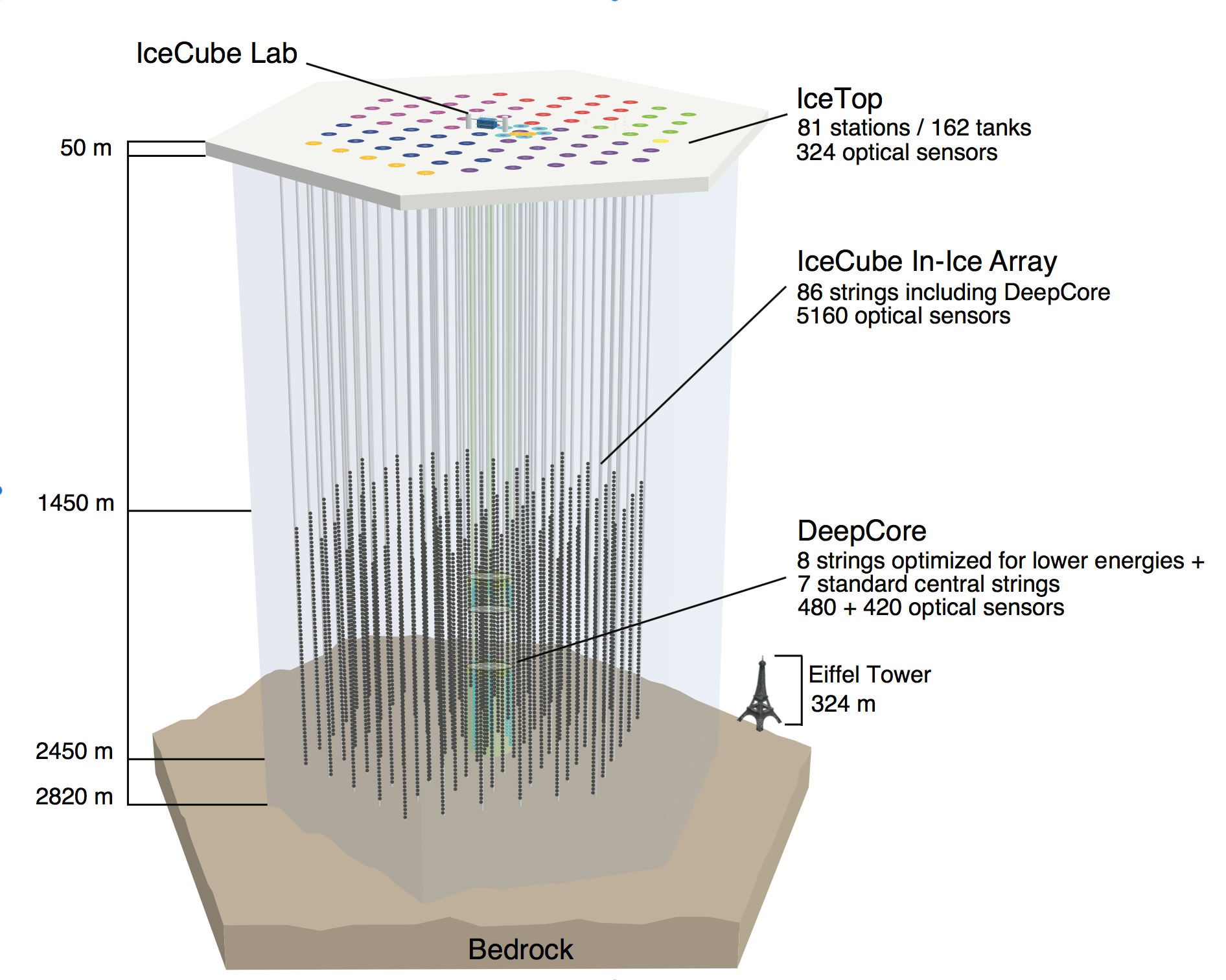}
  \caption{Schematic overview of the \icecube detector. Image source: \cite{instrumentation}}
  \label{fig:aiThai0e}
\end{figure}

Additional to this \textit{in-ice array}, which is designed to measure
neutrinos with energies from the TeV to the PeV scale, the
\textit{DeepCore} sub array hold additional optical modules in order to
lower the detection energy threshold in this region of the detector to
detect neutrinos with energies from \(10\GeV\) to \(100\GeV\).
\cite{instrumentation}

\subsection{Digital Optical Modules (DOMs)}
\label{sec:doms}

\begin{figure}[htbp]
  \subcaptionbox{Components of the optical module. Image source: \cite{instrumentation}}{\halfimage{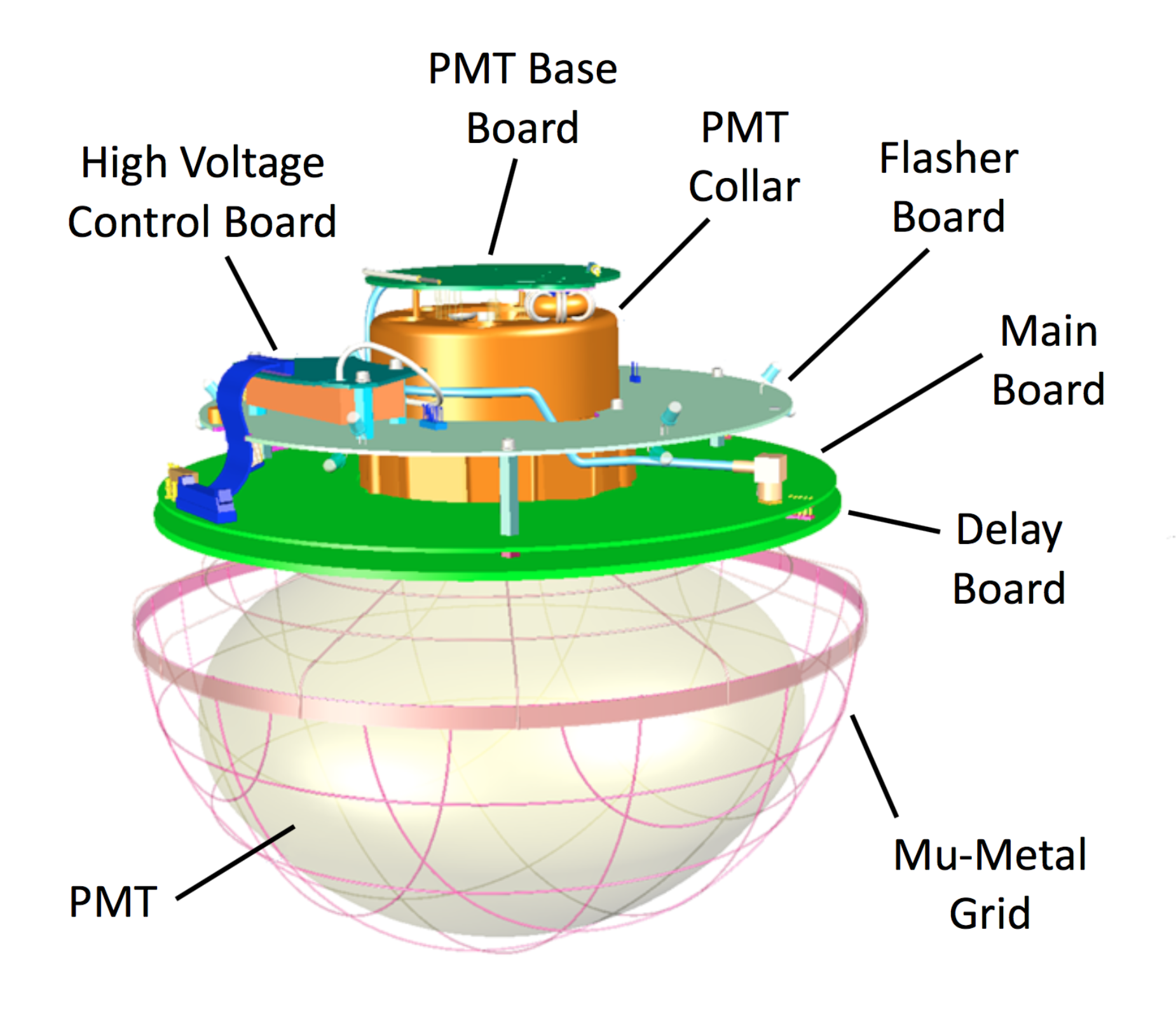}}\hfill
  \subcaptionbox{Glass sphere containing the components of the optical module. Image source: \cite{gallerynoharness}}{\includegraphics[width=0.35\textwidth]{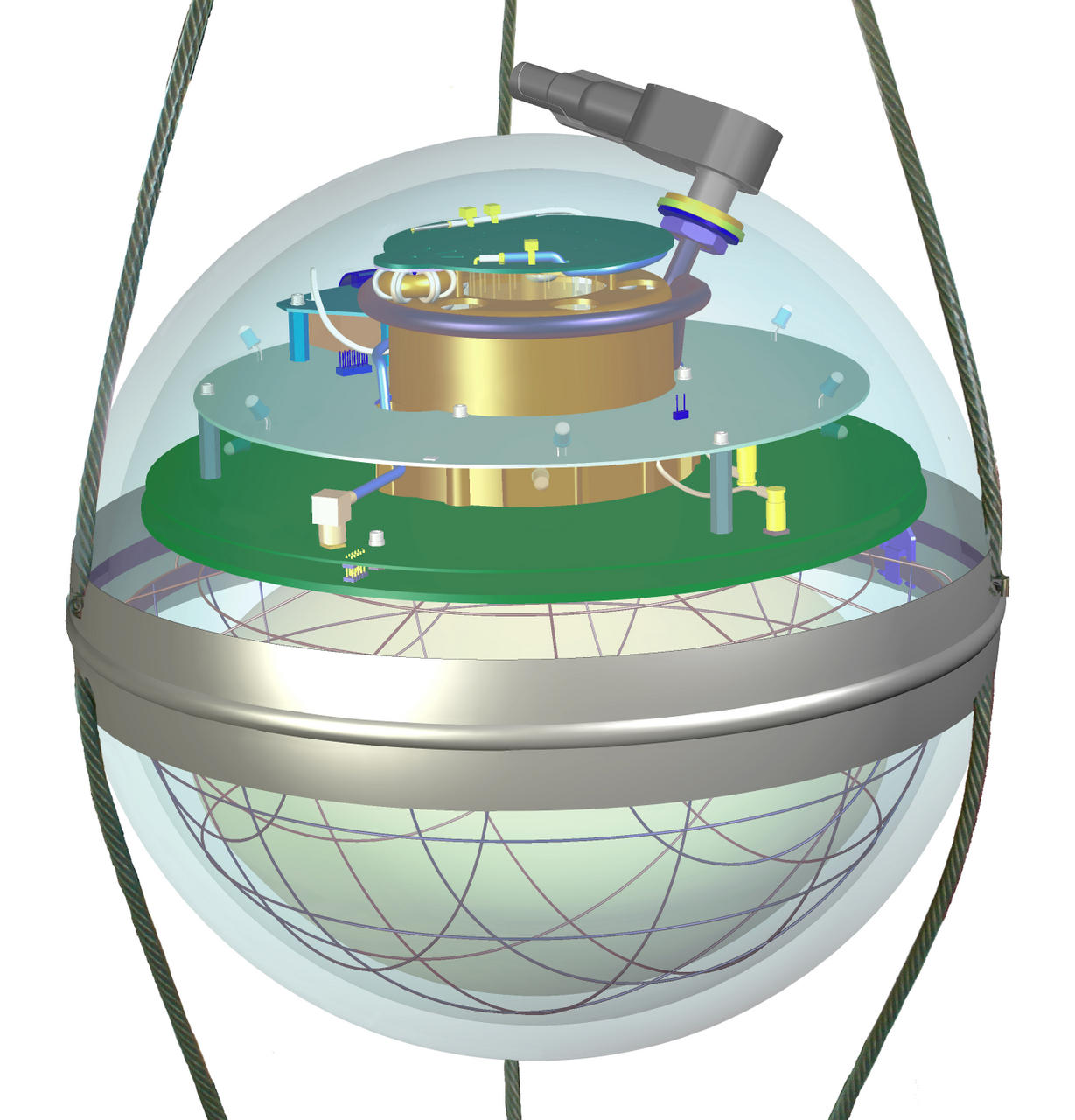}}
  \caption{Schematic display of a digital optical module (DOM), \icecube's basic detection unit.}
  \label{fig:aK4raigh}
\end{figure}

The basic detection unit in \icecube is the
\textit{Digital Optical Module} (DOM). Its main components are a photo
multiplier tube (PMT), which detects impacting photons, a main
electronics board, and a flasher board containing light-emitting diodes
(LEDs) for calibration purposes (figure \ref{fig:aK4raigh} a). The DOM's
components are contained within a glass sphere with an outer diameter of
about \(35\cm\) that can withstand high pressures (figure
\ref{fig:aK4raigh} b). The space in between is filled with a gel to
avoid optical effects at the medium boundaries. \cite{instrumentation}

The recorded signals from the PMT are digitized within the module before
sending the signal to the surface in order to minimize the loss of
information from degradation of analog signals sent over long distances.
\cite{firstyearperformance}

The optical modules are optimized for detecting Cherenkov light emitted
by particles with energies from \(10\GeV\) to \(10\PeV\) up to \(500\m\)
away from the optical module. \cite{instrumentation}

\subsection{Properties of South-Polar Ice}
\label{sec:ice}

The ice at the South Pole has exceptional optical properties, because
air bubbles shrink and vanish under large pressure, forming a so-called
\textit{clathrate hydrate}, where impurities are enclosed inside the ice
crystal structure. \cite{rongenswedishcamera}

Using \icecube's LED flasher calibration system, the properties of
\icecube's glacial ice have been measured: The average distance to
absorption, \(\lambda\abs\), the average distance between successive
scatters \(\lambda\sca\), and the angular distribution of the new
direction after scattering.

To fit these parameters for different depths, the ice has been divided
into \(z\)-layers of an arbitrary thickness of \(10\m\). The ice
parameters have been fitted for each layer such that the properties are
best interpreted as average of their true values over the thickness of
the ice layers. \cite{icepaper}

\paragraph{Scattering}

\begin{figure}[htbp]
  \subcaptionbox{Depth dependence of the effective scattering length~$\lambda\esca$ and the effective scattering coefficient~$b_\text{e}:=\sfrac{1}{\lambda\esca}$.}{\halfimage{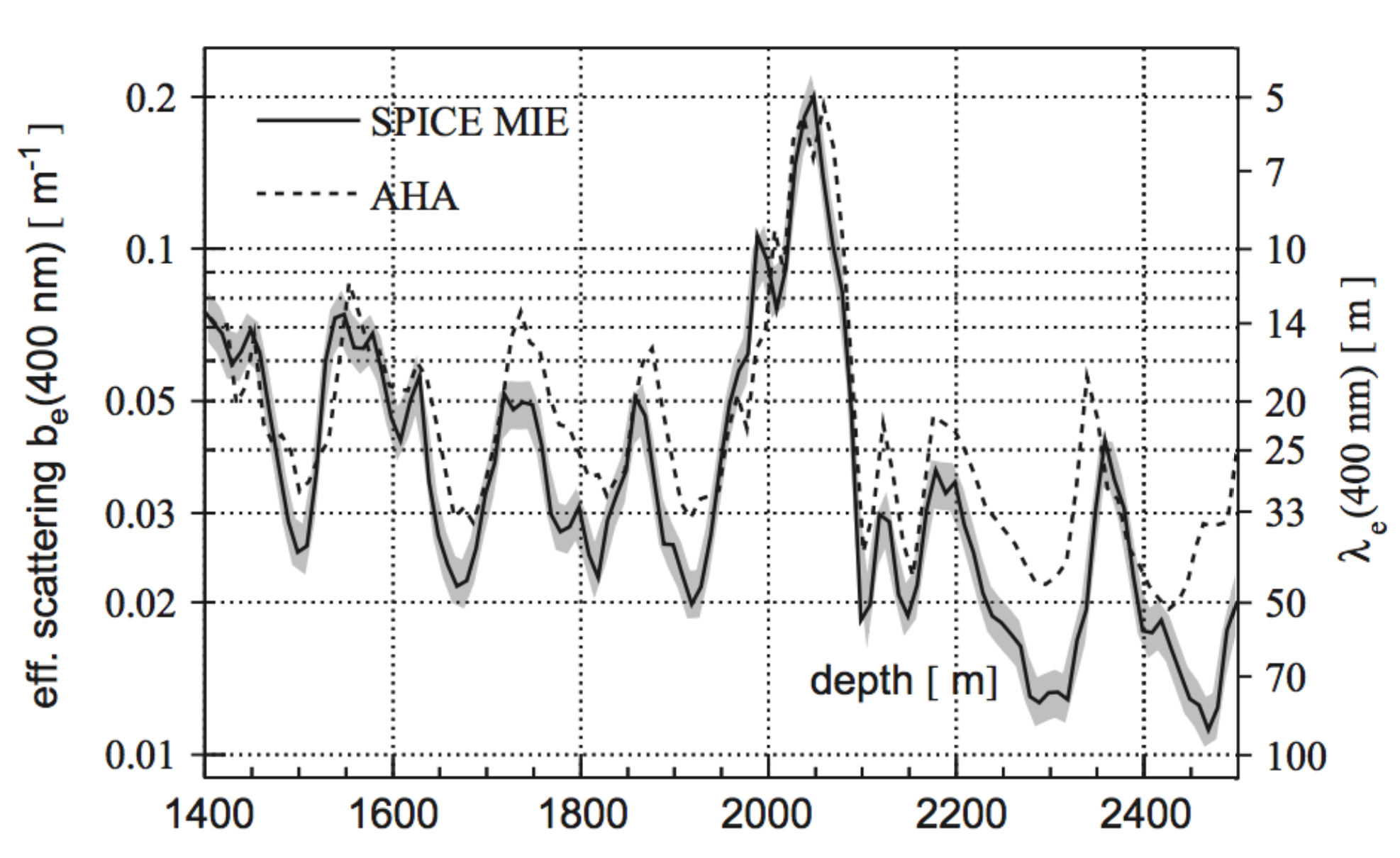}\vspace*{2mm}}\hfill
  \subcaptionbox{Depth dependence of the absorption length~$\lambda\abs$ and the absorption coefficient~$a:= \sfrac{1}{\lambda\abs}$.}{\halfimage{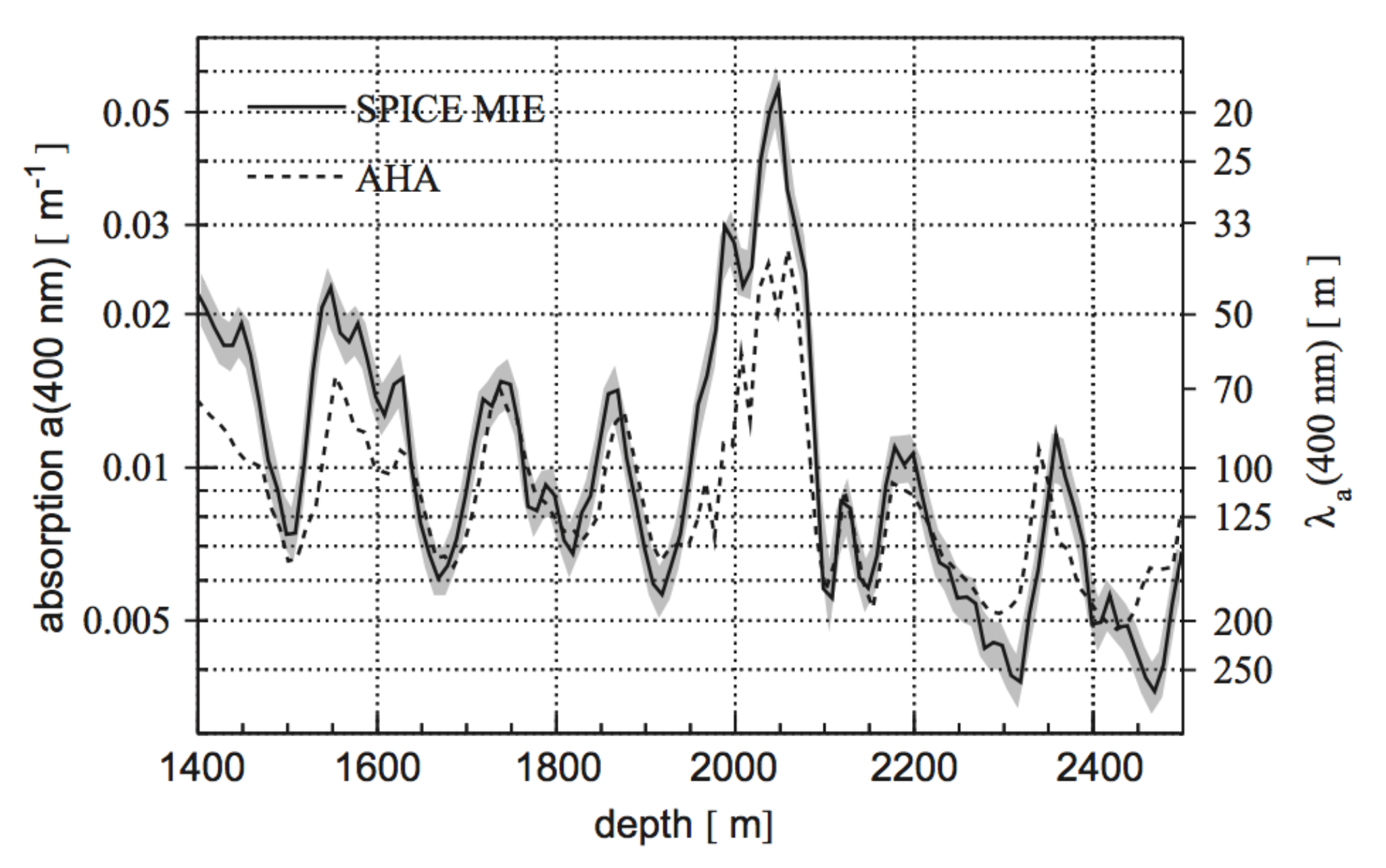}}
  \caption{Values of the absorption length $\lambda\abs$ and the effective scattering length $\lambda\esca$ for different depths, but for a fixed photon wavelength of $400\nm$. Plot taken from \cite[figure 16]{icepaper}. A detailed data table is given in \cite[table C1]{icepaper}.}
  \label{fig:Ahxobai3}
\end{figure}

The measured effective scattering lengths \(\lambda\esca\) range from
\(5\m\) to \(90\m\), corresponding to the geometric scattering length
ranging from \(0.3\m\) to \(5.4\m\). The depth dependence of the
scattering length is shown in figure \ref{fig:Ahxobai3} (a).
\cite{icepaper}

The wavelength dependence is given by equation \ref{eq:escawavelength}
\cite[section 4]{icepaper}, where
\(b_\text{e} := \sfrac{1}{\lambda\esca}\) is the effective scattering
coefficient, \(\nu\) the photon wavelength, and \(\alpha\) a global fit
parameter. \(b_\text{e}(\nu = 400\nm)\) is given by figure
\ref{fig:Ahxobai3} (a) and \cite[table C4]{icepaper}. The global
parameter \(\alpha\) has been fitted to \(\alpha = 0.90 \pm 0.03\)
\cite[section 5.1]{ackermann}.

\begin{equation}
  b_\text{e}(\nu) = b_\text{e}(\nu = 400\nm) \cdot \left(\frac{\nu}{400\nm}\right)^{-\alpha}
  \label{eq:escawavelength}
\end{equation}

The scattering prefers the forward direction, with a mean cosine of the
scattering angle of \(\meancostheta = 0.94\)
\cite[paragraph 9]{ackermann} (or \(\meancostheta = 0.90\)
\cite{icepaper}).

\paragraph{Absorption}

The absorption lengths in the South-Polar ice vary between \(10\m\) in
dusty regions and \(280\m\) in very clear ice layers. Their depth
dependence is given in figure \ref{fig:Ahxobai3} (b). The absorption
length, which is governed by the dust concentration, is especially low
in the so-called ``dust peak'' at a depth of about \(2000\m\).
\cite{ackermann, ppcpaper, icepaper}

The dependence of the absorption coefficient
\(a := \sfrac{1}{\lambda\abs}\) on the photon wavelength \(\nu\), the
temperature difference \(\delta\tau\) is given by equation
\ref{eq:abswavelength} \cite{icepaper}.

\begin{equation}
  a(\nu) = a_\text{dust}(\nu) + A\,\e^{-\sfrac{B}{\nu}}\, (1 + 0.01 \cdot \delta\tau)
  \label{eq:abswavelength}
\end{equation}\begin{equation}
  a_\text{dust}(\nu) = a_\text{dust}(\nu = 400\nm) \left(\frac{\nu}{400\nm}\right)^{-\kappa}
\end{equation}

The global parameters have been fitted to \(A = (6954 \pm 973)\m^{-1}\),
\(B = (6618 \pm 71)\nm\), and \(\kappa = 1.08 \pm 0.01\)
\cite[section 5.2]{ackermann}.\footnote{The quantity $A$ in \cite{icepaper} corresponds to the quantity $A_\text{IR}$ in \cite{ackermann}. $B$ in \cite{icepaper} corresponds to $\lambda_0$ in \cite{ackermann}.}
The temperature difference \(\delta\tau(d) = T(d) - T(1730\m)\) for
depths \(d\) is given by equation \ref{eq:temperature} \cite{icepaper}.

\begin{equation}
  T(d) = 221.5\unit{K} - 0.00045319\,\frac{\text{K}}{\text{m}}\cdot d + 5.822 \cdot 10^{-6}\,\frac{\text{K}}{\text{m}^2} \cdot d^2
  \label{eq:temperature}
\end{equation}

\paragraph{Ice-Layer Tilt and Ice Anisotropy}

\begin{figure}[htbp]
  \centering
  \includegraphics[width=0.5\textwidth]{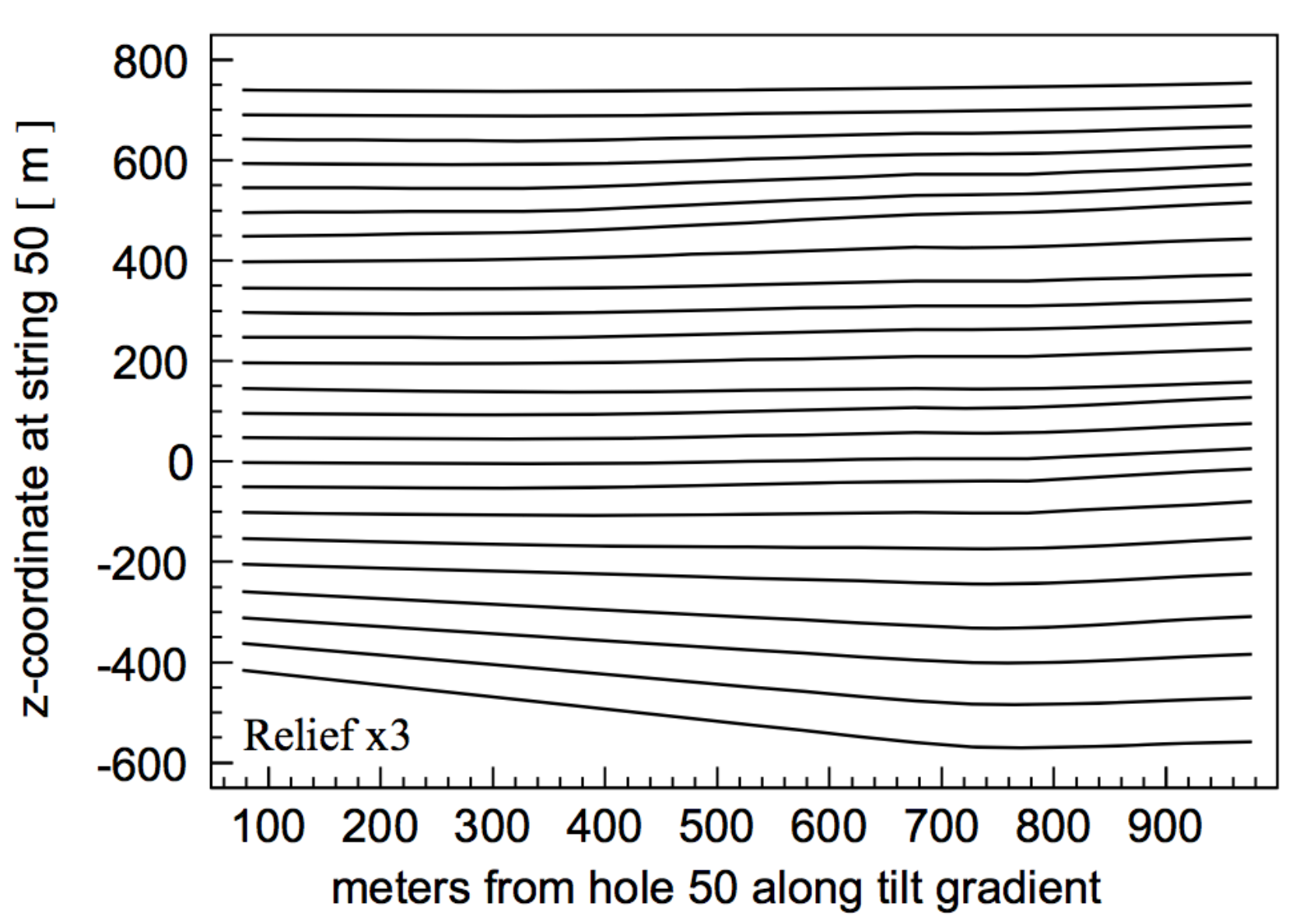}
  \caption{Ice layers along the average gradient direction within the ice. The relief is amplified by a factor of 3 to enhance the clarity of the layer structure. The lowest layer shown exhibits a shift of $56\m$ between its shallowest and deepest points, which is the largest shift of all layers shown in the figure. Plot and caption taken from \cite[figure 14]{icepaper}.}
  \label{fig:wohr8uaY}
\end{figure}

The absorption properties of the ice layers follow the dust
concentration, which does not strictly follow the arbitrary
\(z\)-layers, but is tilted. To model this feature, the ice layers can
also be considered tilted by using an effective-\(z\) coordinate,
\(z_\text{e}(x,y,z) = z + \text{relief}(x,y,z)\), which is shown in
figure \ref{fig:wohr8uaY}. \cite{icepaper}

Furthermore, scattering and absorption show also a slight dependency on
the propagation direction of the photons, aligned with the ice flow
direction of the glacier \cite{icrc17pocam}, which is referred to as
\textit{ice anisotropy}.

This study considers the dependencies of scattering and absorption
length on depth, temperature and wavelength, but does not consider the
ice-layer tilt and the ice anisotropy.

An overview of the different ice models used in \icecube is given in
\cite{flasherdataderivedicemodels}.

\subsection{Hole Ice Around the Detector Strings}
\label{sec:hole_ice}

The so-called \textit{hole ice} is the refrozen water within the drill
holes that were necessary to deploy the detector strings with the
optical modules.

For the \icecube detector, 68 boreholes with an approximate diameter of
\(60\cm\) to a depth of about \(2500\m\) were created using a
\textit{hot-water drilling} technique. Drilling one hole required about
48 hours time. During the deployment, the drill hole was filled with
water. \cite{instrumentation}

When the glacial ice in the drill holes became water, the structures
that were responsible for the specific properties of the bulk-ice layers
were destroyed. Thus, the properties of the hole ice are considered
largely independent of those of the surrounding bulk ice. The deployed
instrumentation become frozen in place and optically coupled to the
surrounding ice sheet when the water in the boreholes became ice again.
\cite{instrumentation}

In order to monitor the freeze-in process, a camera system consisting of
two video cameras in separate spheres, each also equipped with four LEDs
and three lasers, has been deployed along one of the detector strings.
The cameras observed that the drill hole became completely refrozen
within 15 days. \cite{instrumentation}

\begin{figure}[htbp]
  \centering
  \subcaptionbox{}{\thirdimage{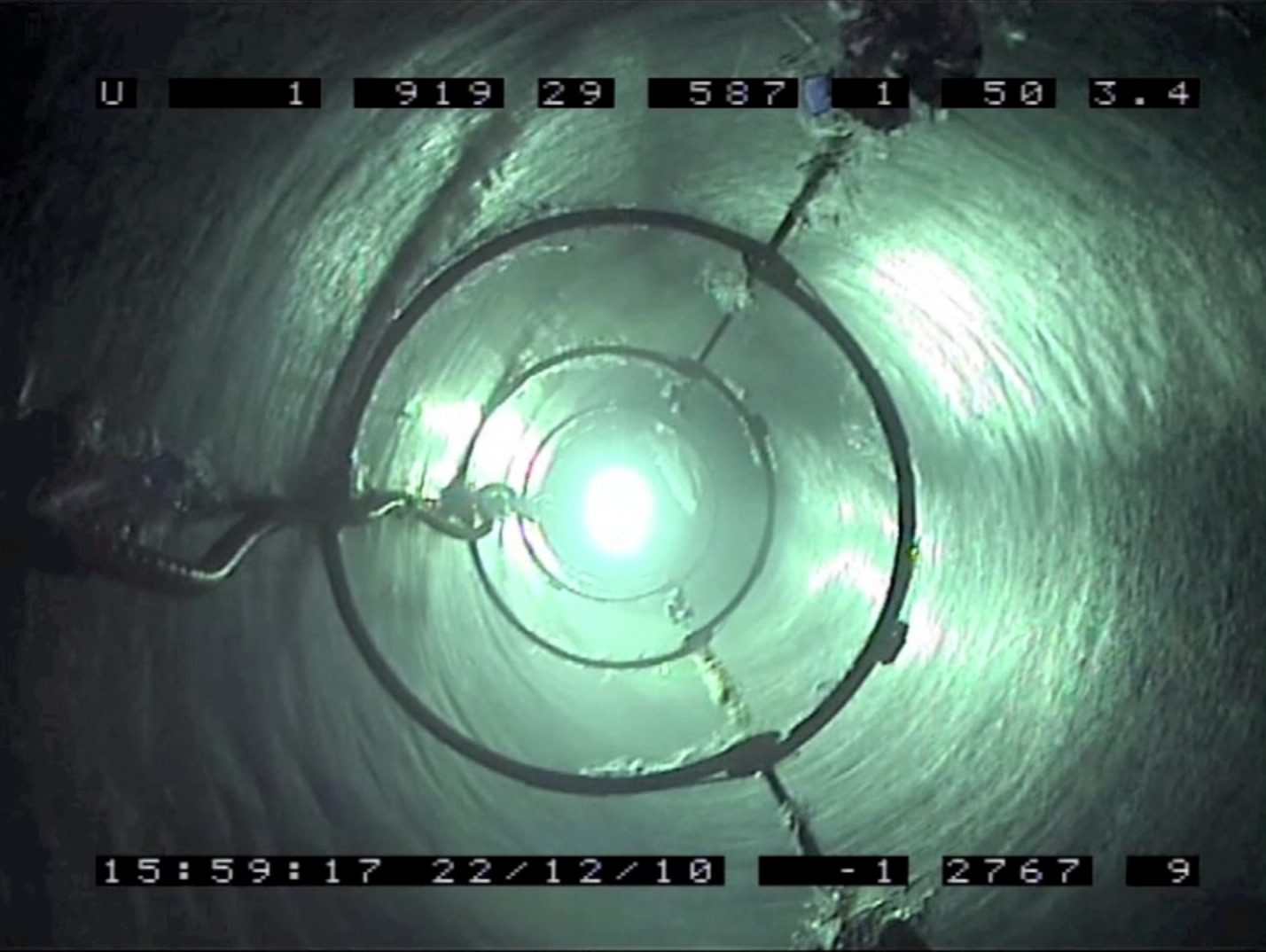}}\hfill
  \subcaptionbox{}{\thirdimage{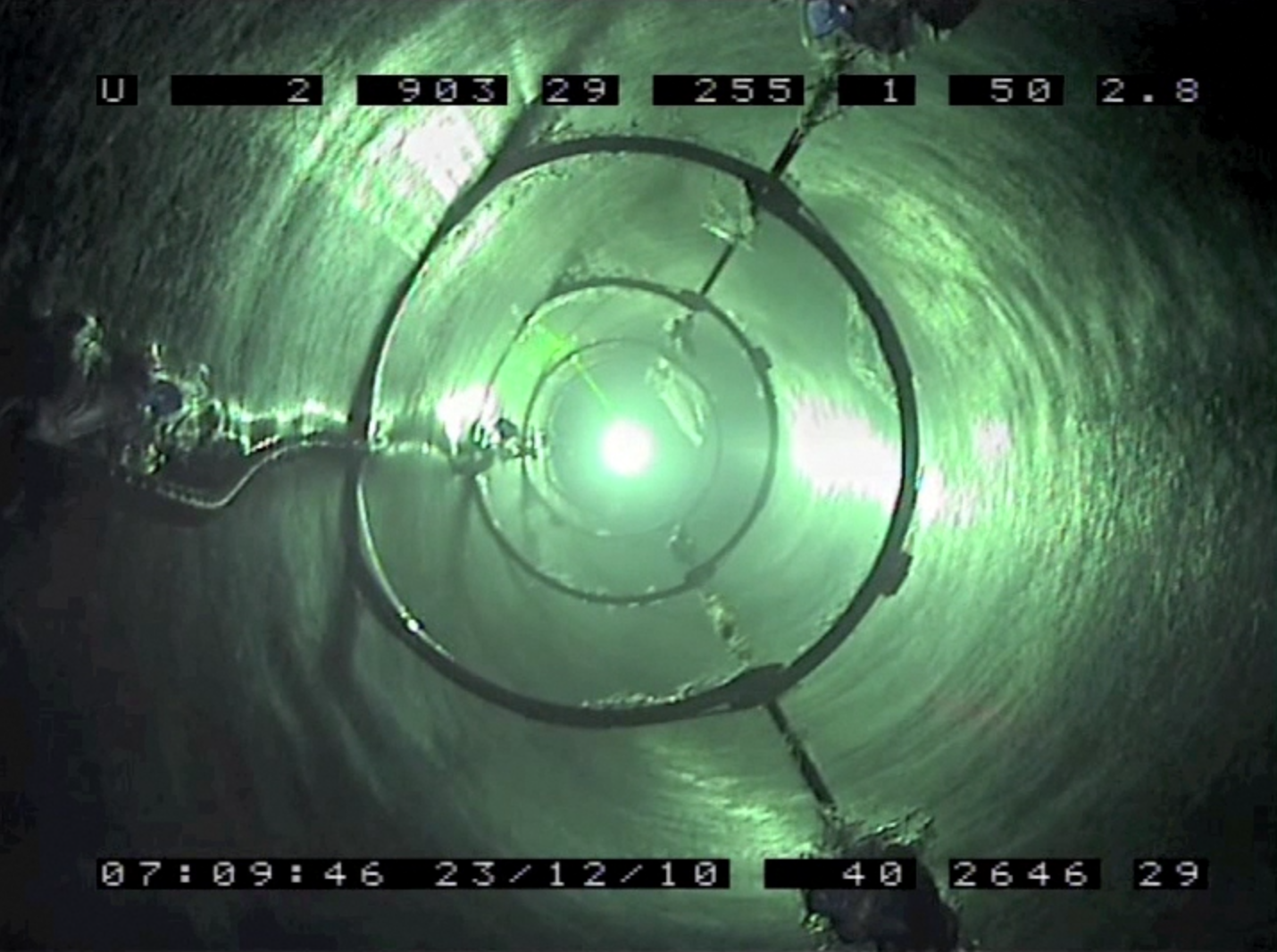}}\hfill
  \subcaptionbox{}{\thirdimage{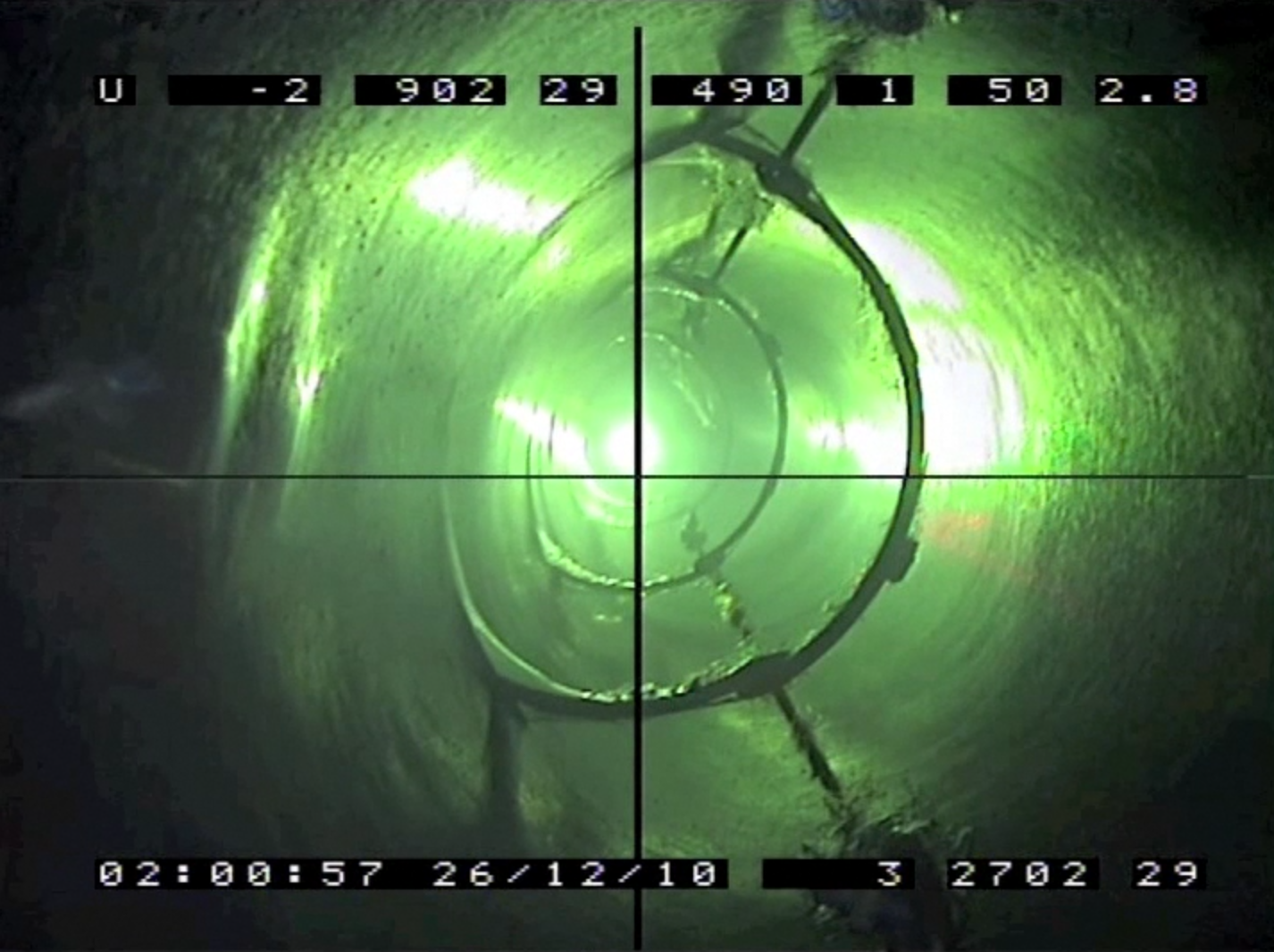}}\hfill
  \subcaptionbox{}{\thirdimage{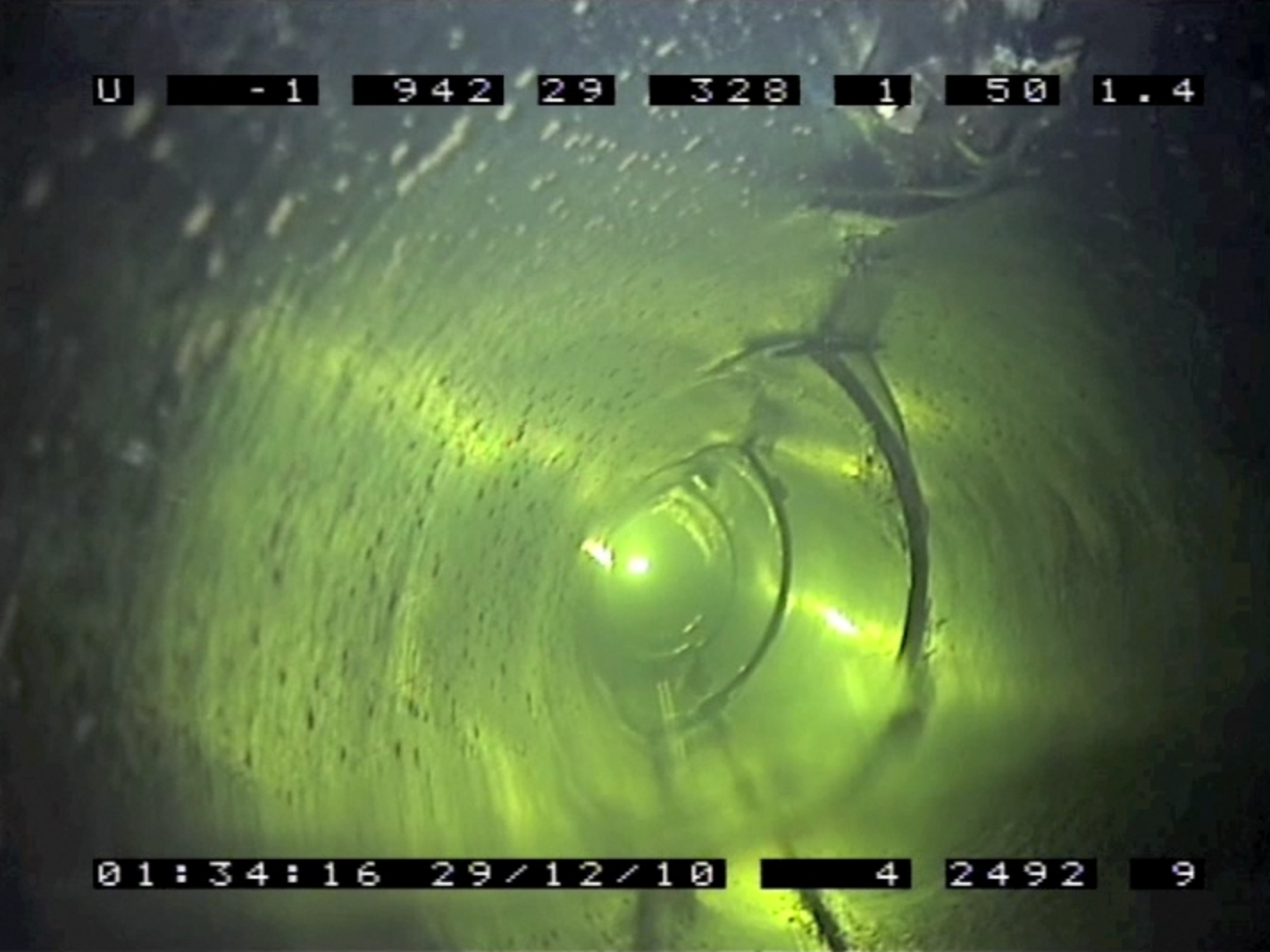}}\hfill
  \subcaptionbox{}{\thirdimage{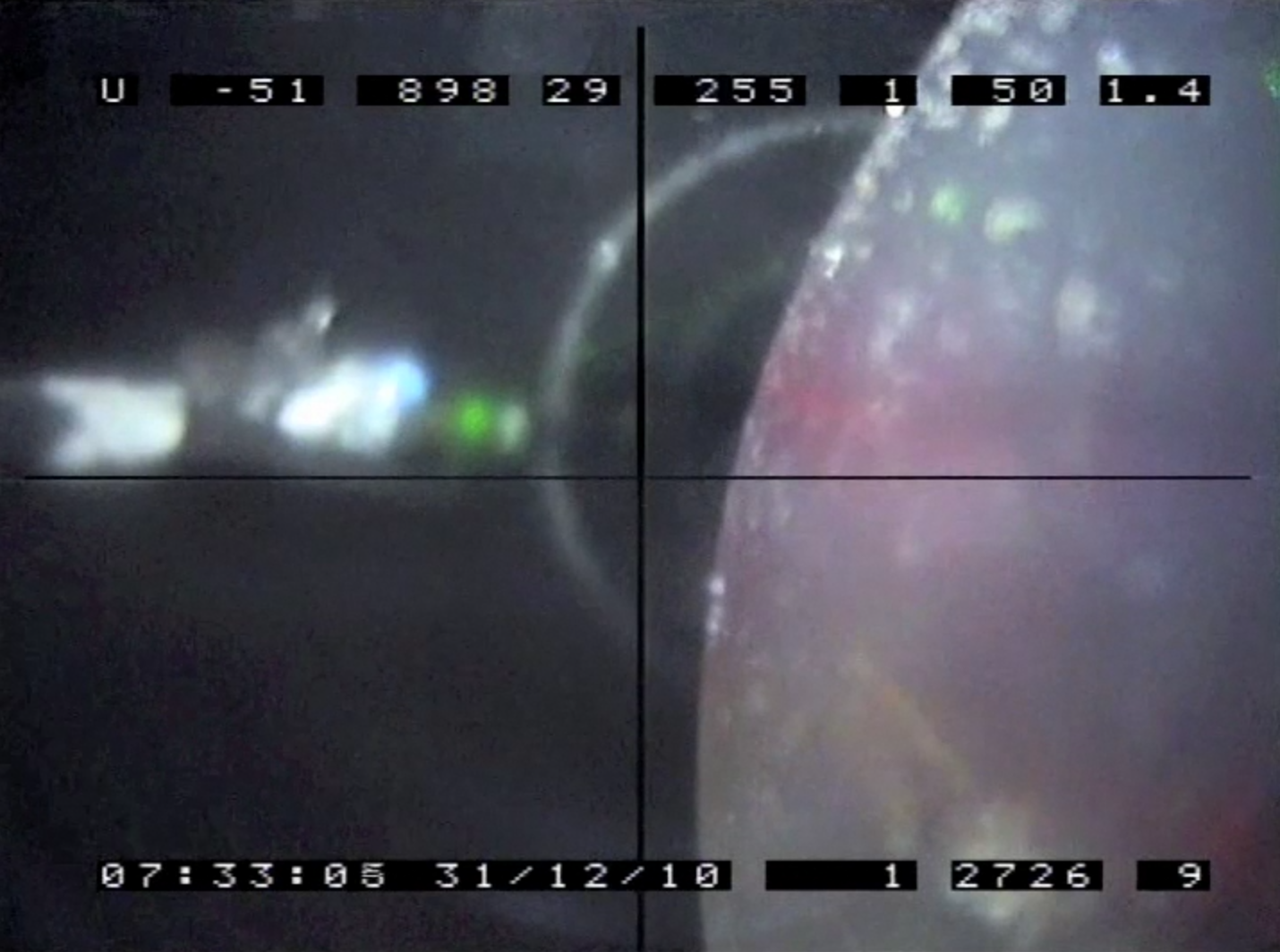}}\hfill
  \subcaptionbox{}{\thirdimage{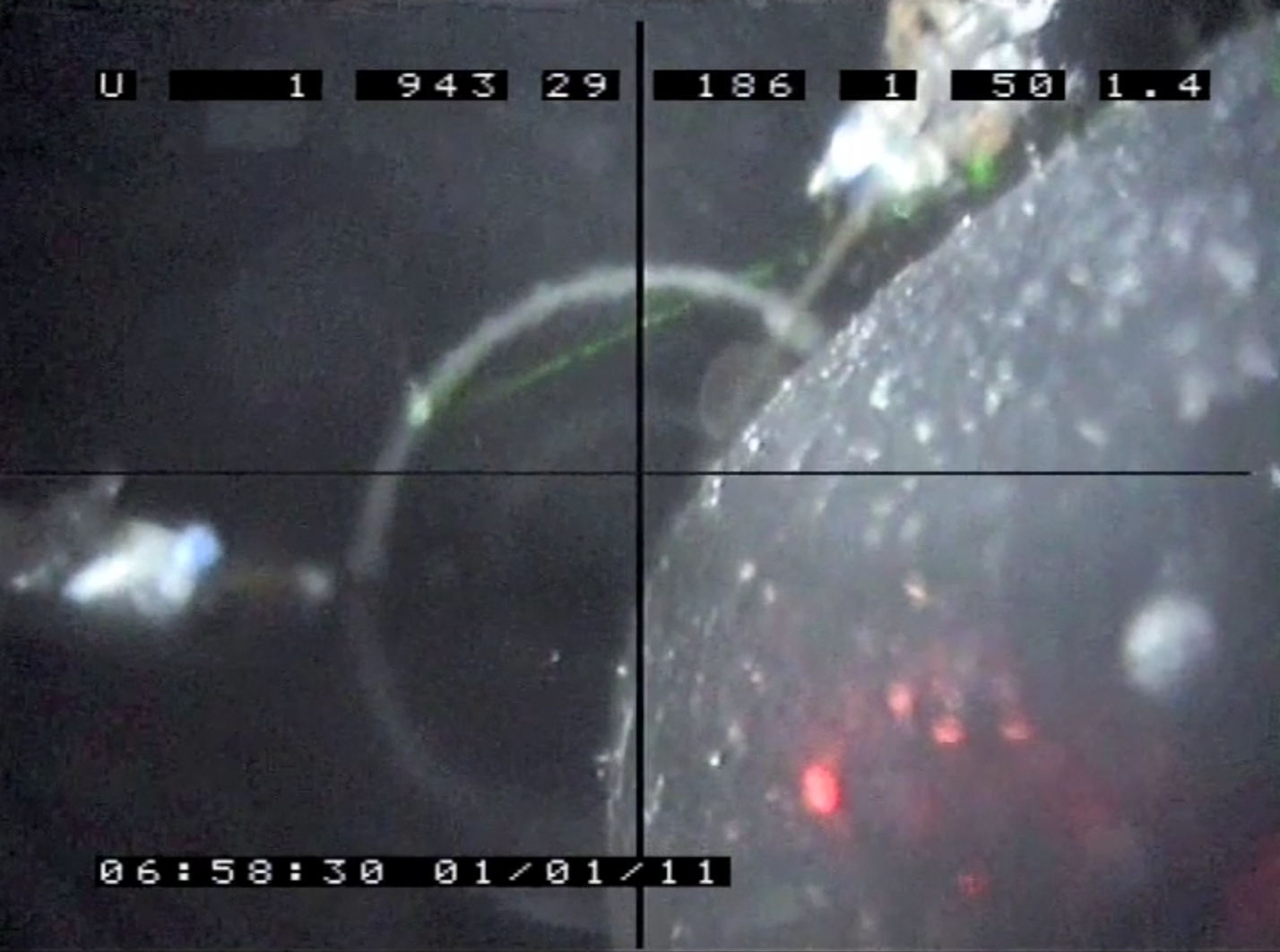}}\hfill
  \subcaptionbox{}{\halfimage{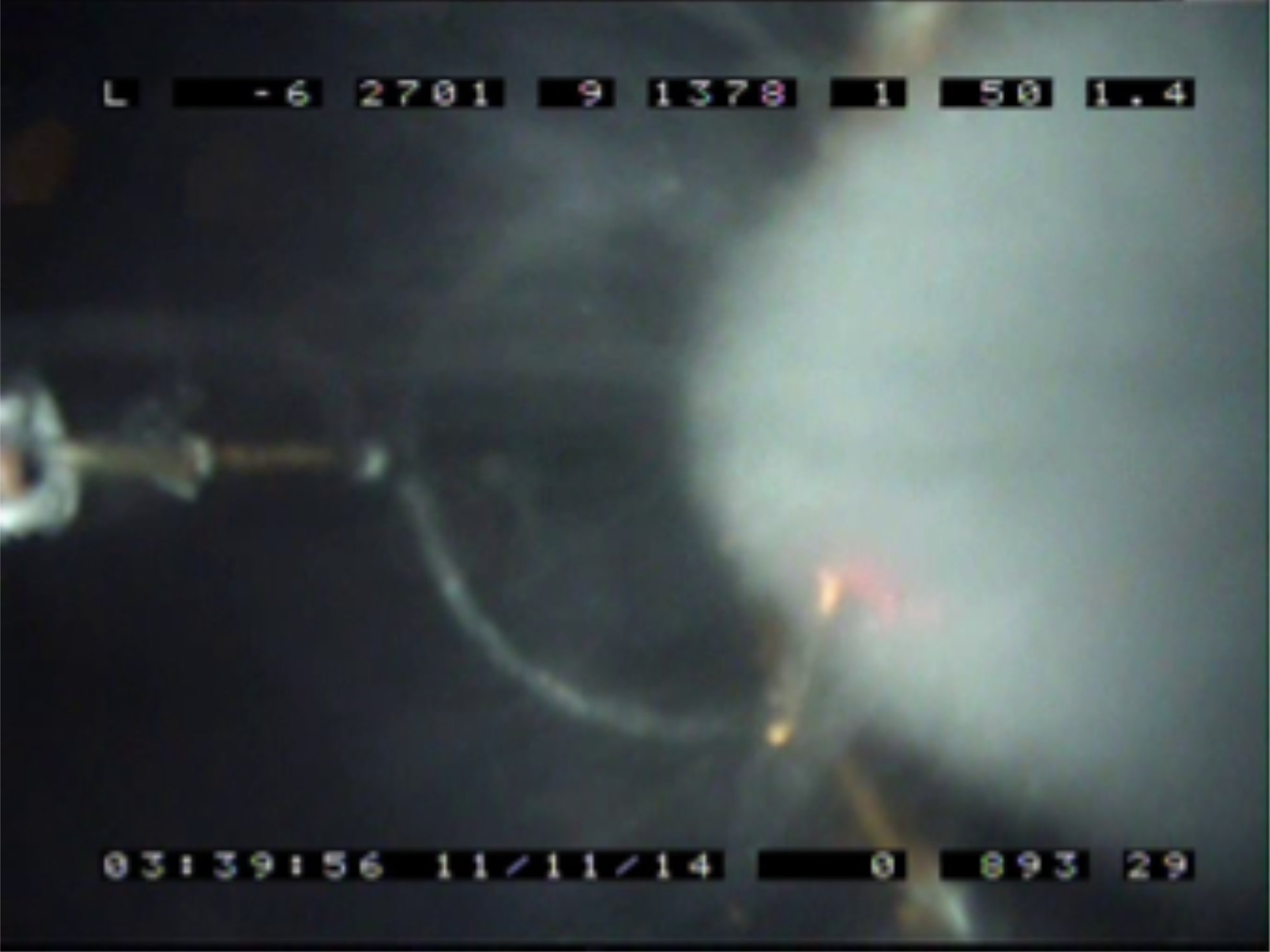}}\hfill
  \subcaptionbox{}{\halfimage{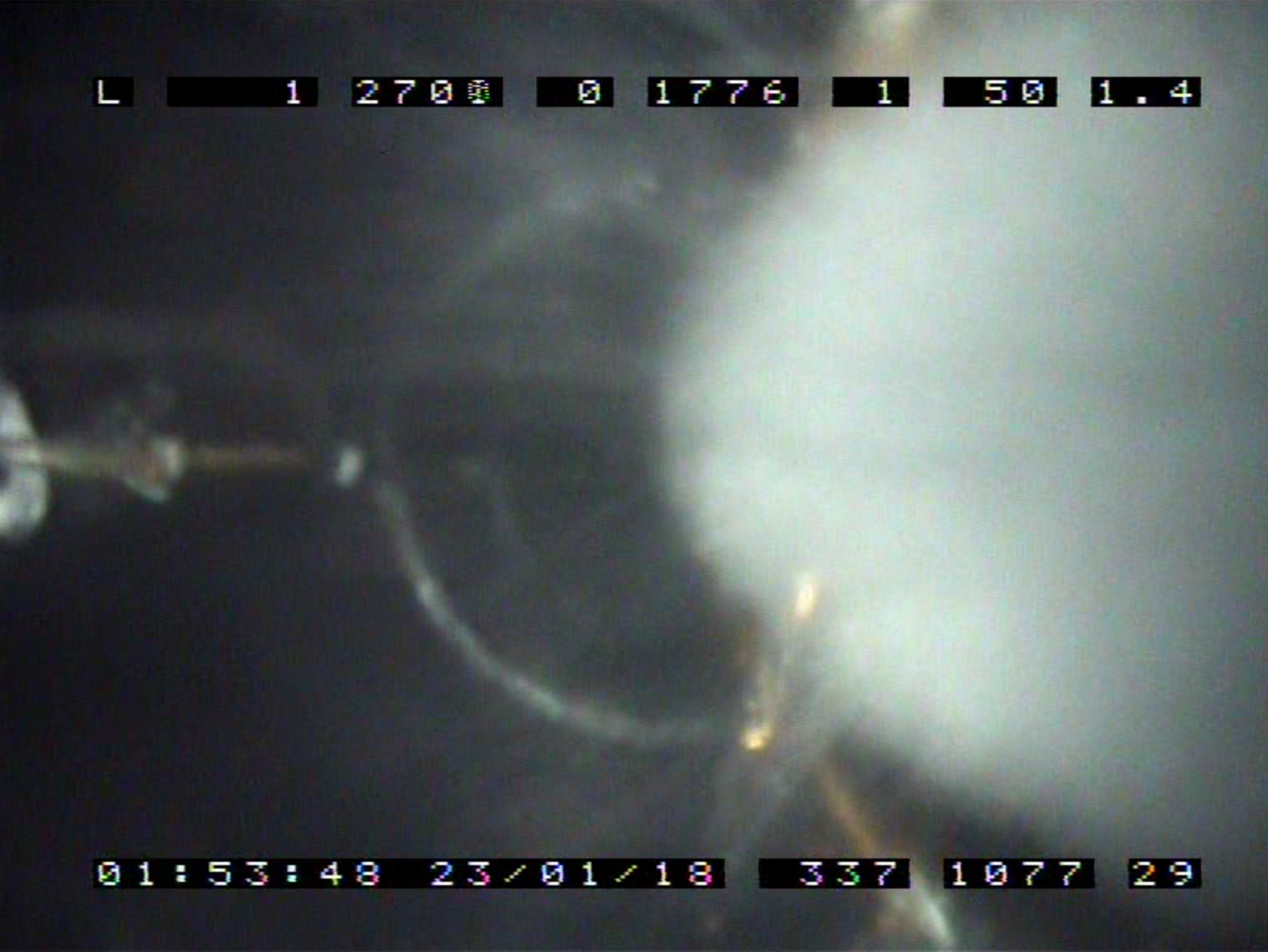}}\hfill
  \caption{Monitoring the freeze-in process using a camera system deployed within string number 80. The drill hole freezes from outside in as seen in (a) to (f). The final configuration that can still be observed in 2018 as seen in (h) still shows a diffuse column, called \enquote{bubble column}, on the right-hand side of images (g) and (h). Image sources: \cite{icrc17pocam, camera2010, camera2018}}
  \label{fig:daeM6yot}
\end{figure}

The camera observations suggest two hole-ice components, a clear outer
region, and an inner column of a smaller scattering length and a
diameter of about \(16\cm\) (figure \ref{fig:daeM6yot}).
\cite{rongenswedishcamera,instrumentation}

The observed freeze-in process from the outside in is consistent with
the model of \textit{cylindrical freezing}, where impurities or air
bubbles are pushed inwards along the freezing boundaries until they
merge in the center. \cite{rongenswedishcamera} After completing the
freeze-in process, no long-term changes have been observed from 2010 to
2018. \cite{instrumentation, camera2010, camera2018}

In this study, when needing to differentiate between the different
components of the hole ice, the outer clear component will be called
\textit{drill-hole ice}, or \textit{drill-hole column}. The inner
component with a shorter scattering length will be called
\textit{bubble column}. Note, however, that there is no established
nomenclature in the context of \icecube publications, yet.

In this study, the hole-ice columns will be modeled as cylinders. More
complicated geometries such as accounting for the inevitable swinging of
the drilling head, or pressure effects that could lead to a vertical
gradient in the properties of the hole ice, are not considered in this
study.\followup

  \section{Computing Concepts}
\label{sec:simulation_background}

\subsection{Monte-Carlo Simulations of Photons}
\label{sec:monte_carlo}

A Monte-Carlo simulation is a computational method that utilizes a large
amount of random numbers. In order to substitute a complex, possibly
unknown probability distribution, samples of random numbers are drawn
from one or several basic probability distributions and processed in
deterministic calculations. The results are evaluated to gain
information about processes or quantities involved. This method is
especially useful for systems with many degrees of freedom.
\cite{physiklexikon}

This study needs to determine whether and when detector modules are hit
by photons from a given source assuming different ice parameters. In
principle, one could devise a mathematical function of input quantities
and random variables that determines whether and when a photon is
detected by an optical module. This task would be disproportionately
complex, however, especially as the function would have to be revised
for every change in the underlying models.

Therefore, this study uses Monte-Carlo simulations that propagate
photons through the ice in several simulation steps. Based on drawing
random numbers from basic probability distributions, the simulation
determines in each step, whether and when to scatter or to absorb the
photon next, and, in which direction the photon is to be scattered.
Checking for DOM collisions in each step, the simulation is able to
determine for each photon and each detector module whether and when the
photon is detected by the module.

To obtain probability statements from Monte-Carlo simulations, the
\textit{law of large numbers} is employed: If an experiment involving
random processes is repeated \(n\) times, the relative frequency
\(h_n(A):=\sfrac{H(A)}{n}\) of an event\nbsp \(A\), which occurs
\(H(A)\) times in total in these \(n\) experiments, approaches the
\textit{probability}\nbsp \(p(A)\) of the event\nbsp \(A\) for large
numbers\nbsp \(n\) with certainty. \cite{physiklexikon}

\[
  \lim_{n \rightarrow \infty} \text{P}(|h_n(A) - p(A)| < \epsilon) = 1, \ \ \ \epsilon \in \reals
\]

Technical progress concerning computational devices, graphics processing
units (GPUs) in particular, allows to perform highly parallelized
Monte-Carlo simulations on large scale. For a random-walk description of
the propagation of photons, see \cite{absorption1997}. A first
implementation of a photon-propagation simulation through ice is
described in \cite{lundberg}. A study of propagation simulations using
GPUs is presented in \cite{ppcpaper}.

\subsection{Parallel Computing on Graphics Processing Units (GPUs)}
\label{sec:parallel_computing}

Graphics processing units (GPUs) are optimized for performing simple
calculations for a large number of values in parallel.

The general procedure for GPU calculations is to allocate memory on the
GPU, to copy input parameters onto the GPU, perform calculations on the
GPU, and then to download the results from the GPU. This procedure is
efficient if the time that is spent for allocating, and copying to and
from the GPU memory is short compared to the time spent with
calculations on the GPU. \cite{cudacourse}

The basic units for parallelization are \textit{computational threads}.
All operations within a thread are sequential, while all those
operations are applied to a set of threads, called
\textit{thread block}, in parallel. Each GPU may have one or several
thread blocks. \cite{cudacourse}

While technological progress is achieved rapidly, GPU memory is still a
considerably limited resource. In particular fast memory, which requires
a small amount of physical time for reading and writing information, is
expensive. Therefore, memory is divided in several categories: The
\textit{local memory} that belongs only to one thread, is fastest but
most limited. The \textit{shared memory}, which is common to all threads
of one thread block, is slower. Next slower is the
\textit{global memory}, which is common to all thread blocks on the GPU.
Slowest, but in comparison to the GPU memory practically unlimited, is
the \textit{host memory}, which is memory not on the GPU but on other
components of the computer. Efficient memory use is one of the key
concepts for performant GPU programming.
\textit{Coalesce memory access}, which means that each thread in a
thread block reads or writes from or to a coherent memory block parallel
to the other threads of the thread block, increases memory access
performance. \cite{cudacourse}

Techniques that utilize the internal optimizations of GPUs allow for
further performance improvements: Using GPU-native
\textit{atomic operations} such as the increment operator that increases
the value of a variable by 1 is more performant than using a generic
mathematical operation. GPUs support vectors with four components as
native data types. Using \textit{native vectorial operations} such as a
dot product is more performant than implementing the operation as
mathematical function manually. \cite{cudacourse}

In parallel computing, the \textit{step complexity} of an algorithm is a
measure of the physical time the parallelized algorithm needs to run.
The \textit{work complexity} is a measure for the summed computational
work that is done by all threads in that time. A pattern to avoid in
this context is \textit{thread divergence} where some threads have to
stay idle and wait for other threads completing their work.
\cite{cudacourse}

  \label{sec:tools}

\subsection{\icecube Simulation Framework}

This study uses the \noun{IceCube Simulation Framework}, short \icesim,
in Version \texttt{V05-00-07}. The framework, written in
\noun{C++}\footnote{C++ programming language, \url{https://isocpp.org}}
and
\noun{Python}\footnote{Python programming language, \url{https://www.python.org}},
provides the software needed to run simulations, to write and to process
\icecube-specific data such as simulated or recorded events and the
detector geometry.

\docframe{
\docparwithoutframe{The documentation of the \noun{IceCube Simulation Framework} can be found at \url{http://software.icecube.wisc.edu/documentation/}.}\medskip

\docparwithoutframe{A guide on how to install the \noun{IceCube Simulation Framework} with all the other tools needed for this study is provided at \url{https://github.com/fiedl/hole-ice-study/blob/master/notes/2018-01-23_Installing_IceSim_in_Zeuthen.md}.}\medskip

\sourceparwithoutframe{The source code of the \noun{IceCube Simulation Framework} can be found at \url{http://code.icecube.wisc.edu/projects/icecube/browser/IceCube/meta-projects/simulation}.}
}

\subsection{Photon Propagation With \clsim}

The main tool of this study is the photon-tracking simulation software
\clsim. This software implements a ray-tracing algorithm described in
section \ref{sec:standardphotonpropagationalgorithm}, modeling
scattering and absorption of light in the deep-glacial ice at the South
Pole or in Mediterranean sea water. \cite{clsimreadme} Written in C++,
\noun{Python} and
\noun{OpenCL C}\footnote{OpenCL, Open Computing Language, \url{https://www.khronos.org/opencl}},
\clsim uses \noun{OpenCL} to simulate the photon propagation in a highly
parallelized way on graphics processing units (GPUs). \cite{clsimsource}
\clsim reads the photon sources from the \icesim data, converts the data
into a format that can be used on GPUs, uploads the data and the
propagation program (``kernel'') onto the GPU, and propagates the
photons there. Then \clsim downloads the hits and tracks of the
propagated photons, performs post-processing operations such as
accepting hits on optical modules based on the acceptance criteria of
the modules, and converts them back into an \icesim-compatible format.

\docframe{
\sourceparwithoutframe{The source code of \clsim, which is released under the Internet Systems Consortium (ISC) license, can be accessed at the \clsim code repository \url{https://github.com/claudiok/clsim}.}\medskip

\sourceparwithoutframe{In order to simulate the propagation through hole ice, \clsim has been modified for this study. Until the modified source code has been merged into the main code repository, the modified \clsim source code can be accessed through the forked code repository at \url{https://github.com/fiedl/clsim}.}\medskip

\docparwithoutframe{A guide on how to install the modified version of \clsim can be found in appendix \ref{sec:howtoclsim} and on \url{https://github.com/fiedl/hole-ice-study/blob/master/notes/2018-01-23_Installing_IceSim_in_Zeuthen.md\#install-patched-clsim}.}
}

\subsection{Photon Visualization With \steamshovel}

\begin{figure}[htbp]
  \image{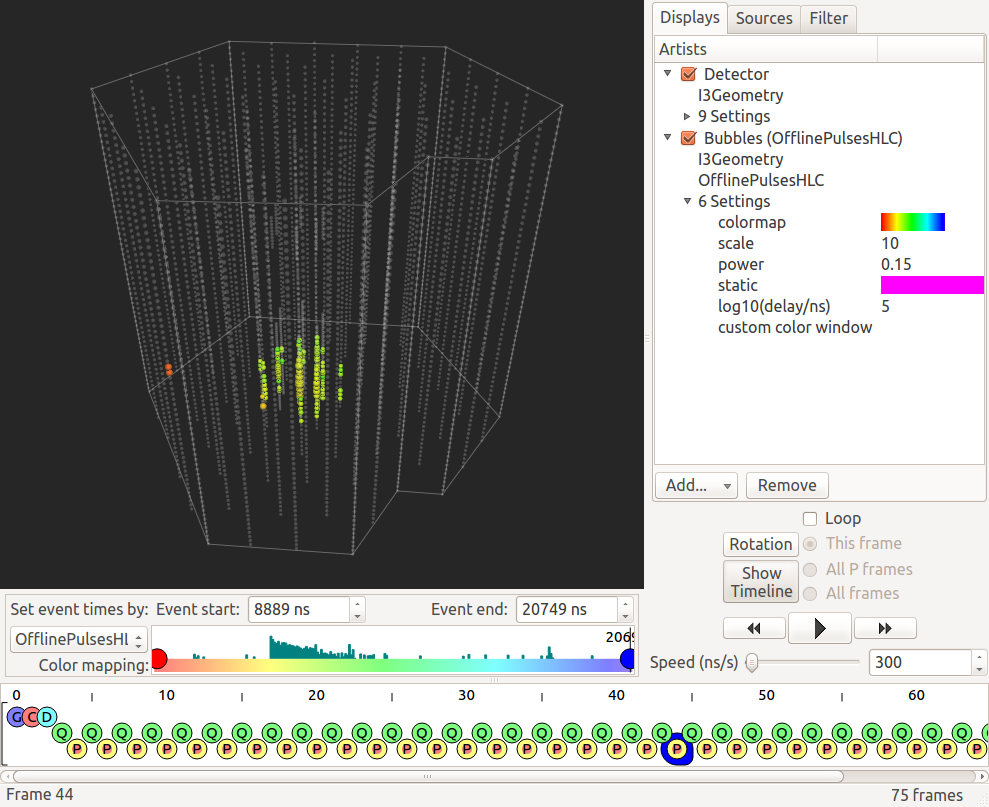}
  \caption{Screenshot of \steamshovel, the \icecube event and data viewer software. The main workspace shows a schematic of the \icecube detector with its 86 strings holding 60 optical detector modules each. Control elements allow to hide and display components and to navigate through the event information. Image source: \cite{steamshoveldocumentation}}
  \label{fig:steamshovel}
\end{figure}

\steamshovel (Figure \ref{fig:steamshovel}) is the event and data viewer
software of \icecube. It allows to visualize the \icecube detector as
well as events that have been recorded, reconstructed or simulated in
the detector. For example, \steamshovel can be used to visualize light
sources such as muons traveling through the detector producing Cherenkov
light, or LED flashes that can be emitted by the optical modules for
calibration purposes. \steamshovel may also visualize the photons
propagating through the ice, or the amount of light detected by the
optical modules. Data containing timing information can be visualized in
an animated manner, watching an event as slow-motion animation.

Most event visualizations in this study are created using \steamshovel.
In order to visualize hole ice and propagating photons for this study,
several patches of the standard \steamshovel code are required.

\sourcepar{The required \steamshovel patches are provided within the code repository of this study: \url{https://github.com/fiedl/hole-ice-study/tree/master/patches/steamshovel}}

\subsection{Other Software Tools}

In order to process data, to perform statistical analyses, and to plot
data, this study uses several supplementary software libraries, such as
\noun{NumPy}\footnote{NumPy package for scientific computing, \url{http://www.numpy.org}},
\noun{Matplotlib}\footnote{Matplotlib plotting library, \url{http://matplotlib.org}}
and
\noun{Pandas}\footnote{Pandas Python Data Analysis Library, \url{http://pandas.pydata.org}}.
For unit tests (section \ref{sec:unit_tests}), this study uses the
\noun{gtest}\footnote{Google Test Framework, gtest, \url{https://github.com/google/googletest}}
testing framework.

\docpar{The necessary scripts and installation instructions can be found in the code repository of this study at \url{https://github.com/fiedl/hole-ice-study}.}

  \section{Development of Algorithms For the Direct Photon-Propagation Through Hole Ice With \clsim}
\label{sec:methods}

\renewcommand\currentsection{Hole-Ice Algorithms}

\subsection{Modeling Hole Ice as Distinct Cylindrical Ice Volumes}

In the \icecube simulation framework, photon propagation simulation
takes several dependencies into account: The photon absorption length
and the photon scattering length may depend on the photon's wavelength.
Absorption and scattering length may also depend on the photon's
\(z\)-coordinate as the South-Polar ice consists of several
\textit{ice layers}. These ice layers may also be tilted. Furthermore,
the absorption length may depend on the photon's direction of motion,
which is called \textit{absorption anisotropy}.

This study adds another ice feature to the simulation: Hole ice may be
modeled by adding cylinder-shaped volumes within the surrounding bulk
ice where the propagation properties, or, to be specific, the photon's
absorption length and scattering length, differ from the propagation
properties of the bulk ice.

\begin{figure}[htb]
  \centering
  \includegraphics[width=0.6\textwidth]{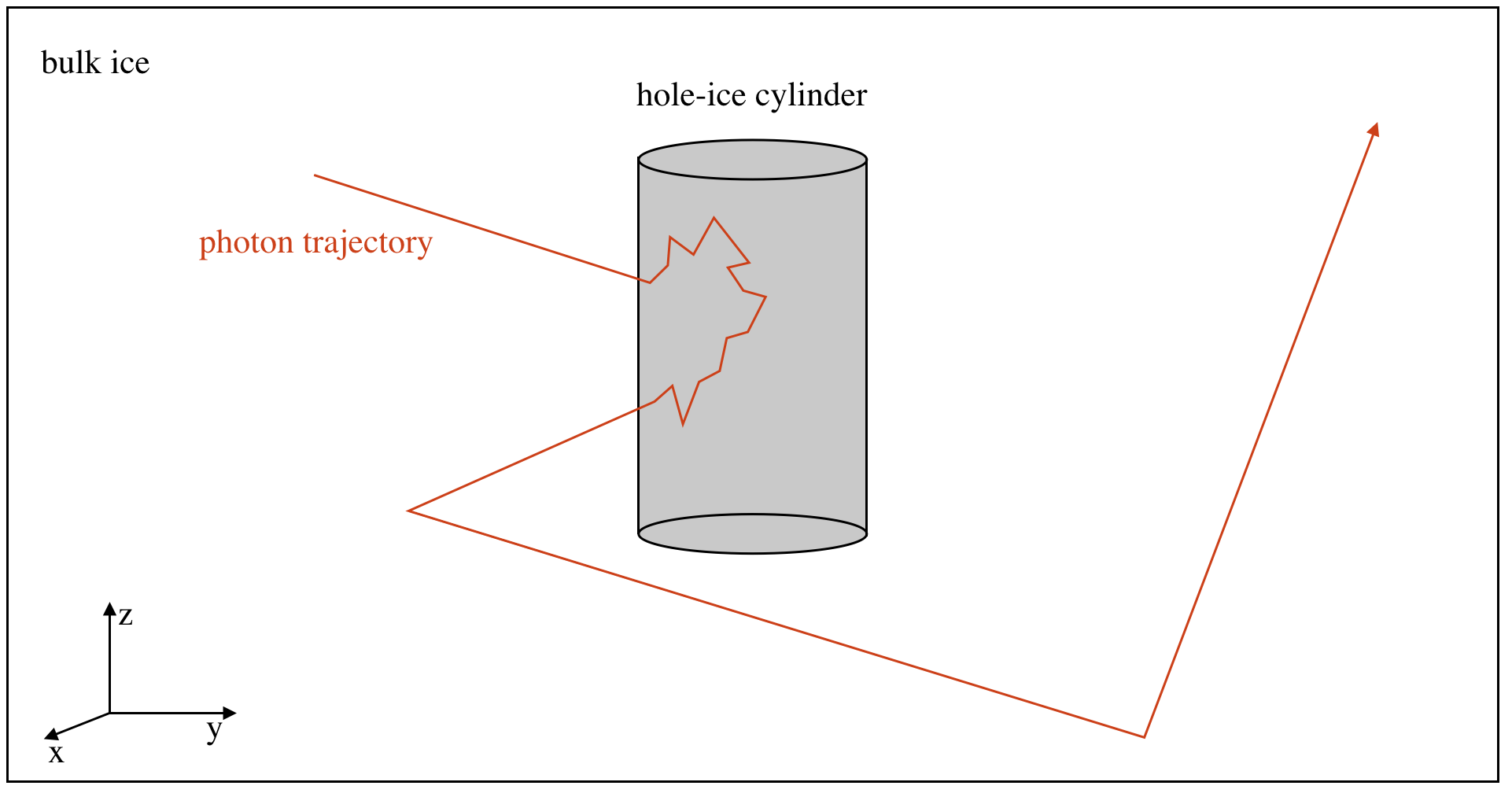}
  \caption{Schematic diagram of a propagating photon. The photon enters a hole-ice cylinder with a scattering length different from the outside ice. When leaving the cylinder the photon assumes the scattering length of the outside ice again.}
  \label{fig:aiw2Thah}
\end{figure}

Figure \ref{fig:aiw2Thah} illustrates such a scenario for one single
photon: The photon's trajectory starts in the bulk ice. The photon
enters the hole-ice cylinder where the scattering length is shorter as
in the bulk ice. Hence the photon scatters more frequently within the
cylinder. When the photon leaves the hole-ice cylinder the propagation
properties of the bulk ice take effect again, resulting in the photon
scattering less frequently after leaving the cylinder.

In this study, cylinders are always defined along the \(z\)-axis. Also,
the scattering angle is assumed to behave the same within the hole ice
as in the bulk ice.

  \subsection{Propagating Photons Through Different Media}

\subsubsection{Very Basic Photon-Propagation Algorithm}

\begin{figure}[htb]
  \smallerimage{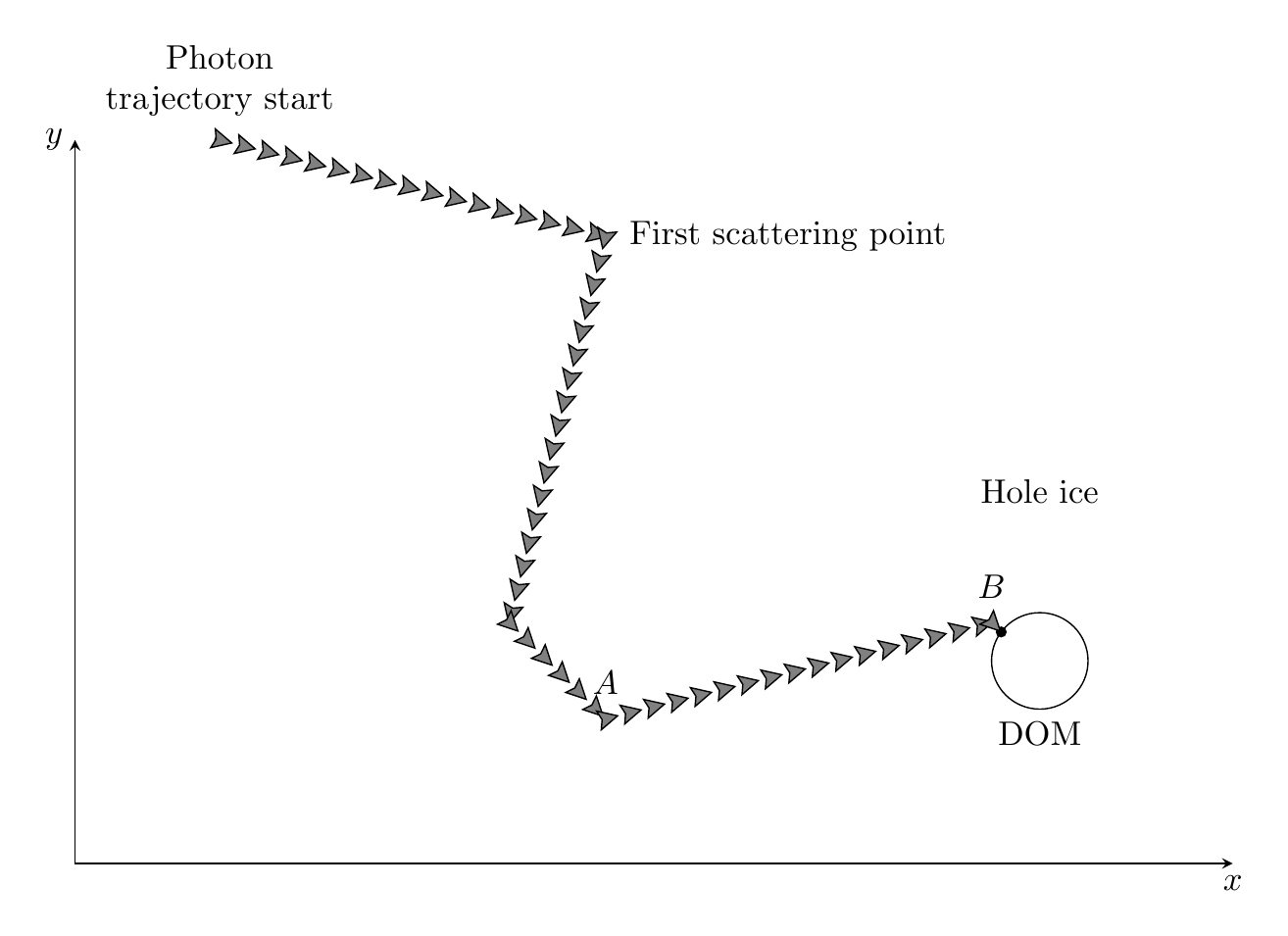}
  \caption{Illustration of a basic photon-propagation algorithm in a two-dimensional coordinate system. The photon is propagated by a small distance in each propagation step. At each position, the algorithm checks for absorption, scattering and whether an optical module has been hit.}
  \label{fig:ieph6Bie}
\end{figure}

\begin{figure}[p]
  \image{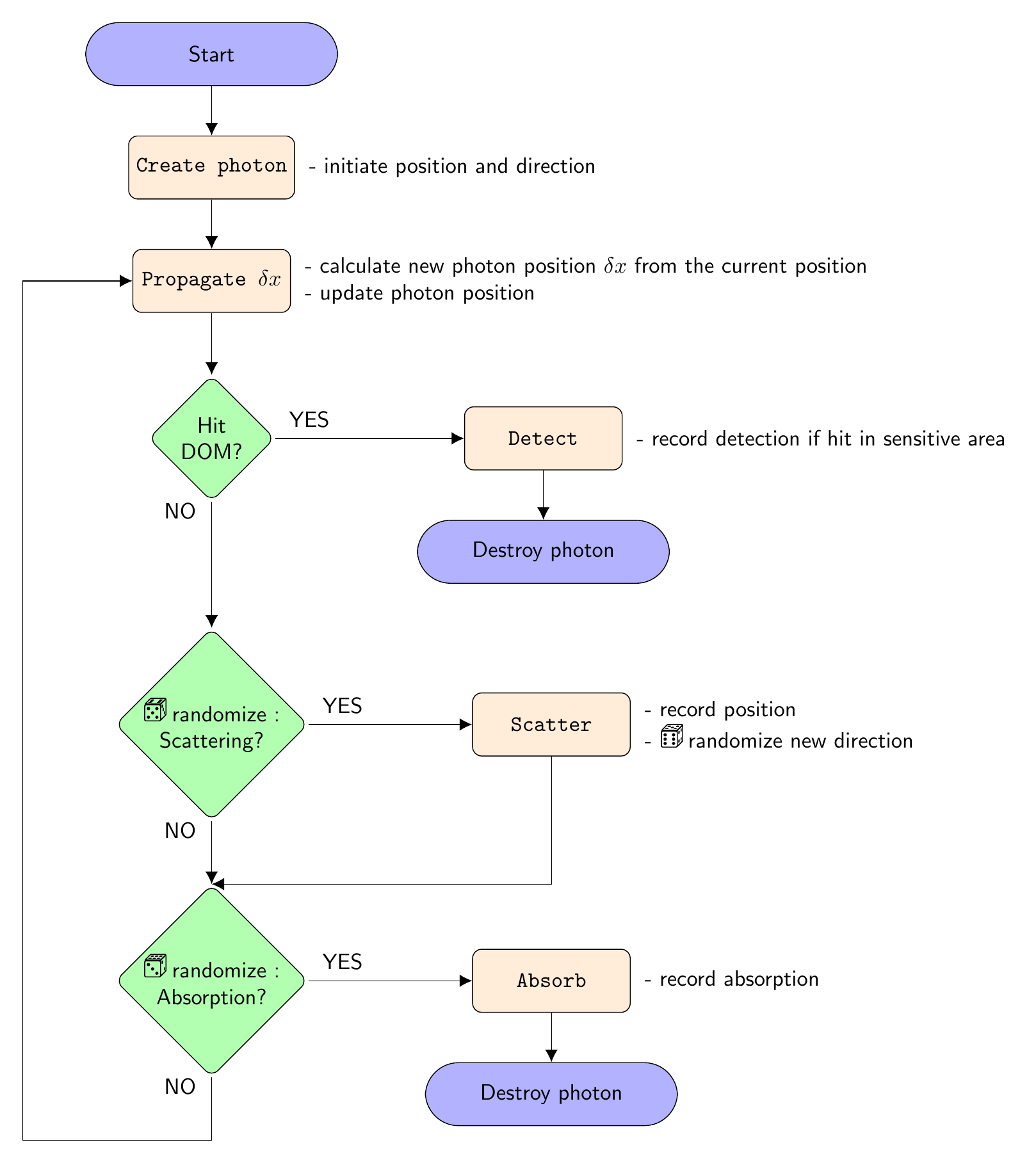}
  \caption{Flow chart of a basic photon-propagation algorithm. Interfaces, where the algorithm begins or ends, are displayed as violet pill shapes. Processes, where the algorithm performs an operation or a calculation, are displayed as brown rounded boxes. Decisions are displayed as green rounded diamond shapes. One propagation step consists of moving the photon by a fixed small distance $\delta x$, checking for detection as well as randomizing scattering and absorption within the ice.}
  \label{fig:ohsa0miG}
\end{figure}

A first, very basic photon-propagation algorithm, which is not actually
implemented in \clsim, but is presented as comparative example, moves
the photon by a small distance \(\delta x\). At the new photon position,
the algorithm checks for detection at an optical module and randomizes
as a function of the scattering and absorption lengths whether the
photon should be scattered or absorbed by the ice at this position.
Then, the photon is propagated again by the same small distance
\(\delta x\). The same loop repeats until the photon either hits an
optical module or is absorbed by the ice. Figure \ref{fig:ieph6Bie}
illustrates a propagation scenario in a two-dimensional coordinate
system. Figure \ref{fig:ohsa0miG} presents the algorithm as flow chart.

This very basic propagation algorithm can handle propagation through
different media just by making the scattering and absorption
probabilities depend on the current photon position and direction. Ice
layers and ice layer tilt can be implemented by making the scattering
and absorption probabilities depend on the current photon position,
absorption anisotropy by making the absorption probability depend on the
current photon direction as well. Hole-ice cylinders can be implemented
by checking whether the current photon position is within any of a list
of cylinders defined in any way, either by supplying a list of cylinder
coordinates and radii, or by re-using the coordinates of the detector
strings.

This basic propagation algorithm, however, is very inefficient regarding
computational performance: The algorithm moves the photon over long
distances in small steps without changing the direction. For a typical
geometric scattering length of two meters, moving the photon in steps of
\(\delta x = 1\mm\) per propagation step would mean performing 2000
propagation steps before changing the direction of motion.

The propagation algorithm can be made more efficient by moving the
photon in each propagation step not by a small distance \(\delta x\) but
by the whole distance to the next interaction point. This way, the above
example would take only one propagation step rather than 2000. This
performance improvement, however, comes at the cost that propagating
through different media requires a different, more involved
computational approach. This more efficient propagation algorithm, which
is the standard photon-propagation algorithm in \icecube, will be
described in the next section.

\subsubsection{Standard Photon Propagation Algorithm}
\label{sec:standardphotonpropagationalgorithm}

\label{sec:standard_photon_propagation_algorithm}
\label{sec:standard_clsim}

In \icecube's standard photon-propagation algorithm, a propagation step
moves the photon not just by a small, fixed distance \(\delta x\) but to
the next interaction point at once. An interaction may be the photon
scattering within the ice, the photon being absorbed by the ice, or the
photon hitting an optical module. \cite{clsimsource, ppcpaper}

At each scattering point, the algorithm randomizes the new photon
direction based on the scattering angle distribution, and how far the
photon will travel until it is scattered again based on the scattering
length. How far the photon will travel until it is absorbed is
randomized just once when the photon is created.

For each propagation step, the algorithm checks whether the photon will
hit an optical module on the path between two scattering points. Also,
the algorithm checks whether the destined distance to absorption will be
reached before the next scattering point. If the photon will be
destroyed in this step, either by being absorbed in the ice, or by
hitting an optical module, the final position is recorded. Otherwise,
the photon proceeds to the next scattering position. This loop is
repeated until the photon is either absorbed or hits an optical module.
\cite{ppcpaper}

\begin{figure}[htbp]
  \smallerimage{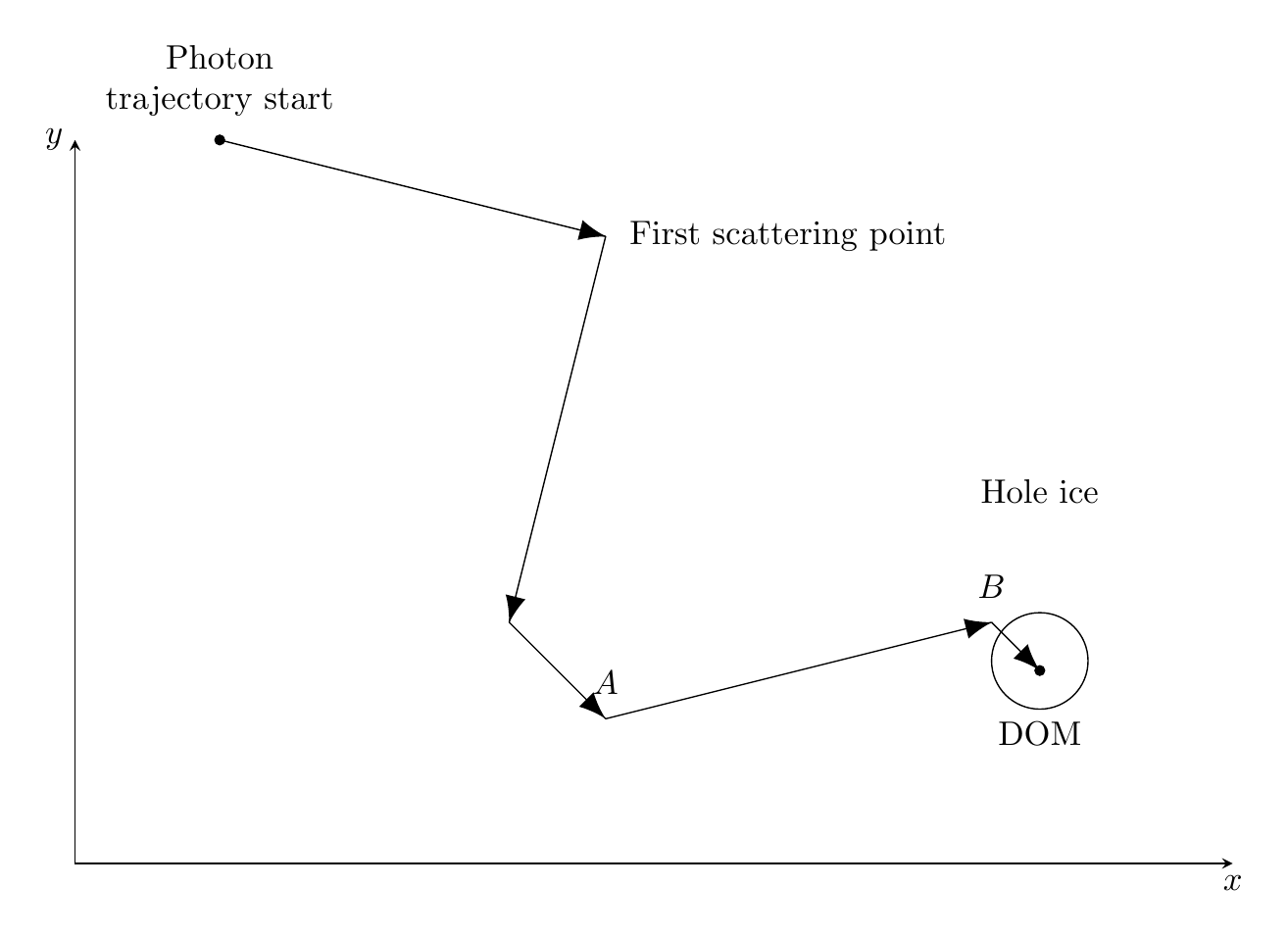}
  \caption{Illustration of \icecube's standard photon-propagation algorithm in a two-dimensional coordinate system. The photon is propagated from one scattering point to the next scattering point in each propagation step. In each step, the algorithm checks for absorption and whether an optical module (DOM) is hit in between the two scattering points.}
  \label{fig:oheeL3ai}
\end{figure}

\begin{figure}[p]
  \image{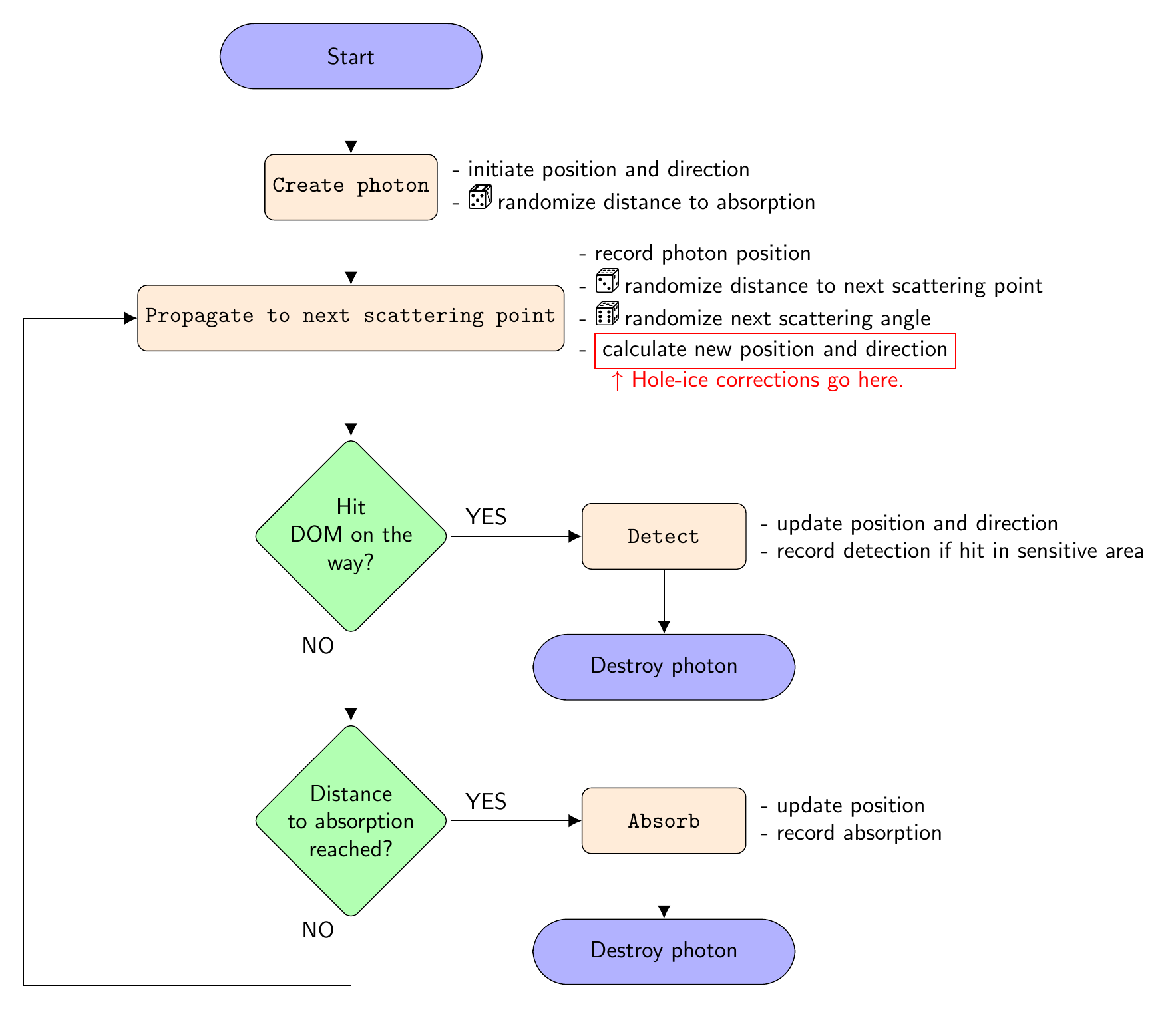}
  \caption{Flow chart of the standard photon-propagation algorithm. One propagation step consists of moving the photon from one scattering point to the next scattering point. If the photon hits an optical module in between the two scattering points, the algorithm records a hit and destroys the photon. If the photon is absorbed in the ice in between the two scattering points, it will only be propagated to the position of absorption and it will not reach the next scattering point. When adding propagation through hole ice with a different scattering length to this algorithm, the calculation of the next scattering point needs to be modified accordingly.}
  \label{fig:Ar0vai8u}
\end{figure}

Figure \ref{fig:Ar0vai8u} shows a flow chart of this algorithm. Figure
\ref{fig:oheeL3ai} shows the same two-dimensional scenario as figure
\ref{fig:ieph6Bie} in the previous section, but illustrates how the same
photon trajectory is modeled by this algorithm with significantly less
simulation steps.

Assuming the number of calculations in each scattering step is of the
same order of magnitude in both described algorithms, the number of
simulation steps translates to the total number of calculations the
processing unit needs to perform for each photon. Thus, the algorithm
that needs much less simulation steps is also much more efficient.

Propagating a photon through several media with different optical
properties is more involved in this algorithm than in the basic one. If
the algorithm calculates the distance to the next interaction point
considering only the interaction properties of the ice at the current
position of the photon as suggested by the basic algorithm, then the
next interaction point would be calculated inaccurately if the medium
properties change between two interaction points. The larger the
scattering length gets compared to the size of the ice volumes with
constant interaction properties, the worse the inaccuracies become. This
can be illustrated by an extreme example where the scattering length and
the absorption length within the bulk ice are several meters long, but
the photon would hit a cable with a diameter of only a couple of
centimeters between two scattering points. The cable would absorb the
photon at once. But if the next interaction point is calculated
evaluating only the interaction properties at the position of the
current scattering point, then the photon ignores the cable and
continues on its path without being absorbed by the cable.

The solution to this problem chosen for the standard propagation
algorithm in \icecube is to implement medium transitions in the
following manner: Rather than randomizing the geometrical distance
\(X:=\overline{AB}\) from the current interaction point \(A\) to the
next interaction point \(B\) itself, the algorithm randomizes the number
\(N :\in \reals^+\) of interaction lengths the photon will travel to the
interaction point, like a budget. This budget is spent by considering
the different media on the way from the current interaction point along
the photon's direction, and converting the number of interaction lengths
into a geometrical distance according to the shares of the different
media in the path of the photon between the two interaction points.

Suppose, starting at interaction point \(A\), the photon travels in
\(m :\in \naturals\) media \(M_i\) with interaction lengths
\(\lambda_i\) for a distance of \(x_i\) respectively until it reaches
the next interaction point \(B\). In each medium \(M_i\), the algorithm
spends \(n_i: x_i = n_i\,\lambda_i\) of the budget \(N:=\sum_1^m n_i\)
that has been randomized at interaction point \(A\) until it reaches
interaction point \(B\) in a distance \(X:=\overline{AB}\).

\begin{equation}
  X = \sum_{i=1}^m x_i = \sum_{i=1}^m n_i\,\lambda_i, \ \ \ \ N = \sum_{i=1}^m n_i
  \label{eq:convertbudgettodistance}
\end{equation}

Each distance \(x_i\) is the length of the trajectory the photon spends
in medium \(M_i\). The first distance \(x_1\) is the distance from the
starting point \(A\) to the medium border of \(M_1\) and \(M_2\) along
the photon direction. The subsequent distances \(x_i\) are the distances
of the first medium border (of \(M_{i-1}\) and \(M_i\)) and the second
medium border (of \(M_i\) and \(M_{i+1}\)) respectively along the photon
direction. The last distance \(x_m\) is determined by how much of the
budget \(N\) is left in the last medium, \(x_m = n_m\,\lambda_m\), such
that the overall budget \(N:=\sum_{i=1}^m n_i\) is spent.

Coming back to the extreme example where a cable lies ahead on the
photon path, the algorithm now considers all media on the path along the
photon direction in order to convert the number of interaction lengths
into a geometrical distance to the interaction point. As the absorption
length within the cable's medium is set to zero, all of the
absorption-length budget is spent at the medium border to the cable,
resulting in the final interaction point of the photon being right at
the point where the photon enters the cable. Thus, this algorithm lets
the photon be absorbed by the cable as intended.

Ice layers, the tilt of the ice layers, and the absorption anisotropy
are modeled in this algorithm by implementing the rules on how the
interaction budget is spent accordingly. Adding cylinder-shaped volumes
to model hole ice or cables means to further extend these rules on how
the scattering and absorption budgets are spent along the photon's
trajectory.

Two different algorithms for the photon propagation through
cylinder-shaped volumes are described in the following sections: Section
\ref{sec:algorithm_a} describes a first approach where the propagation
through the hole-ice cylinders is added as subsequent correction for the
existing medium-propagation algorithm. Section \ref{sec:algorithm_b}
describes a second approach where the existing medium-propagation
algorithm is rewritten in order to support the propagation through
hole-ice cylinders the same way as other medium transitions.

  \subsection{Algorithm A: Hole-Ice Propagation as Subsequent Correction of the Propagation Without Hole Ice}
\label{sec:algorithm_a}

A first approach to adding propagation through hole-ice cylinders to the
standard propagation algorithm (section
\ref{sec:standardphotonpropagationalgorithm}) implemented in \clsim is
to add a subsequent correction to each simulation step without rewriting
the existing algorithm.

\sourcepar{The source code of the implementation of this first approach can be found in appendix \ref{sec:algorithm_a_source}, in the folder \texttt{algorithm\_a} on the CD-ROM, as well as in the code repository at \url{https://github.com/fiedl/clsim/tree/sf/hole-ice-2017/resources/kernels/lib/hole_ice}.}

This approach assumes that the scattering length \(\lambda\hi\sca\) and
the absorption length \(\lambda\hi\abs\) within the hole ice can be
expressed as multiple of the corresponding scattering length
\(\lambda\sca\) and absorption length \(\lambda\abs\) within the
surrounding bulk ice.

\[
  \lambda\sca\hi = f\sca \, \lambda\sca, \ \ \ \lambda\abs\hi = f\abs \, \lambda\abs
\]

The factor \(f\sca :\in \reals^+_0\) will be called
\textbf{scattering length factor}, \(f\abs :\in \reals^+_0\)
\textbf{absorption length factor}. These factors can be implemented as
properties of the individual hole-ice cylinders, or as common property
of all cylinders.

If, for example a cylinder has an absorption length factor of
\(f\abs = 1\), the absorption length within the hole ice is the same as
in the surrounding bulk ice. If the absorption length factor is
\(f\abs = 0\), a photon will be absorbed instantly when entering the
hole-ice cylinder. A scattering factor of \(f\sca=0.1\) means that the
scattering length within the hole ice is one tenth of the scattering
length within the surrounding bulk ice.

\paragraph{Task}

The task of this \textbf{hole-ice-correction algorithm} is to modify the
quantities that are effected by the hole ice in each simulation step.

\begin{figure}[htbp]
  \smallerimage{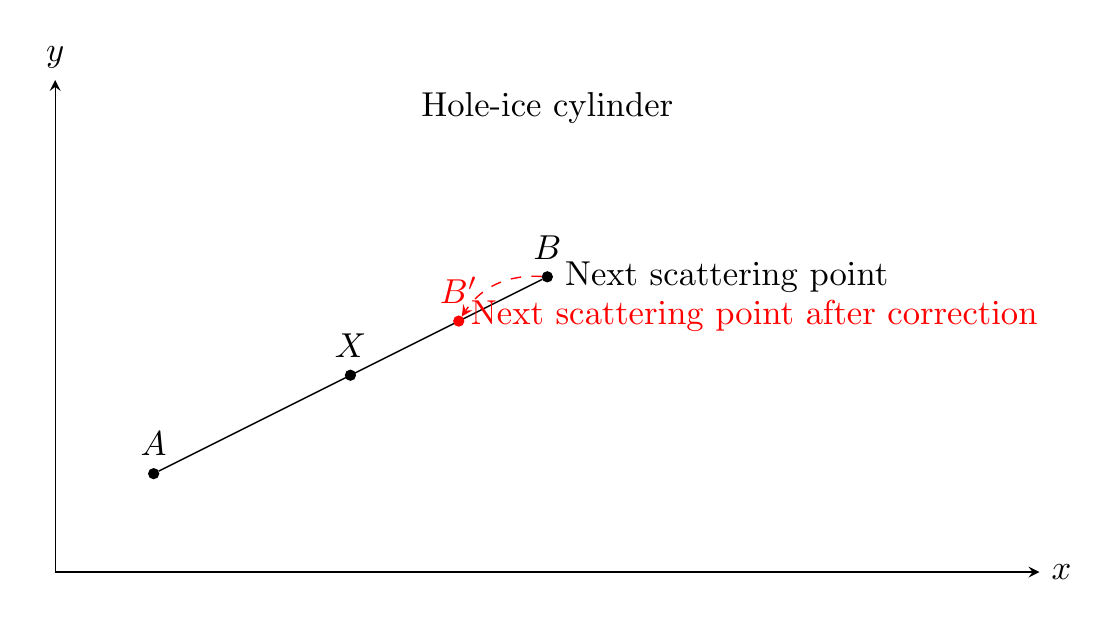}
  \caption{Illustration of the task of the hole-ice correction algorithm. In this two-dimensional scenario, the hole-ice cylinder is represented by a circle. When the photon scatters at point $A$, the standard-\clsim algorithm calculates the next scattering point $B$ without knowledge of the hole ice. The hole-ice correction algorithm calculates what fraction of the trajectory $AB$ runs through the hole ice and determines a correction, resulting in a hole-ice-corrected next scattering point $B'$.}
  \label{fig:Edahi9sh}
\end{figure}

Figure \ref{fig:Edahi9sh} illustrates this task: The standard
\clsim propagation algorithm calculates the next scattering point \(B\)
without any knowledge of the hole ice. The hole-ice algorithm takes the
properties of the hole ice into account and calculates a correction for
the next scattering point. \clsim will use the corrected next scattering
point \(B'\) rather than the originally calculated point \(B\).

\paragraph{Context}

The hole-ice-correction algorithm is inserted into the
\clsim propagation algorithm's simulation step right after \clsim has
calculated the distance to the next scattering point and the remaining
distance to the absorption point without knowledge of the hole ice.
After the hole-ice-correction algorithm follows the check whether the
photon hits an optical module on the way from \(A\) to \(B'\).

The hole-ice-correction algorithm takes the following input parameters:
Current photon position at point \(A\), photon direction after
scattering at point \(A\), a list of the hole-ice cylinders with their
coordinates and radii, the hole-ice scattering length factor \(f\sca\)
and absorption length factor \(f\abs\) as common properties of all
hole-ice cylinders, the distance the next scattering point calculated by
the standard algorithm, and the remaining distance to the absorption
point calculated by the standard algorithm. As output parameters, the
hole-ice correction algorithm returns the hole-ice-corrected distance to
the next scattering point and the hole-ice-corrected remaining distance
to the absorption point.

\paragraph{Procedure}

For each propagation from one scattering point \(A\) to the next
scattering point \(B\), the hole-ice algorithm calculates the
intersection points of the photon trajectory \(AB\) and the hole-ice
cylinders in range. Based on what portion of the distance \(AB\) runs
through the hole ice, the algorithm calculates a correction for the
distance to the next scattering point and a correction for the remaining
distance to the absorption point.

Both calculations, scattering correction and absorption correction,
depend on each other: If the photon scatters earlier within the hole ice
than determined by the standard algorithm, then the corrected path
within the hole ice is shorter than the original one, which means that
the remaining distance to the absorption point will be longer as less of
the absorption budget is spent, yet. If, on the other hand, the hole ice
absorbs the photon instantly when entering the cylinder, then the
corrected final position \(B'\) of this simulation step won't be the
point determined by the scattering correction but the point where the
photon enters the cylinder.

\begin{figure}[htbp]
  \image{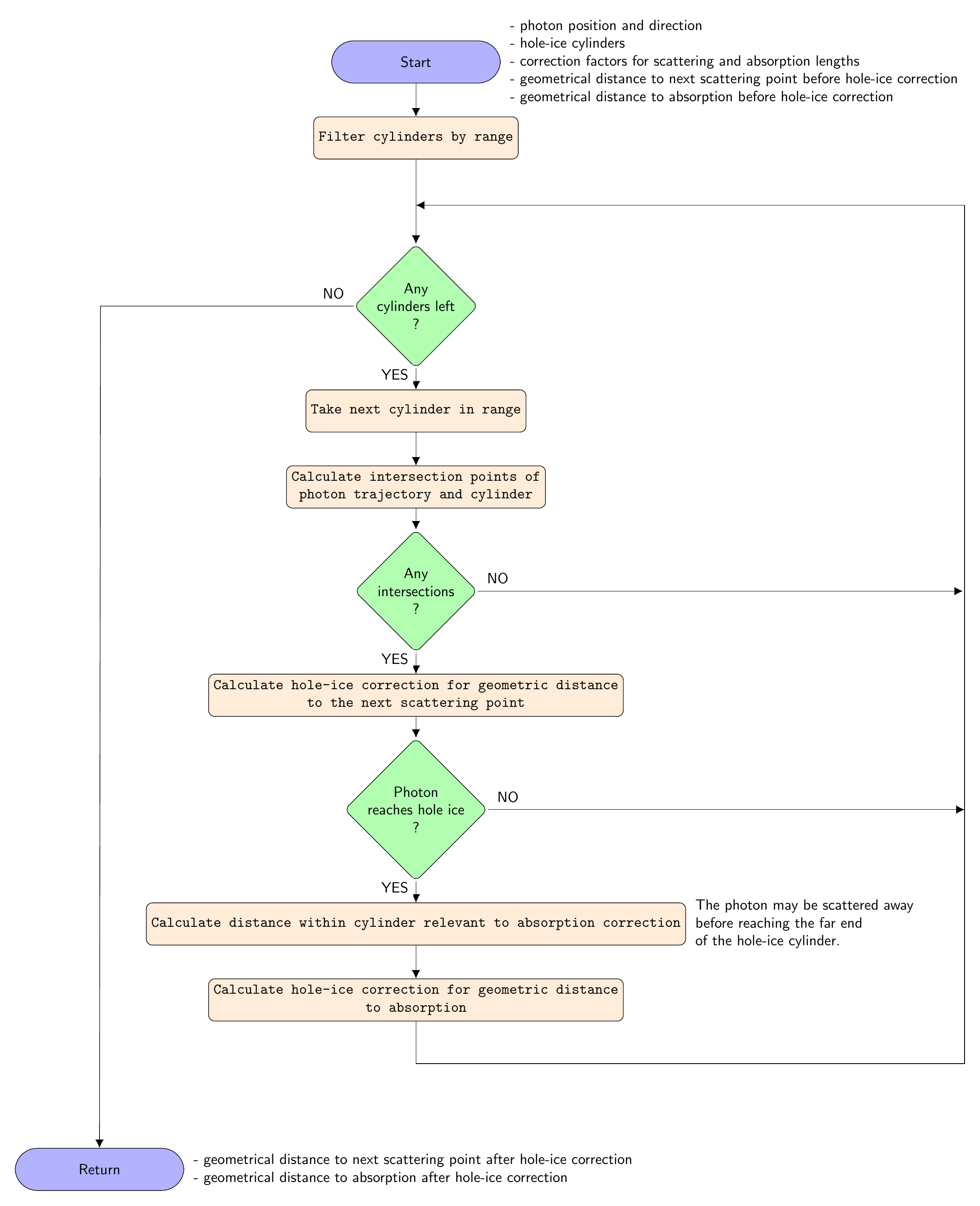}
  \caption{Flow chart of the hole-ice-correction algorithm, which calculates subsequent corrections for the quantities that are effected by the hole ice: the distance to the next scattering point and the remaining distance to absorption of the photon. Both corrections depend on each other: If the photon scatters earlier, the absorption correction will be smaller. The algorithm returns the corrected quantities to the standard algorithm, which uses the corrected quantities from that point on.}
  \label{fig:bahxug7O}
\end{figure}

Figure \ref{fig:bahxug7O} shows a flow chart of this algorithm including
the loop over the hole-ice cylinders in range, the calculation of the
intersection points of photon path and hole-ice cylinder, the correction
of the next scattering point and the correction of the remaining
distance to absorption.

\paragraph{Intersection Calculations}

In order to determine the fraction of the path \(AB\) of the photon that
runs through the hole-ice cylinder, the intersection points of the
photon path and the cylinder need to be calculated. This can either be
done solving the geometric equations coordinate-wise for the coordinates
of the zero, one or two intersection points, or by treating the same
scenario as vectorial problem, calculating the coordinate vectors of the
intersection points using other vectorial quantities rather than
separate coordinates. The latter approach turns out to be more efficient
as it utilizes the support of native vector operations of the graphics
processing units. Both approaches are presented in appendix
\ref{sec:intersections}.

\paragraph{3D-2D Projection}

Intersection points can be calculated in two dimensions by projecting
all coordinates from the three-dimensional coordinate system onto the
\(x\)-\(y\) plane along the \(z\)-axis, and then calculating the
intersection of a line and a circle rather than the intersections of a
line and a cylinder. The fraction of the photon path that runs through
the hole ice is the same as in three dimensions due to the intercept
theorem, because both, the distance to the intersection point,
\(\len{AX}_\text{2D}:=\xi \len{AX}_\text{3D}\), and the distance to the
next scattering point, \(\len{AB}_\text{2D}:=\xi \len{AB}_\text{3D}\),
use the same projection factor \(\xi \in \reals^+\), which itself
depends on the photon direction.

\[
  \frac{\len{AX}_\text{2D}}{\len{AB}_\text{2D}} = \frac{\len{AX}_\text{3D}}{\len{AB}_\text{3D}}
\]

The relative distance corrections, \(\sfrac{\Delta x}{\len{AB}}\), are
the same in two and three dimensions due to the same reason. But the
absolute distance correction
\(\Delta x : \len{AB'} = \len{AB} + \Delta x\) needs to be expressed in
three dimensions when adding the correction to the distances used in the
standard algorithm as they are defined for a three-dimensional
coordinate system. Keeping this conversion requirement in mind, the
geometric cases of the photon path \(AB\) running through a hole-ice
cylinder can be examined in two dimensions.

\paragraph{Geometric Cases}

When the hole-ice algorithm calculates the distance correction
\(\Delta x: \len{AB'} = \len{AB} + \Delta x\), there are several
geometric cases to consider, depending on the number \(N\) of
intersections of the photon path \(AB\) with the hole-ice cylinder and
whether the path starts within the cylinder or outside the cylinder.

\begin{figure}[htbp]
  \smallerimage{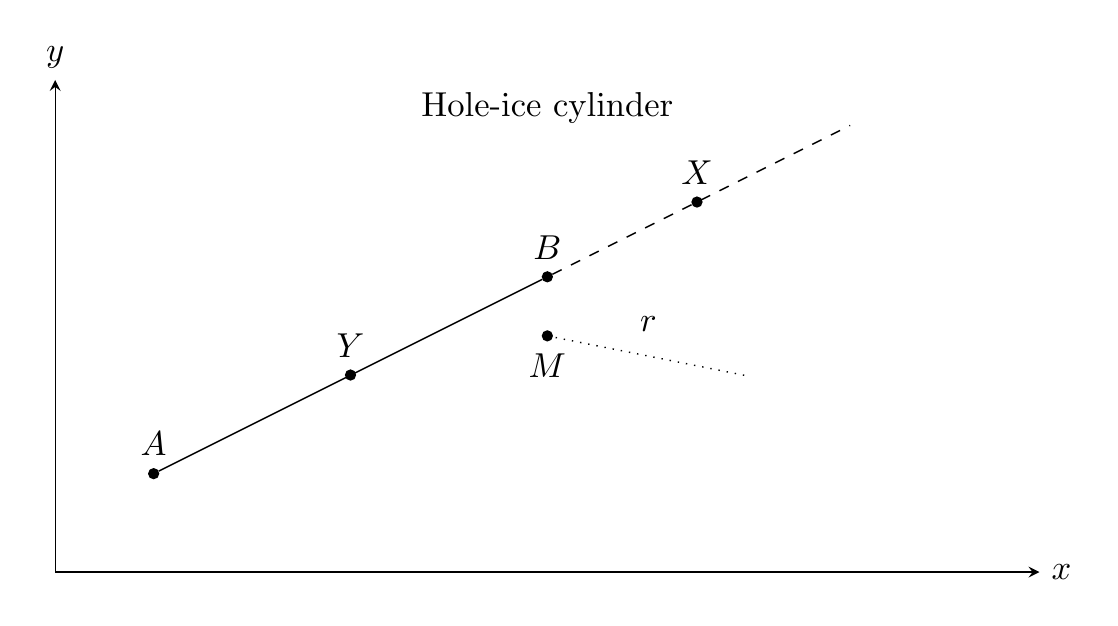}
  \caption{Intersection of the line along the photon path $AB$ with the hole-ice cylinder in two dimensions, where the cylinder is represented by a circle of radius $r$ around the center $M$. The first intersection point is $Y$, the second intersection point $X$. The second intersection point $X$ is beyond the path's end point $B$. Thus, the number $N$ of intersections is considered to be $N=1$ in this case.}
  \label{fig:iefai4iV}
\end{figure}

The distance correction depends on the fraction \(w :\in \reals^+_0\) of
the path length \(\len{AB}\) that runs within the hole-ice cylinder.

\begin{description}
  \item[Case 1] If the photon starts outside the cylinder and the path has no intersection with the cylinder ($N = 0$), then the path does not run through the cylinder, $w = 0$, and the distance correction $\Delta x$ needs to be zero: $\Delta x = 0$.
  \item[Case 2] If the photon starts inside the cylinder and the path has no intersection with the cylinder ($N = 0$), then the whole path is within the cylinder, $w = 1$. A photon that would travel a distance of $\lambda$ within the bulk ice, travels a distance of $\lambda\hi$ within the hole ice, $\lambda\hi = f\,\lambda$. The distance correction $\Delta x$ for this photon would be $\Delta x = \lambda\hi - \lambda = (f - 1)\lambda$, or more generally, $\Delta x = (f - 1) \len{AB}$.
  \item[Case 3] If the photon starts outside the cylinder, but intersects the cylinder once ($N = 1$), then the first part of the trajectory, $AY$, stays the same, and only the path from the first intersection point $Y$ to the destination point $B$ runs within the hole ice, $w = \len{YB} / \len{AB}$. As only this fraction needs to be corrected, the distance correction is $\Delta x = (f - 1)\len{YB}$.
  \item[Case 4] If the photon intersects the cylinder once ($N = 1$), but starts within the cylinder, only the first part, $AC$, of the path needs to be corrected.

  $C$ is the \textbf{termination point} of the trajectory inside the cylinder. In cases where the photon reaches the far end of the cylinder, $C$ is just the second intersection point, $C = X$. If the trajectory within the cylinder is terminated, however, by the other interaction respectively, for example if the photon is scattered away before reaching the far end when calculating the absorption correction, $C$ is the point where the trajectory is terminated by the other interaction.

  In contrast to Case 3 where the outside distance $\len{AY}$ has been fixed, in this case the outside trajectory part $\len{CB}$ comes after the hole-ice part $\len{YC}$ and will be the shorter the stronger the hole-ice correction is. Therefore, one needs to use the scaled distance $\len{AC} / f$ to determine how much of the path will remain outside the cylinder, resulting in a distance correction of $\Delta x = (1 - \frac{1}{f})\len{AC}$.
  \item[Case 5] If the photon starts outside the cylinder and has two intersections ($N = 2$) with the cylinder, this is a combination of Case 3 and Case 4, resulting in a distance correction of $\Delta x = (1 - \frac{1}{f})\len{YC}$.
\end{description}

\begin{figure}[htbp]
  \includegraphics[height=\textheight]{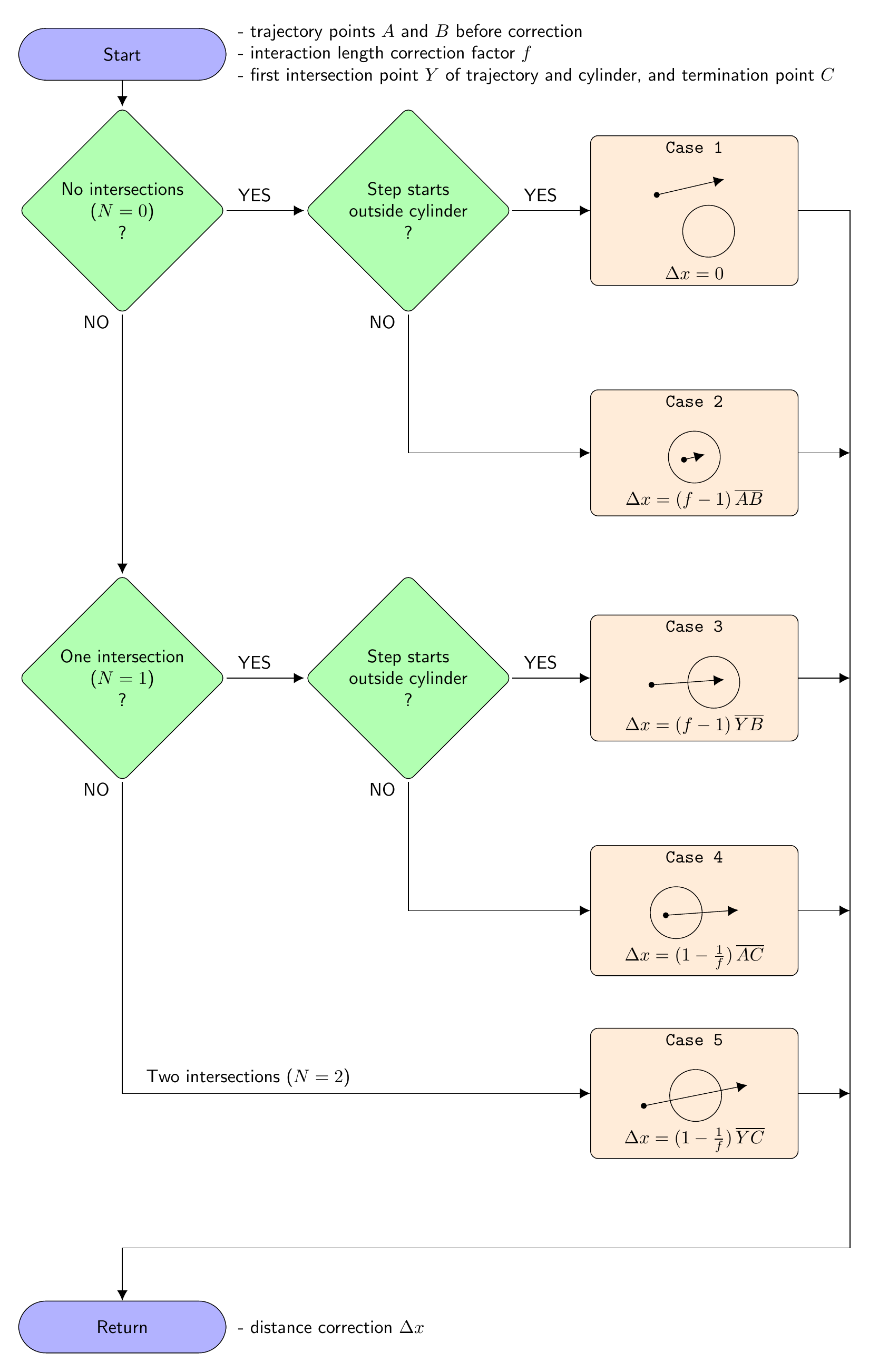}
  \caption{Flow chart for calculating the hole-ice correction for the geometric distance to the next interaction point.}
  \label{fig:Eeshi4Oh}
\end{figure}

A flow chart of the distance-correction algorithm is shown in figure
\ref{fig:Eeshi4Oh}.

\paragraph{Pros and Cons}

As this first hole-ice algorithm only adds subsequent corrections for
the standard-\clsim algorithm, it leaves the standard-\clsim code, which
is already well tested, almost untouched. The additions of the hole-ice
algorithm interface with the \clsim algorithm only through a small
surface area, corresponding to a small number of variables passed from
one algorithm to the other. This allows to test the hole-ice algorithm
with so-called \textit{unit tests} (section \ref{sec:unit_tests}) where
specific examples with fixed input variables are provided and tested
whether the algorithm produces the expected results for the example.

The hole-ice correction algorithm, however, has important limitations.
First, the current understanding of the hole ice suggests that the
interaction properties of the hole ice are independent of the properties
of the surrounding bulk ice (see section \ref{sec:hole_ice}). But with
this hole-ice correction algorithm, it is only possible to define the
interaction lengths of the hole ice relative to the interaction
properties of the surrounding bulk ice. While one could workaround this
problem locally by adjusting the correction factors \(f\) in relation to
the local properties of the bulk ice, large jumps over several layers of
ice with different properties would cause inevitable errors.

Secondly, the algorithm does work only correct in scenarios where the
photon crosses only one cylinder between two scattering points. In
scenarios where the photon crosses several cylinders, for example when
nested cylinders are used to model radially changing properties of the
hole ice, or when instantly-absorbing cylinders are used to model cables
within a hole-ice cylinder, this algorithm is not able to calculate the
corrections properly because it would interpret the interaction factors
\(f\) not only relative to the surrounding bulk ice but also relative to
the cylinders the photon has already crossed in this simulation step.

A different approach to adding propagation through hole-ice cylinders to
the standard-propagation algorithm is to treat hole-ice cylinders the
same way as other media such as ice layers rather than accounting for
the hole-ice effects using subsequent corrections. This second approach
gets rid of the above shortcomings, but at the cost that the
standard-\clsim algorithm that handles the propagation through different
media needs to be re-written. This second approach is described in the
following section.

  \subsection{Algorithm B: Hole-Ice Propagation by Generalizing the Previous Medium-Propagation Algorithm}
\label{sec:algorithm_b}

A second approach to adding propagation through hole-ice cylinders to
the standard propagation algorithm (section
\ref{sec:standardphotonpropagationalgorithm}) implemented in \clsim is
to rewrite the part of the standard algorithm that propagates the photon
through different media, aiming to make this algorithm more generic and
support hole-ice cylinders at the same level as other structures such as
ice layers.

\sourcepar{The source code of the implementation of this second approach can be found in appendix \ref{sec:algorithm_b_source}, in the folder \texttt{algorithm\_b} on the CD-ROM, as well as in the code repository at \url{https://github.com/fiedl/clsim/tree/sf/hole-ice-2018/resources/kernels/lib}.}

In order to propagate a photon through different media, the standard
propagation algorithm converts the randomized number of interaction
lengths \(N\) into a geometrical distance
\(X:=\sum_i n_i\,\lambda_i,\ N = \sum_i n_i\) (equation
\ref{eq:convertbudgettodistance} on page
\pageref{eq:convertbudgettodistance}) based on the interaction lengths
\(\lambda_i\) of the different media. In its current implementation in
\clsim, this conversion is hard-coded in a way that does support medium
changes in the form of tilted ice layers but does not support other
shapes such as cylinders of hole
ice.\footnote{A comparison of the flow charts of standard \clsim's media-propagation algorithm and the new media-propagation algorithm is given in appendix \ref{sec:flow_charts}.}
In the approach described in this section, the medium-propagation
algorithm is re-implemented in a way that does support cylinder-shaped
media.

\paragraph{Task}

The task of this new \textbf{medium-propagation algorithm} is to
determine in each simulation step, which media, including ice layers as
well as hole-ice cylinders, are on the photon's path along its current
direction and to convert the randomized scattering and absorption length
budgets to geometrical distances according to the interaction lengths of
the different media.

\paragraph{Context}

The new medium-propagation algorithm is inserted into the
\clsim propagation algorithm's simulation step right after \clsim has
randomized the number of scattering lengths to the next scattering
point, replacing \clsim's original medium-propagation algorithm. After
the medium-propagation algorithm follows the check whether the photon
hits an optical module on the way from the current scattering point
\(A\) to the next scattering point \(B\).

The medium-propagation algorithm takes the following input parameters:
Current photon position at point \(A\), photon direction after
scattering at point \(A\), a list of the hole-ice cylinders with their
coordinates, radii, scattering lengths and absorption lengths, a list of
ice layers with their coordinates, scattering lengths and absorption
lengths, the randomized number of scattering lengths to the next
scattering point, the remaining number of absorption lengths to the
absorption point, and the remaining geometric distance to absorption. As
output parameters, the medium-propagation algorithm returns the updated
number of remaining absorption lengths to the absorption point, the
geometric distance to the next scattering point, and the updated
remaining geometric distance to absorption.

\paragraph{Procedure}

The new algorithm works with a generic list of media. For each medium,
the algorithm stores the scattering length, the absorption length, and
the distance along the photon's direction from the current photon
position to the medium's border. This way, the algorithm knows in which
distance from the current position which medium properties take effect.
Due to this generic structure, homogeneous media with all kinds of
shapes could be added to this list. In this first implementation, the
algorithm adds un-tilted ice layers and cylinders, which model hole ice
and cables. Tilted ice layers as well as absorption anisotropy are not
re-implemented in this first implementation, but may be added
later.\footnote{At the time of writing, ice tilt and absorption anisotropy are not re-implemented, yet. To check the current state of this issue, see \url{https://github.com/fiedl/hole-ice-study/issues/48}.}
After adding ice layers and hole-ice cylinders to the media list, the
algorithm sorts the list by ascending distance from the current photon
position in order to have the list in the proper order for the following
calculations. Finally, the algorithm loops over the media list and
calculates the geometrical distances to the next scattering point and to
the absorption point just like the previous media-propagation algorithm
has done for ice layers only.

\paragraph{Media Loop}

\begin{figure}[p]
  \resizebox{\textwidth}{!}{%
    \begin{tikzpicture}
  [
    node distance = 1.8cm,
    every node/.style = {fill=white, font=\sffamily},
    align = center
  ]

  \newcommand\process[3]{\node(#1)[process, #3]{#2}}
  \newcommand\decision[3]{\node(#1)[decision, #3]{#2}}
  \node(start)[start]{Start};
  \decision{boundaryleft}{Medium\\boundary left\\?}{below = 2cm of start};

  \process{next}{Take next medium boundary}{right = 4cm of boundaryleft};
  \process{scadistance}{Calculate contribution of this medium\\up to the next medium boundary\\to the geometrical distance to the\\next \textbf{scattering} point}{below of = next};
  \process{absdistance}{Calculate contribution of this medium\\up to the next medium boundary\\to the geometrical distance to \textbf{absorption}}{below = 5mm of scadistance};

  \process{spendrestsca}{Calculate contribution of the last medium\\to the geometrical distance to the next scattering point}{below = 5cm of boundaryleft};
  \process{spendrestabs}{Calculate contribution of the last medium\\to the geometrical distance to absorption}{below of = spendrestsca};

  \decision{ifabsorbed}{Distance to\\absorption reached\\?}{below = 5mm of spendrestabs};
  \process{propagateonlytoabs}{Only propagate to the absorption point}{right = 2.5cm of ifabsorbed};

  \node(return)[stop, below = 2cm of ifabsorbed]{Return};

  \newcommand\comment[2]{\node ()[substeps, right = 1mm of #1]{#2}}
  \comment{start}{
    \substep number of scattering lengths left to next scattering point\\
    \substep number of absorption lengths left to absorption\\
    \substep list of media. Each: distance to medium boundary, local scattering lenght, local absorption length
  };
  \comment{return}{
    \substep geometrical distance to next scattering point\\
    \substep geometrical distance to absorption
  };

  \makeatletter
    \@ifundefined{r@fig:eewoo3Be}{}{
      \draw [decorate, decoration={brace,amplitude=10pt,raise=4pt}, yshift=0pt]
        (absdistance.west) -- +(0,2.5) node [midway,xshift=-1.8cm] {See figure \ref{fig:eewoo3Be}};
    }
  \makeatother

  \newcommand\connect[2]{\draw[->](#1)--(#2)}
  \connect{start}{boundaryleft};
  \connect{boundaryleft}{next};
  \connect{next}{scadistance};
  \connect{scadistance}{absdistance};
  \draw [->] (absdistance) |- +(5,-1.5) |- (0,-1.5);

  \connect{boundaryleft}{spendrestsca};

  \connect{spendrestsca}{spendrestabs};
  \connect{spendrestabs}{ifabsorbed};
  \connect{ifabsorbed}{propagateonlytoabs};
  \draw [->] (propagateonlytoabs) |- (0,-20.0);

  \connect{ifabsorbed}{return};

  \node () [right = 1mm of boundaryleft, yshift = 3mm] {YES};
  \node () [below = -1mm of boundaryleft, xshift = -5mm] {NO};

  \node () [right = 1mm of ifabsorbed, yshift = 3mm] {YES};
  \node () [below = -1mm of ifabsorbed, xshift = -5mm] {NO};

\end{tikzpicture}
  }
  \vspace*{5mm}
  \caption{Flow chart of the part of the new media-propagation algorithm that loops over list of media and calculates geometrical distances to the next scattering point and to the absorption point.}
  \label{fig:nimuriX4}
\end{figure}
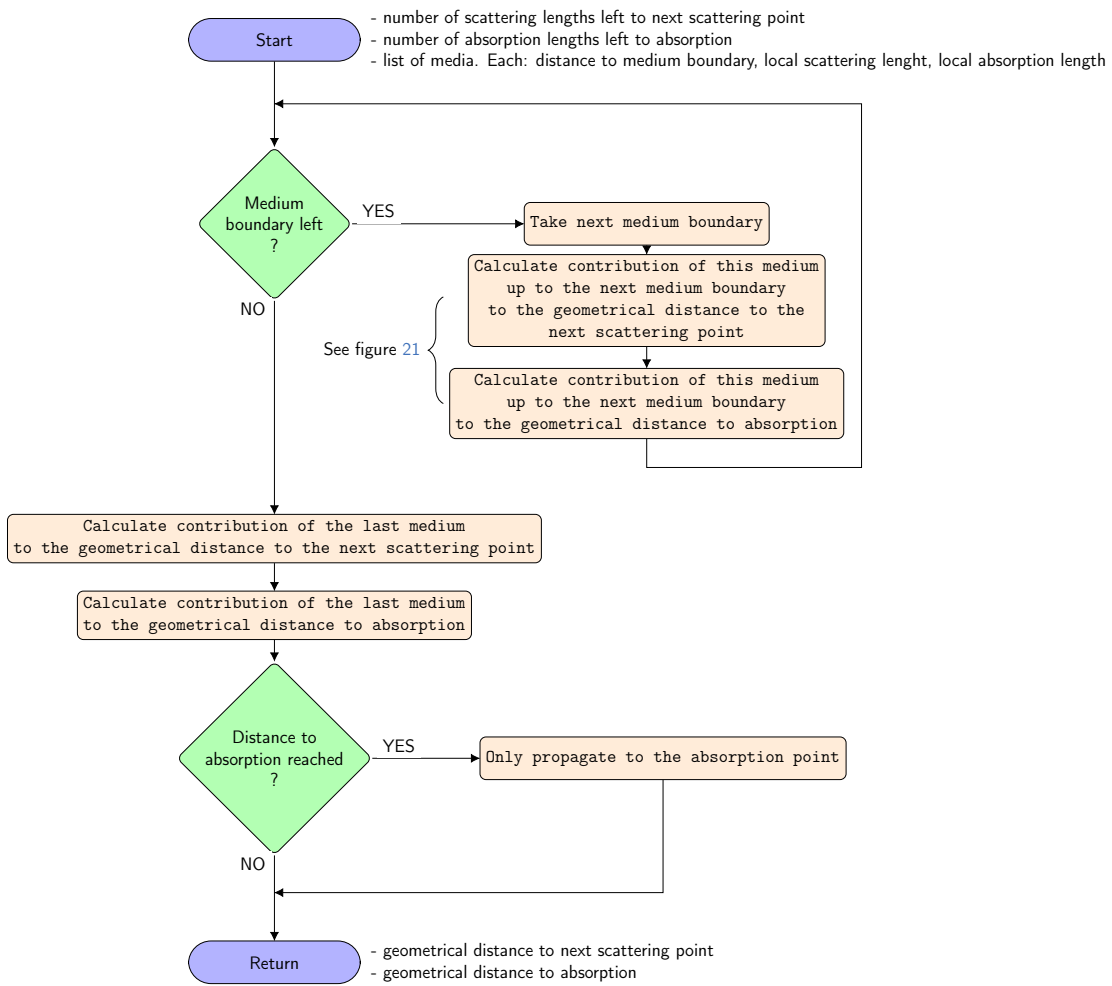

\begin{figure}[htbp]
  \image{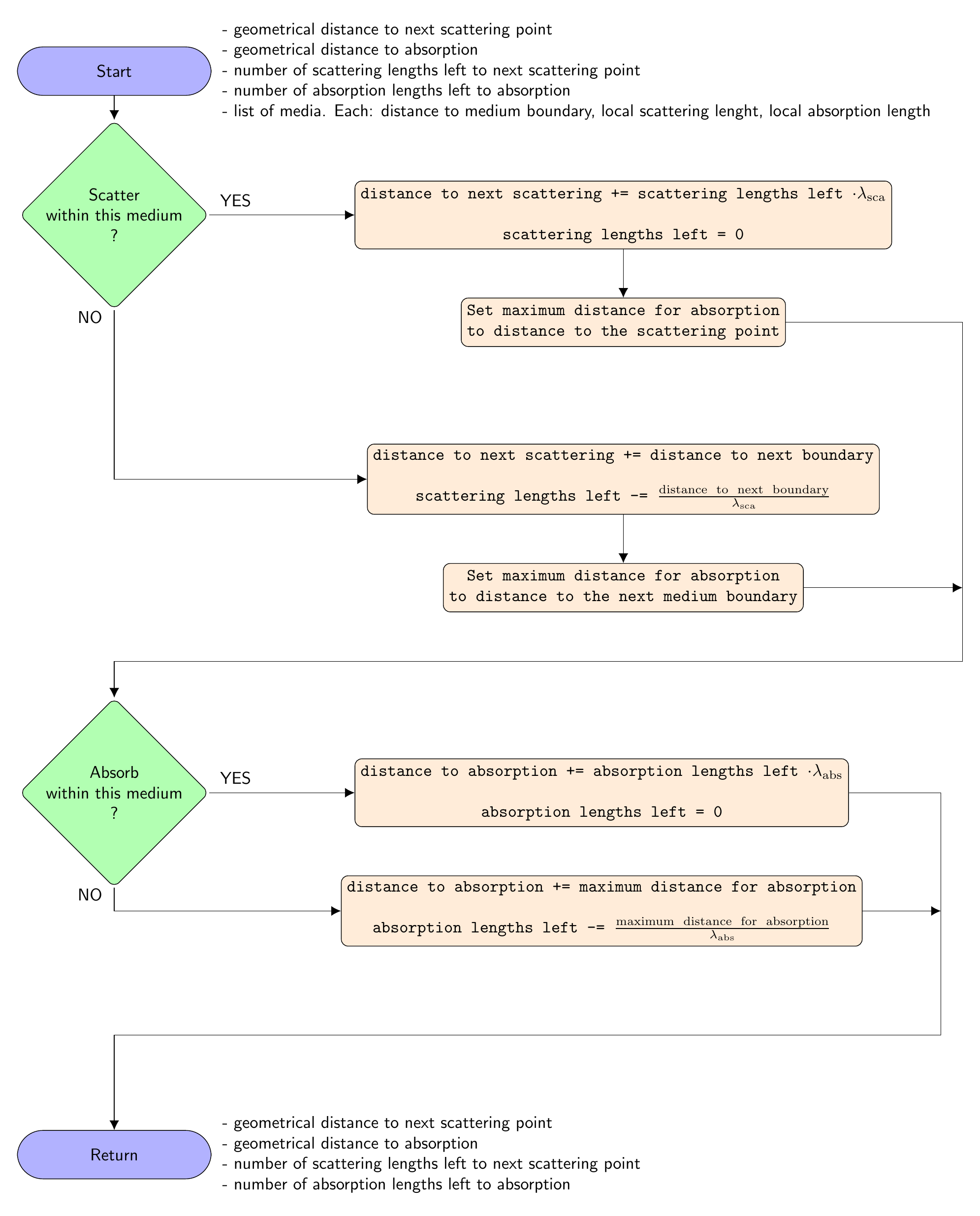}
  \caption{Flow chart of the part of the new media-propagation algorithm that calculates the contribution of one specific medium to the geometrical distances to next scattering point and to the absorption point.}
  \label{fig:eewoo3Be}
\end{figure}

In the loop over the different media along the photon's direction, the
algorithm calculates the contribution of the media to the geometrical
distances to the next scattering point and to the absorption point. The
flow chart of the whole media loop is shown in figure
\ref{fig:nimuriX4}. The flow chart for the calculation of one medium's
contribution is shown separately in figure \ref{fig:eewoo3Be}.

\paragraph{Contribution of the Last Medium}

As shown in figure \ref{fig:nimuriX4}, the last medium is handled
separately as the contribution of the last medium is not determined by
the size of the medium but by the amount of scattering and absorption
budget left. The contributions of the first \((m-1)\) media to the
geometrical distance sum up to \(\sum_{i=1}^{m-1} x_i\) where each
contribution \(x_i := n_i\,\lambda_i\) determines how much of the budget
\(N:=\sum_{i=1}^m n_i\) is spent. After the first \((m-1)\) media have
been crossed, \(n_m := N - \sum_{i=1}^{m-1} n_i\) of the budget is left
for the last medium. Thus, the photon travels a distance of
\(x_m := n_m\,\lambda_m\) in the last medium where it reaches the
interaction point.

\paragraph{Absorption Before Reaching the Scattering Point}

When calculating the distance to the next scattering point based on the
scattering length budget, if the photon's absorption length budget is
spent before reaching the next scattering point, or equivalently, if the
distance from the current scattering point \(A\) to the absorption point
is shorter than the distance from \(A\) to the next scattering point,
then the next trajectory point \(B\) is set to be the absorption point
as shown in figure \ref{fig:nimuriX4}. The final trajectory point of the
photon is its point of absorption.

\paragraph{Scattering Before Reaching the Absorption Point}

When calculating the contribution of a medium to the remaining distance
to the absorption point as shown in the lower half of figure
\ref{fig:eewoo3Be}, the algorithm needs to know what distance \(\gamma\)
within that medium contributes to spending the absorption length budget.
If the photon does not scatter within this medium, the distance
\(\gamma\) is just the distance from one medium border to the next
medium border along the photon
direction.\footnote{This corresponds to the termination point being set to the second intersection point, $C = X$, in Case 4 and Case 5 of the hole-ice-correction algorithm described in section \ref{sec:algorithm_a}.}
If the photon does scatter within this medium, the distance \(\gamma\)
is set to the distance to the scattering
point\footnote{This corresponds to the termination point being set to the point of the other interaction in Case 4 and Case 5 of the hole-ice-correction algorithm described in section \ref{sec:algorithm_a}.}
as shown in the upper half of figure \ref{fig:eewoo3Be}.

\paragraph{Order and Precedence of Cylinders}

\begin{figure}[htbp]
  \subcaptionbox{In a case where the inner cylinder is smaller, intuition suggests that the inner cylinder represents a special area within a more general area and should take precedence for a photon located within both cylinders.}{\includegraphics[width=0.48\textwidth]{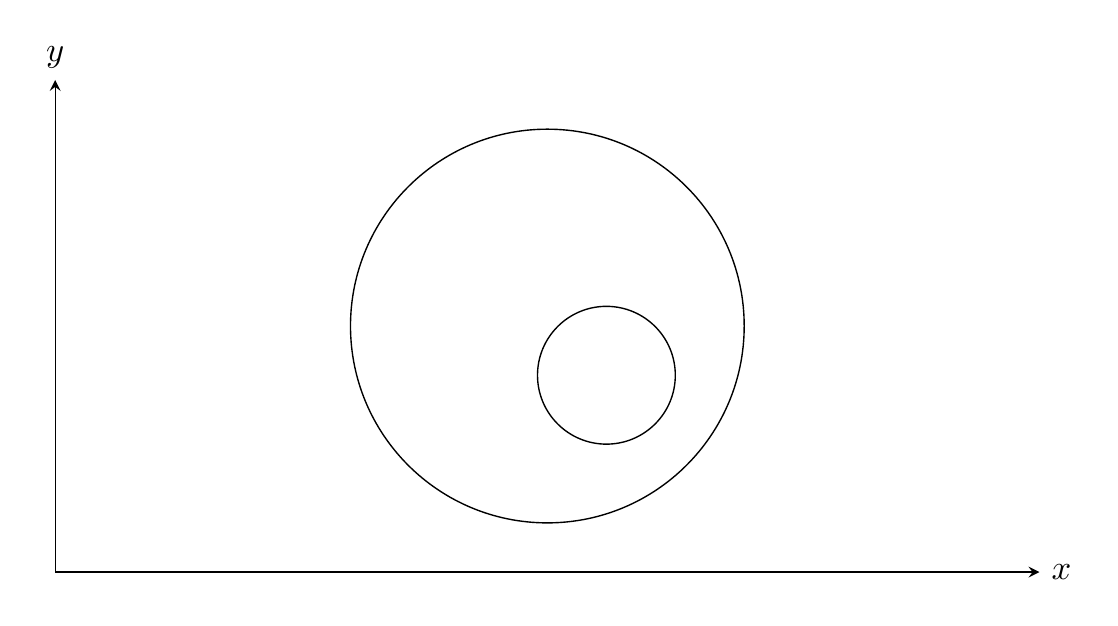}}\hfill
  \subcaptionbox{If both overlapping cylinders have the same size, the choice which properties should be applied to a photon at a position within the overlap of both cylinders, is arbitrary. }{\includegraphics[width=0.48\textwidth]{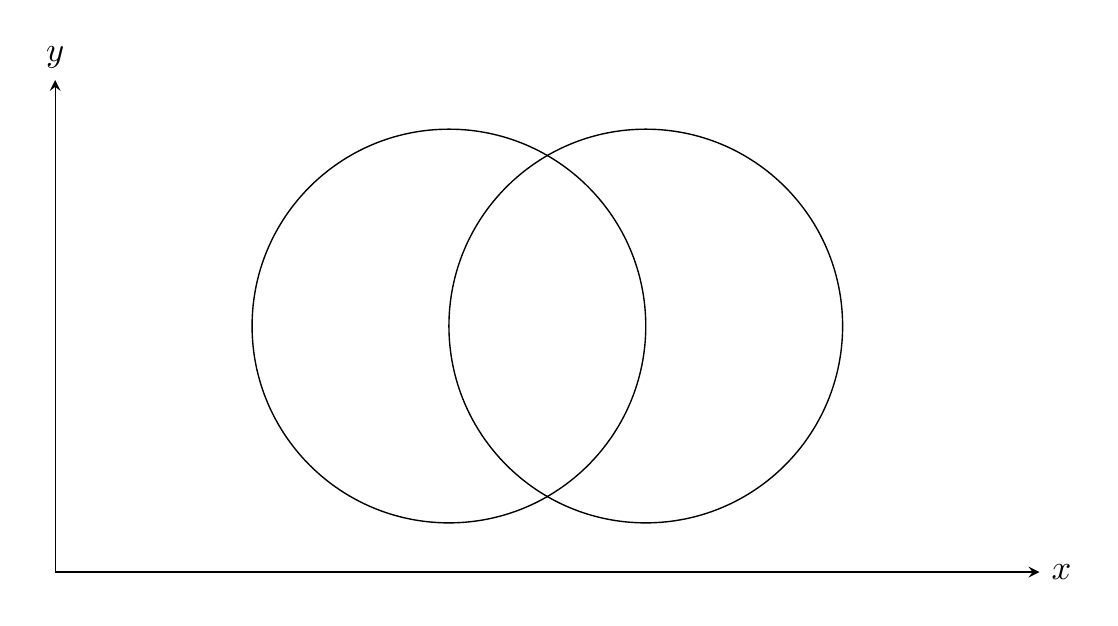}}
  \caption{Two-dimensional projection of nested or overlapping hole-ice cylinders with different ice properties. If a photon is at a position within the overlap of both cylinders, which ice properties should be applied for the propagation? The algorithm defines that the cylinder added last takes precedence.}
  \label{fig:kuZ8deek}
\end{figure}

When nesting or overlapping cylinders as shown in figure
\ref{fig:kuZ8deek}, the algorithm needs to determine which ice
properties to apply for a photon at a position within both cylinders. In
the left part (a) of the figure, the intuition might be: The smaller
cylinder models a special area within the more general outer cylinder
and, therefore, should take precedence. In the right part (b) of the
figure, however, both cylinders have the same size. Here it is unclear
which ice properties to apply for a position within the cylinders'
overlap. To make this situation unambiguous, the algorithm defines that
the media added to the list later take precedence over media added
earlier. Therefore, when modeling nested cylinders, the algorithm
requires to add the larger cylinder before adding the smaller cylinder
to the definition list.

\paragraph{Pros and Cons}

This second approach allows to define cylinder-shaped volumes with
absolute scattering and absorption lengths, which meets the requirements
of the current understanding of the hole-ice phenomenon. Simulation
steps over large \(z\)-distances that cross hole-ice cylinders are
handled well by this algorithm in contrast to the previously described
hole-ice-correction algorithm (section \ref{sec:algorithm_a}).
Additionally, this algorithm supports nested or overlapping cylinders,
which can be utilized to model bubble column together with the
drill-hole column as well as the string's cable. Ice tilt and absorption
anisotropy have not been ported to this algorithm, yet, but can be added
later
on.\footnote{See: \url{https://github.com/fiedl/hole-ice-study/issues/48}.}

The down side to this approach is that an important part of
standard-\clsim's media-propagation algorithm needed to be
re-implemented, requiring evidence that the new algorithm is still doing
what is expected. In order to provide this evidence, section
\ref{sec:unit_tests_and_cross_checks} describes a series of cross checks
to make sure the new algorithm results in the expected scattering and
absorption behavior. Also, standard-\clsim's medium-propagation has been
ported to the new interfaces as drop-in replacement for the new
medium-propagation algorithm, such that one can switch between both
algorithms without a code
rollback.\footnote{A guide on how to switch the medium-propagation algorithm is presented in appendix \ref{sec:how_to_switch_media_propagation}.}

\begin{table}
  \begin{tabelle}{l|L|L}
    & \textbf{Algorithm (a)} & \textbf{Algorithm (b)} \\
    \hline
    Approach
      & Leave \clsim medium propagation as it is and add \textbf{hole-ice effects as correction} afterwards
      & Unify \clsim medium propagation through layers and hole ice: Treat them as \textbf{generic medium changes} \\
    Hole-ice properties
      & defined relative to bulk-ice properties
      & defined absolute \\
    Pros
      &
        \begin{itemize}
          \item[+] Small surface area of hole-ice code, well testable through unit tests
          \item[+] Standard \clsim almost untouched
        \end{itemize}
      &
        \begin{itemize}
          \item[+] Supports nested cylinders and cables
        \end{itemize}
      \\
    Cons
      &
        \begin{itemize}
          \item[--] Current understanding of hole-ice suggests defining hole-ice properties absolute rather than relative
        \end{itemize}
      &
        \begin{itemize}
          \item[--] Needed rewrite of \clsim's medium-propagation code
          \item[--] Ice tilt and ice anisotropy not re-implemented, yet
        \end{itemize}
  \end{tabelle}
  \caption{Comparison of the hole-ice-correction algorithm (a) presented in section \ref{sec:algorithm_a} and the new generic medium-propagation algorithm (b) presented in section \ref{sec:algorithm_b}.}
\end{table}

A comparison to other methods to account for the hole-ice effect is
presented in section \ref{sec:comparison_methods}.

  \subsection{Technical Issues and Optimizations}
\label{sec:technical_issues_and_optimizations}

Running a simulation with millions of photons propagating and
scattering, has to meet certain performance requirements that make sure
that the simulations of interest can be run in a reasonable time on
current hardware. On the other hand, current graphics processing units
(GPUs) afford the opportunity to utilize parallelization capabilities.
Both impose technical challenges. To deal with the necessary complexity
and, in particular, to make sure that a complex algorithm does what it
is supposed to do, unit tests and cross checks are used as will be
described in the next section \ref{sec:unit_tests_and_cross_checks}.
This section focuses on dealing with computational limits, introducing
the techniques of GPU parallelization, cluster parallelization,
thinning, and performance optimizations.

\paragraph{Simulation-Step Parallelization}

\begin{figure}[htbp]
  \hspace{1.5cm}
  \includegraphics[width=0.7\textwidth, trim={7.7cm 0 0 0}, clip]{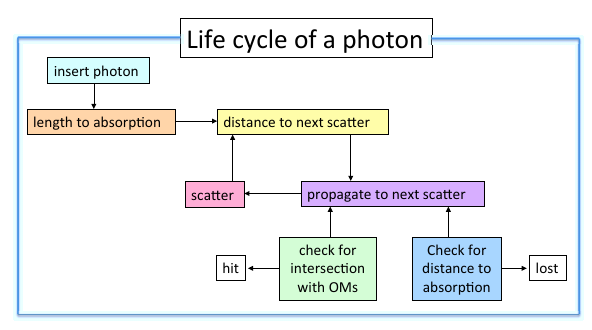}
  \newline\hspace*{1.5cm}
  \includegraphics[width=0.9\textwidth, trim={0 0 6.2cm 0}, clip]{img/dima-parallel-photons-Id3ioyie}
  \caption{Parallelization of simulation steps: Tracking of photons entails the computationally identical steps: Propagation to the next scatter, calculation of the new direction after scatter, and evaluation of intersection points of the photon track segment with the detector array. These same steps are computed simultaneously for thousands of photons. (Image and caption taken from \cite{ppcpaper}, figure 1.)}
  \label{fig:Id3ioyie}
\end{figure}

Each simulation step consists of only a few operations: Scattering the
photon, checking for absorption, checking for collision with a detector
module, and propagating the photon to the next scattering position (see
\ref{sec:standard_photon_propagation_algorithm}). This small set of
operations can be parallelized, performing the same calculations and
operations simultaneously for several thousands of simulation steps,
each with different values such as coordinates and directions, on
graphics processing units, which are designed for exactly this kind of
work. Figure \ref{fig:Id3ioyie} illustrates this parallelization of
simulation steps.

This parallelization leads to a performance improvement by factors of
150 and more when running the propagation simulation on a single GPU
compared to running the same simulation on a single CPU core
\cite{ppcpaper}. Standard \clsim as well as the hole-ice extended
\clsim can be run on CPUs, GPUs, or both using \noun{OpenCL} as
abstraction layer. In practice, simulating 1000 photons for
visualization purposes and recording their paths, which requires a lot
of memory, works best on a CPU. Simulating millions of photons, but only
recording their final position, where the photon is absorbed within the
ice or has it an optical module, works best on GPUs.

\paragraph{Cluster Parallelization}
\label{sec:cluster_parallelization}

A different technique, which adds another layer of parallelization on
top of the simulation-step parallelization, is to run the same
simulation on a cluster of machines, each equipped with one ore several
GPUs. Each machine runs the same simulation, but with different
parameters, such as different radii and scattering lengths of the
hole-ice cylinders. This study uses this technique for parameter scans
(sections \ref{sec:parameter_scan} and \ref{sec:flasher}) where the same
simulation is performed for different parameter sets in order to find a
specific parameter set which best fits some kind of external
requirement.

\paragraph{Issues Concerning Graphics Processing Units}

Running and developing a simulation software for graphics processing
units poses challenges specific to this architecture. A list of
technical issues to watch out for in follow-up studies is provided in
appendix \ref{sec:gpu_technical_issues}.

\paragraph{Parallel-Programming Optimizations}

Applying parallel-programming optimization techniques listed in section
\ref{sec:parallel_computing} may help to improve the performance of the
algorithm that runs on the graphics processing units.

When optimizing code, \noun{Ahmdal's Law} gives the theoretical limit on
how much a process can be sped up by optimizing one of its components.
\cite{raytracingtips}

\[ s_\text{total} = \frac{1}{(1 - p) + \sfrac{p}{s_\text{comp}}}, \ \ \ s = \frac{t_\text{before}}{t_\text{after}}, \ \ \
\lim_{s_\text{comp}\rightarrow\infty} s_\text{total} = \frac{1}{1-p} \]

Here, \(s_\text{total}\) is the resulting speedup of the whole process,
which is defined as the fraction of the time the process takes before
the optimization and the time the same process takes after the
optimization has been applied. \(p\) is the portion of execution time
that the component requires before optimizing. \(s_\text{comp}\) is the
speedup the the component that is optimized achieved by the
optimization. Even in the limit that the component that is optimized
takes zero time after applying the optimization,
\(s_\text{comp}\rightarrow\infty\), the speedup of the whole process is
limited. Also, \authorname{House} and \authorname{Wyman}
\cite{raytracingtips} recommend to first write working code, and then
optimize in a second step.

\begin{figure}[htbp]
  \subcaptionbox{Checking if a cylinder is in range in the same loop as propagating through that cylinder. The cylinders are processed successively. Lots of photons are idle.}{\includegraphics[width=0.48\textwidth, clip, trim = {4cm 2cm 6cm 2cm}]{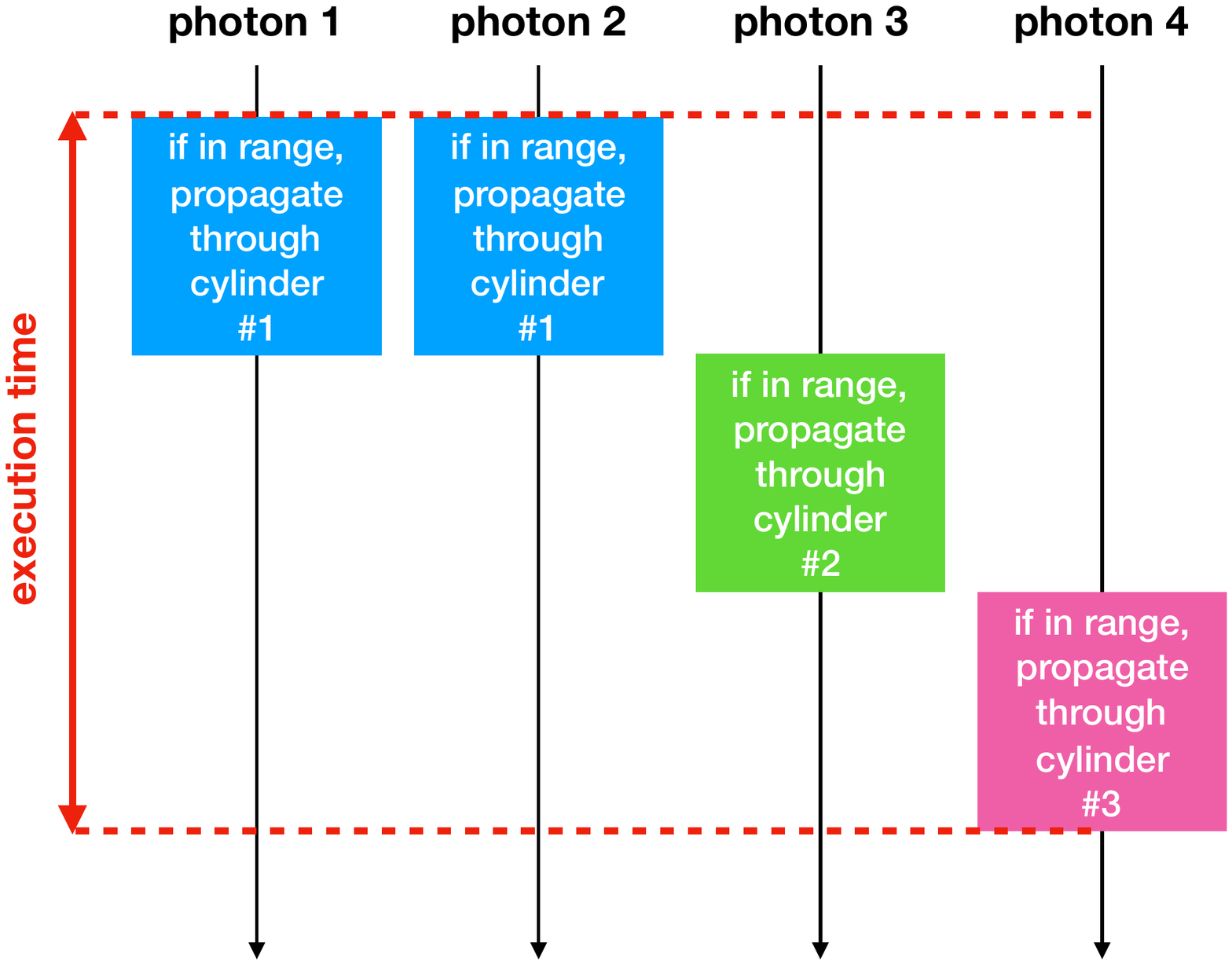}}\hfill
  \subcaptionbox{Checking if a cylinder is in range in a separate loop. The main loop where the photons are propagated through the cylinders does not loop over all cylinders but only over cylinders marked as in range for each photon respectively. }{\includegraphics[width=0.48\textwidth, clip, trim = {4cm 2cm 6cm 2cm}]{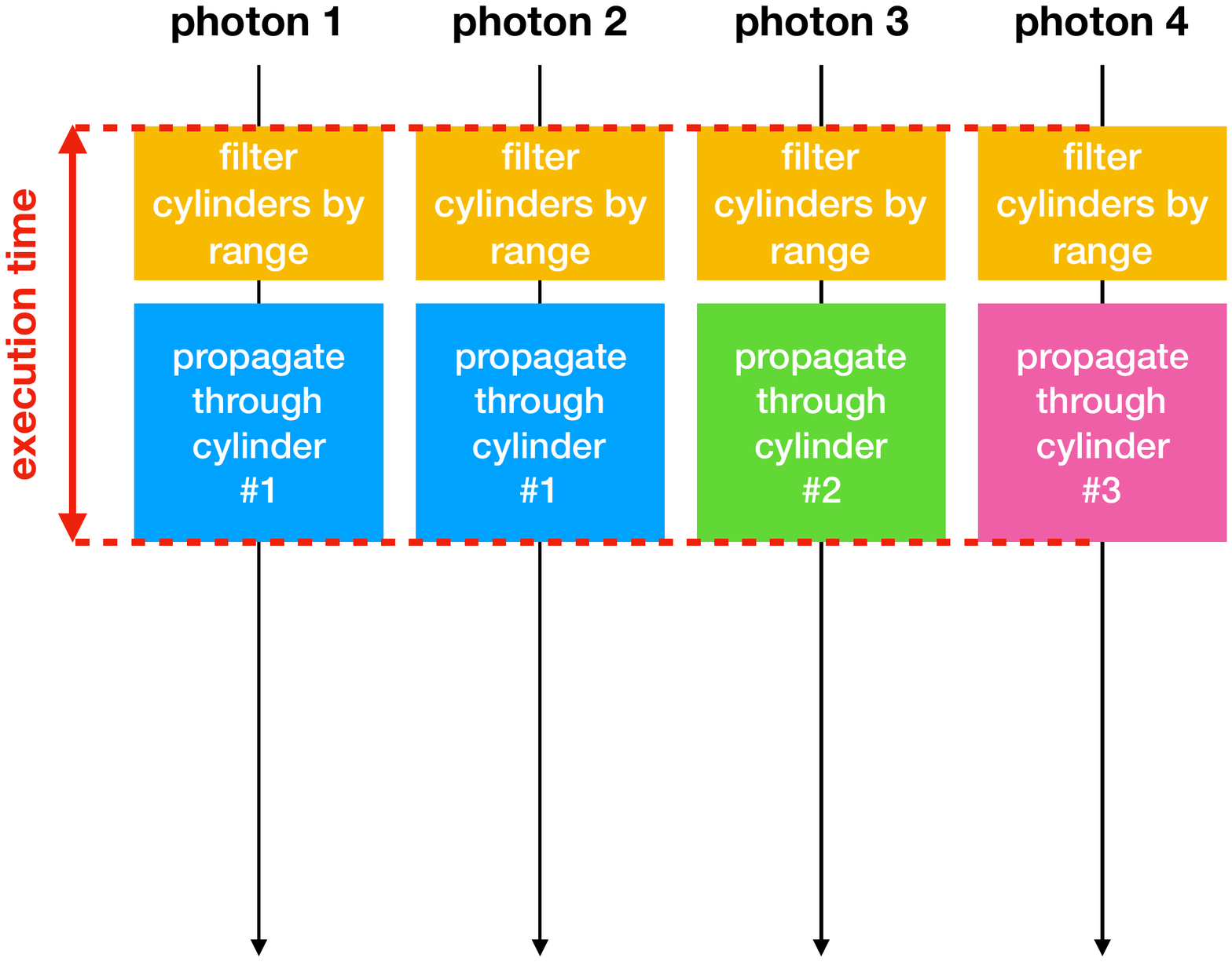}}
  \caption{Processing the propagation of photons through hole-ice cylinders in parallel. While filtering cylinders by range separately adds additional executional time, it improves parallelization and thereby saves execution time overall.}
  \label{fig:ceiV8Yai}
\end{figure}

One key concept to efficient parallel programming is to avoid
executional threads being idle, waiting for other threads to finish
before resuming. One application of this concept in the hole-ice
algorithm is to \textbf{filter hole-ice cylinders by range} in a
\textbf{separate loop}. The algorithm needs to check in each simulation
step, which hole-ice cylinders are in range of the photon and therefore
should be considered when calculating the propagation through the hole
ice. Without this optimization, the cylinders are processed one by one
in each simulation step. While, in each iteration of the cylinder loop,
all photons in range of one specific cylinder are propagated through
that cylinder, all other photons not in range of the cylinder, wait idly
and thereby waste computation time. This is illustrated in figure
\ref{fig:ceiV8Yai} (a).

Even though the optimization adds time for the additional loop where the
indices of the cylinders in range are saved into a local array, the main
loop where the propagation through each cylinder is handled is much
faster now, because it does not need to iterate over all cylinders in
the detector but only over the cylinders in range of the photon and
improves parallelization as illustrated in figure \ref{fig:ceiV8Yai}
(b).

\docpar{The implementation of this optimization is documented in \issue{30}.}

Another optimization used in the hole-ice algorithm is to
\textbf{order mathematical cases by their frequency of application}. In
a case-by-case analysis, each executional thread skips the additional
case checks when its case is determined. If the checks for common cases
are handled first, there is a good chance that all of the parallel
threads fall into common cases such that the checks for the rare cases
can be skipped for the whole thread block. This technique is used in the
implementation of the geometric cases for both the hole-ice-correction
algorithm (\ref{sec:algorithm_a}) and the new media-propagation
algorithm (\ref{sec:algorithm_b}).

\paragraph{GPU-Specific Optimizations}

In addition to the general principles of parallel programming (section
\ref{sec:parallel_computing}), \authorname{House} and \authorname{Wyman}
\cite{raytracingtips} list practical tips for optimizing GPU-ray-tracing
code. Most notably, passing data structures by reference rather than by
value proved to have significant performance impact because
\textbf{allocating memory on a GPU is expensive}. This also includes
simple structures like arrays of numbers, and even re-using the same
array for several calculations rather than re-defining it. This
technique has caused a performance improvement on GPUs by a factor of
six in this study.

\docpar{This optimization of re-using arrays is documented in \issue{70}.}

\textbf{Optimizing mathematical operations} on paper is an important
technique as each calculation operation that can be cut down on in a
procedure that saves time when executing the procedure many times. In
addition to simplifying equations mathematically, avoiding numerically
expensive operations such as square roots improves performance.

GPUs support 4-dimensional vectors as native data types as well as
native vector operations such as the dot product. Expressing equations
in terms of vectors rather than simplifying equations on the coordinate
level, resulted in improved performance and numerical accuracy.

\docpar{This optimization of using vector types rather than scalar types is documented in \issue{28}.}

\paragraph{Performance Profiling}

\begin{figure}[htbp]
  \image{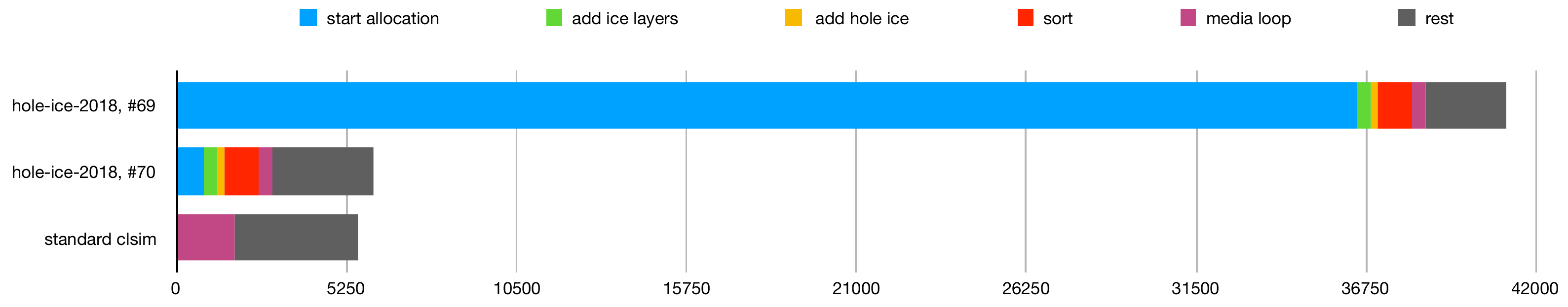}
  \caption{Performance profile of the average simulation step in arbitrary units (GPU clock cycles). Each row of the chart represents a different algorithm. In the first row, which shows an early stage of the new medium-propagation algorithm, the initial memory allocation within each simulation step takes a lot of time, especially when comparing to standard \clsim (row 3). The second row shows the profile of the new medium-propagation algorithm after a memory issue has been fixed. Notable operations in each simulation step are the initial memory allocation for local variables, adding ice layers, adding hole-ice cylinders, sorting medium boundaries by distance to the photon and looping over the media to convert the interaction budget into geometrical distances.}
  \label{fig:profiling-paz4Eig6}
\end{figure}

\authorname{Owens} recommends profiling tools like \noun{gprof},
\noun{VTune}, or \noun{VerySleepy} \cite{cudacourse}. But also basic
techniques like printing time differences between executional blocks can
help to isolate performance problems. Figure
\ref{fig:profiling-paz4Eig6} shows the execution time profile of the
simulation step of the new medium-propagation algorithm.

Note that printing a profiling result may be more expensive as the
computation that is to be profiled itself. In this study, printing the
time difference of the beginning and the end of a simulation step takes
five times as long as the simulation step itself.

\docpar{Profiling the simulation step is documented in \issue{69}.}

\paragraph{Performance Optimizations for Production Use}

After implementing and debugging a new algorithm, a lot of debug outputs
and other development tools are no longer required when running
simulations. For those production simulations, the debug features can be
turned off by
\textbf{compiling the \icecube simulation framework for production} use
rather than for debugging. Thereby, simulations run significantly faster
as debug outputs often cost more time than the actual simulation itself.

Also, activating \textbf{kernel caching} on the production machines may
improve simulation performance. However, this should only be used when
always using the same propagation kernel, and should be avoided when
switching between kernel versions or turning kernel features on and off,
because the kernel cache will not be reliably reset automatically when
changing kernels.

\label{sec:thinning} Another way to save computational time when
performing simulations, especially large cluster scans, is to scale down
the number of photons using a \textbf{thinning factor}. A representative
fraction of the photons is propagated in the simulation, and then the
simulation results, for example the numbers of hits in each detector
module, are scaled up accordingly.

\docframe{
  \docparwithoutframe{How to switch between a debug build and a production build is documented in \issue{73}.}\medskip

  \docparwithoutframe{How to enable or disable kernel caching is documented in \issue{15}.}\medskip

  \docparwithoutframe{Using a thinning technique in flasher parameter scans is documented in \issue{59}.}
}

  \subsection{Unit Tests and Cross Checks}
\label{sec:unit_tests_and_cross_checks}

There are several techniques to ensure that the propagation algorithm
produces results that meet the expectations. First, \clsim itself
implements a number of tests that run in each simulation and check
computational quantities for consistency. \cite{clsimsource}

Secondly, one can define scenarios with known quantities for single
photons and check externally whether the algorithm produces the expected
results for those single photons. This is done with \textit{unit tests}
and is presented in section \ref{sec:unit_tests}.

Thirdly, checks for a sample of photons can be devised where all photons
should behave the same way, for example all photons should be absorbed
by a medium that is configured for instant absorption. This check is
presented in section \ref{sec:instant_absorption_tests}.

Finally, a sample of propagated photons can be examined regarding their
statistical properties, for example the sample's distributions of
quantities like arrival time and path length. These tests are presented
in sections \ref{sec:arrival_time} to \ref{sec:cross_check_71}.

\subsubsection{Unit Tests With Single
Photons}\label{unit-tests-with-single-photons}

\label{sec:unit_tests}

In order to verify that individual components (``units'') of the
implementation produce results as expected, unit tests were implemented
for the algorithm that calculates intersections as well as for the
hole-ice-correction algorithm.

\docframe{
  \sourceparwithoutframe{The unit tests for the intersection algorithm can be found in the folder \texttt{unit\_tests} on the CD-ROM and at \url{https://github.com/fiedl/clsim/blob/sf/hole-ice-2017/resources/kernels/lib/intersection/intersection_test.c}.}\medskip

  \sourceparwithoutframe{The unit tests for the hole-ice-correction algorithm can be found in the folder \texttt{unit\_tests} on the CD-ROM and at \url{https://github.com/fiedl/clsim/blob/sf/hole-ice-2017/resources/kernels/lib/hole_ice/hole_ice_test.c}.}
}

\paragraph{Task}

The task of unit tests is to test individual components of a software.
In this case, the tests execute separate functions of the new algorithms
with fixed input parameters and check whether the functions return the
expected results, which, in preparation of the test, have been obtained
by other means, either by a separate program, using a separate
programming language, a separate algorithm, or via calculations by hand.

In this study, unit tests define single photons that cross hole-ice
cylinders in a pre-defined way and check whether the intersection
algorithm determines the correct intersection points, whether the 2d-3d
projections are handled correctly, and whether the hole-ice-correction
algorithm determines the expected corrections for the geometric
distances according to the hole-ice parameters provided by the test
scenario. Extreme examples can be designed to produce simple results
that can be calculated by hand and verified by intuition. For example,
for hole-ice cylinders configured for instant absorption, the absorption
point is identical to the intersection point of the photon trajectory
and the cylinder. More complex examples require more involved
calculations by hand or using other software like \noun{Python} scripts
or computer algebra tools for calculating intersections.

\paragraph{Testing Framework}

This study uses the
\noun{gtest}\footnote{Google Test Framework, gtest, \url{https://github.com/google/googletest}}
testing framework. This framework has been chosen due to its good
documentation, wide adoption and slim architecture that made it possible
to use the framework to test individual components of the new source
code without interfering with the rest of the \icesim framework.

\docpar{The introductory documentation of \noun{gtest} can be found at \url{https://github.com/google/googletest/blob/master/googletest/docs/primer.md}.}

\paragraph{Limitations of Unit Tests}

Unit tests are most useful to ensure stability when adjusting,
refactoring or rewriting software components. Even after replacing a
large part of the source code, unit tests can make sure that the
software still produces the same results as before. Tests also allow for
so-called test-driven development where the expected results are
specified first, and then the software is built or changed iteratively
until it produces the expected results. This technique works best when
the components to be tested have small interfaces because the technique
requires the tests to provide all input parameters for the components to
be tested.

Unexpected issues may arise when unit tests are run on another
architecture as the production code. For example, the unit tests used in
this study were insensitive to certain driver issues and numerical
issues\footnote{See section \ref{sec:gpu_technical_issues}.} that did
only occur when running the software on the GPUs of the computing
cluster and did not occur when running the unit tests on the local CPU
of the development machine.

Unit tests are, by design, also insensitive to high-level problems. Each
individual component may produce the results expected from this
component while still issues may arise when components are not tied
together correctly, or because the expectations for an individual
component may be wrong, which becomes apparent only when adopting a
larger perspective rather than the focused perspective on individual
components.

To increase confidence in the new software, therefore, in addition to
unit testing, high-level cross checks need to be performed as described
in the following subsections.

\subsubsection{Instant-Absorption Tests}\label{instant-absorption-tests}

\label{sec:instant_absorption_tests}

\begin{figure}[htbp]
  \subcaptionbox{Distribution of the distance to the hole-ice center, evaluated for each scattering step.}{\includegraphics[width=0.32\textwidth, trim={1cm 0 25cm 2cm}, clip]{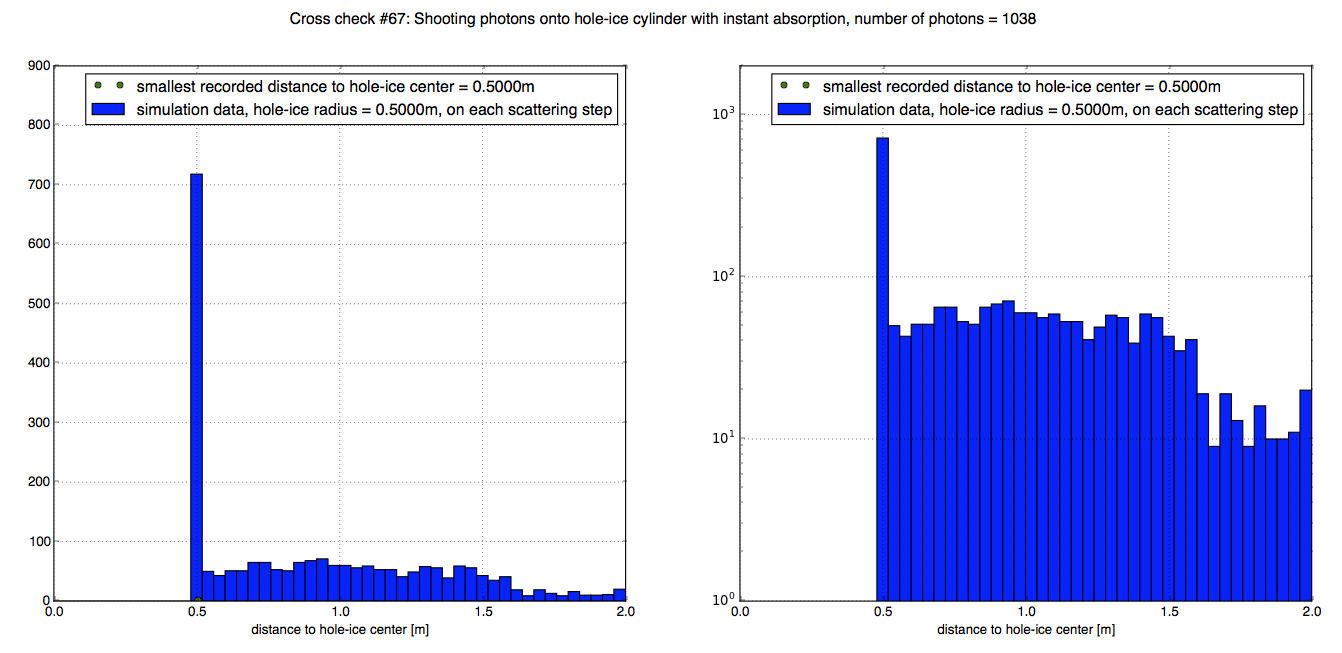}}\hfill
  \subcaptionbox{Distance to the next scattering point and distance to hole-ice center for each scattering step.}{\includegraphics[width=0.65\textwidth, trim={0 0 1cm 0}, clip]{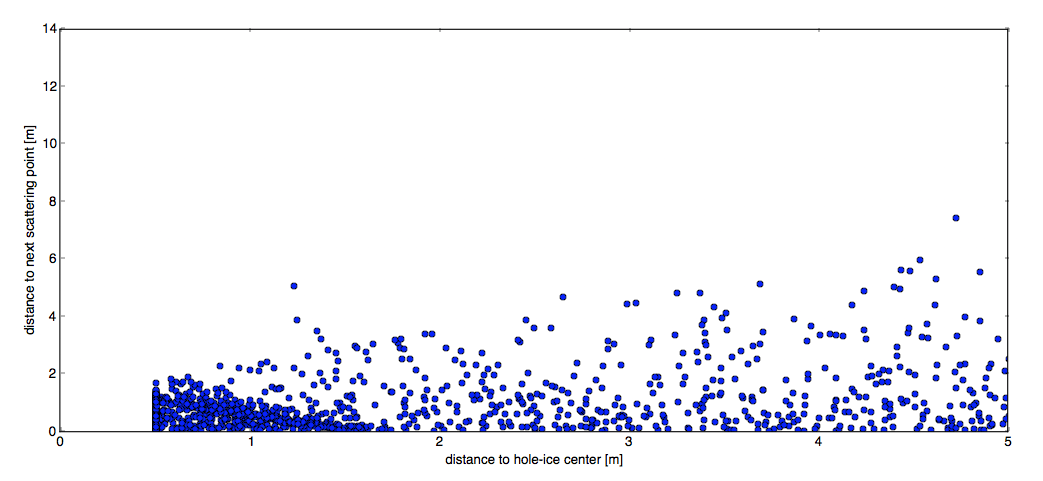}}
  \caption{Analyzing the simulation of photons propagating towards a hole-ice cylinder configured for instant absorption. No photon can be recorded within the hole-ice cylinder's radius.}
  \label{fig:chiep7Is}
\end{figure}

\begin{figure}[htbp]
  \subcaptionbox{Starting from different positions into the same direction. View from above.}{\includegraphics[width=0.49\textwidth]{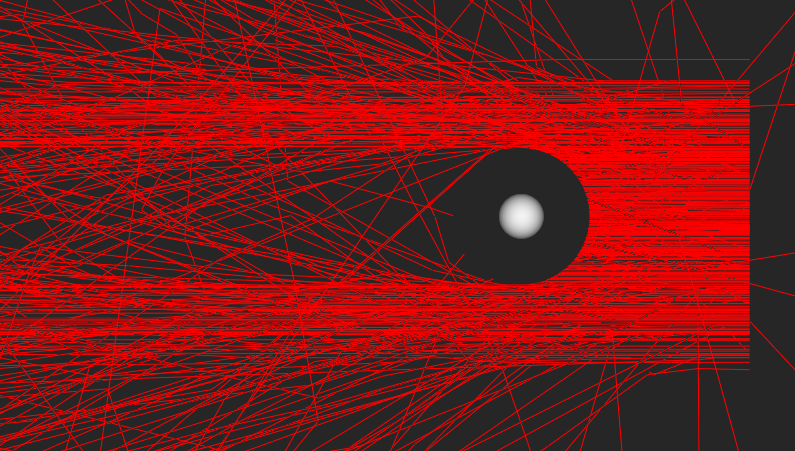}}\hfill
  \subcaptionbox{Starting from the same position into different directions.}{\includegraphics[width=.49\textwidth, trim={0 0 40mm 0}, clip]{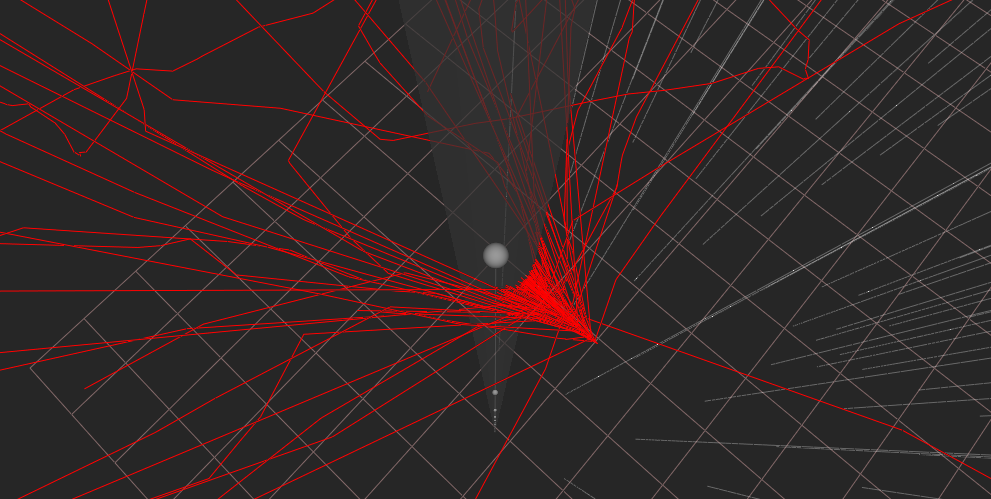}}
  \caption{Visualizing an instant-absorption test using the \steamshovel event viewer. In the simulation, photons are started towards a cylinder configured for instant absorption. If the medium-propagation algorithm works as expected, no photon can get inside the cylinder.}
  \label{fig:moo9Eiqu}
\end{figure}

A simple high-level test can be performed by starting several photons
towards a cylinder, which is configured for instant absorption. After
propagating the photons using the new media-propagation algorithm and
recording the path of each individual photon, one can verify the results
making sure than no photon position is recorded inside the
instant-absorption cylinder (figure \ref{fig:chiep7Is}). Additionally,
one can visualize the simulation using the \steamshovel event viewer to
verify that no photon enters the instant-absorption cylinder as shown in
figure \ref{fig:moo9Eiqu}.

\sourcepar{The implementation of this cross check can be found in \issue{67}, \issue{47}, and \issue{22}.}

\begin{figure}[htbp]
  \subcaptionbox{View from above.}{\includegraphics[width=.32\textwidth, trim={0 3cm 16cm 0}, clip]{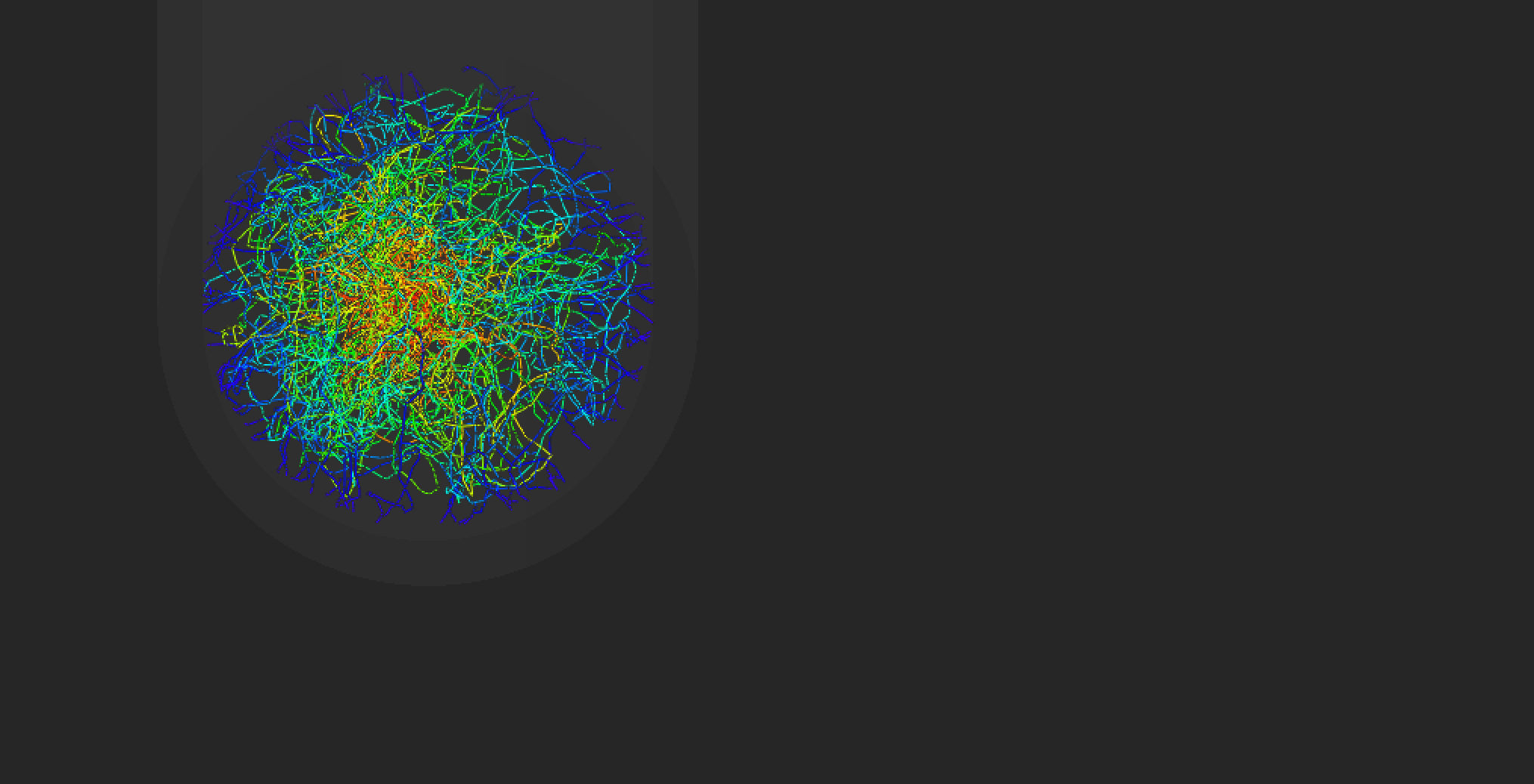}}\hfill
  \subcaptionbox{View from the side.}{\includegraphics[width=.32\textwidth, trim={6cm 4cm 13cm 1.5cm}, clip]{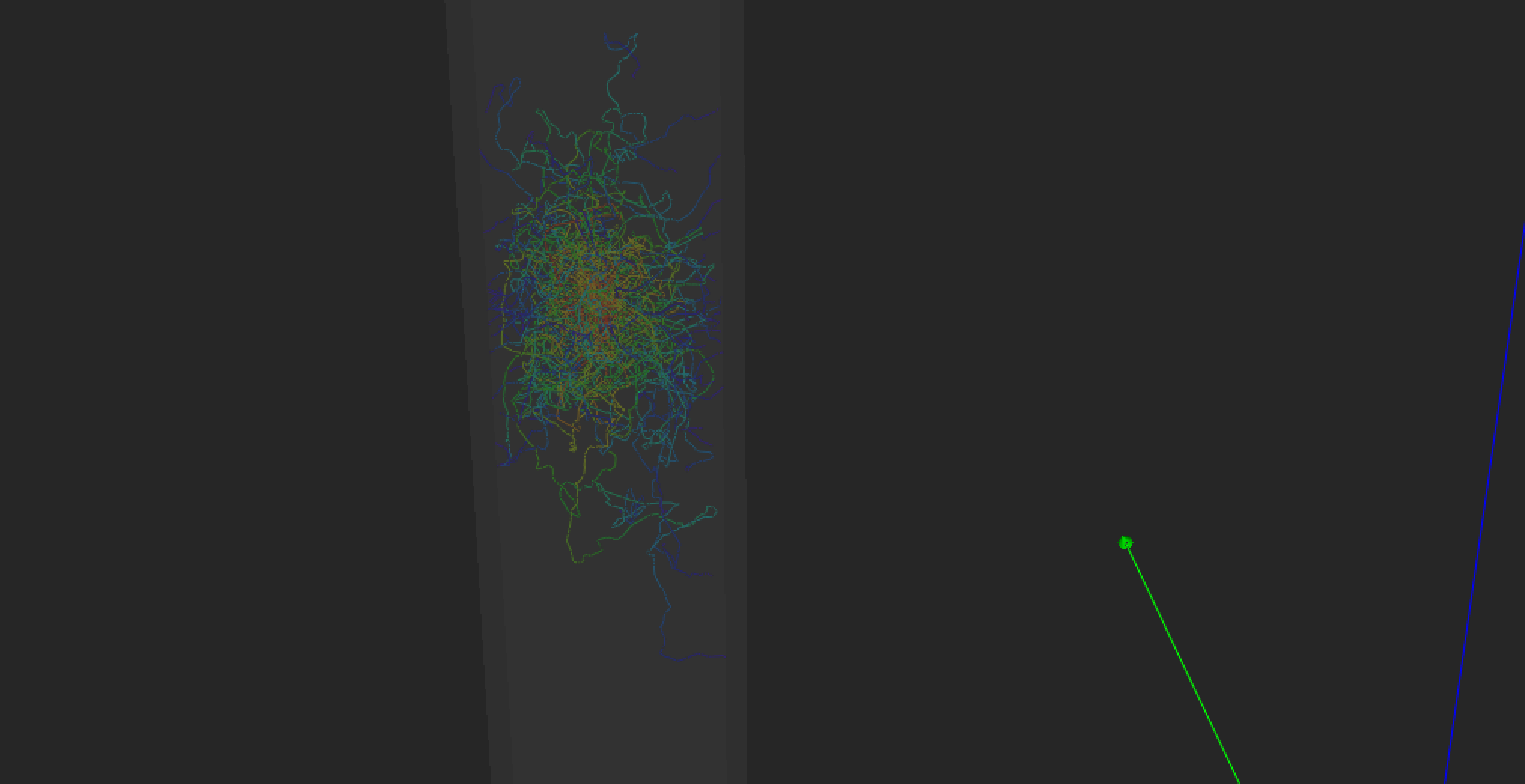}}\hfill
  \subcaptionbox{Instant absorption of the outer cylinder turned off.}{\includegraphics[width=.32\textwidth, trim={6cm 4cm 13cm 1.5cm}, clip]{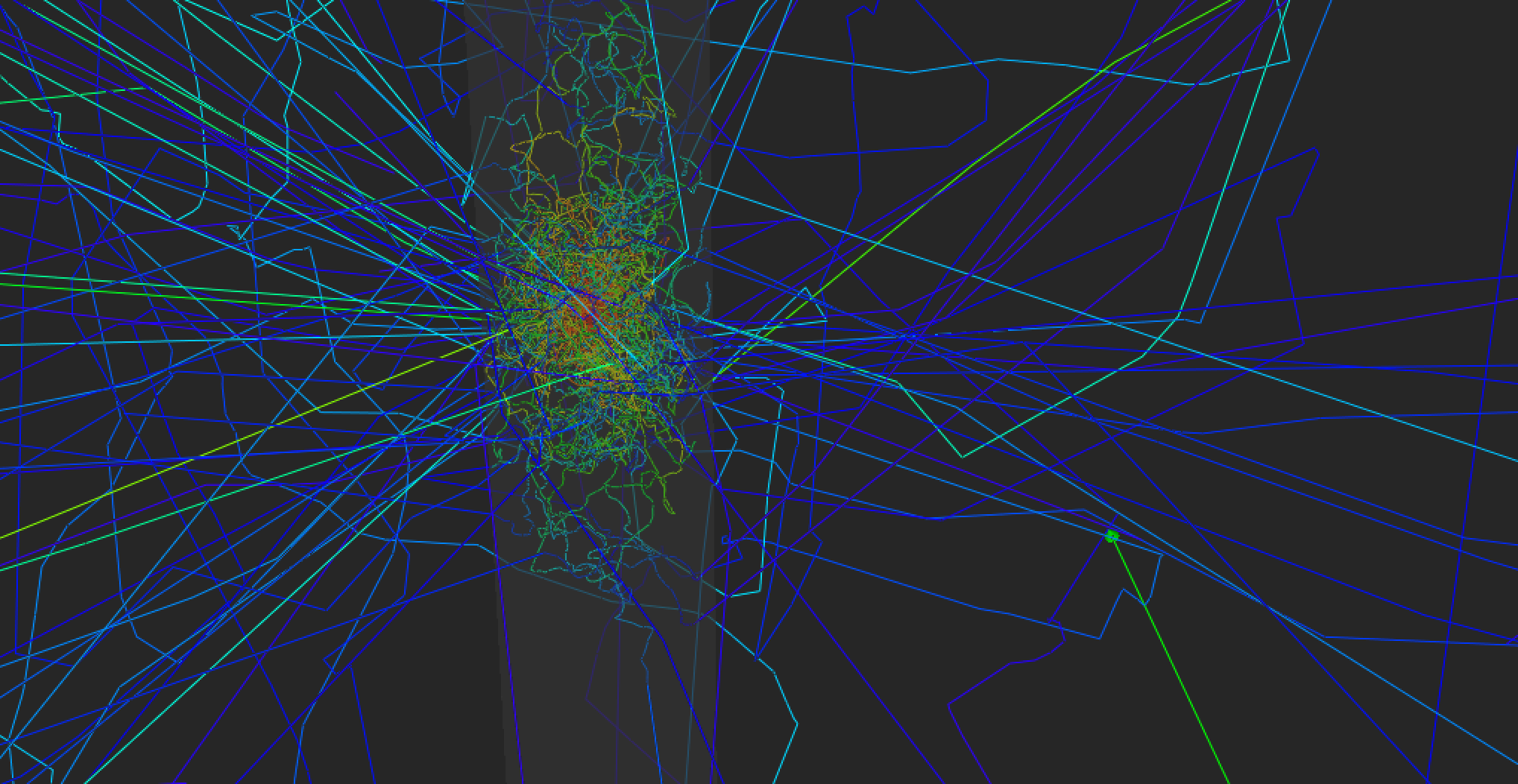}}
  \caption{Visualizing an instant-absorption test where photons are started within two nested cylinders. The outer cylinder is configured for instant absorption. No photon can pass through the area between both cylinders unless the instant absorption is turned off.}
  \label{fig:sahmoo8O}
\end{figure}

A related, but more complex scenario is starting photons within two
nested cylinders where the inner cylinder is configured for a short
scattering length and the outer cylinder is configured for instant
absorption (figure \ref{fig:sahmoo8O}).

\FloatBarrier\newpage
\subsubsection{Arrival-Time Distributions}
\label{sec:arrival_time}

\begin{figure}[htbp]
  \subcaptionbox{Top view of the detector strings in this simulation. The photons are started at the middle optical module of string 63, and are received by the optical modules of the surrounding strings 70, 71, 64, 55, 54, and 62.}{\includegraphics[width=.48\linewidth]{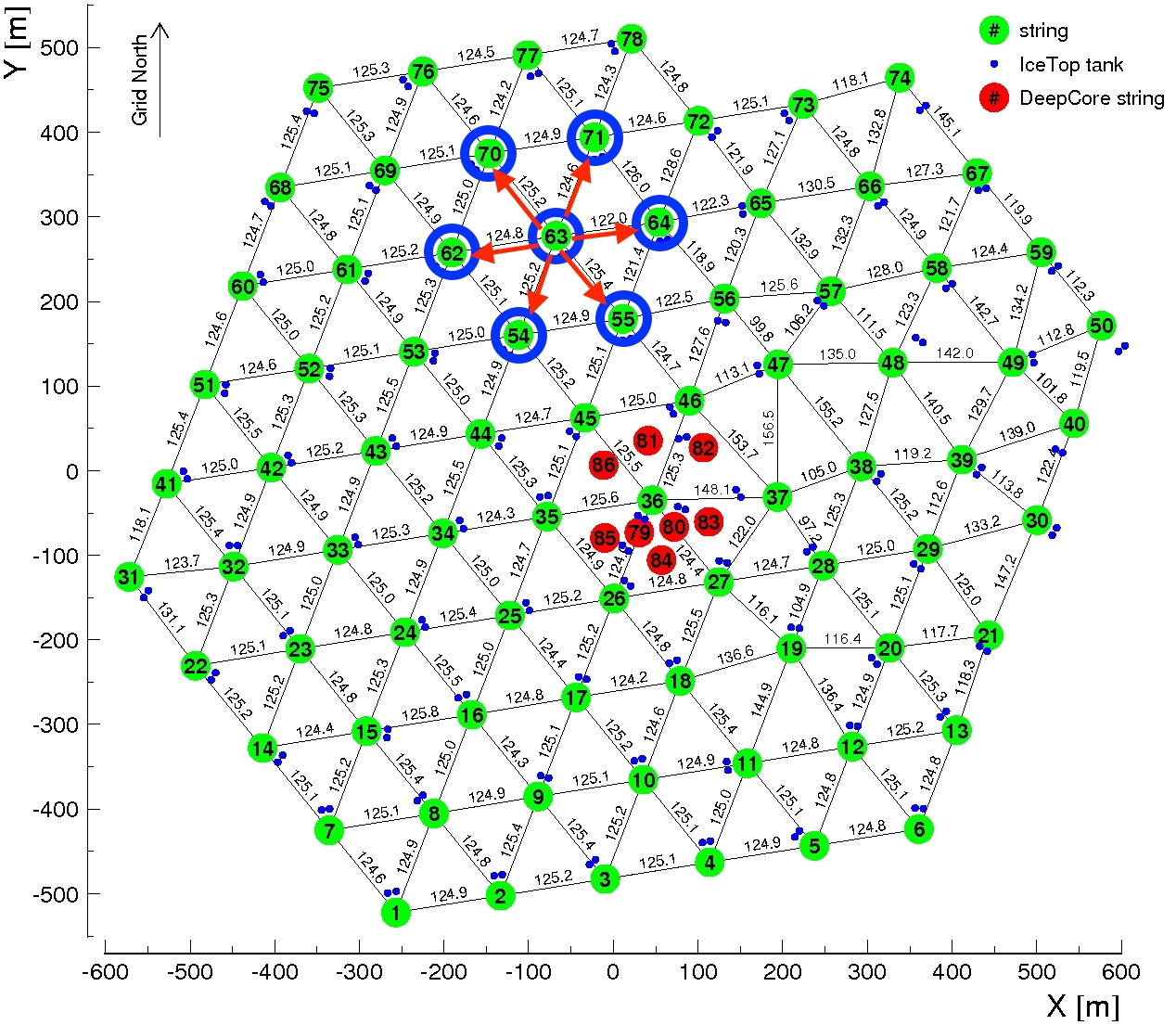}}
  \hfill
  \subcaptionbox{Photon-arrival-time distributions for different hole-ice configurations. The dashed lines indicate the mean arrival times.}{\includegraphics[width=.48\linewidth]{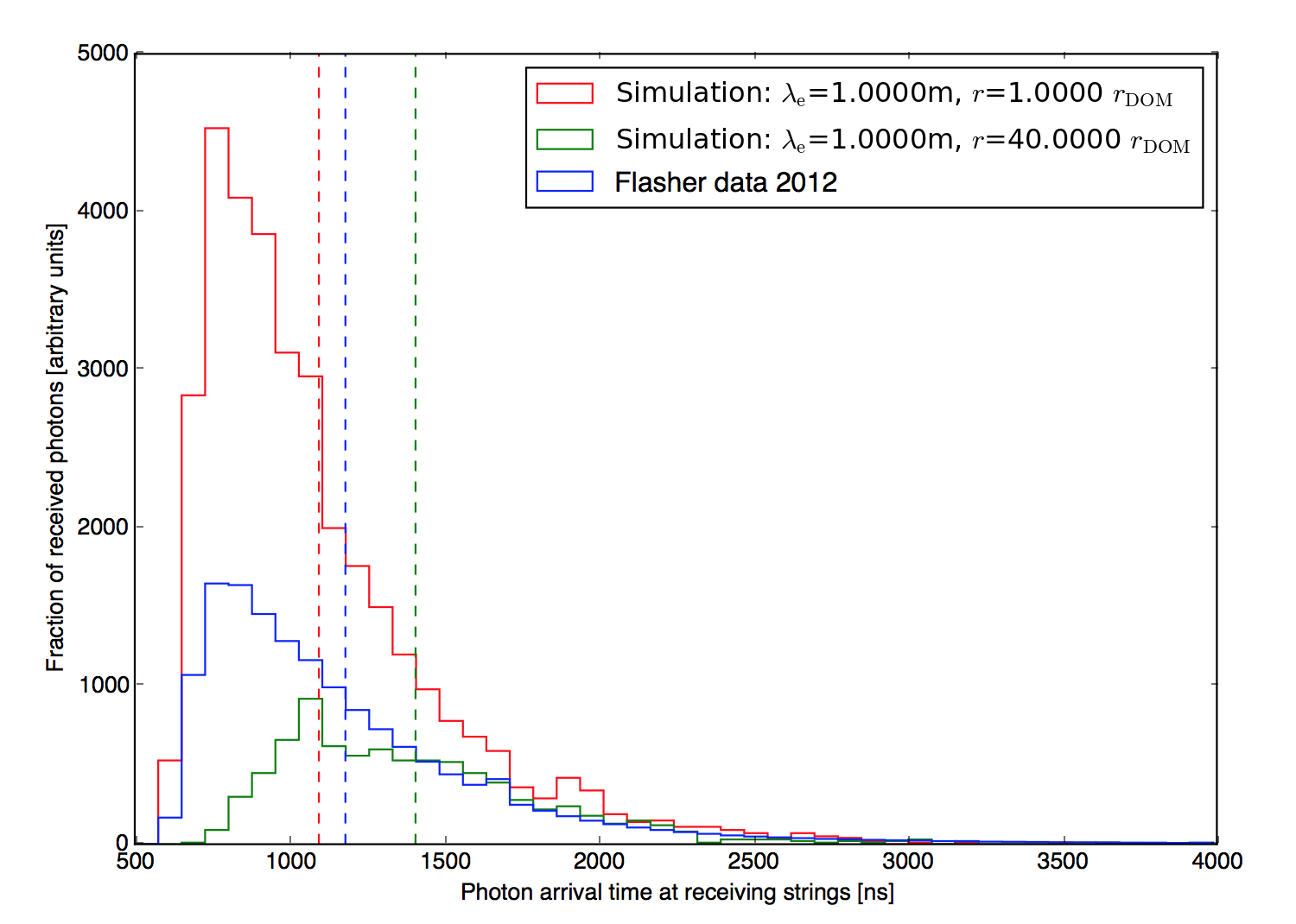}}
  \caption{Creating arrival-time distributions for photons propagating through hole ice. The photons are started at a central position and detected by optical modules around the starting position.}
  \label{fig:eipau6Ag}
\end{figure}

One way to test the behavior of statistical properties of a sample of
photons is to plot the arrival-time distribution of photons traveling
from a central position to receiving optical modules around the starting
position. Figure \ref{fig:eipau6Ag} shows arrival-time distributions for
this scenario being carried out with a flasher experiment compared to
two simulations with different hole-ice configurations each.

\sourcepar{The source for the simulations and for creating these histograms can be found in \issue{91}.}

An estimation based on the mean scattering angle and the radius of the
hole-ice cylinder suggests that one would need a difference of the
hole-ice cylinder's radius of several meters to cause a difference in
the mean arrival time on the order of \(100\ns\). While this is no
realistic scenario as the drill hole's radius is only about \(50\cm\),
the simulation allows to use extreme scenarios to observe the effects
more noticeable.

From the comparison of the arrival-time distributions, note that the
distribution that is based on real data is comparable to the ones based
on simulations, which suggests that the new algorithm does not introduce
unexpected, unphysical effects to the photon propagation.

In the simulation where the hole-ice cylinders have a much larger
radius, the effects of the hole ice should be stronger. Indeed, the
comparison of the distributions shows three effects to expect from this
assumption: First, in the simulation with stronger hole-ice effect, less
photons arrive at the receiving optical modules in total as more photons
are scattered away by the hole ice. Secondly, for stronger hole ice, the
photons arrive later at the receiving optical modules, because they
spend more time scattering randomly within the hole-ice cylinder before
reaching the receiving optical module. Thirdly, for stronger hole ice,
the left-hand side of the distribution histogram is less steep, because
with more scattering points on the photon's path, there is a larger
number of possible paths, which leads to the arrival time being more
distributed.

While these effects confirm the expectations qualitatively, examining
other high-level observables allows to compare expectations to the
simulations' results even quantitatively. This is the subject of the
following sections, beginning with observing the photons' path-length
distributions.

\FloatBarrier\newpage
\subsubsection{Exponential Distribution of the Total Path Length}
\label{sec:total_path_length_distribution}

Let the \textbf{Total path length} of a photon be the summed distance
from the position where the photon is created, along the photon's path,
to the position where the photon is absorbed. The total path length
\(X\) of photons that propagate in a medium with an absorption
probability that is the same everywhere in the medium, for example
within a confined volume within the bulk ice, is expected to follow an
exponential distribution with a rate parameter corresponding to the
absorption probability, or equivalently the absorption length
\(\lambda\) within the
medium.\footnote{If the photons do not depend on each other, the number of photons that are absorbed at a certain path length does only depend on the absorption probability, or equivalently on the absorption length $\lambda$, and the number of photons remaining at this path length. When a photon is absorbed, the number of remaining photons decreases. Thus, the change in the number of photons is proportional to the number of photons, causing the number of photons absorbed at a certain path length following an exponential distribution. See also appendix \ref{sec:exponential_distribution}.}

\begin{equation}
  X: \ \ P(X \in [x; x + \dx[) \propto e^{\sfrac{-x}{\lambda}}, \ \ x \in \reals^+_0
\end{equation}

When propagating photons within a hole-ice cylinder, their path length
should also follow an exponential distribution, but with different
absorption length \(\lambda\hi\), as long as the absorption length is
sufficiently small such that most photons do not leave the cylinder
before being absorbed.

This behavior can be verified using a simulation. In the simulation, a
pencil beam of \(10^4\) photons is started within a hole-ice cylinder
with a radius of \(1.0\m\). The photons are started with a distance of
\(0.9\m\) to the cylinder center towards the cylinder center. Let the
effective scattering length within the cylinder be \(100.0\m\) and the
absorption length be \(0.1\m\). For each simulated photon, the total
path length is recorded.

\begin{figure}[htbp]
  \caption{\steamshovel event display of the simulation of photons propagating and being absorbed within the hole-ice cylinder without medium transition. The photons are started within the hole-ice cylinder, which is represented by the grey cylinder, inwards. The detector module, which is represented by the white sphere in the center of the figure, is shown only to illustrate the size of the scenery and is configured not to interact with the simulated photons.}
\end{figure}

\sourcepar{The implementation of this cross check can be found in \issue{64}.}

The distribution of the total path length should follow an exponential
decay curve. From the rate of the decay, one should be able to read the
absorption length \(\lambda\) (see appendix
\ref{sec:exponential_distribution}).

\begin{figure}[htbp]
  \image{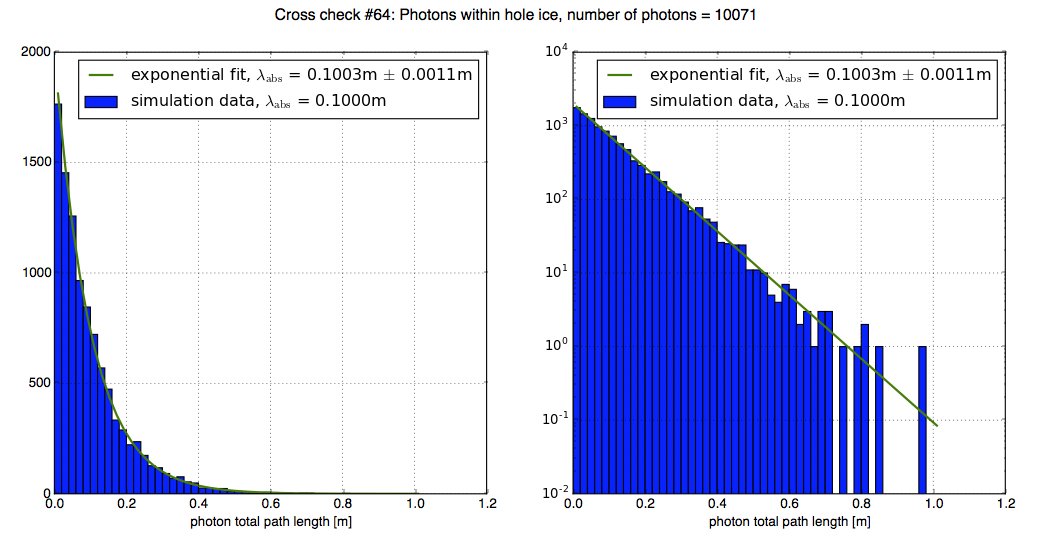}
  \caption{Distribution of the total path length of simulated photons both started and absorbed within a hole-ice cylinder using the new medium-propagation algorithm. The distribution follows an exponential curve. The fitted absorption length is $\lambda_\text{abs}=0.1003\m \pm 0.0011\m$. The true absorption length used in the simulation is $0.1000\m$.}
  \label{fig:gieNa2Th}
\end{figure}

Indeed, the simulation yields the expected distribution of the total
path length (figure \ref{fig:gieNa2Th}). Via a curve fit, the absorption
length \(\lambdaabs\) can be determined for this simulation to be
\(\lambdaabs = 0.1003\m \pm 0.0011\m\), which is in accordance with the
true absorption length of \(0.1000\m\) used in the simulation.

\FloatBarrier\newpage
\subsubsection{Piecewise Exponential Distribution of the Total Path Length for One Medium Boundary}

This cross check aims to verify that the medium-boundary transition is
handled correctly when photons enter a hole-ice cylinder from outside.

In a simulation, a pencil beam of \(10^5\) photons is started in a
distance of \(2.0\m\) to cylinder center towards the hole-ice cylinder
with a radius of \(1.0\m\). The effective scattering length outside is
set to \(10^6\m\), inside to \(100\m\). The absorption length outside is
set to \(1.0\m\), inside to \(0.10\m\). As before, for each simulated
photon, the total path length, from the starting point of the photon
along the photon's trajectory to the point where the photon is absorbed,
is recorded.

\sourcepar{The implementation of this cross check can be found in \issue{65}.}

\begin{figure}[htbp]
  \smallerimage{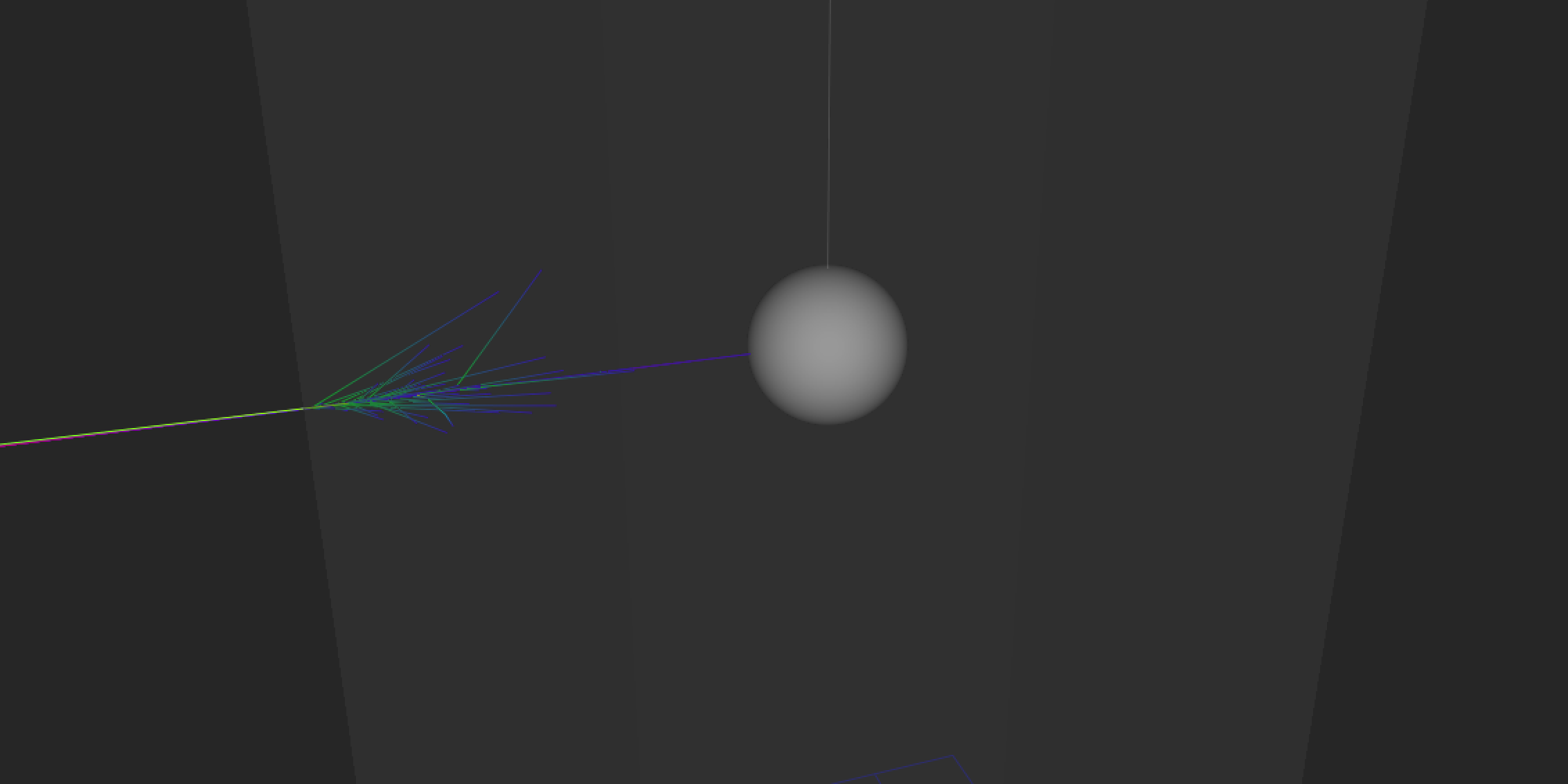}
  \caption{\steamshovel event display of the simulation of photons propagating through the hole-ice medium boundary. The photons are started outside and towards the hole-ice cylinder on the left hand side. As the scattering length within the cylinder is smaller, the photons scatter a lot more within the cylinder. The detector module is shown as a white sphere only to illustrate the size of the scenery and is configured not to interact with the simulated photons.}
\end{figure}

As the absorption lengths are different outside and inside the cylinder,
the histogram should now follow two separate exponential curves: The
left hand side of the histogram, which corresponds to the area outside
the cylinder, should follow an exponential curve governed by the
absorption length outside the cylinder. The right hand side, which
corresponds to the area within the cylinder, should follow an
exponential curve governed by the absorption length within the hole-ice
cylinder.

Note that the histogram's bins are proportional to the number
\(m(x):=\sfrac{-\d n(x)}{\dx}\) of photons that are absorbed within the
path length interval \([x; x+\dx[\), not the number \(n(x)\) of
remaining photons. Thus, the histogram is expected to show a jumping
discontinuity at the position of the medium
border.\footnote{See also appendix \ref{sec:exponential_distribution}.}

\begin{figure}[htbp]
  \image{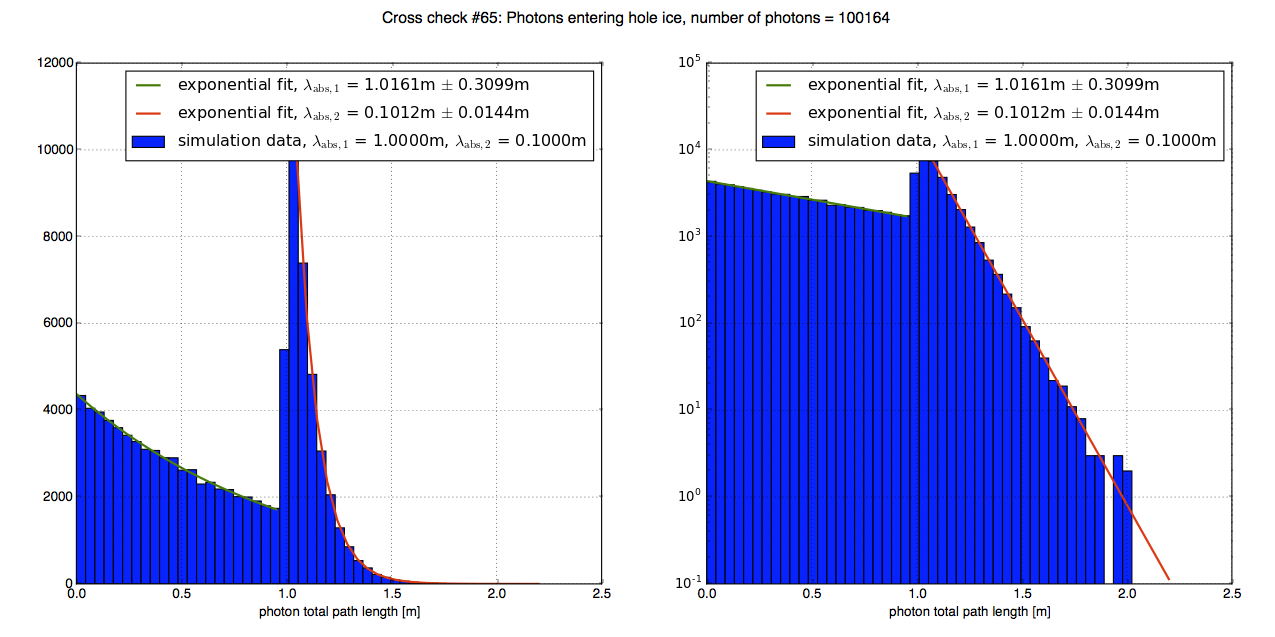}
  \caption{Distribution of the total path length of simulated photons started outside and absorbed within a hole-ice cylinder using the new medium-propagation algorithm. The distribution follows two exponential curves. The fitted absorption lengths are $\lambda_\text{abs,1} = 1.016\m \pm 0.310\m$ and $\lambda_\text{abs,2} = 0.101\m \pm 0.014\m$. The true absorption length in the simulation is set to $1.000\m$ outside, and to $0.100\m$ within the hole-ice cylinder.}
  \label{fig:iquo3Ou3}
\end{figure}

The simulation yields the expected distribution of the total path length
(figure \ref{fig:iquo3Ou3}). Via a curve fit, the absorption lengths
\(\lambda_\text{abs,1} = 1.016\m \pm 0.310\m\) and
\(\lambda_\text{abs,2} = 0.101\m \pm 0.014\m\) are in determined in
accordance with the true absorption length outside of \(1.000\m\) and
within the hole-ice cylinder of \(0.100\m\) set in the simulation.

\FloatBarrier\newpage
\subsubsection{Piecewise Exponential Distribution of the Total Path Length for Two Medium Boundaries}

This cross check aims to verify that the medium-boundary transition is
handled correctly when photons enter and leave a hole-ice cylinder.

In a simulation, a pencil beam of \(10^5\) photons is started in a
distance of \(1.5\m\) to the cylinder center towards the hole-ice
cylinder with a radius of \(0.5\m\). The effective scattering length
outside and inside is set to \(10^6\m\). The absorption length outside
is set to \(1.0\m\), inside to \(0.75\m\). As before, for each simulated
photon, the total path length is recorded.

\sourcepar{The implementation of this cross check can be found in \issue{66}.}

\begin{figure}[htbp]
  \smallerimage{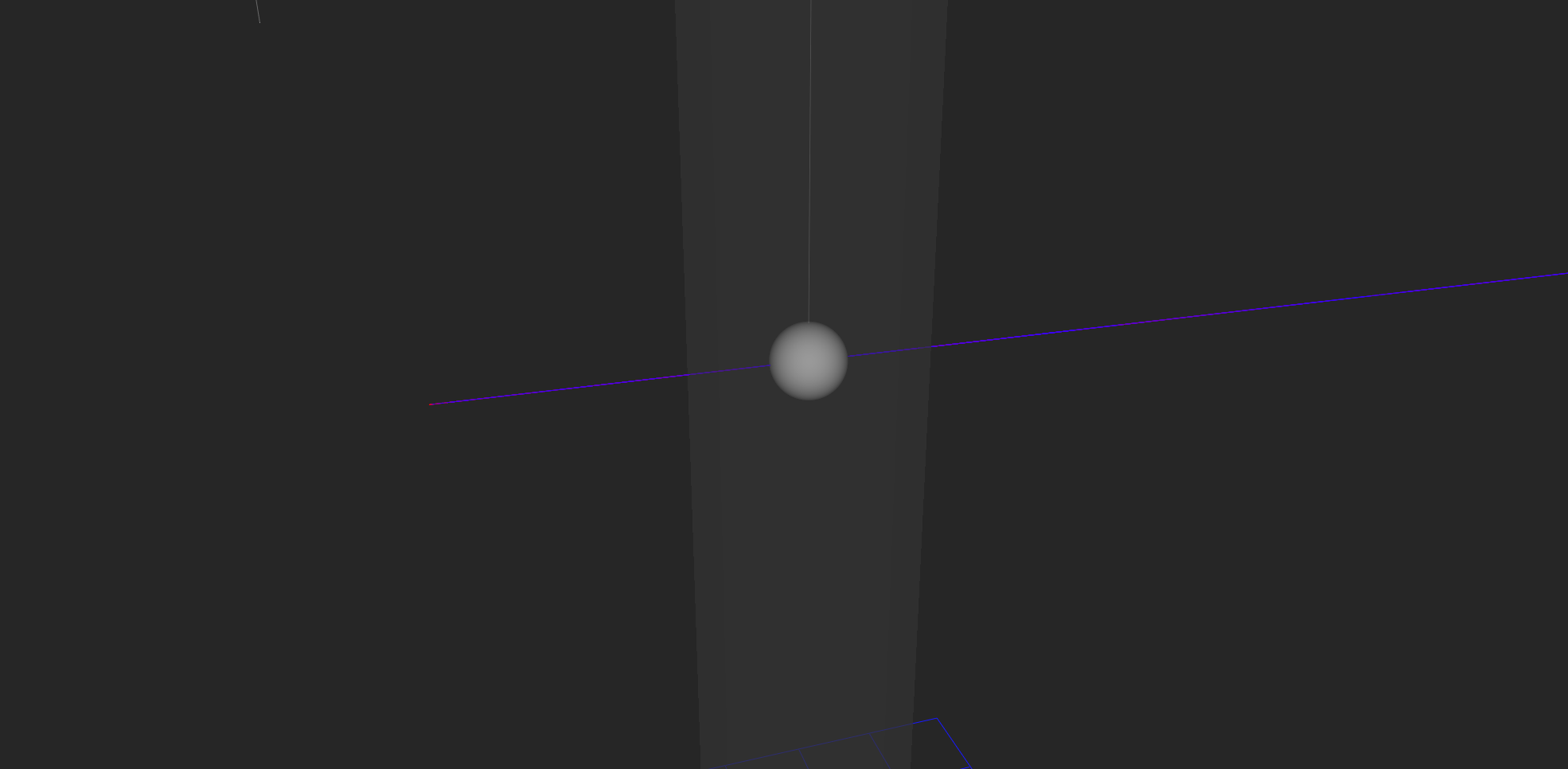}
  \caption{\steamshovel event display of the simulation of photons propagating through a hole-ice cylinder. As the scattering length is chosen to be very large, the photon path appears as a single line.  The photons are started outside the hole-ice cylinder on the left hand side. Some of them are absorbed before entering the cylinder, some within the cylinder and some after leaving the cylinder on the right hand side. The detector module is shown as a white sphere only to illustrate the size of the scenery and is configured not to interact with the simulated photons.}
\end{figure}

As the simulated photons may now cross two different medium boundaries,
the distribution of the path lengths should follow three exponential
curves: On the left hand side of the histogram, which corresponds to the
photons that are absorbed before entering the hole ice, the histogram
should follow an exponential curve governed by the absorption length
outside. In the middle, which corresponds to the photons that are
absorbed inside the cylinder, the histogram should follow an exponential
curve governed by the absorption length within the hole ice. On the
right hand side, which corresponds to the photons that are absorbed
after leaving the hole-ice cylinder, the histogram should follow an
exponential curve, again, governed by the absorption length outside the
cylinder.

\begin{figure}[htbp]
  \image{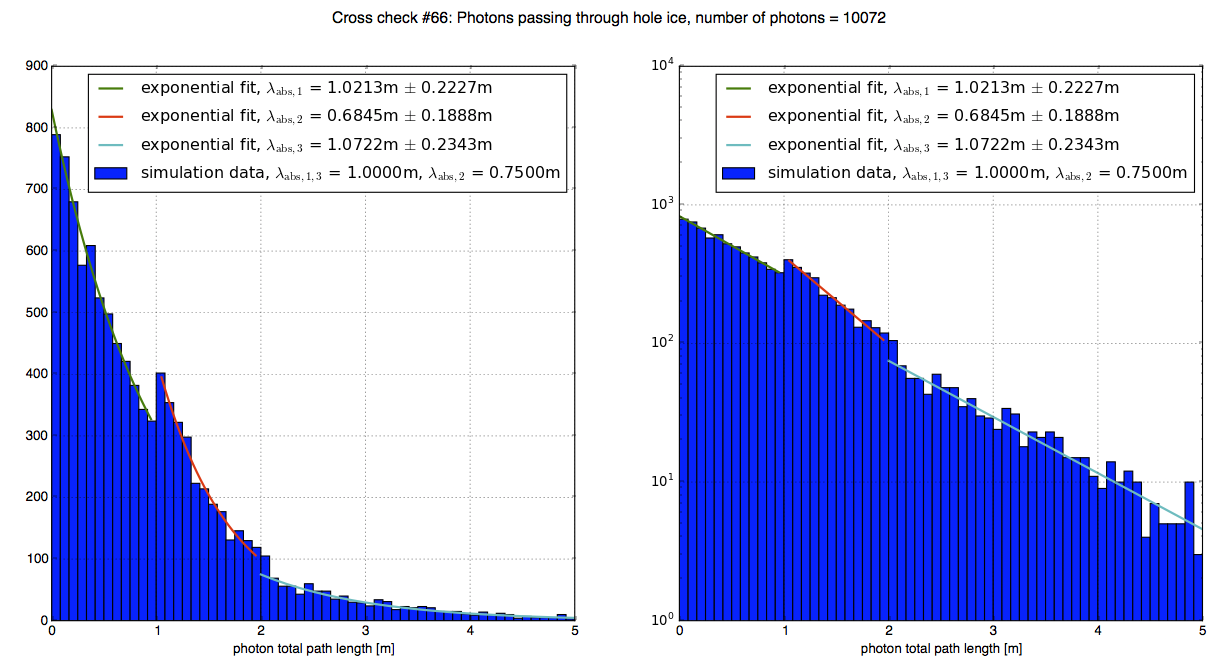}
  \caption{Distribution of the total path length of simulated photons, which start outside the cylinder, and may be absorbed before, within or after passing the cylinder, using the new medium-propagation algorithm. The distribution follows three exponential curves. The fitted absorption lengths are $\lambda_\text{abs,1} = 1.02\m \pm 0.22\m$, $\lambda_\text{abs,2} = 0.68\m \pm 0.19\m$, and $\lambda_\text{abs,3} = 1.07\m \pm 0.23\m$. The true absorption length in the simulation is set to $1.00\m$ outside, and to $0.75\m$ within the hole-ice cylinder.}
  \label{fig:io9eiDee}
\end{figure}

The simulation yields the expected distribution of the total path length
(figure \ref{fig:io9eiDee}). The absorption lengths
\(\lambda_\text{abs,1} = 1.02\m \pm 0.22\m\),
\(\lambda_\text{abs,2} = 0.68\m \pm 0.19\m\) and
\(\lambda_\text{abs,3} = 1.07\m \pm 0.23\m\) determined via a curve fit
are in accordance with the true absorption length outside of \(1.00\m\)
and within the hole-ice cylinder of \(0.75\m\) set in the simulation.

\FloatBarrier
\subsubsection{Distance to Next Scattering Point in Relation to the Distance From the Hole-Ice Center}
\label{sec:cross_check_71}

The cross checks of the previous sections refer to the absorption
length. To examine the scattering-length behavior of the photons
simulated using the new media-propagation algorithm, this cross check
observes the distance to the next scattering point in relation to the
distance from the hole-ice center at each scattering point.

In a simulation, a plane wave of 100 photons is started towards a
hole-ice cylinder. Let the hole-ice radius be \(0.50\m\). The scattering
length within the hole-ice cylinder is \(0.06\m\), the scattering length
in the surrounding bulk ice is \(1.48\m\). For each scattering step, the
distance to the next scattering point and the current distance to the
center of the hole-ice cylinder are recorded.

\sourcepar{The implementation of this cross check can be found in \issue{71}.}

\begin{figure}[htbp]
  \smallerimage{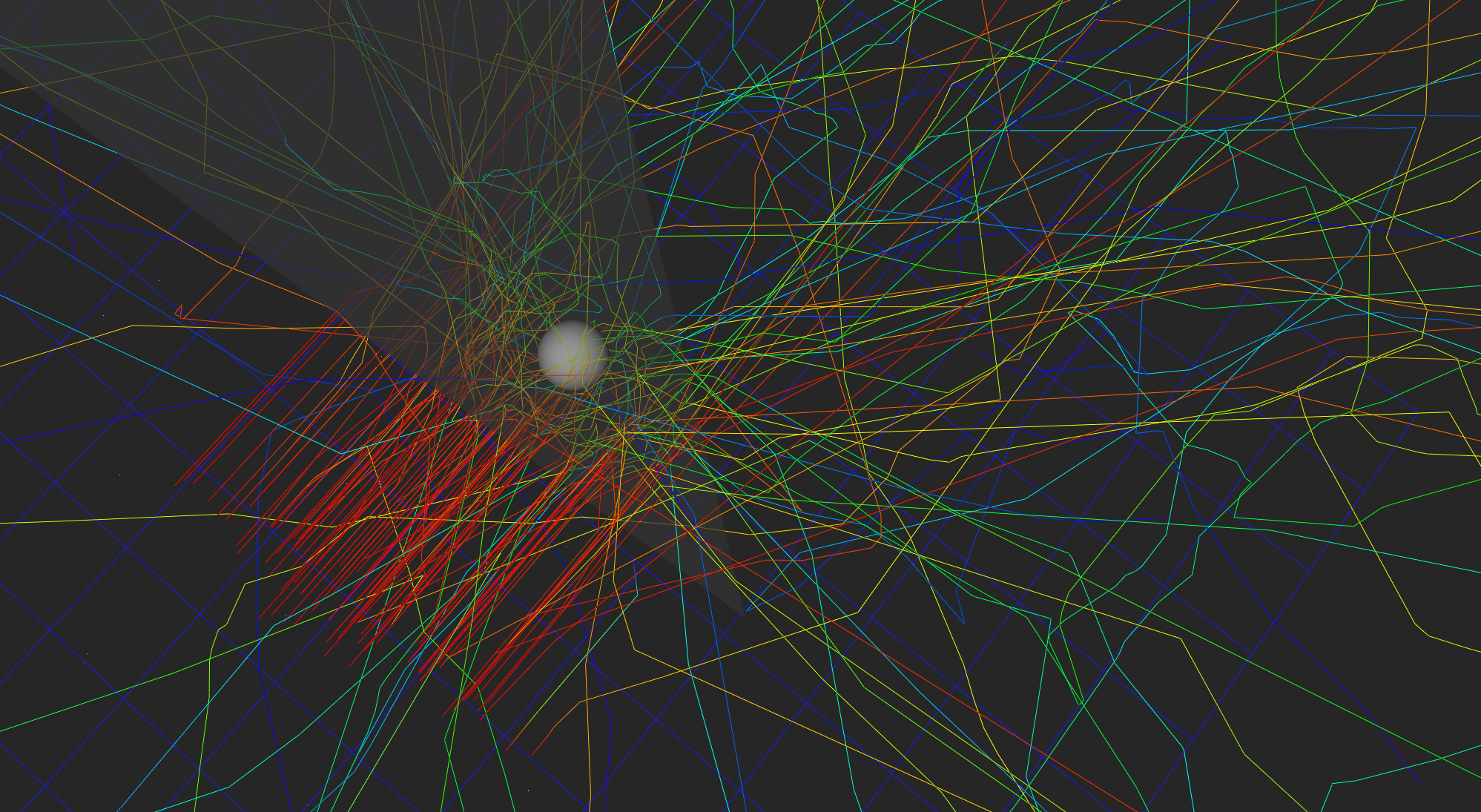}
  \caption{\steamshovel event display of a simulation of a plane wave of photons propagating towards a hole-ice cylinder. The scattering length within the hole-ice cylinder is different from the scattering length within the surrounding bulk ice.}
  \label{fig:An7ik8pu}
\end{figure}

As the distance to the next scattering point should be exponentially
distributed, governed by the scattering lengths within and without the
hole-ice cylinder, the distance to the next scattering point should
differ in the following three regions:

\begin{enumerate}
  \item Well within the hole-ice radius, the distance to the next scattering point should be in the range of the scattering length of the hole ice.
  \item Well outside the hole-ice radius, the distance to the next scattering point should be in the range of the scattering length of the surrounding bulk ice.
  \item In the vicinity of the hole-ice-cylinder radius, there should be some kind of transition due to the complex geometrical simulation.
\end{enumerate}

\begin{figure}[htbp]
  \centering
  \subcaptionbox{All data points}{\includegraphics[width=0.48\linewidth]{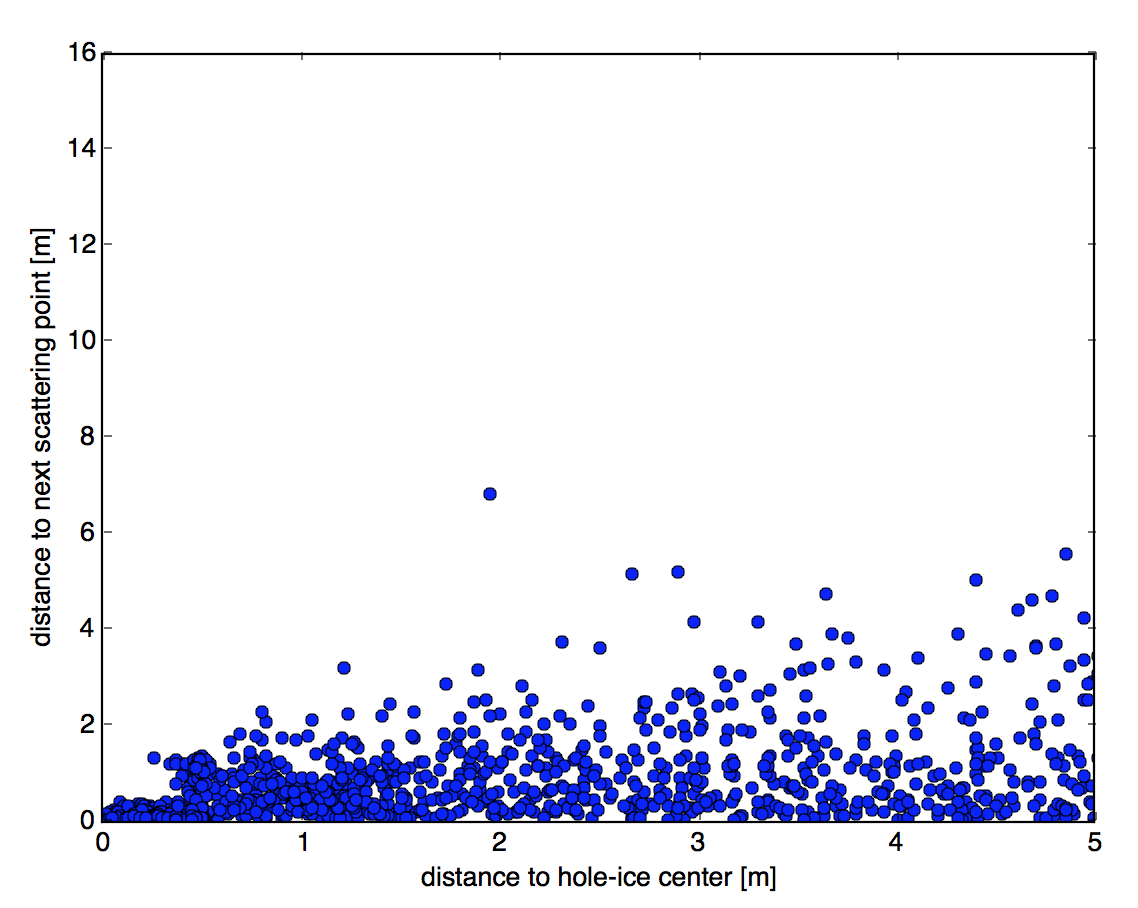}}\hfill
  \subcaptionbox{Averaged for bins of a width of  $10\cm$}{\includegraphics[width=0.48\linewidth]{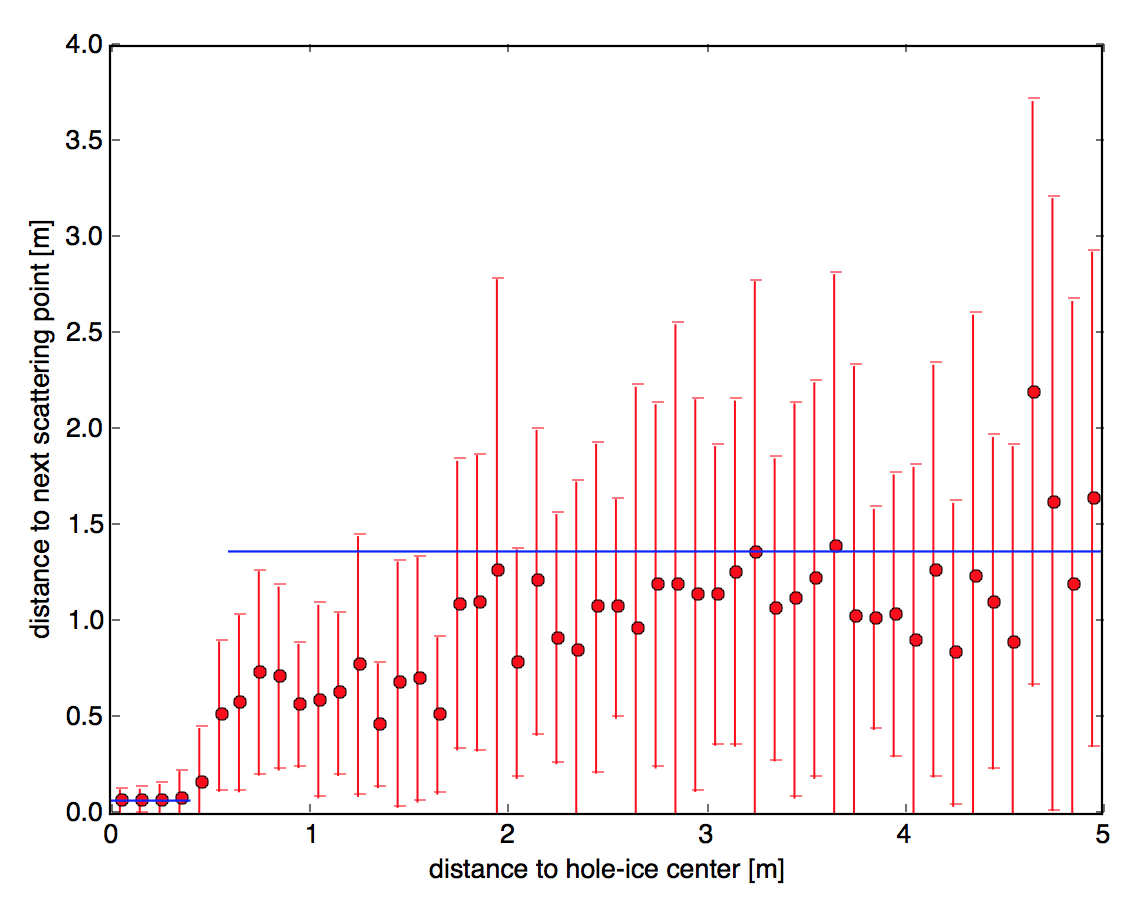}}\hfill
  \subcaptionbox{Magnification of the vicinity of the hole-ice border}{\includegraphics[width=0.48\linewidth]{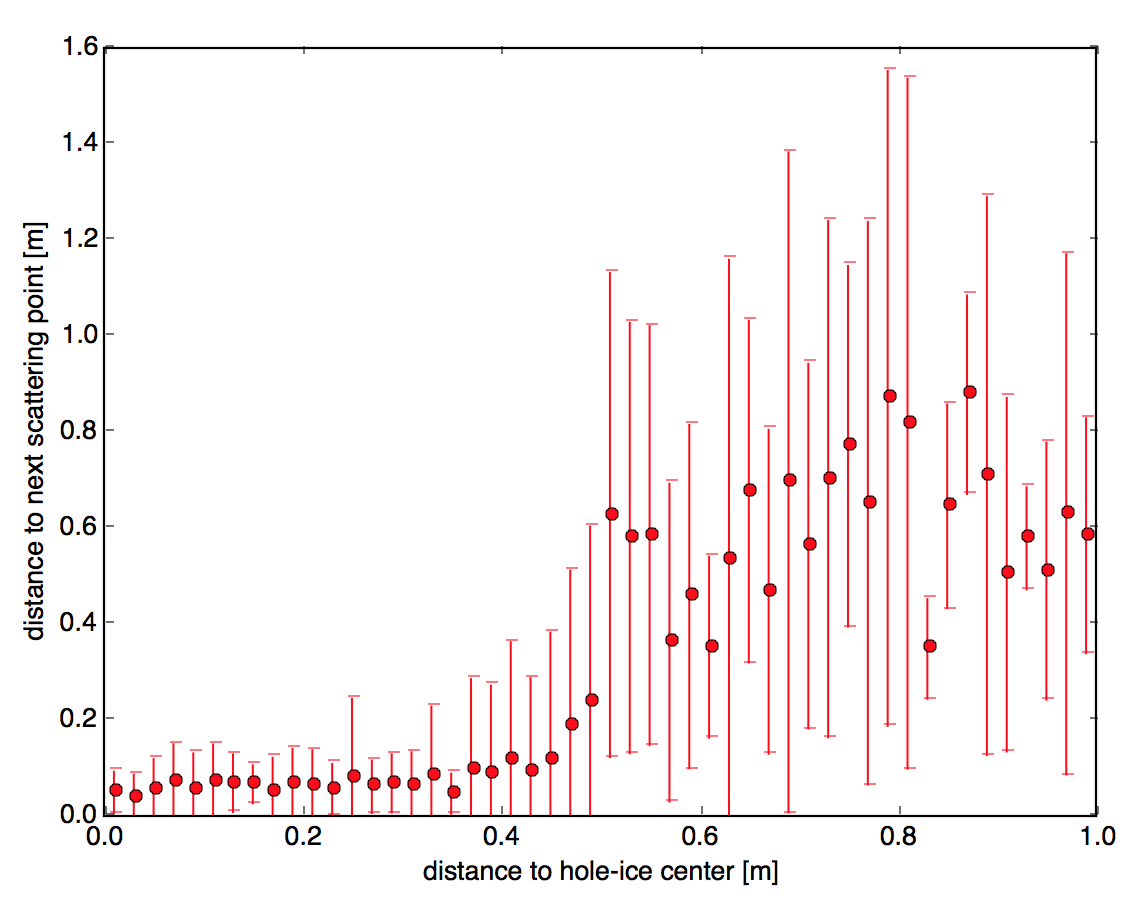}}
  \caption{For each scattering step of a simulation of photons entering a hole-ice cylinder, plot the distance to the next scattering step in relation to the current distance to the cylinder center. The mean distance to the next scattering point within $80\,\%$ of the hole-ice radius is fitted to $0.07\m \pm 0.07\m$. The mean distance to the next scattering point outside of $120\,\%$ of the hole-ice radius is fitted to $1.37\m\pm 1.37\m$. The true scattering length in the simulation is set to $0.06\m$ inside, and to $1.48\m$ outside the hole-ice cylinder.}
  \label{fig:eeYoid2p}
\end{figure}

Assuming an exponential distribution of the distance to the next
scattering point, the mean distance to the next scattering point within
\(80\,\%\) of the hole-ice radius is fitted to \(0.07\m \pm 0.07\m\),
the mean distance to the next scattering point outside of \(120\,\%\) of
the hole-ice radius is fitted to \(1.37\m\pm 1.37\m\) (figure
\ref{fig:eeYoid2p} b), which is in the range of the true scattering
lengths of \(0.06\m\) and \(1.48\m\) inside and outside the hole ice
respectively.

When plotting the local average of the distance to the next scattering
point for the vicinity of the hole-ice border (figure \ref{fig:eeYoid2p}
c), the curve shows no abrupt jump but a transition between both domains
as expected.

  \section{Examples of Application}
\label{sec:applications}

This section shows examples of what can be done with the new hole-ice
algorithms. Section \ref{sec:scattering_simulation} visualizes the
scattering of photons within hole-ice cylinders with different
scattering lengths. Sections \ref{sec:angular_acceptance_scan} and
\ref{sec:parameter_scan} demonstrate scanning the effective angular
acceptance of optical modules for different hole-ice parameters.
Sections \ref{sec:cylinder_shift} to \ref{sec:cables} show examples of
shifting and nesting hole-ice cylinders as well as adding cables as
cylinders of instant absorption. Section \ref{sec:flasher} shows an
example of a flasher-calibration study. Section
\ref{sec:flasher_with_cable} demonstrates a flasher simulation with
shadowing cable.

\subsection{Visualizing a Hole-Ice Column With Different Scattering Lengths}
\label{sec:scattering_simulation}

One of the first things one can use the new propagation algorithms for,
is to visualize the scattering behavior of photons entering a hole-ice
cylinder.

\paragraph{Simulation}

\begin{figure}[htbp]
  \subcaptionbox{$\lambda\hi\sca = 1.0 \cdot \lambda\sca$, 3D view. The hole-ice is represented by the grey cylinder around the optical module represented by a white circle.}{\halfcropimage{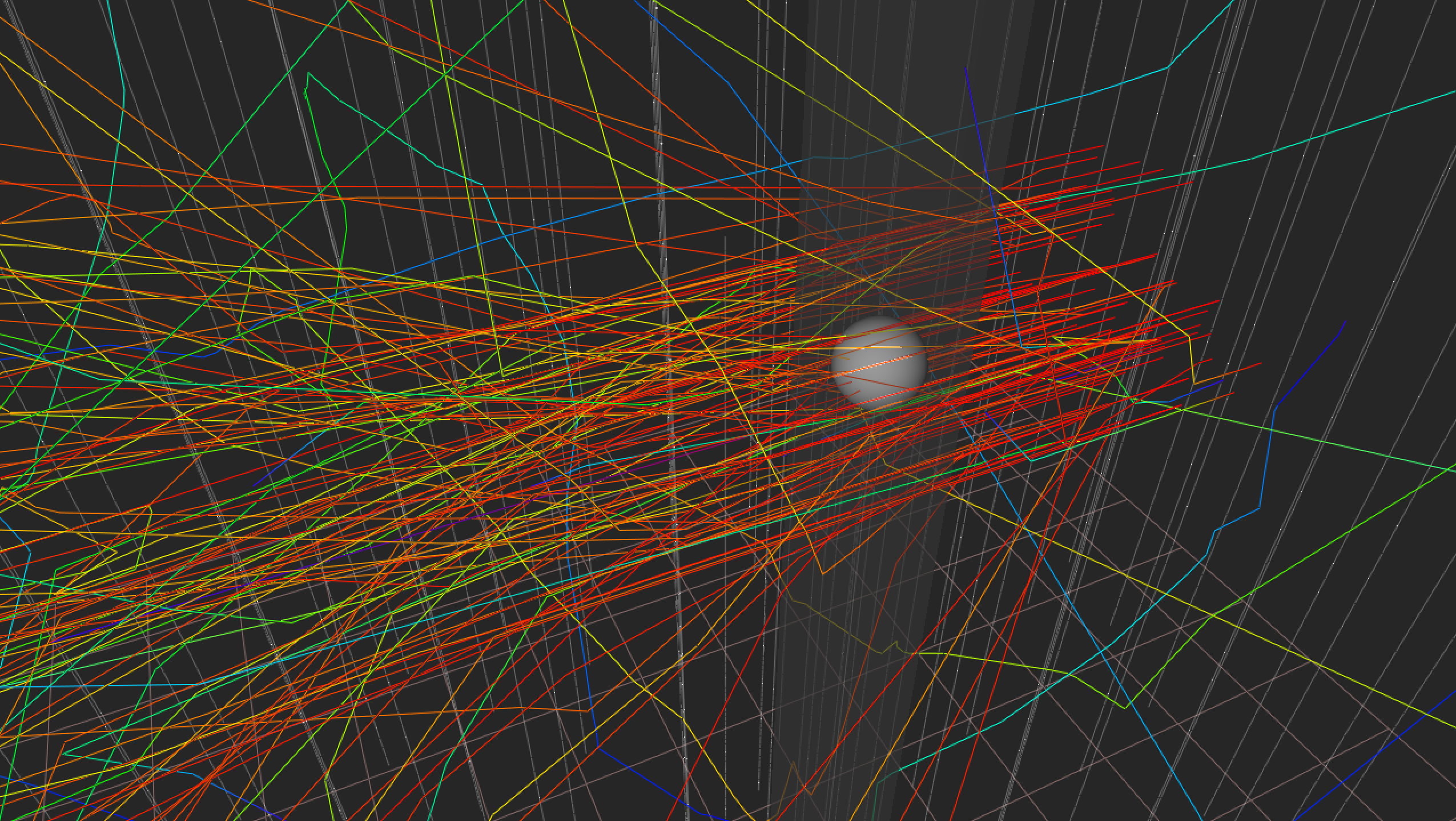}{0 0 0 2.2cm}}\hfill
  \subcaptionbox{$\lambda\hi\sca = 1.0 \cdot \lambda\sca$, view from above}{\halfimage{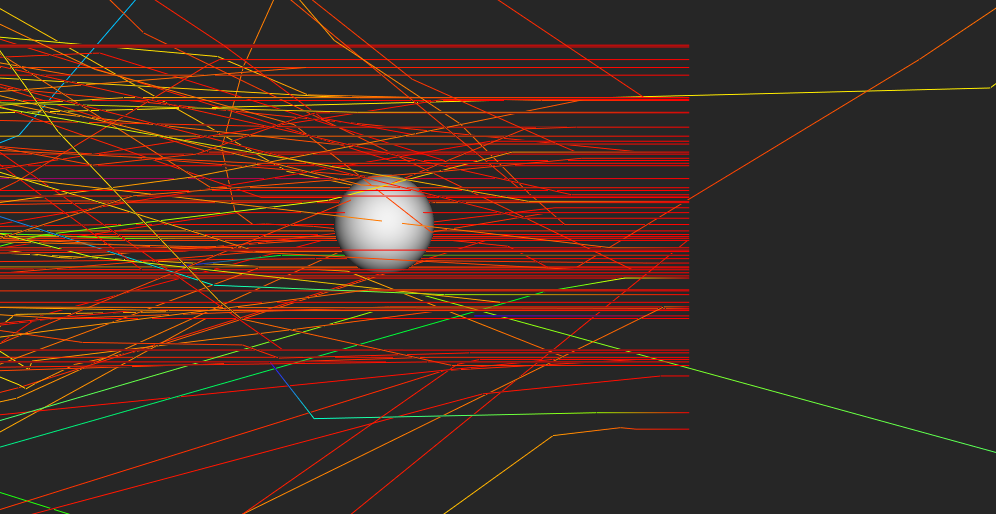}}\hfill
  \subcaptionbox{$\lambda\hi\sca = 0.1 \cdot \lambda\sca$}{\halfimage{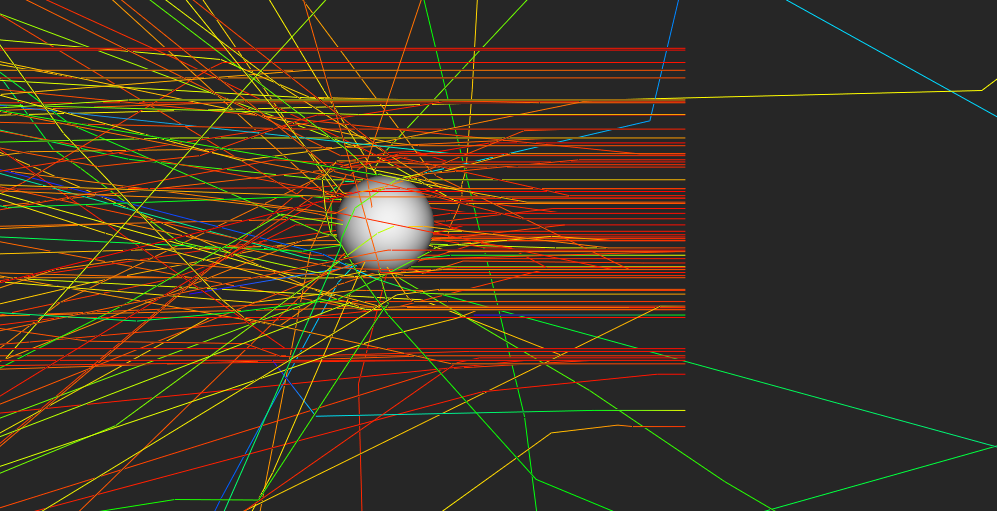}}\hfill
  \subcaptionbox{$\lambda\hi\sca = 0.01 \cdot \lambda\sca$}{\halfimage{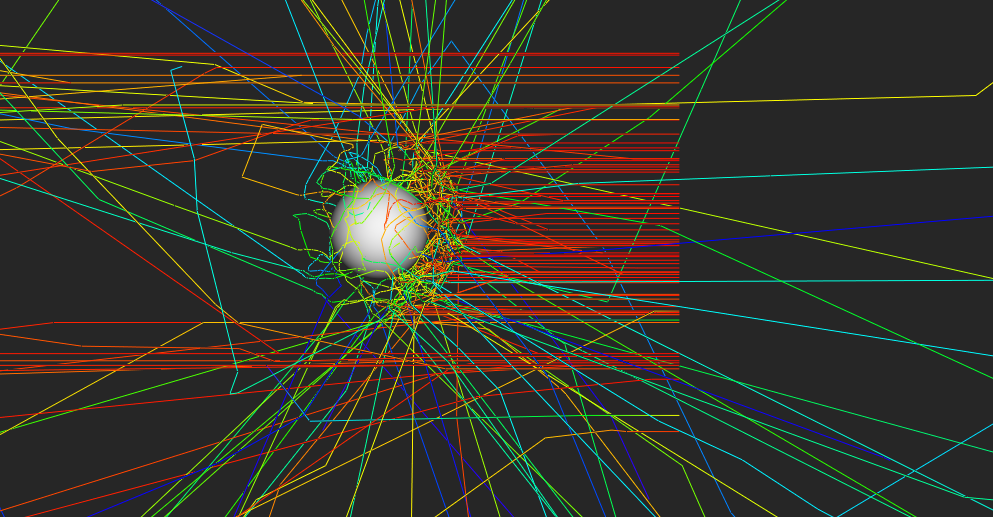}}\hfill
  \subcaptionbox{$\lambda\hi\sca = 0.001 \cdot \lambda\sca$}{\halfimage{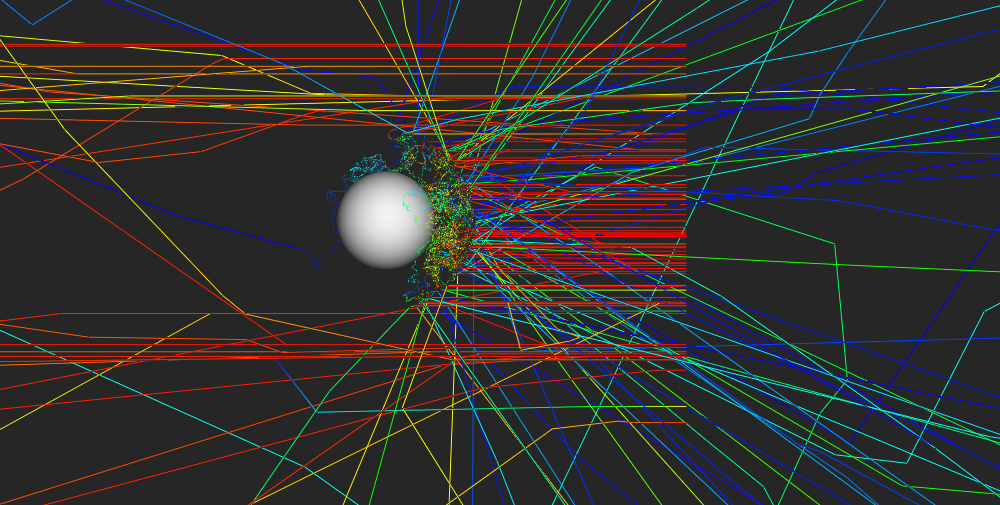}}\hfill
  \subcaptionbox{$\lambda\hi\sca = 0.0001 \cdot \lambda\sca$}{\halfimage{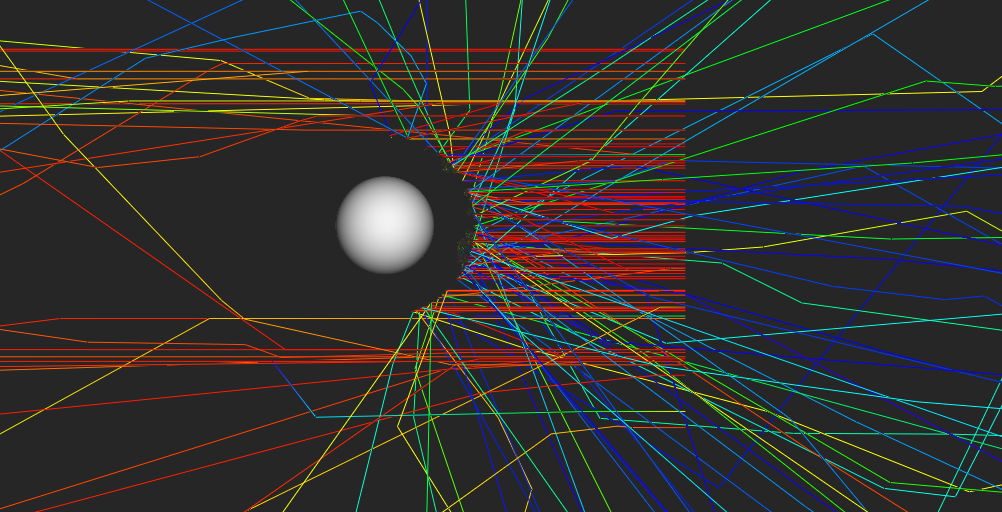}}\hfill
  \subcaptionbox{$\lambda\hi\sca = 0.0001 \cdot \lambda\sca$, zoom}{\halfimage{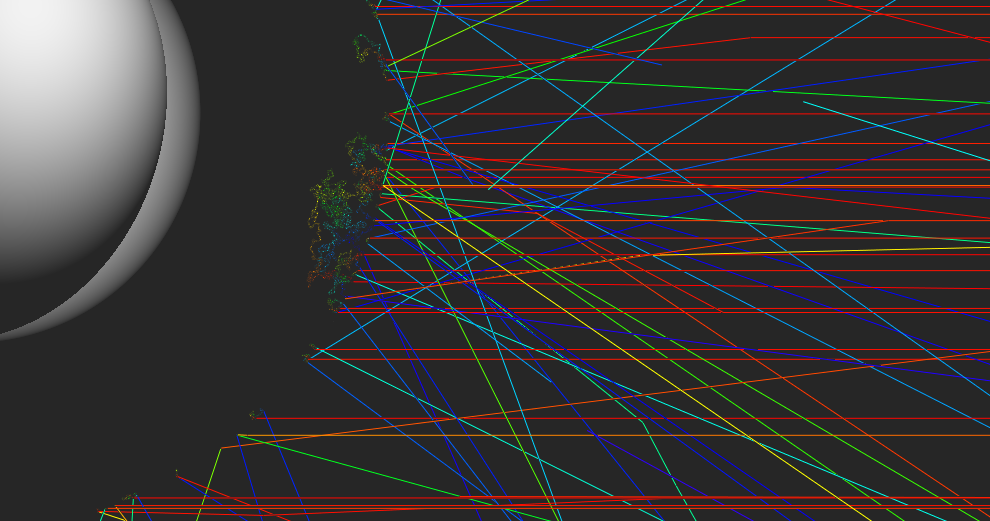}}\hfill
  \subcaptionbox{$\lambda\hi\sca = 0.0001 \cdot \lambda\sca$, zoom 2}{\halfimage{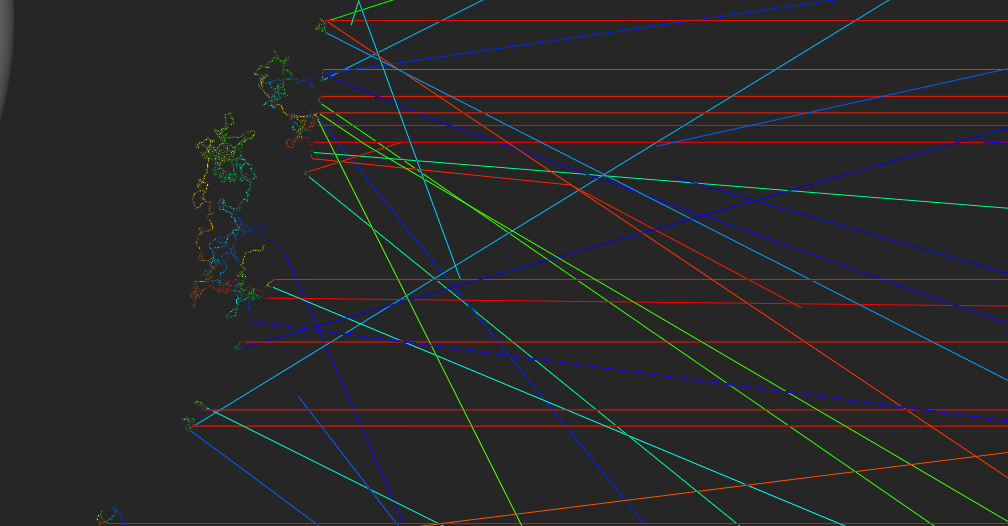}}\hfill
  \caption{\steamshovel visualization of a photon-propagation simulation where photons are started from a plane on the right hand side onto a hole-ice cylinder with a radius of $30\cm$ from $1\m$ distance. The scattering length $\lambda\sca\hi$ within the hole-ice cylinder is given relative to the scattering length $\lambda\sca$ of the surrounding bulk ice. The absorption length is kept the same within the cylinder as in the bulk ice. The geometric scattering length of the bulk ice is $\lambda\sca = 1.3\m$, the absorption length is $\lambda\abs = 48.0\m$. Colors of the photon trajectories indicate the number of scatterings relative to the total number of scatterings of the photon until absorption. Red: photon just created, blue: photon about to be absorbed. The optical module is shown as a white sphere only to indicate the scale of the scenery. Interaction with the optical module is turned off in the simulation. Animation on \noun{YouTube}: \protect\url{https://youtu.be/BhJ6F3B-I1s}}
  \label{fig:Uo8kuo2z}
\end{figure}

In this example, 100 simulated photons are propagated towards and into a
hole-ice cylinder. The photons are started from random positions within
a 1m-by-1m plane in a distance of \(1\m\) from the cylinder center in
parallel towards the cylinder. (See figure \ref{fig:Uo8kuo2z} a.)

The cylinder radius is \(30\cm\). The scattering length within the bulk
ice is \(\lambda\sca = 1.3\m\), corresponding to an effective scattering
length of \(\lambda\esca = 21.7\m\). The absorption length within the
bulk ice is \(\lambda\abs = 48.0\m\).

The scattering length \(\lambda\hi\sca\) within the hole ice is defined
relative to the bulk-ice scattering length:
\(\lambda\hi\sca = f\,\lambda\sca, \ f \in \reals_0^+\), for example
\(f = \sfrac{1}{10}\) in figure \ref{fig:Uo8kuo2z} (c),
\(f = \sfrac{1}{100}\) in figure \ref{fig:Uo8kuo2z} (d), and
\(f = \sfrac{1}{1000}\) in figure \ref{fig:Uo8kuo2z} (d).

This simulation uses the hole-ice-correction algorithm described in
section \ref{sec:algorithm_a}. The interaction with the optical detector
module is deactivated in this simulation.

\docframe{
\docparwithoutframe{The configuration and implementation of this simulation is documented in \issue{40}.}\medskip

\sourceparwithoutframe{A script for configuring and running this and other simulations of this kind is provided at \url{https://github.com/fiedl/hole-ice-study/tree/master/scripts/FiringRange}.}
}

\paragraph{Visualization}

The photon-propagation simulation records the starting point, each
scattering point, and the final trajectory point of each photon. The
photon trajectories can be visualized using the \steamshovel event
display software.

Photon trajectories are represented as lines connecting the scattering
points. The colors of the trajectory segments indicate simulation steps.
A red trajectory segment represents a photon that has just been created.
The blue end of the spectrum represents a photon that is about to be
absorbed.

Figure \ref{fig:Uo8kuo2z} shows the \steamshovel visualizations of the
simulation described above for different hole-ice scattering lengths.

\youtubepar{An animated visualization of this photon-propagation simulation can be found at\newline \url{https://youtu.be/BhJ6F3B-I1s}, and in the folder \texttt{video} on the CD-ROM.}

\paragraph{Observations}

If the scattering length of the hole ice is the same as the scattering
length of the bulk ice (figure \ref{fig:Uo8kuo2z} a and b), the photons
pass right through the cylinder as if it were not there.

The mean scattering angle in the bulk ice as well as in the hole ice is
assumed as \ang{20} \cite{escawiki}. Therefore, for a weak hole ice with
a large scattering length, the photons are deflected within the hole
ice. See figure \ref{fig:Uo8kuo2z} (c) in comparison to (b).

For a stronger hole ice with smaller scattering length, several
scatterings occur within the hole ice, changing the direction of the
scattered photons more drastically. This makes the hole-ice cylinder
cause an effective reflection of the incoming photons. See figure
\ref{fig:Uo8kuo2z} (d).

In figure \ref{fig:Uo8kuo2z} (e), the scattering length is so small that
the space between the cylinder border and the optical module is large
enough to show the typical ``random walk'' \cite{randomwalk} of the
randomly scattering photons. This random scattering delays the photons
on their way to the optical module (compare section
\ref{sec:arrival_time} regarding arrival-time distributions).

For even stronger hole ice, which is shown in figures \ref{fig:Uo8kuo2z}
(f), (g), and (h), the hole-ice cylinder around the optical module is
shielding the module from the incoming photons. The small scattering
length confines the random walk of the photons to the outer region of
the cylinder as it is unlikely for a photon to be propagated further
into the cylinder. The photons walk in the outer region until they are
absorbed, or scattered out of the cylinder and escape.

\begin{figure}[htbp]
  \smallerimage{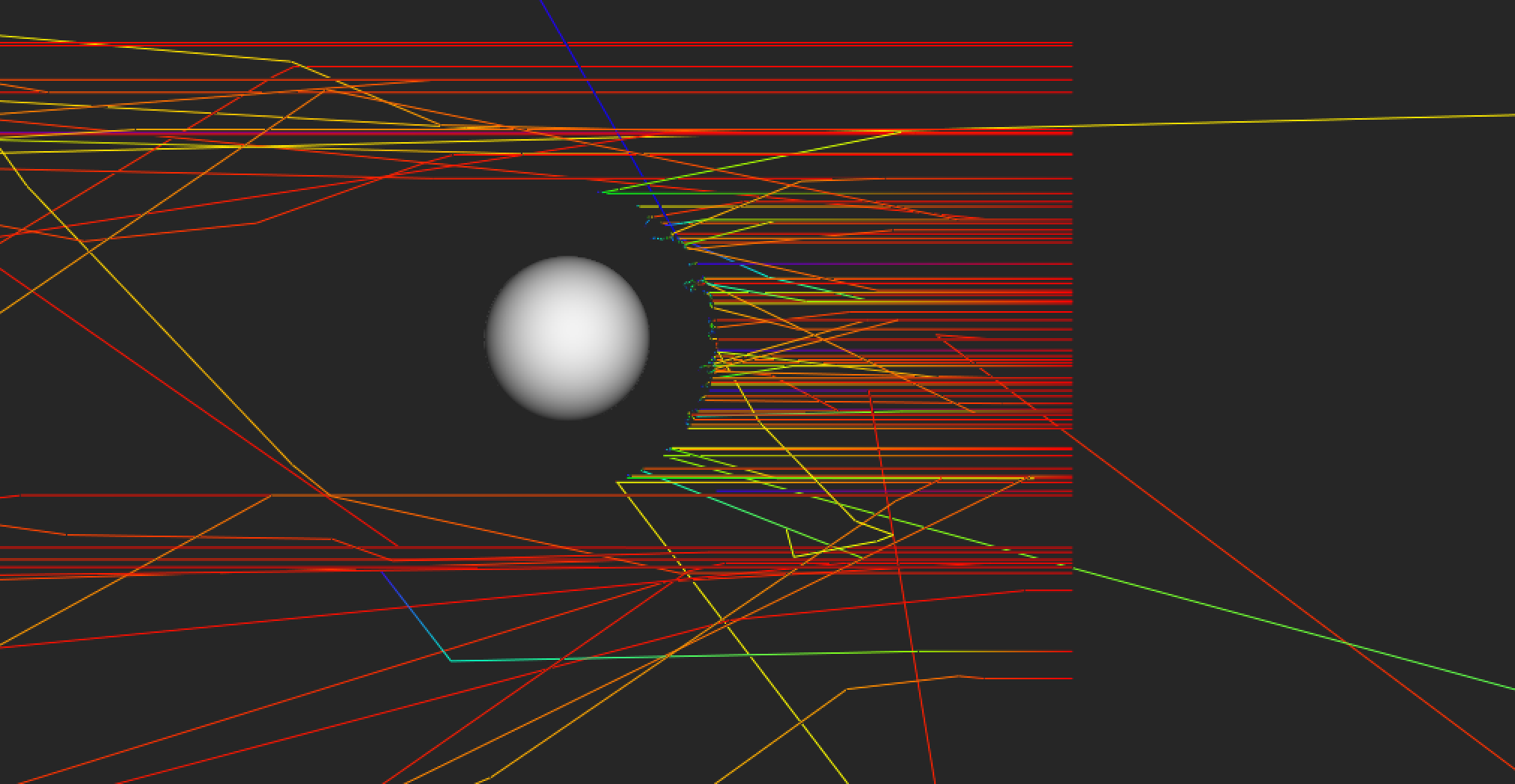}
  \caption{\steamshovel visualization of a simulation where the absorption length within the cylinder is set to $\lambda\abs\hi \approx 5\cm$. The scattering length factor is $f\sca=0.001$. No photon can reach more than $5\cm$ into the cylinder. Figure \ref{fig:Uo8kuo2z} (e) on page \pageref{fig:Uo8kuo2z} shows the same simulation without setting the hole-ice absorption length.}
  \label{fig:Lie7laxa}
\end{figure}

The same effect, confining the photons to the outer region of the
cylinder, can be achieved by defining the absorption length within the
cylinder to \(\lambda\abs\hi = 5\cm\). Then, no photon can reach more
than \(5\cm\) into the cylinder (figure \ref{fig:Lie7laxa}).

  \subsection{Scanning the Angular Acceptance of an Optical Module}
\label{sec:angular_acceptance_scan}

\begin{figure}[htbp]
  \subcaptionbox{Side view of an \icecube digital optical module (DOM). The photomultiplier tube (PMT) in each DOM is facing downwards. Incoming photons are shown in the left upper corner of the illustration. They are moving towards the DOM under a polar angle $\eta$ measured from below. Image taken from \cite{martindardupdate}.}{\halfimage{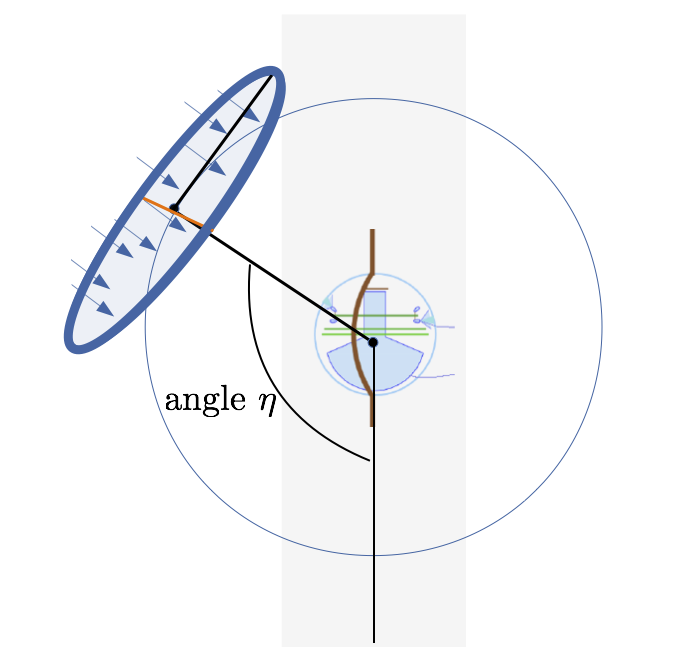}}\hfill
  \subcaptionbox{Angular acceptance $a(\eta)$: Relative sensitivity of the DOM to incoming photons depending on the polar angle $\eta$. The ``nominal'' curve $a\dom(\eta)$ does not consider hole-ice effects and is based on lab measurements. The ``hole ice'' curve $a\domhi(\eta)$ effectively contains hole-ice effects and is based on simulations (see footnote \ref{footnote:hole_ice_h2_curve}, page \pageref{footnote:hole_ice_h2_curve}). Plot taken from \cite{icepaper}, figure 7.}{\halfimage{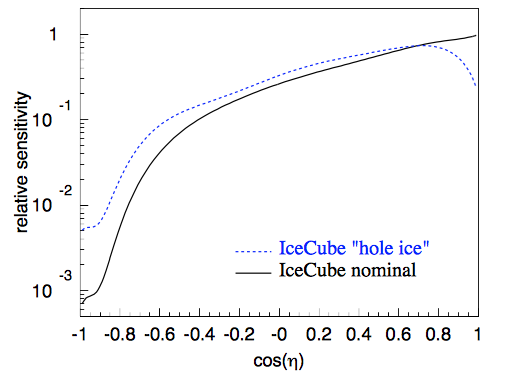}}
  \caption{Angular acceptance: The sensitivity of \icecube's digital optical modules (DOMs) depends on the polar angle $\eta$ of the direction of incoming photons relative to the DOM.}
  \label{fig:weihee6X}\label{fig:icepaper}
\end{figure}

The angular acceptance, or angular sensitivity to incoming photons, of
\icecube's digital optical modules (DOMs) depends on the direction of
the incoming photons. This is due to the detector design as the
photomultiplier tube (PMT) is facing downwards as illustrated in figure
\ref{fig:weihee6X} (a). Therefore, a photon coming from below is more
likely to hit the PMT and be detected than a photon coming from above.
Also, losses due to glass and gel transmission effects as well as the
quantum and collection efficiencies of the PMT depend on the impact
angle of the incoming photon \cite{icepaper}.

Figure \ref{fig:weihee6X} (b) shows the measured
\textbf{angular acceptance} \(a\dom(\eta)\) of the \icecube DOMs by
plotting the \textbf{relative sensitivity} from lab measurements for
each polar angle representing the direction of incoming photons.
\cite{icepaper} The \textbf{sensitivity} is the fraction of the incoming
photons that are registered by the detector. The sensitivity is
normalized \textbf{relative} to the optimal incoming angle, which is
from below, where the relative sensitivity is defined to be 1. Note that
the polar angle \(\eta\) is measured from below: For photons coming from
below, \(\eta = 0\). For photons coming from above, \(\eta = \pi\).

\FloatBarrier
\subsubsection{Expected Effect of Hole Ice on the Angular Sensitivity}
\label{sec:hole_ice_effects}

The effect of hole ice on the detection of incoming photons is expected
to depend on the polar angle \(\eta\) of the incoming photons.
\cite{icepaper} When the photons are coming from the side,
\(\eta = \sfrac{\pi}{2}, \cos \eta = 0\), the distance that incoming
photons need to travel through the hole ice is minimal. Therefore, the
effect of the hole ice is expected to be minimal for photons from this
direction.

For photons approaching the optical module from below,
\(\eta = 0, \cos \eta = 1\), the hole-ice effect should be strong as the
distance that photons need to travel through the hole ice is maximal.
The photons are likely to be scattered away before reaching the optical
module, effectively reducing the sensitivity for photons approaching the
optical module from below as shown in figure \ref{fig:weihee6X} (b) by
the blue curve in the region of \(\cos \eta = 1\).

For photons approaching the optical module from above,
\(\eta = \pi, \cos \eta = -1\), the hole-ice effect should also be
strong as the distance that photons need to travel through the hole ice
is maximal. The optical module is very insensitive to registering
photons coming from above. But propagating through the hole ice, photons
approaching the optical module from above are likely to be scattered
away before reaching the optical module. There is a chance that those
photons travel around the optical module and finally hit the module at a
lower position in the sensitive area. This effectively increases the
sensitivity of the module to photons approaching the optical module from
above as shown in figure \ref{fig:weihee6X} (b) by the blue curve in the
region of
\(\cos \eta = -1\).\footnote{The effective ``hole-ice'' angular-acceptance curve (blue curve in figure \ref{fig:weihee6X} b), has been obtained with simulations using the \noun{Photonics} software \cite{lundberg}, assuming a hole-ice cylinder of $30\cm$ radius and a geometric scattering length of $50\cm$. These are the so-called ``H2 parameters'' from an \noun{Amanda} calibration study \cite{holeicestudieswithyag,icemodelsdata}. See also section \ref{sec:a_priori_curve} and appendix \ref{sec:angular_acceptance_simulation_for_h2_parameters}. \label{footnote:hole_ice_h2_curve}}

\newpage

\subsubsection{Measuring the Effective Angular Acceptance With Simulations}
\label{sec:measuring_angular_acceptance_with_simulations}

\label{sec:gauging}

In order to quantify the hole-ice effect on the angular acceptance of
\icecube's optical modules for different hole-ice models, this study
performs a series of simulations, starting photons from different angles
\(\eta_i\) towards an optical module and counting the photons registered
by the optical module.

The simulation records the number \(N(\eta_i):= N\) of started photons
for each angle \(\eta_i\), as well as the number \(k(\eta_i)\) of
registered hits for each angle. The relative hit frequency is
\(h(\eta_i)\).

\begin{equation}
  h(\eta_i) = \frac{k(\eta_i)}{N}
\end{equation}

To make this relative frequency \(h(\eta_i)\) comparable to the DOM's
angular acceptance \(a\dom(\eta)\), which is defined such that
\(a\dom(\eta = 0) = 1\), this study often uses a renormalized relative
frequency \(\tilde{h}(\eta_i):=g\,h(\eta_i)\) with a scaling factor
\(g\in\reals\) such that \(\tilde{h}(\eta = 0) = 1\).

\begin{equation}
  \tilde{h}(\eta_i) = g\,h(\eta_i), \ \ \
  g = \frac{1}{h(\eta = 0)}
  \label{eq:gauging_factor}
\end{equation}

\sourcepar{A script to configure and perform these kinds of simulations is provided in \script{AngularAcceptance}.}

\newpage

\subsubsection{Photon Sources: Pencil Beams and Plane Waves}

\begin{figure}[htbp]
  \subcaptionbox{Pencil beams: Start photons at fixed positions.}{\halfimage{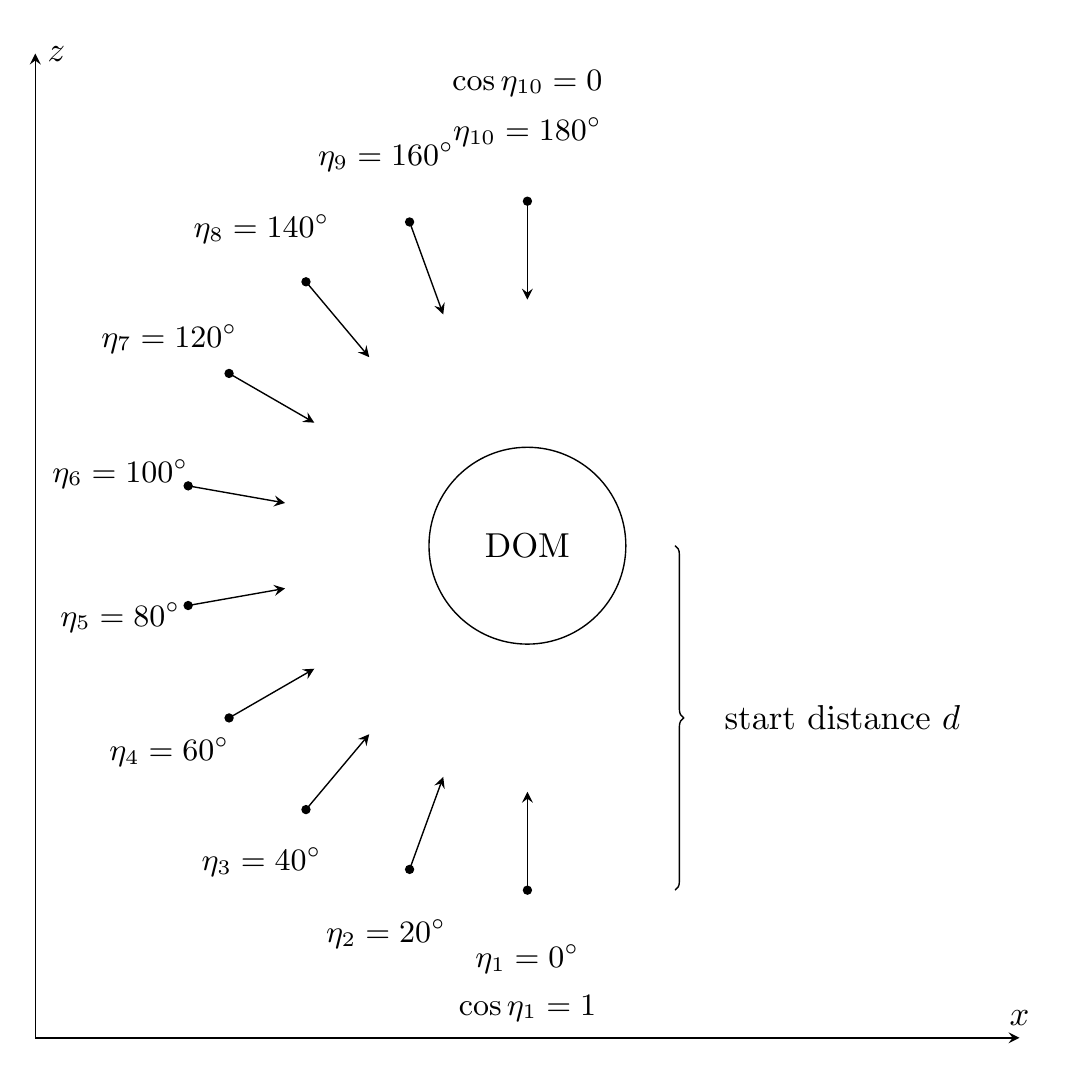}}\hfill
  \subcaptionbox{Plane waves: Start photons from random positions within planes with extent $e$ approximating plane waves approaching the DOM. The two-dimensional planes are represented by lines in this $x$-$z$-projected view.}{\halfimage{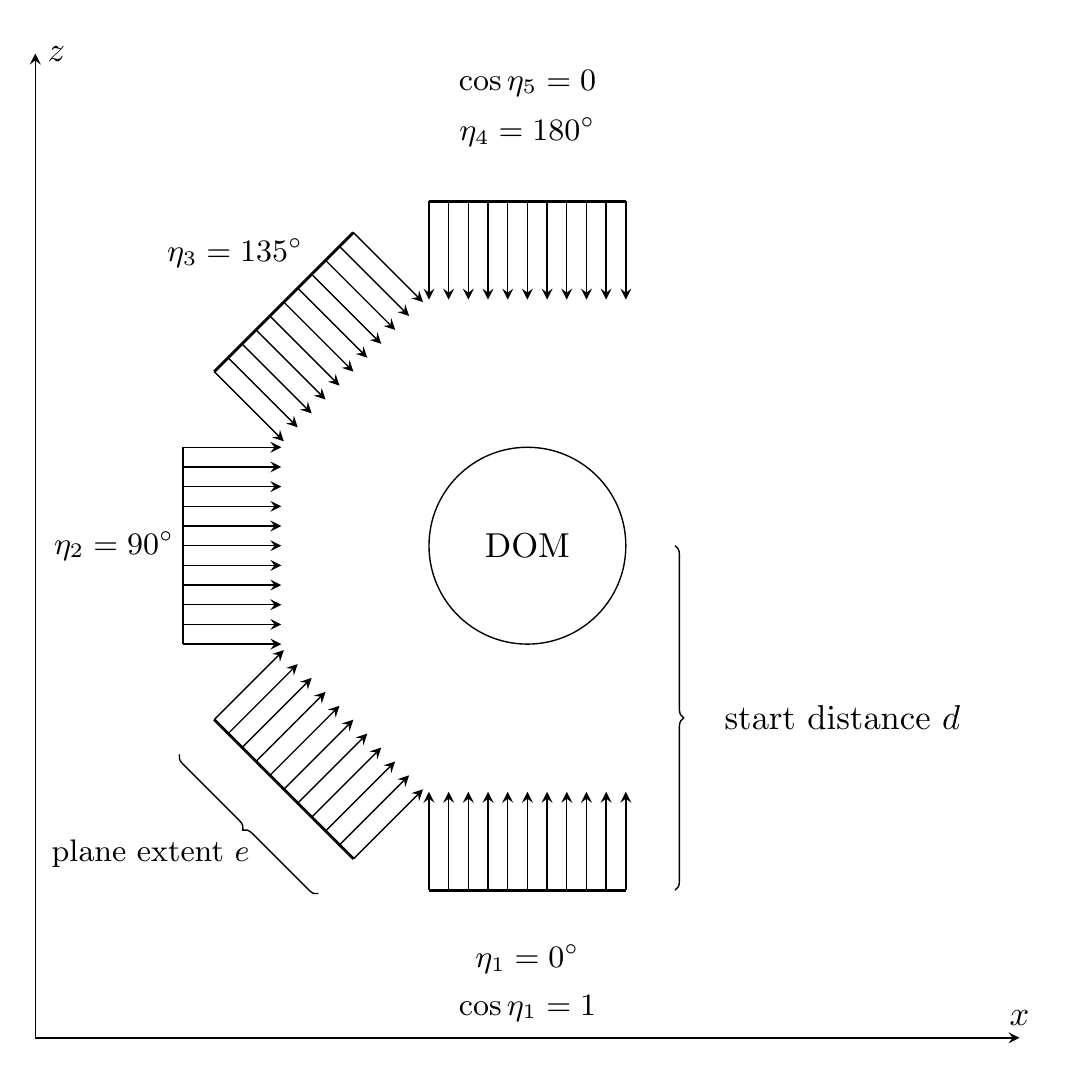}}
  \caption{Angular-acceptance scan: In each simulation, start photons from a different polar angle $\eta_i$ from a distance $d$ towards the digital optical module (DOM) and record the number of photons registered by the DOM. In this illustration, the DOM is viewed from the side. For optical modules aligned along the $z$-axis, the acceptance is symmetrical with respect to the azimuth angle.}
  \label{fig:quie8Oof}
\end{figure}

This study considers two types of photon sources for angular-acceptance
studies, pencil beams and plane waves. Both are illustrated in figure
\ref{fig:quie8Oof}.

One simulation is performed for each angle \(\eta_i\). Photons are
started from a distance \(d\) from the center of the optical module. In
the case of pencil beams, photons are started at a fixed position for
each angle \(\eta_i\) such that it approaches the optical module under
this angle as shown in figure \ref{fig:quie8Oof} (a).

In order to approximate plane waves as photon sources, photons are not
started from fixed positions but from random positions within a plane
that is located in a distance \(d\) from the center of the optical
module and oriented perpendicular to the distance vector as shown in
figure \ref{fig:quie8Oof} (b).

Due to the limitation of computational resources, rather than simulating
plane waves with infinite extent, an arbitrary extent \(e\) of the plane
is chosen. \rongen \cite{martindardupdate,rongenswedishcamera} compares
the influence of the plane extent \(e\) on the angular-acceptance curves
(\cite{martindardupdate}, slides 6 and 9). The angular-acceptance
simulations described in the section arbitrarily choose a plane extent
\(e = 1\m\) and a start distance
\(d = 1\m\).\footnote{For follow-up studies, an extent of $e=2\m$ and a distance of $d=2.5\m$ should be chosen as photon-source parameters, because the angular-acceptance curves become stable for these or larger parameters. \cite{rongenswedishcamera}}

Note that the scaling factor \(g\) (equation \ref{eq:gauging_factor})
depends on the photon source and needs to be redetermined when switching
between pencil beams and plane waves, or when changing the starting
distance \(d\), or the plane extent \(e\).

\begin{figure}[htbp]
  \subcaptionbox{$2\ns$}{\halfimage{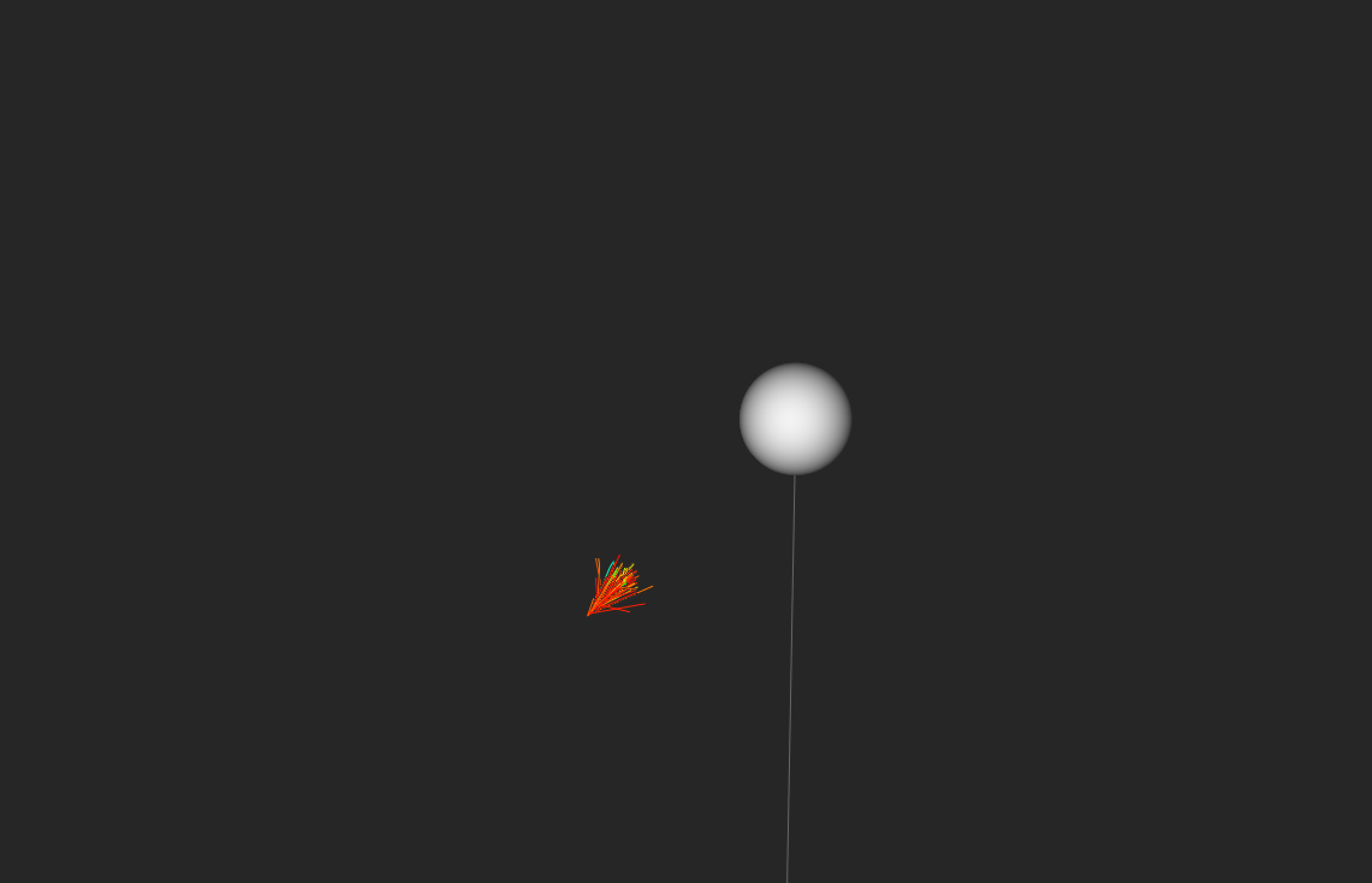}}\hfill
  \subcaptionbox{$4\ns$}{\halfimage{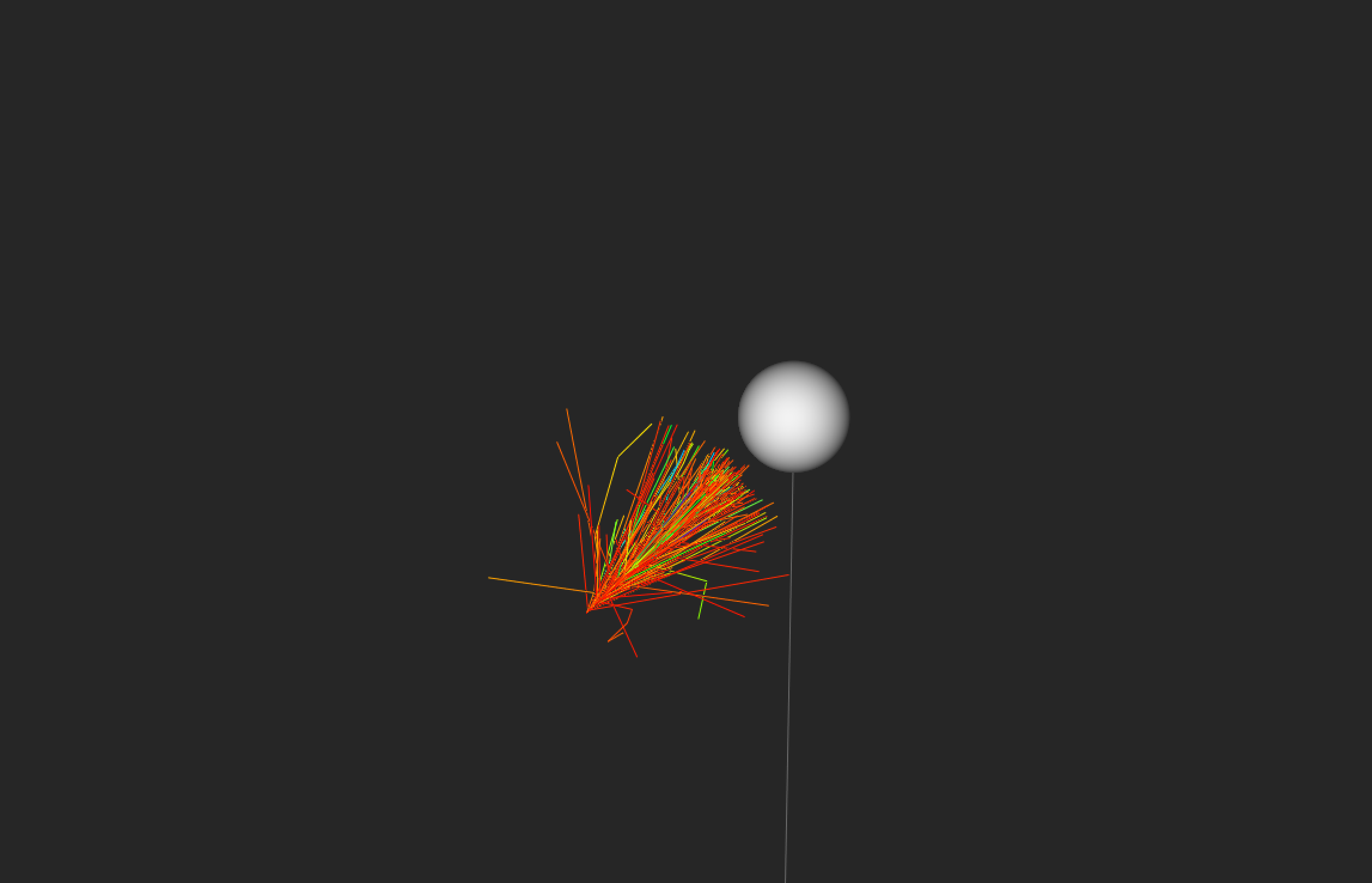}}\hfill
  \subcaptionbox{$9\ns$}{\halfimage{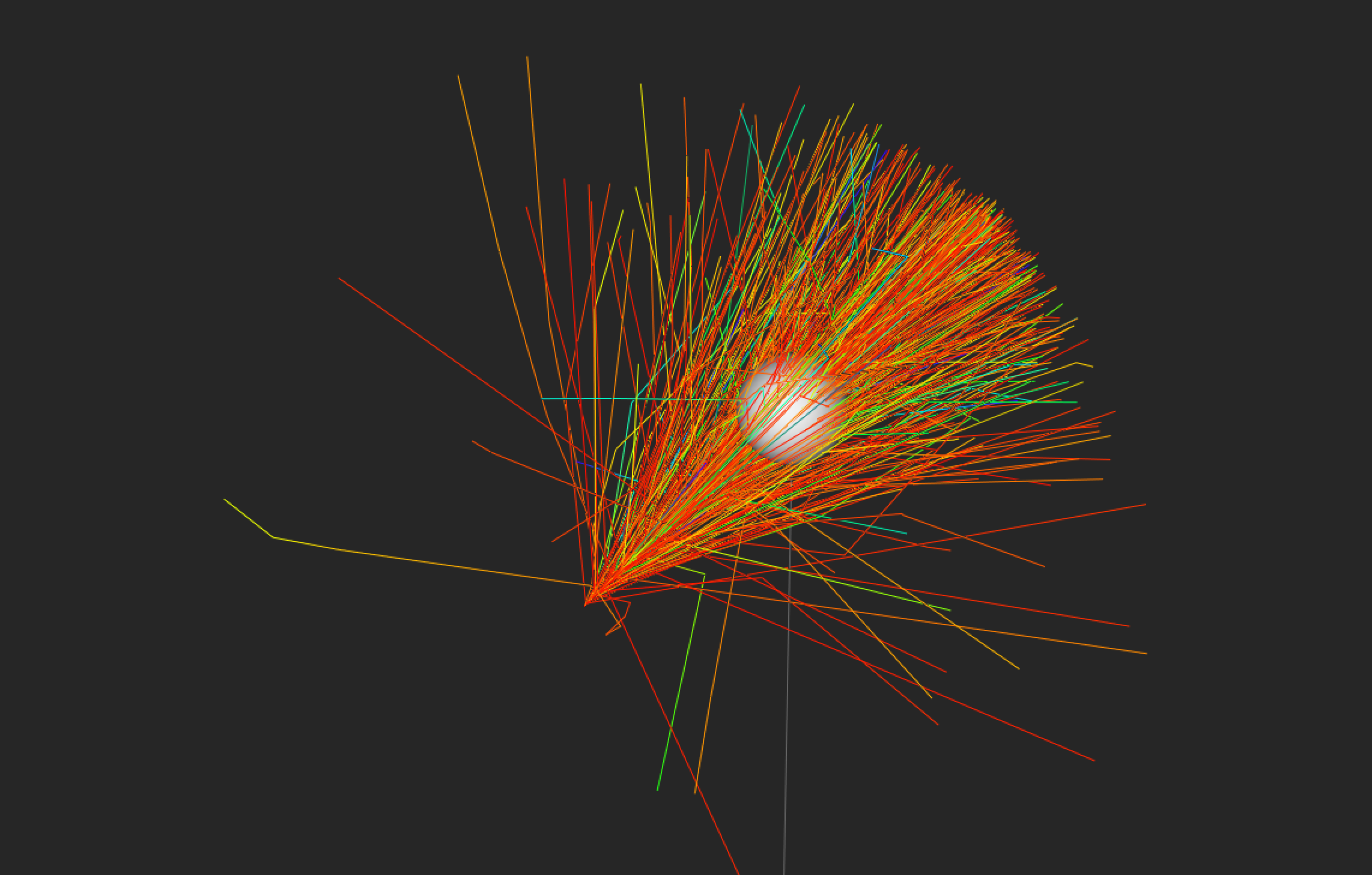}}\hfill
  \subcaptionbox{$100\ns$}{\halfimage{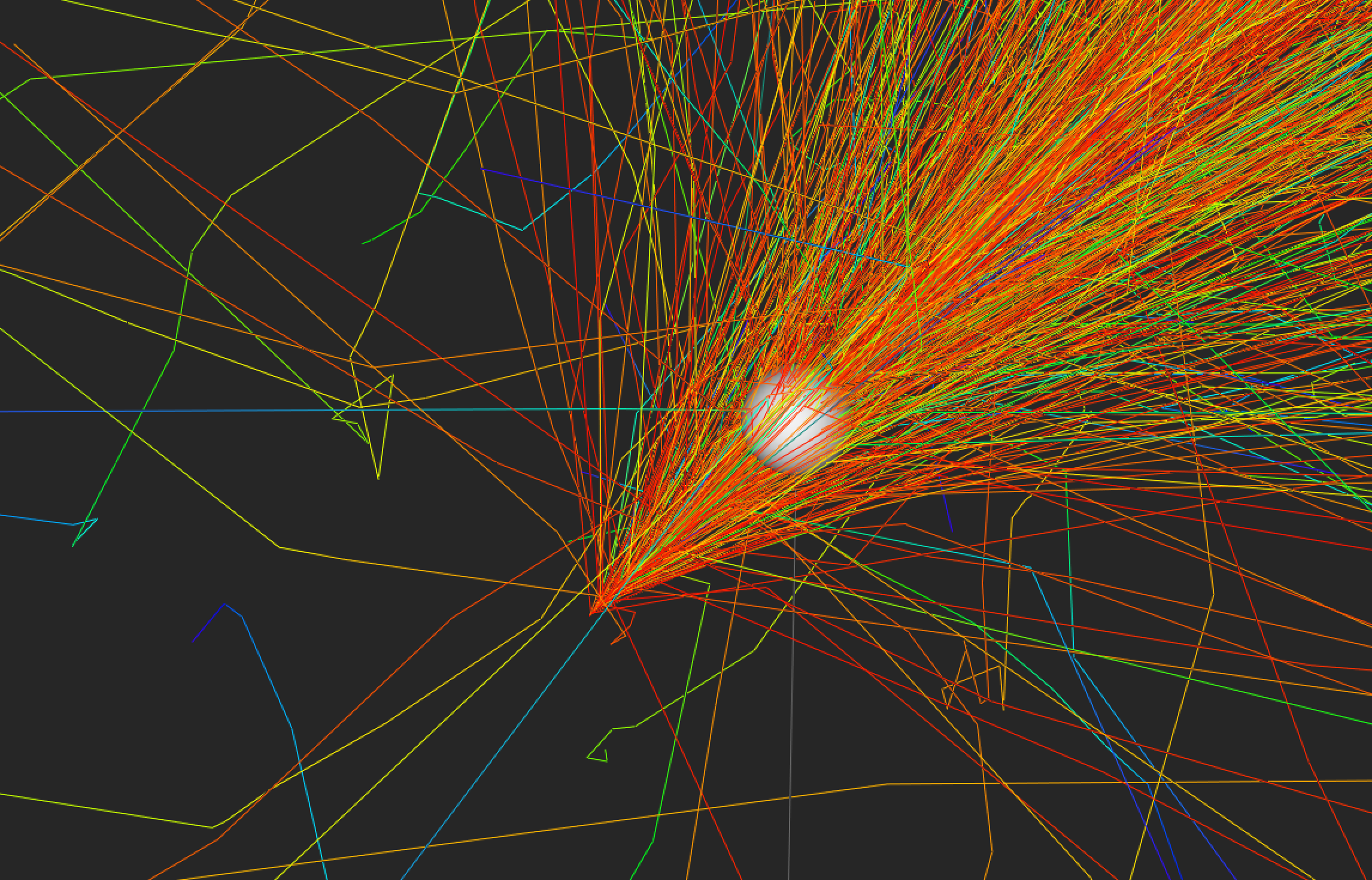}}
  \caption{\steamshovel visualization of photons started as pencil beam under an angle of $\eta = \ang{45}$ from a distance of $d = 1\m$ towards the upmost optical module of a detector string. Snapshots are taken at $2\ns$, $4\ns$, $9\ns$, and $100\ns$ after starting the photons. The opening angle of the beam is $\ang{0.001}$. The photon spread seen in the event display is due to scattering along the photon trajectory.}
  \label{fig:Paihah7h}
\end{figure}

\begin{figure}[htbp]
  \subcaptionbox{$2\ns$}{\halfimage{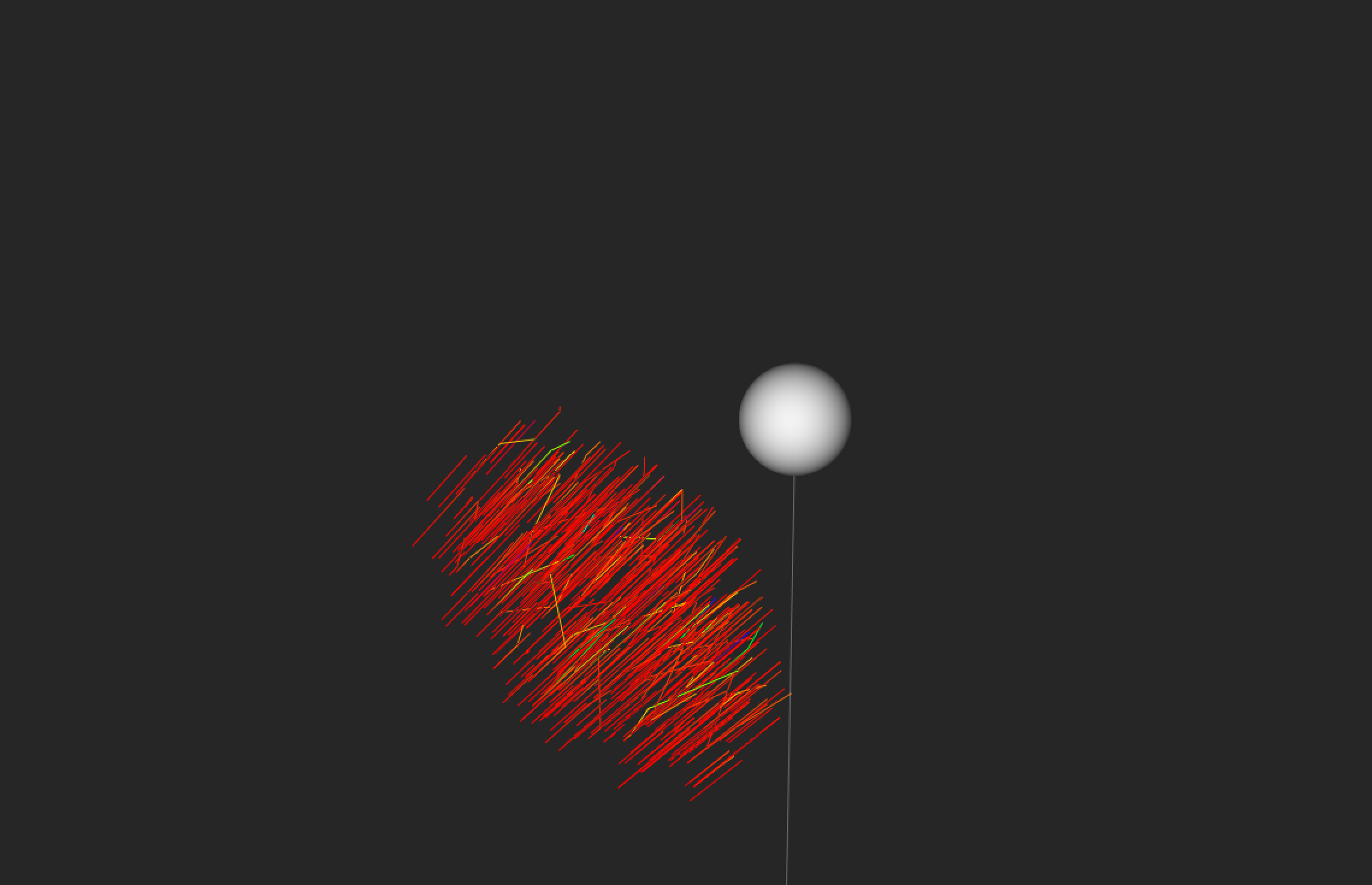}}\hfill
  \subcaptionbox{$4\ns$}{\halfimage{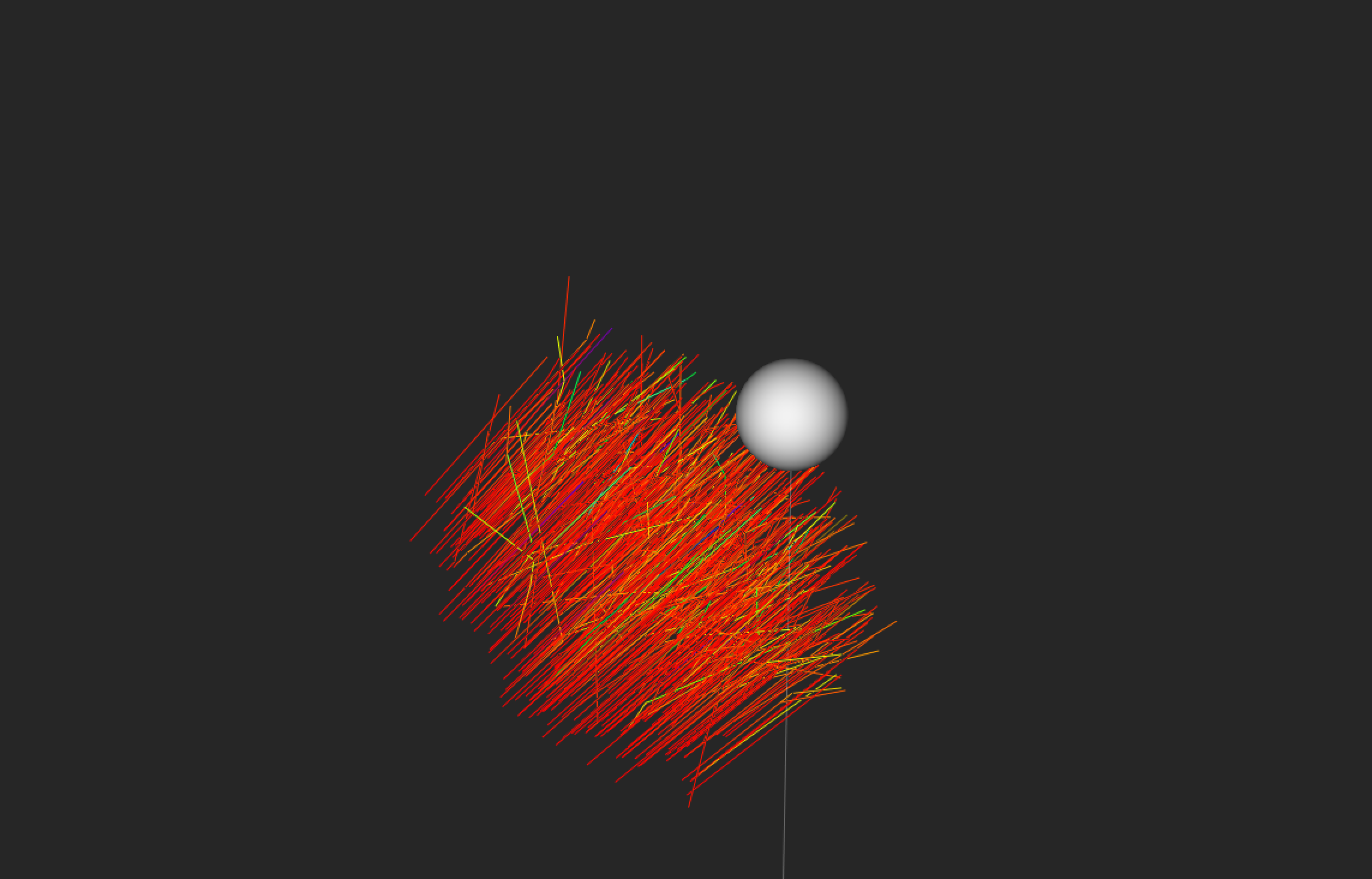}}\hfill
  \subcaptionbox{$9\ns$}{\halfimage{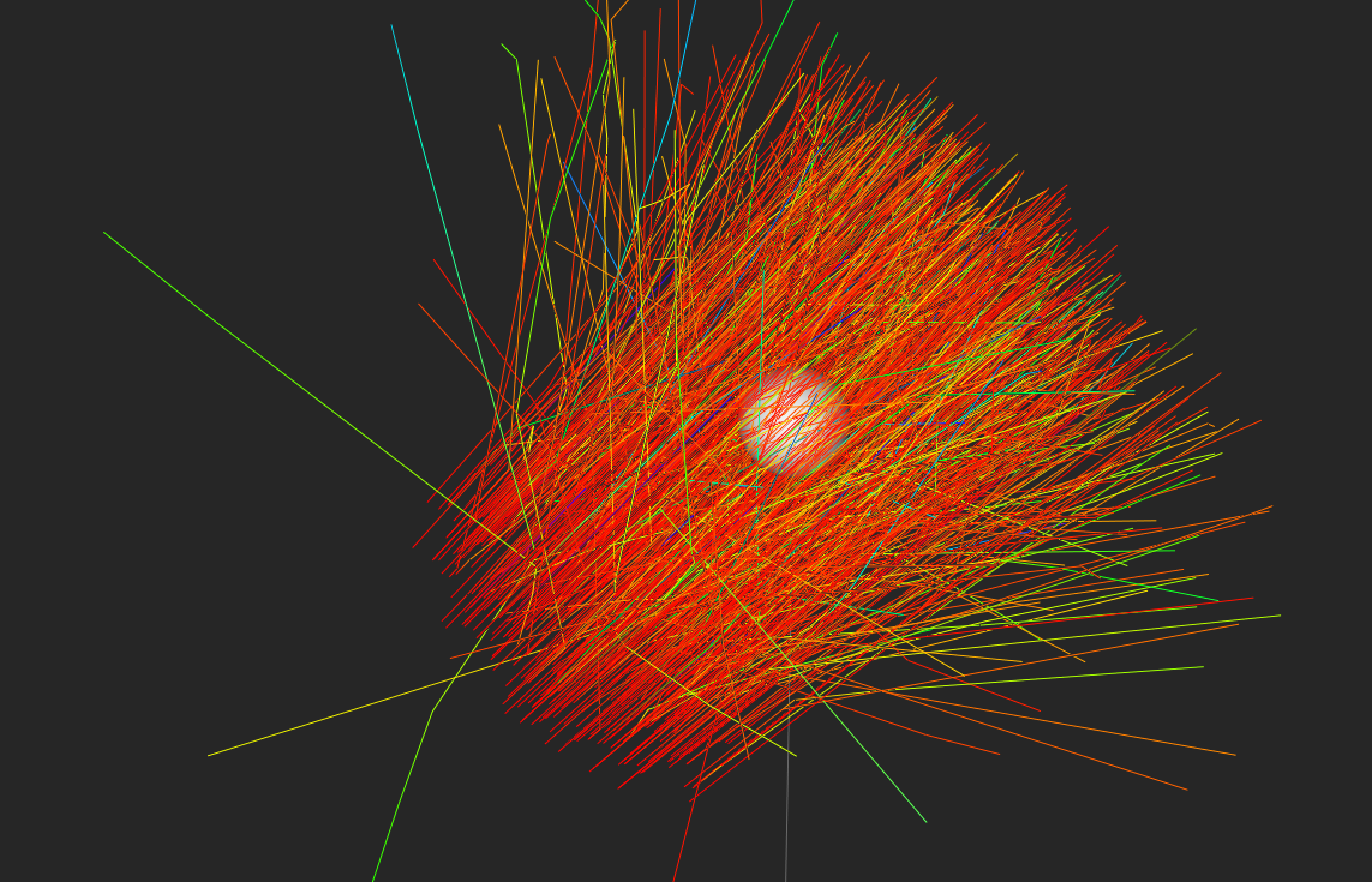}}\hfill
  \subcaptionbox{$100\ns$}{\halfimage{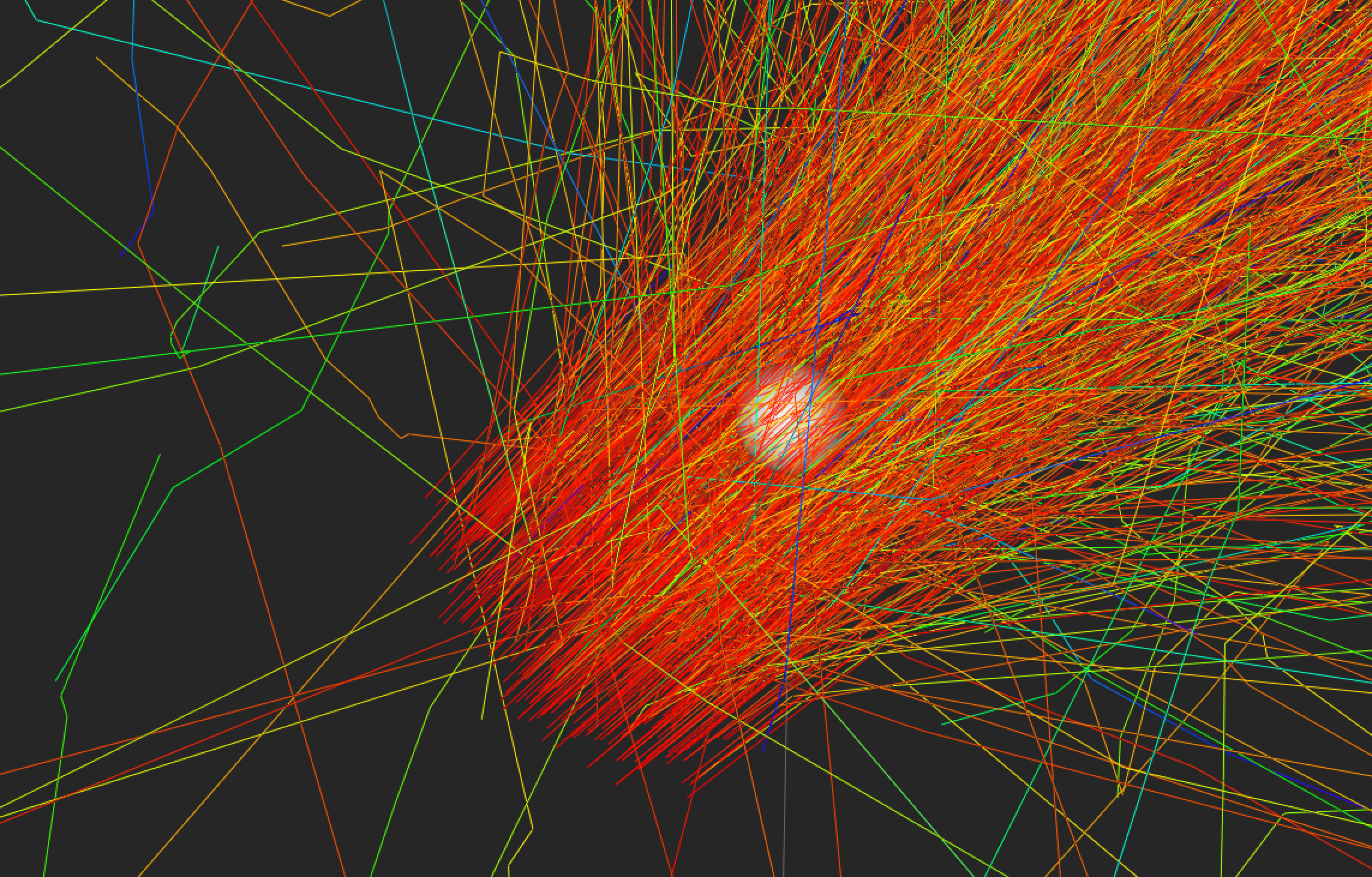}}
  \caption{\steamshovel visualization of photons started as plane wave with an extent of $e = 1\m$ under an angle of $\eta = \ang{45}$ from a distance of $d = 1\m$ towards the detector module. Snapshots are taken at $2\ns$, $4\ns$, $9\ns$, and $100\ns$ after starting the photons.}
  \label{fig:Aehi7kae}
\end{figure}

Figure \ref{fig:Paihah7h} shows a \steamshovel visualization of
simulated photons started from a pencil beam, figure \ref{fig:Aehi7kae}
for photons started from a plane.

\FloatBarrier
\subsubsection{Acceptance Criterion: A Priori Angular Acceptance or Direct Detection}
\label{sec:acception_criterion}\label{sec:a_priori_angular_acceptance}
\label{sec:direct_detection}

In each simulation step of the photon-propagation simulation, the
algorithm checks whether the photon intersects the sphere representing a
optical module between two scattering points (see section
\ref{sec:standard_photon_propagation_algorithm}). At this time, the
intersection position and the direction of the photon are known. Based
on that information, in order to model the sensitivity of the optical
module, the simulation needs to decide whether to consider the incoming
photon as detected, or to ignore the hit.

\begin{figure}[htbp]
  \centering
  \halfimage{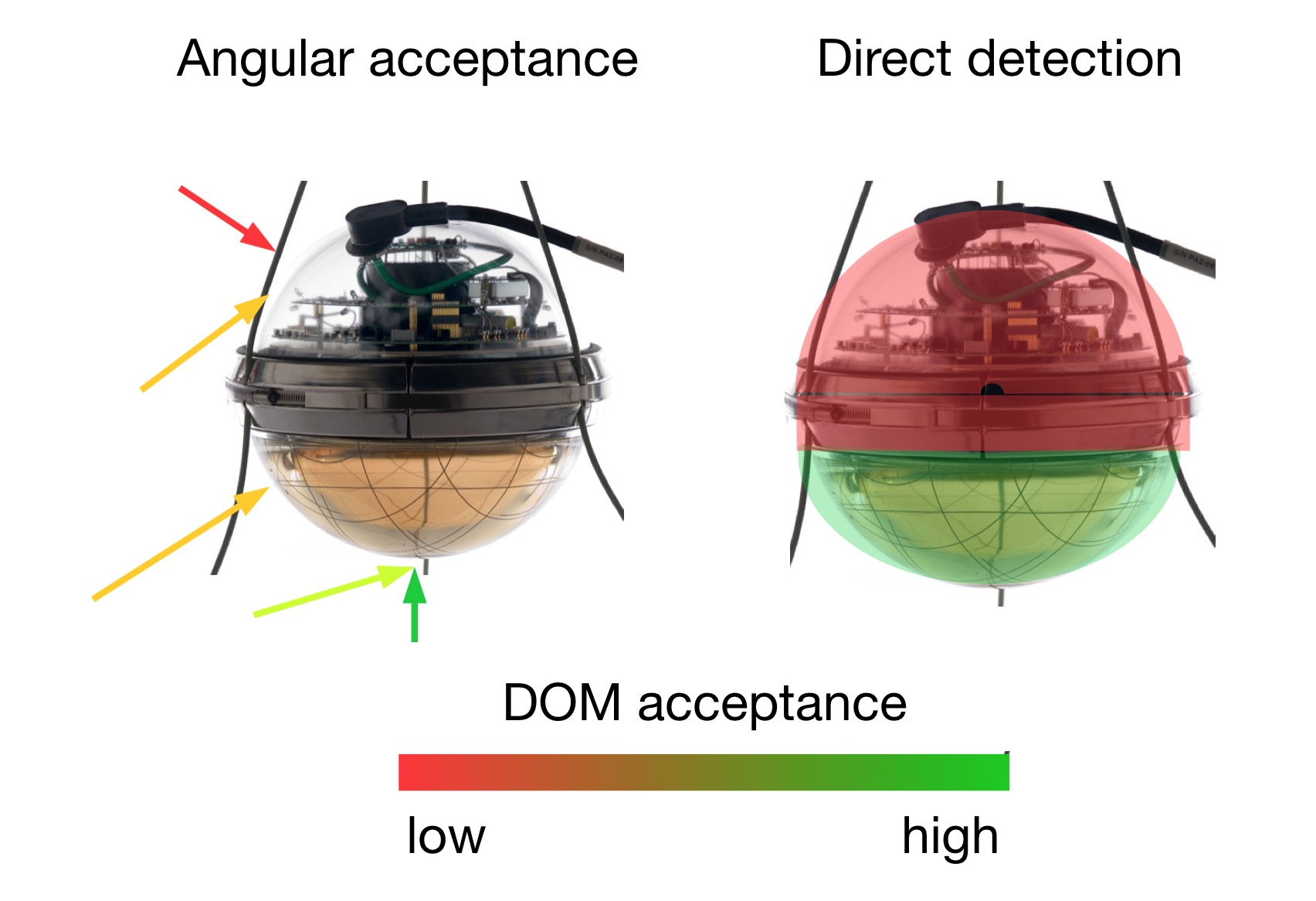}
  \caption{Acceptance criteria: The DOM on the left-hand side accepts or rejects incoming photons as hits based only on the direction of the photon (``angular acceptance''). The DOM on the right-hand side accepts or rejects incoming photons only based on the impact position that determines whether the sensitive area of the photomultiplier tube has been hit (``direct detection''). Image taken from \cite[slide 17]{martinspicehddard}.}
  \label{fig:kieQuoh1}
\end{figure}

This study considers two approaches to this decision-making process:
Accept the hits based only on the measured \textbf{angular acceptance}
\(a\dom(\eta)\) of the optical module and the directions of the incoming
photons, or accept the hits based only on the location of the sensitive
photomultiplier area and the positions of the hits. The latter method is
called \textbf{direct detection} \cite{martinspicehddard}. Both
approaches are illustrated in figure \ref{fig:kieQuoh1}.

Using only the photon direction as acceptance criterion, is the default
approach in \clsim. This method has the advantage that both, the
location of the photomultiplier tube, and other angular-dependent
characteristics such as its quantum efficiency, can be abstracted into a
single angular-acceptance function that only depends on one parameter,
the photon direction.

In \clsim, the angular acceptance \(a\dom(\eta)\) of \icecube's optical
modules has been implemented as polynomial.

\begin{equation}
  a\dom(\eta) = \sum_{j = 0}^{10} b_j\,\cos(\eta)^j, \ \ \ \eta \in [0; \pi]
  \label{eq:anominal}
\end{equation}

\begin{gather*}
   b_0 =  0.26266,  \ \  b_1    =  0.47659,   \ \ b_2 =  0.15480,  \\
   b_3 = -0.14588,  \ \  b_4    =  0.17316,   \ \ b_5 =  1.3070,   \\
   b_6 =  0.44441,  \ \  b_7    = -2.3538,    \ \ b_8 = -1.3564,   \\
   b_9 =  1.2098,   \ \  b_{10} =  0.81569
\end{gather*}

\sourcepar{The polynomial parameters for different angular-sensitivity models like $a\dom(\eta)$, also called \enquote{nominal} or \enquote{h0} model, and other models that include hole-ice approximations, called \enquote{h1}, \enquote{h2}, and \enquote{h3}, can be found in \icecube's source repository within the \noun{ice-models} project: \url{http://code.icecube.wisc.edu/projects/icecube/browser/IceCube/projects/ice-models/trunk/resources/models/angsens}.}

In this study that focuses on the propagation through hole ice, the
photons are expected to frequently scatter, and thereby to change their
direction in close proximity to the optical module. In this scenario,
using the position of a hit rather than only the direction, is of
interest, in particular as, according to \cite{martinspicehddard},
direct detection is needed to distinguish different hole-ice models, and
has been implemented for this study in \clsim as alternative to the
standard angular-acceptance method.

\docpar{The implementation of direct detection is documented in \issue{32}.}

The most accurate approach to modeling the sensitivity of the optical
modules to registering incoming photons would be to take both, the hit
position, and the photon direction into account. The position would
account for whether the photon would hit the sensitive area of the
photomultiplier tube. The impact angle would account for
angular-dependent transmission effects and for the quantum efficiency of
the tube. Combining both approaches in \clsim, however, is considered
out of scope of this study.\followup

Also, this study does not consider inclined orientations of the optical
modules, but always assumes that the modules are oriented along the
\(z\)-axis.\footnote{To check if DOM orientations have been implemented at the time of reading, check \url{https://github.com/fiedl/hole-ice-study/issues/53}.}

\subsubsection{Angular-Acceptance Simulations Without Hole Ice}
\label{sec:angular_acceptance_simulations_without_hole_ice}

In order to compare the different approaches regarding photon sources
and acceptance criteria, angular-acceptance simulations have been
conducted using the new propagation algorithm, but without any hole-ice
cylinders.

\docpar{These angular-acceptance simulations without hole ice have been documented in \issue{98}.}

\begin{figure}[htbp]
  \subcaptionbox{pencil beams, a priori angular acceptance}{\halfimage{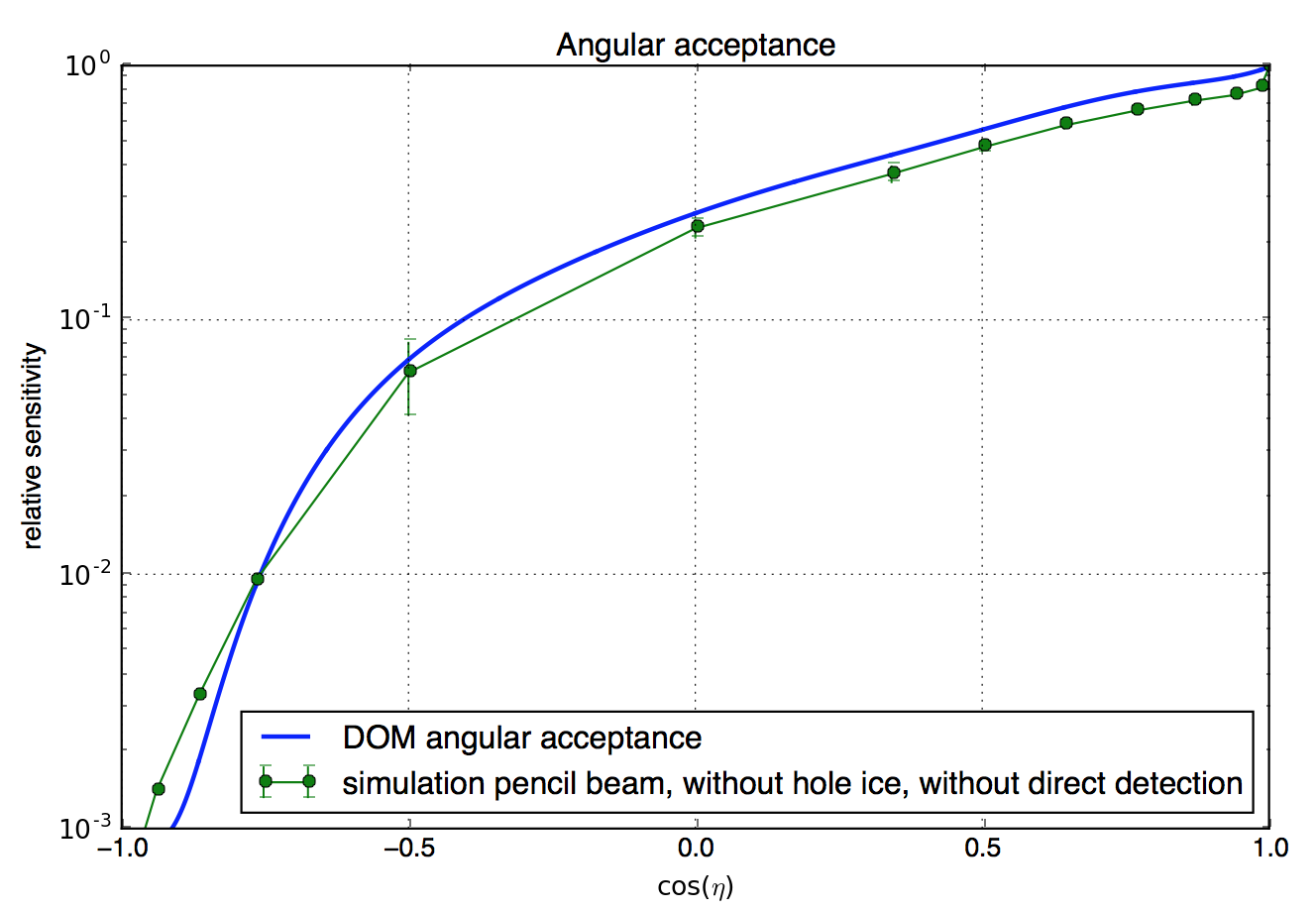}}\hfill
  \subcaptionbox{plane waves, a priori angular acceptance}{\halfimage{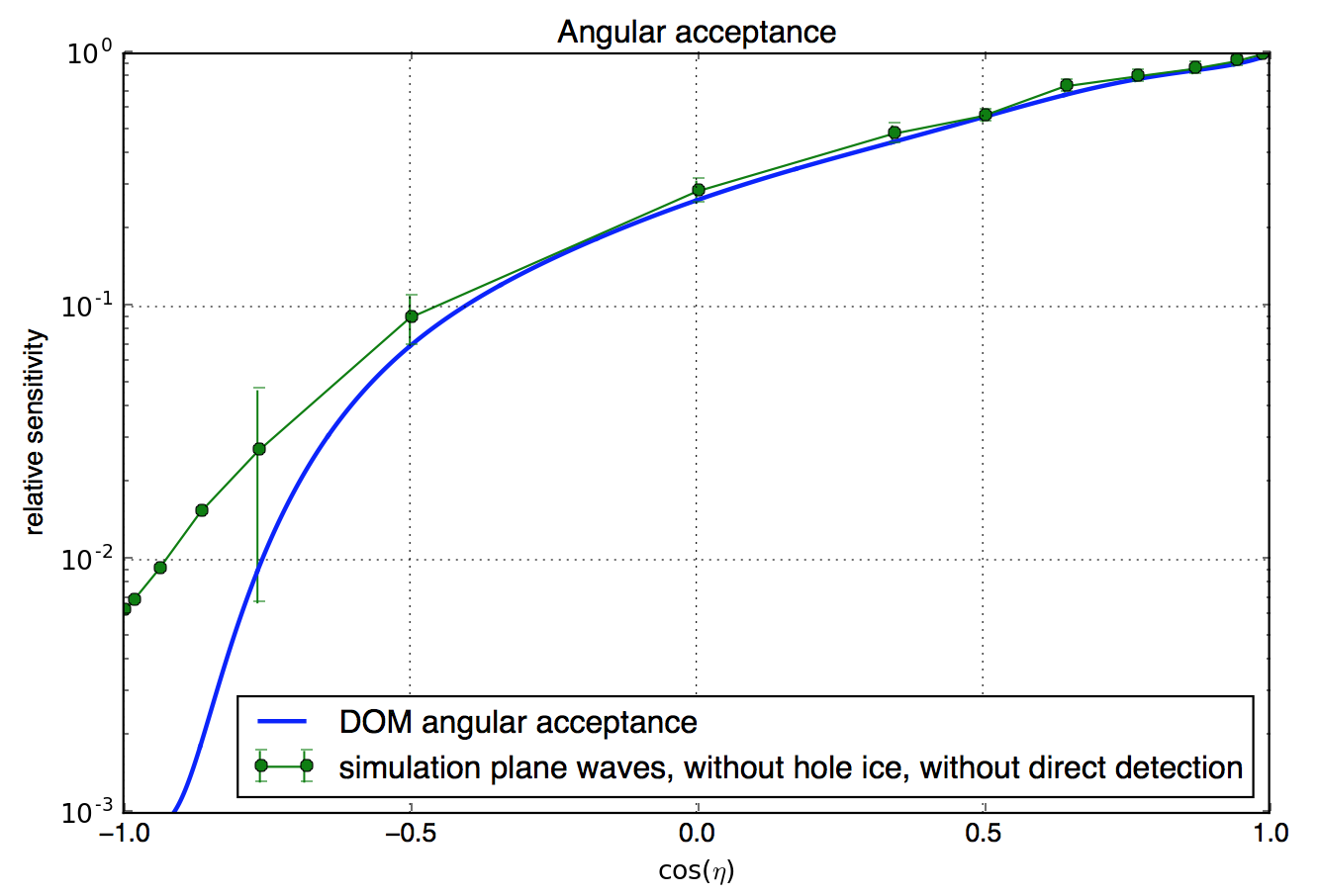}}
  \subcaptionbox{pencil beams, direct detection}{\halfimage{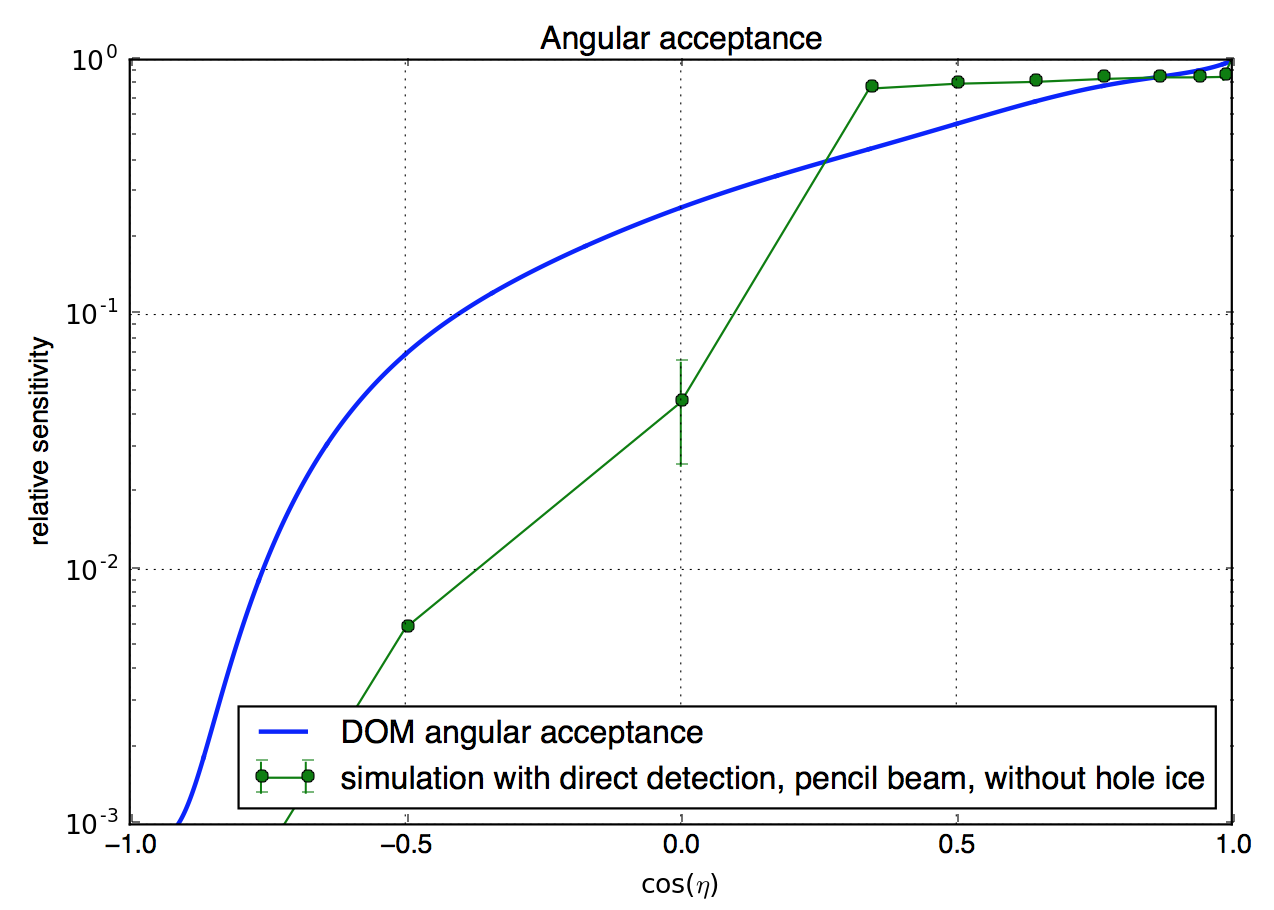}}\hfill
  \subcaptionbox{plane waves, direct detection}{\halfimage{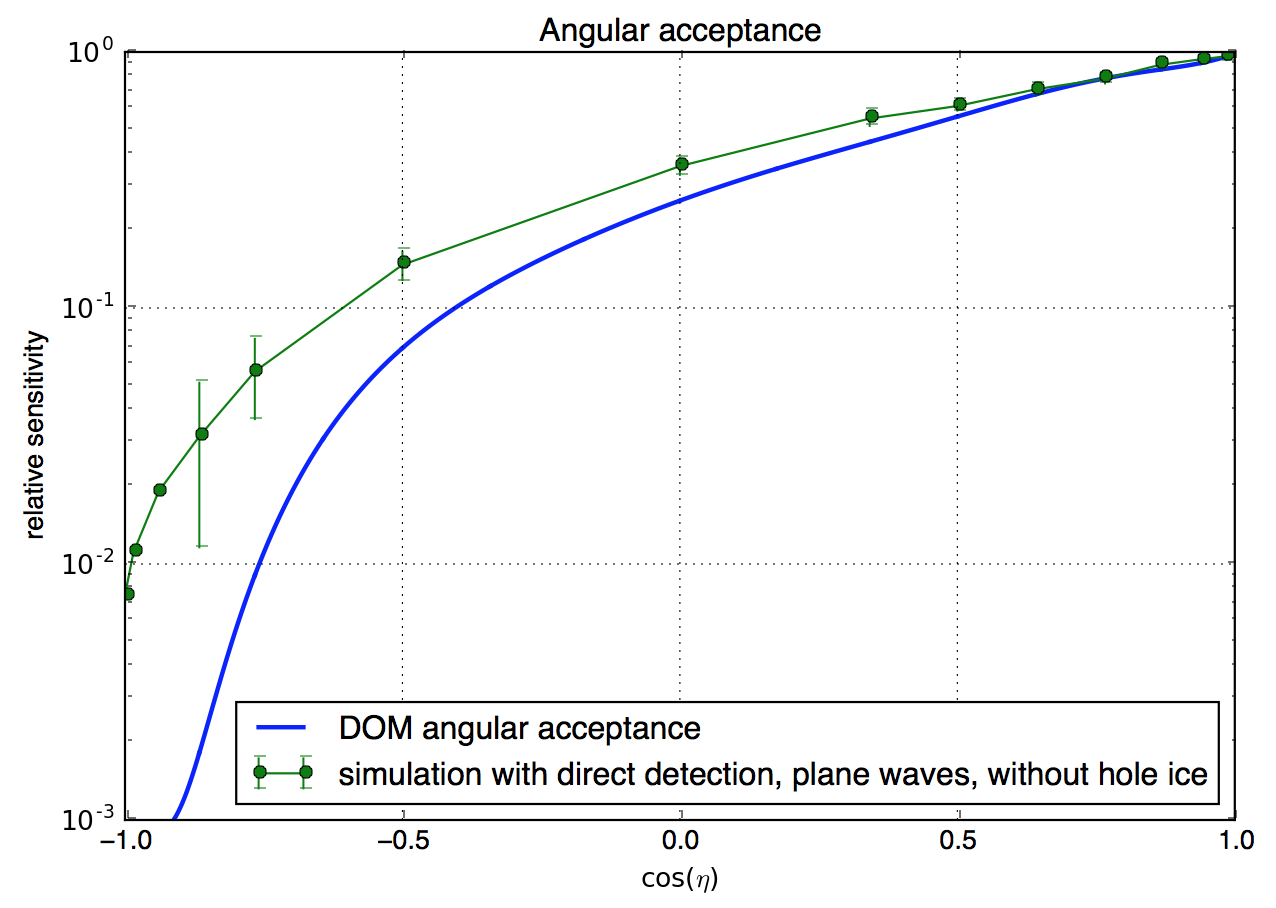}}\hfill
  \caption{Comparison of angular-acceptance-scan simulations with different approaches, all without hole ice. In (a), the simulation curve follows the a priori curve by design as the a priori curve is used as DOM acceptance criterion. With direct detection, (c) and (d), the a priori curve is not put into the simulation. The error bars are based on a binomial likelihood (section \ref{sec:parameter_scan}).}
  \label{fig:Shai8yah}
\end{figure}

Figure \ref{fig:Shai8yah} shows the results of the angular-acceptance
scans, comparing the above approaches, each also in comparison to the a
priori angular-acceptance curve \(a\dom(\eta)\) from \cite{icepaper}
(``nominal'' curve in figure \ref{fig:weihee6X}).

The first simulation, using pencil beams and the a priori angular
acceptance as acceptance criterion, figure \ref{fig:Shai8yah} (a),
follows the a priori angular acceptance curve \(a\dom(\eta)\) by design.
When deactivating scattering and absorption in the bulk ice entirely,
all photons would hit the optical module, and all photons would have
still the original direction. All non-angular acceptance criteria would
be absorbed in the scaling factor \(g\) and both curves, the simulation
curve and the a priori curve, would match exactly in the limit of many
started photons, \(N\rightarrow\infty\).

When switching from pencil beams to plane waves, figure
\ref{fig:Shai8yah} (b), more photons approaching the optical module from
above, in the region of \(\cos \eta = -1\), are accepted because there
are photons that start from above, but from a \(x\)-\(y\) position such
that they will fly by the DOM and then be scattered and hit the DOM with
an angle that is accepted by the DOM. This effect increases when
increasing the extent \(e\) of the plane waves.

When switching to direct detection as acceptance criterion, figure
\ref{fig:Shai8yah} (d), the a priori angular acceptance \(a\dom(\eta)\)
is no longer put into the simulation. Nevertheless, the simulation curve
and the a priori curve have roughly the same shape.
\rongen \cite{martindardupdate} argues that the dominant factor for the
shape of the angular-acceptance curve is determined by geometrical
considerations, which are reproduced by the simulation with direct
detection.

Further studies could investigate whether finding the proper plane
extent \(e\) and starting distance \(d\) would suffice to make both
curves match. Also, the bulk-ice scattering length, which is not
infinite in these simulations, could lead to a systematic error, which
should be investigated in follow-up
studies.\footnote{To check for progress on that matter, see \url{https://github.com/fiedl/hole-ice-study/issues/108}.}\followup

When switching to pencil beams with direct detection, figure
\ref{fig:Shai8yah} (c), the simulation essentially becomes a scan for
where the sensitive area of the photomultiplier tube ends in the optical
module (compare figure \ref{fig:kieQuoh1}).

\subsubsection{Angular-Acceptance Simulations With Hole Ice}
\label{sec:angular_acceptance_simulations_with_hole_ice}

\label{sec:hole_ice_approximation} With the new propagation algorithm,
angular-acceptance simulations can be performed, gathering the relative
frequencies \(\tilde{h}(\eta_i)\) for photons started under an angle
\(\eta_i\) being accepted as hits by the optical module, but this with a
hole-ice cylinder surrounding the target optical module.

In a first attempt, arbitrary properties for the hole-ice cylinder are
used, assuming a cylinder radius \(r:=r\dom\) being the same as the
radius \(r\dom\) of the optical module, and assuming an arbitrary
effective scattering length \(\lambda\hi\esca = 1\m\).

\docpar{These angular-acceptance simulations with hole ice are documented in \issue{99}.}

\begin{figure}[htbp]
  \subcaptionbox{pencil beams, a priori angular acceptance}{\halfimage{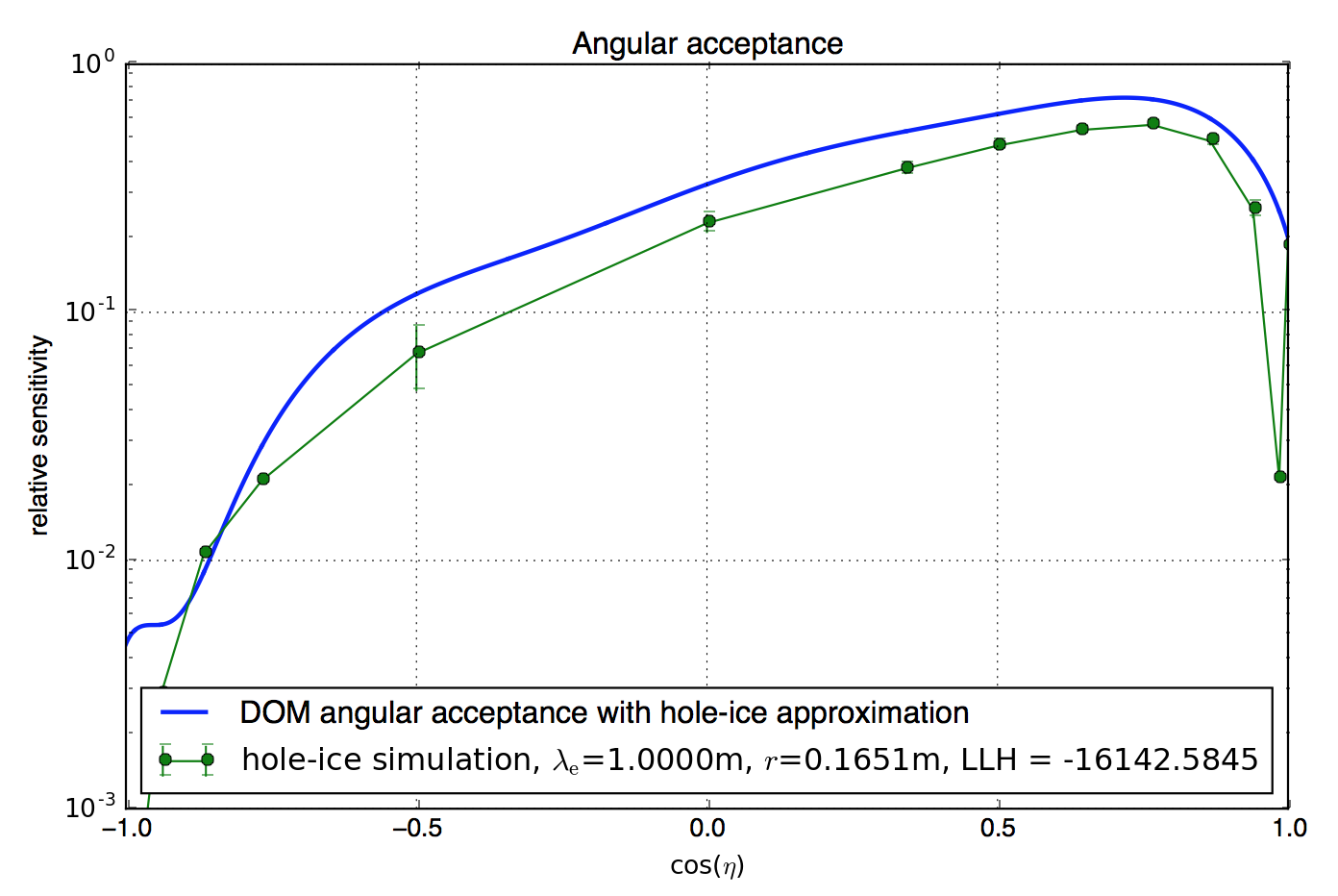}}\hfill
  \subcaptionbox{plane waves, a priori angular acceptance}{\halfimage{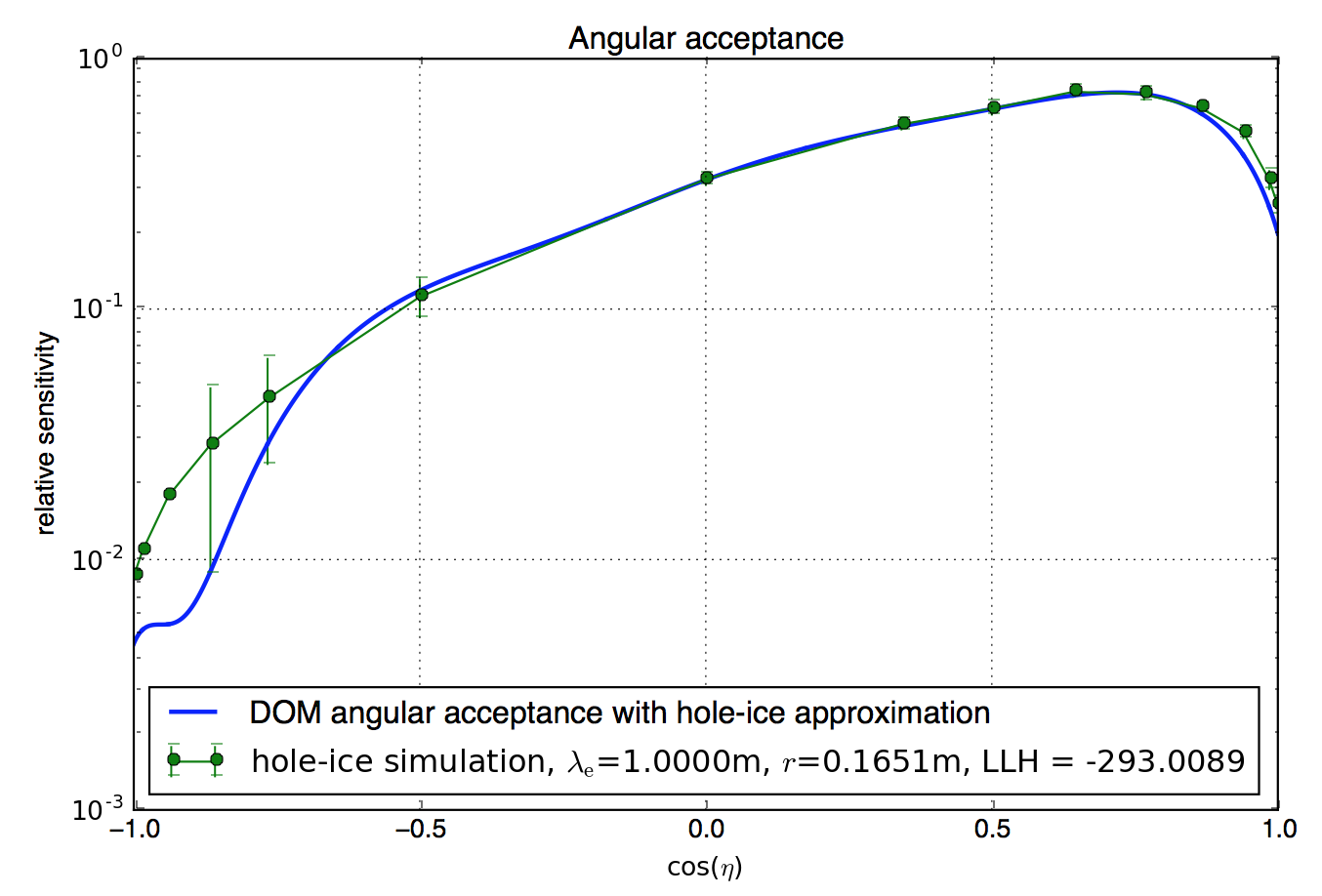}}\hfill
  \subcaptionbox{pencil beams, direct detection}{\halfimage{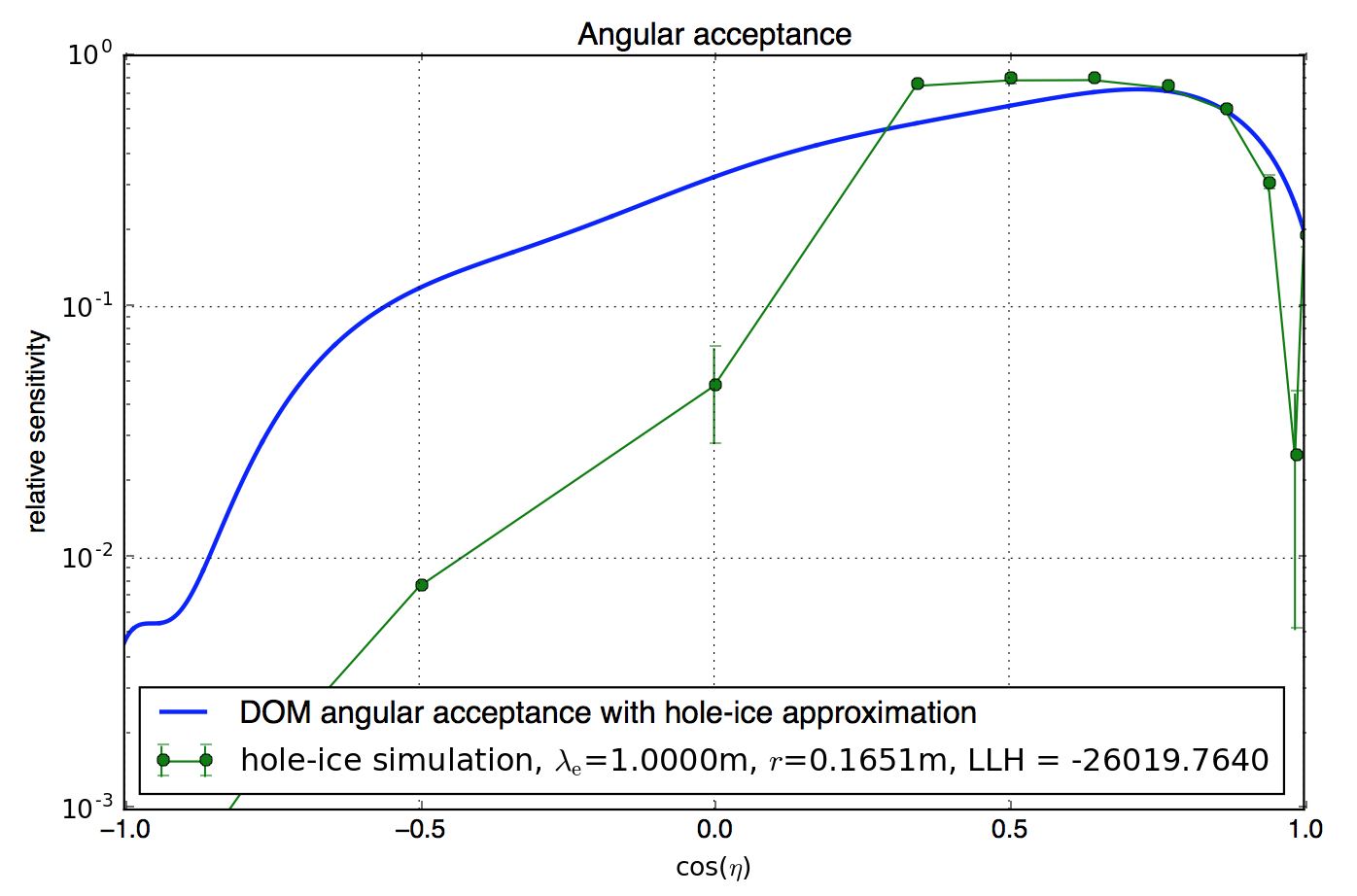}}\hfill
  \subcaptionbox{plane waves, direct detection}{\halfimage{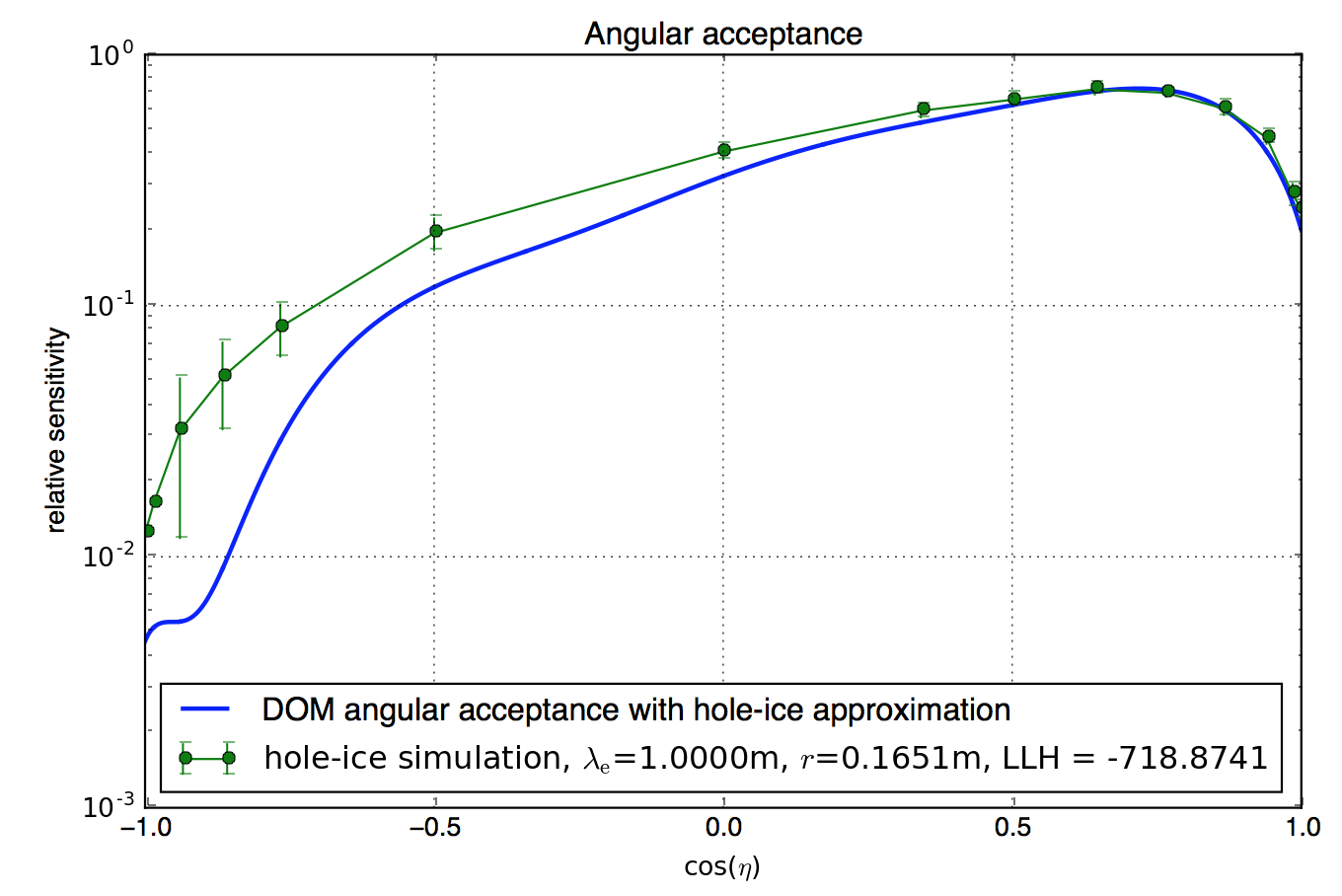}}\hfill
  \caption{Comparison of angular-acceptance-scan simulations with different approaches, all with a hole-ice cylinder with arbitrary properties, assuming an effective hole-ice scattering length of $\lambda\hi_\e=1\m$ and a hole-ice-cylinder radius of $r=r\dom$.}
  \label{fig:eVapie9t}
\end{figure}

Figure \ref{fig:eVapie9t} shows the results of the angular-acceptance
scans with hole ice, comparing the different approaches (direct
detection vs.~angular acceptance, and plane waves vs.~pencil beams),
each in comparison to the a priori effective angular-acceptance curve
\(a\domhi(\eta)\) from \cite{icepaper} that approximates the effect of
hole ice.

\begin{equation}
  a\domhi(\eta) = \sum_{j = 0}^{10} b_j\,\cos(\eta)^j, \ \ \ \eta \in [0; \pi]
  \label{eq:aholeice}
\end{equation}

\begin{gather*}
  b_0 = 0.32813, \ \ b_1 = 0.63899, \ \ b_2 = 0.20049, \\
  b_3 = -1.2250, \ \ b_4 = -0.14470, \ \ b_5 = 4.1695, \\
  b_6 = 0.76898, \ \ b_7 = -5.8690, \ \ b_8 = -2.0939, \\
  b_9 = 2.3834, \ \ b_{10} = 1.0435
\end{gather*}

When using plane waves as photon sources, figure \ref{fig:eVapie9t} (b)
and (d), the behavior of photons approaching the optical module from
below matches the a priori curve \(a\domhi(\eta)\) \cite{icepaper}
already reasonably well for these hole-ice parameters.

The next sections \ref{sec:vary_radius} and \ref{sec:vary_sca} show
effective angular-acceptance curves from simulations with different
hole-ice parameters.

\subsubsection{Varying the Hole-Ice-Cylinder Radius in Simulations}
\label{sec:vary_radius}

With the new medium-propagation algorithm (section
\ref{sec:algorithm_b}), the hole-ice parameters can be varied.

\docpar{Implementing angular-acceptance simulations for different hole-ice-cylinder radii is documented in \issue{82}.}

\begin{figure}[htbp]
  \subcaptionbox{Varying the radius of the hole-ice cylinder.}{\halfimage{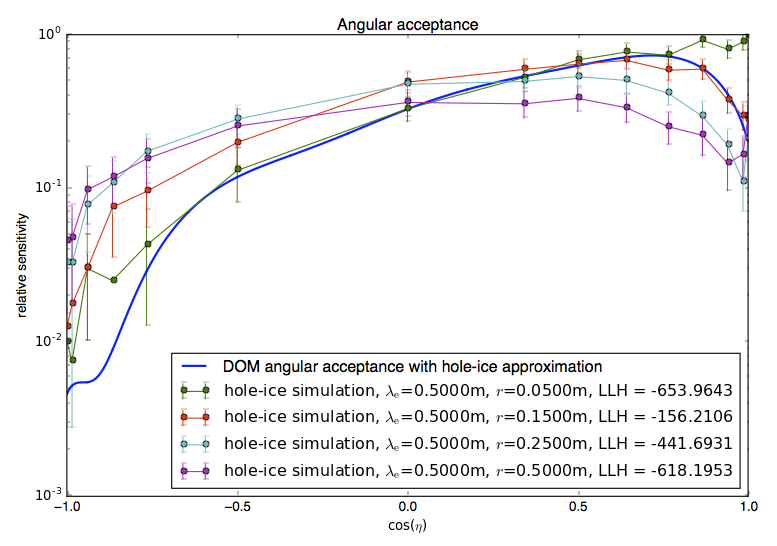}}\hfill
  \subcaptionbox{Varying the scattering length of the hole ice.}{\halfimage{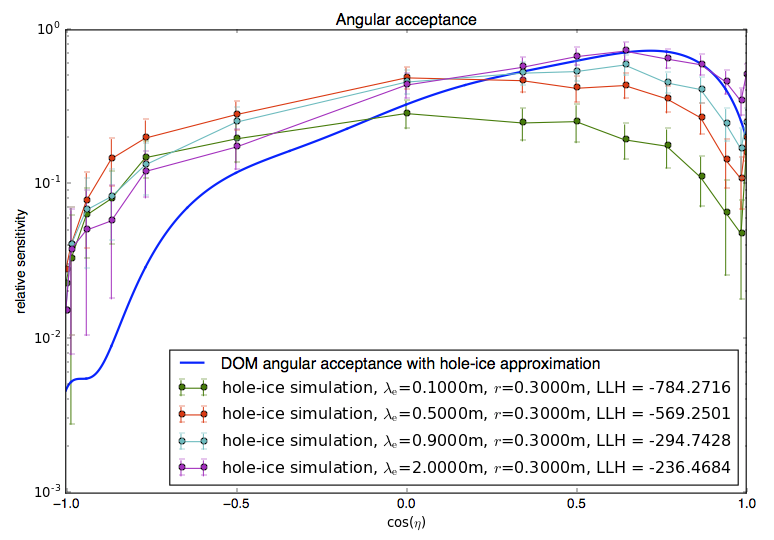}}\hfill
  \caption{Comparison of angular-acceptance simulations with different hole-ice parameters. The blue curve shows the a priori angular-acceptance curve $a\domhi(\eta)$ from \cite{icepaper} that approximates the effect of the hole ice. LLH gives the log-likelihood value of comparing the simulation curve to the a priori curve using a binomial likelihood function (see section \ref{sec:parameter_scan}).}
  \label{fig:neiyi3Al}
\end{figure}

Figure \ref{fig:neiyi3Al} (a) shows angular-acceptance curves from
simulations with different hole-ice-cylinder radii. In these
simulations, the hole-ice cylinder's center is set to be the center of
the optical module. The effective scattering length of the hole ice is
fixed, \(\lambda\esca\hi = 50\cm\).

When the hole-ice-cylinder radius is larger in the simulations, the
hole-ice effect is stronger because the ice volume occupied by the
hole-ice cylinder is larger, making it more likely that photons
approaching the detector module scatter within the hole ice. When
increasing the hole-ice-cylinder radius, photons approaching the optical
module from above (left-hand side of figure \ref{fig:neiyi3Al} a) are
increasingly scattered ``around'' the optical module such that they hit
the optical module in the sensitive area at the bottom, resulting in
more accepted hits. Photons coming from below (right-hand side of figure
\ref{fig:neiyi3Al} a) are increasingly scattered away before reaching
the optical module when increasing the hole-ice-cylinder radius,
resulting in less hits.

\subsubsection{Varying the Hole-Ice Scattering Length in Simulations}
\label{sec:vary_sca}

In another series of simulations, the scattering length of photons
propagating through the hole-ice cylinder is varied.

\docpar{Implementing angular-acceptance simulations for different hole-ice scattering lengths is documented in \issue{83}.}

Figure \ref{fig:neiyi3Al} (b) shows angular-acceptance curves from
simulations with different hole-ice scattering lengths. In these
simulations, the hole-ice cylinder's center is set to be the center of
the optical module. The radius of the hole-ice cylinder is fixed to
\(r = 30\cm\).

When the scattering length within the hole-ice cylinder is shorter,
photons propagating through the cylinder scatter more often. For smaller
scattering lengths, photons approaching the optical module from below
(right-hand side of figure \ref{fig:neiyi3Al} b) are more likely to be
scattered away in the hole ice before reaching the optical module,
resulting in less hits. On the left-hand side of figure
\ref{fig:neiyi3Al} (b), where the photons are approaching the optical
module from above, the effect of varying the scattering length is less
prominent as compared to varying the cylinder radius (\ref{fig:neiyi3Al}
a). The detection of photons approaching the optical module from above
requires that photons flying by the optical module are scattered into
the sensitive area of the module. For a plane wave of photons
approaching the module from above, this is much more likely when
increasing the hole-ice radius as compared to shortening the scattering
length.

Both series of angular-acceptance simulations, varying the hole-ice
scattering length and varying the hole-ice-cylinder radius, confirm
qualitatively the expected hole-ice effects (section
\ref{sec:hole_ice_effects}).

In agreement with figure \ref{fig:neiyi3Al}, \rongen \cite{pocam}
suggests that the scattering length of the hole ice determines the
position of the maximum of the angular-sensitivity curve, the hole-ice
radius dominantly determines the strength of the reduction of the
sensitivity in the region of \(\cos \eta \approx 1\).

Additional angular-acceptance simulations using hole-ice parameters
suggested by other studies, will be presented in section
\ref{sec:angular_acceptance_comparison}.

  \subsection{Determining the Hole-Ice Parameters Corresponding to the Current Hole-Ice Approximation}
\label{sec:parameter_scan}

In current \clsim simulations that do not use the new medium-propagation
algorithm introduced by this study, an effective angular-acceptance
curve \(a\domhi(\eta)\) for optical modules is used to approximate the
effect of the hole ice on the detection of photons (section
\ref{sec:hole_ice_approximation}).

The hole-ice properties that have been assumed to create the effective
angular-acceptance curve \(a\domhi(\eta)\) are a hole-ice-cylinder
radius of \(30\cm\) and a geometric hole-ice scattering length of
\(\lambda\sca\hi = 50\cm\) (\textit{H2 model})
\cite{holeicestudieswithyag}. These properties can be used in a
\clsim simulation using the new medium-propagation algorithm and direct
propagation through the hole ice in order to compare the effective
angular-acceptance curve resulting from the simulation to the
approximation curve currently in use.

\docpar{This simulation using the H2 parameters is documented in \issue{80}.}

\begin{figure}[htbp]
  \centering
  \includegraphics[width=0.75\textwidth]{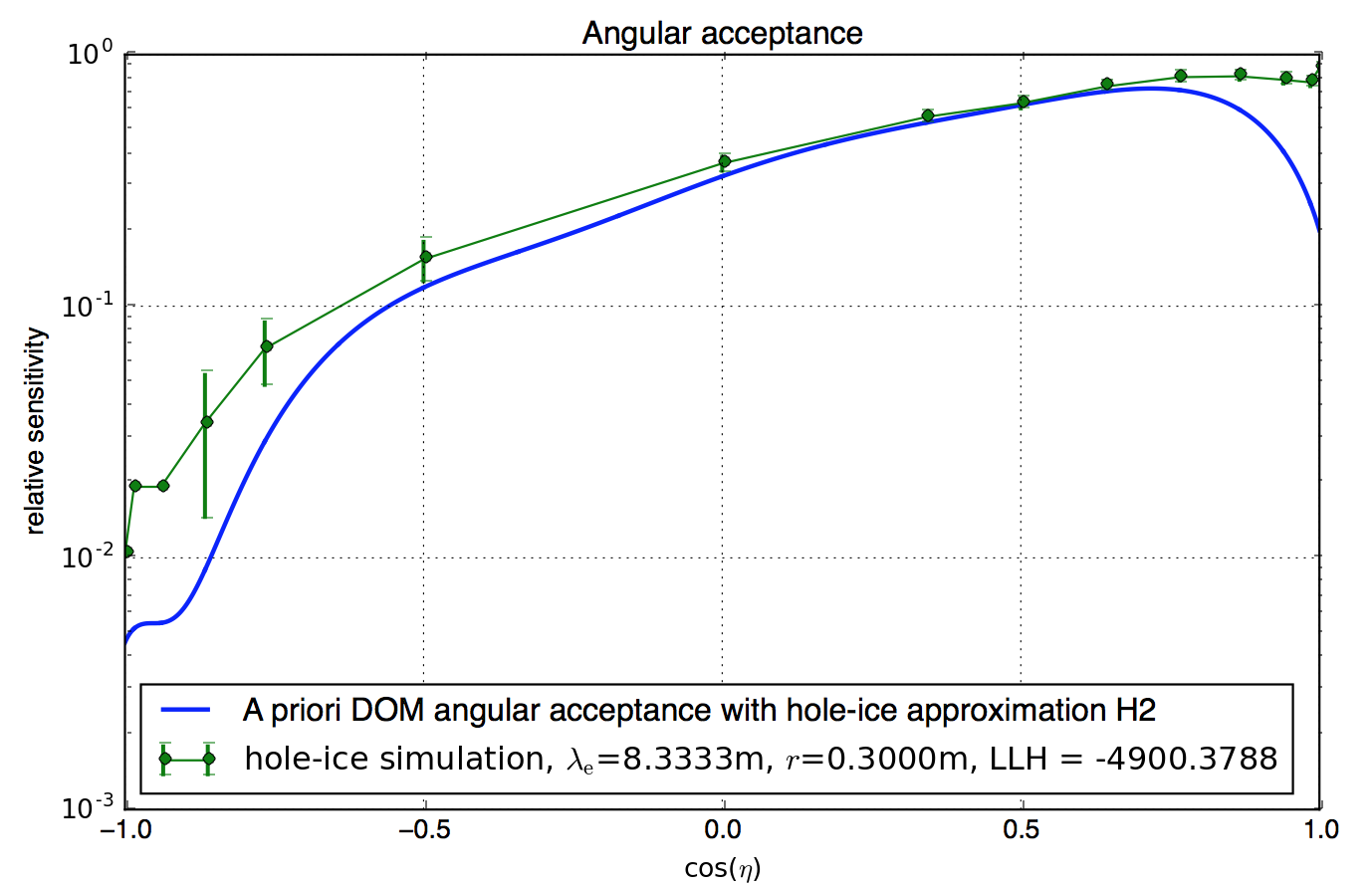}
  \caption{Comparing the effective angular-acceptance curve currently in use for approximating the effect of hole ice on the detection of photons in simulations to a simulation with direct photon propagation through hole ice using the same hole-ice parameters, hole-ice-cylinder radius $r=30\cm$, geometric hole-ice scattering length $\lambda\sca\hi = 50\cm$. LLH is the log binomial likelihood comparing both curves.}
  \label{fig:chie4Ite}
\end{figure}

In the simulation, direct detection is used as acceptance criterion,
plane waves with an extend of \(e=1\m\) and a starting distance of
\(d=1\m\) are used. The hole-ice absorption length is set to a high
fixed value, \(\lambda\abs\hi = 100\m\). Figure \ref{fig:chie4Ite} shows
the result of this simulation. The angular-acceptance curve resulting
from the simulation does not match the approximation curve currently in
use very
well.\footnote{To make sure this is no matter of confusing effective scattering length and geometric scattering length, the same simulation has been performed with an effective scattering length of $50\cm$ rather than a geometric scattering length of $50\cm$. These curves match even worse. See appendix \ref{sec:angular_acceptance_simulation_for_h2_parameters}.}

By performing a grid scan (see section
\ref{sec:cluster_parallelization}) over a range of hole-ice parameters,
one can find the best match with the approximation curve currently in
use, in order to understand which kind of direct-propagation hole ice
best matches the hole ice currently assumed in all simulations using the
approximation curve.

\docframe{
\docparwithoutframe{This grad scan is documented in \issue{12}.}\medskip

\sourceparwithoutframe{A script to configure and perform this kind of parameter scan is provided in \script{ParameterScan}.}
}

In principle, one could scan over a wide variety of parameters,
including the scattering and absorption lengths of the hole ice, the
hole-ice-cylinder radius, the distance from which the photons are
started towards the optical module, the kind of the photon source
(pencil beams, or plane waves), and the extent of the photon source
plane. In order to keep the computational effort minimal, however, all
conditions are kept the same as above, and only hole-ice-cylinder radius
and hole-ice scattering length are varied in the grid scan.

To compare the simulation results to the approximation curve, a binomial
likelihood \(L\) is used.

\begin{equation}
  L = \prod_{i=1}^N \binom{n}{k_i}\,p(\eta_i)^{k_i}\,(1-p(\eta_i))^{n-k_i}
\end{equation}

This likelihood can be interpreted as the probability that for a number
of \(N\) simulations, one simulation for each angle \(\eta_i\), starting
\(n\) photons from the angle \(\eta_i\) towards the optical module,
\(k_i\) photons will be accepted as hit by the optical module if the
acceptance probability \(p(\eta_i)\) for the angle \(\eta_i\) is given
by the approximation curve \(a\domhi(\eta)\),
\(p(\eta_i) = \sfrac{1}{g}\,a\domhi(\eta_i)\). \(g\) is a scaling factor
(see section \ref{sec:gauging}), which is needed due to the
normalization of \(a\domhi(\eta)\).

As one series of simulations is performed for each set \(\HH\) of
hole-ice parameters, there is a likelihood \(L_\HH\) for each parameter
set \(\HH\), which depends on the numbers \(k_{i,\HH}\) of photon hits
for photons started under the angle \(\eta_i\), propagated with hole-ice
parameters \(\HH\).

\[
  L_\HH = \prod_{i=1}^N \binom{n}{k_{i,\HH}}\,p(\eta_i)^{k_{i,\HH}}\,(1-p(\eta_i))^{n-k_{i,\HH}}
\]

The simulation resulting in the maximal likelihood \(L_\HHbest\)
corresponds to the set of hole-ice parameters \(\HHbest\) that best
describe the approximation curve \(a\domhi(\eta)\).

\begin{figure}[htbp]
  \smallerimage{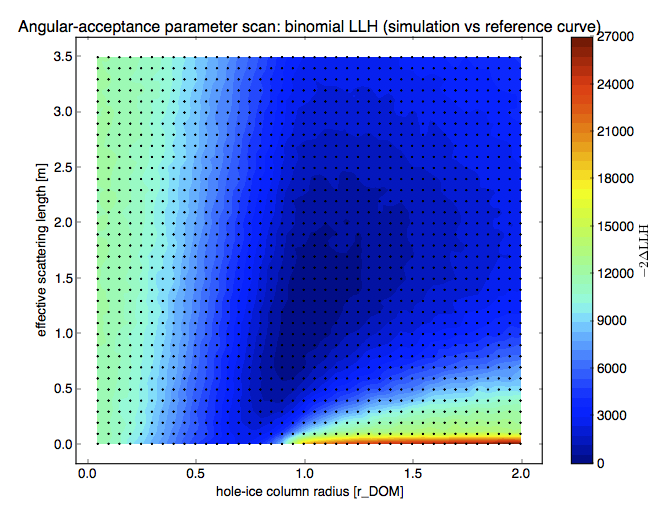}
  \caption{Agreement of the fixed hole-ice-approximation angular-acceptance curve $a\domhi(\eta)$ and direct hole-ice simulations using the new \clsim medium-propagation algorithm with hole-ice-cylinder radii $r$ given in units of the radius $r\dom$ of the optical module, and effective hole-ice scattering lengths $\lambda\esca\hi$. Each dot in the plot represents one parameter set $\HH$ and one corresponding set of simulations.}
  \label{fig:AWa5aiCh}
\end{figure}

Figure \ref{fig:AWa5aiCh} shows the result of the parameter grid scan,
plotting the likelihood, or rather
\(-2\Delta\llh:= -2(\ln L_\HH - \ln L_\HHbest) = -2 \ln \left(\sfrac{L_\HH}{L_\HHbest}\right)\)
as a more common measure of agreement, against the effective scattering
length \(\lambda\esca\hi\) of the hole ice on the one axis, and the
radius \(r\) of the hole-ice cylinder on the other axis. For the optimal
parameters \(\HHbest\), \(\Delta\llh\) is zero.

\begin{figure}[htbp]
  \subcaptionbox{Simulation with hole-ice radius $r=1.0\,r\dom$ and effective hole-ice scattering length $\lambda\esca\hi = 1.3\m$.}{\halfimage{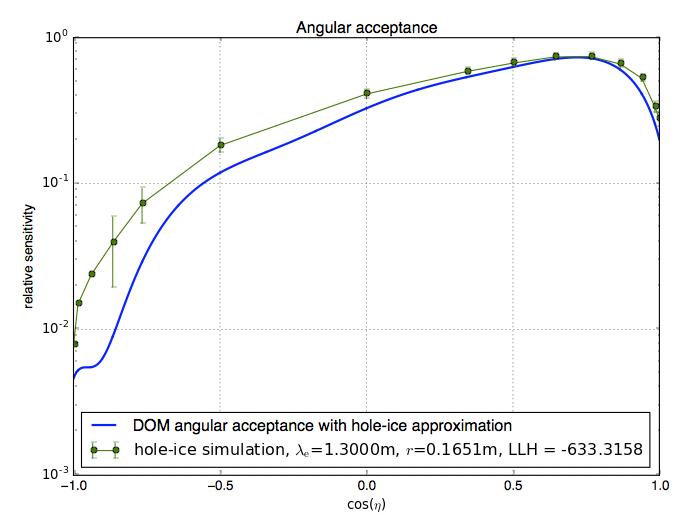}}\hfill
  \subcaptionbox{Simulation with hole-ice radius $r=1.0\,r\dom$ and effective hole-ice scattering length $\lambda\esca\hi = 1.0\m$.}{\halfimage{parameter-scan-1-1-angular}}
  \caption{Angular-acceptance curves from simulations best matching the hole-ice-approximation angular-acceptance curve $a\domhi(\eta)$. The LLH is the log binomial likelihood for the agreement of the simulation curve and the reference curve.}
  \label{fig:weShir8i}
\end{figure}

The hole-ice parameters \(\HHbest\) that lead to a simulation, which
results in an angular-acceptance curve that best agrees with the
approximation curve \(a\domhi(\eta)\), are a hole-ice-cylinder radius
\(r = 1.0\,r\dom\) where \(r\dom\) is the radius of an optical module,
and an effective hole-ice scattering length \(\lambda\esca\hi = 1.3\m\).
Figure \ref{fig:weShir8i} shows an angular-acceptance curve for this
parameter set, as well as for the nearby parameter set of
\(r=1.0\,r\dom, \lambda\esca\hi = 1.0\m\).

For photons approaching the optical module from below (right-hand side
of the plots), the simulations follow the approximation curve. For lower
angles, the simulation sees more photon hits, which, however, is the
case for all simulations in this grid scan.

To summarize this result, choosing a hole-ice cylinder of the same size
as the optical module, and an effective scattering length of about
\(1\m\) for a direct hole-ice simulation, comes closest to using the
hole-ice-approximation angular-acceptance curve, which is currently
default for \clsim simulations.

  \subsection{Simulating the Displacement of Optical Modules Relative to the Hole-Ice Columns}
\label{sec:cylinder_shift}

Optical modules do not necessarily need to be positioned well-centered
relative to the hole ice. In practice, when the drill hole is
refreezing, the optical module may be deposited displaced both relative
to the drill hole as well as to the bubble column.

\begin{figure}[htbp]
  \subcaptionbox{\steamshovel display of the simulation scenario for $\eta = \pi$. The photons are started from a plane on the left-hand side of the image.}{\halfimage{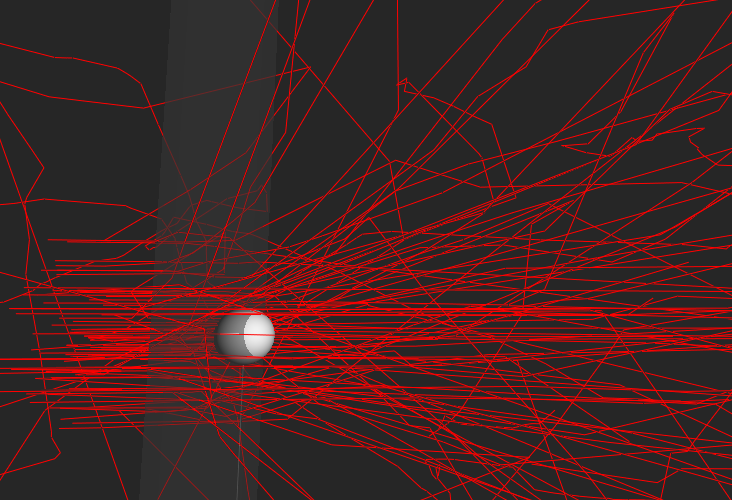}}\hfill
  \subcaptionbox{Effective angular acceptance resulting from this simulation. One simulation curve shows the acceptance of photons arriving from a polar angle $\eta \in [0;\pi[$, the other simulation curve shows the acceptance of photons arriving from $\eta \in [\pi; 2\pi[$. The red curve shows the H2 hole-ice-approximation angular-acceptance curve from \cite{icepaper}.}{\halfimage{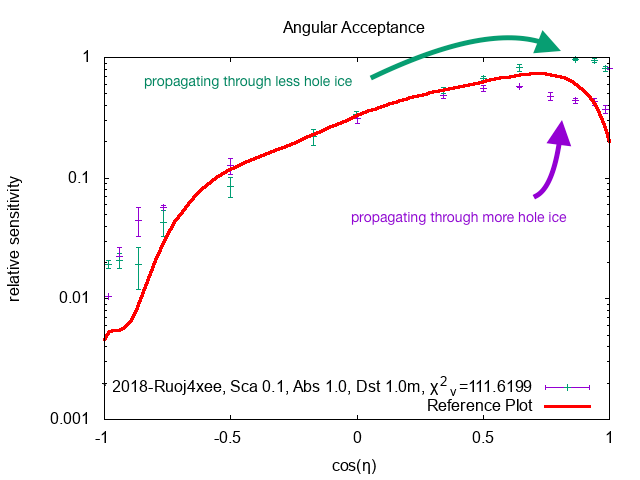}}
  \caption{Simulation of a hole-ice cylinder, which is shifted relative to the position of the optical module. The optical module is shifted beyond the cylinder border such that it is partly within and partly outside of the cylinder. Photons approaching the optical module from one direction, $\eta = \pi$, need to travel through the maximal distance through the hole ice, photons from the opposite direction, $\eta = -\pi$, hit the optical module before reaching the hole-ice cylinder.}
  \label{fig:egieNg5l}
\end{figure}

The hole-ice effect on the detection of photons by an optical module
that is not symmetrically positioned relative to the hole ice cannot be
modeled by an effective angular-acceptance curve because these curves
always assume azimuthal symmetry. Nevertheless, the the asymmetry of the
effect can be visualized using an angular-acceptance simulation,
plotting the acceptance for polar angles \(\eta \in [0;\pi[\) and for
polar angles \(\eta \in [\pi; 2\pi[\) in different colors in the same
plot (figure \ref{fig:egieNg5l} b).

To model the displacement of optical modules relative to the hole ice in
production simulations, both the positions of the optical modules and
the positions of the hole-ice cylinders, or even cylinder sections for
specific \(z\)-ranges can be configured independently.

In this example simulation, the cylinder is shifted relative to the
optical module while the photon sources, which start photons from
different directions towards the optical module, are rotated around the
optical module, not around the position of the cylinder.

\docpar{The simulation with a shifted hole-ice cylinder is documented in \issue{8}.}

In this example, which uses the hole-ice-correction algorithm (section
\ref{sec:algorithm_a}), the hole ice is modeled as cylinder with
\(30\cm\) radius and a hole-ice scattering length of \(\sfrac{1}{10}\)
of the scattering length of the surrounding bulk ice. The optical module
is shifted by \(20\cm\) from the central position, and therefore is
shifted beyond the border of the hole ice in order to demonstrate an
extreme effect. In practice, the optical module can only be shifted by a
maximum of about \(15\cm\) from the central position, because the drill
hole is only about \(60\cm\) in diameter.

Figure \ref{fig:egieNg5l} (a) shows the simulation scenario in
\steamshovel, figure \ref{fig:egieNg5l} (b) shows the resulting
angular-acceptance curves from the simulation, one for the angular range
\(\eta \in [0;\pi[\) and one for \(\eta \in [\pi; 2\pi[\). One of these
curves shows a more extreme hole-ice effect as more photons need to
propagate through the hole ice to hit the optical module in comparison
to the other curve.

\begin{figure}[htbp]
  \smallerimage{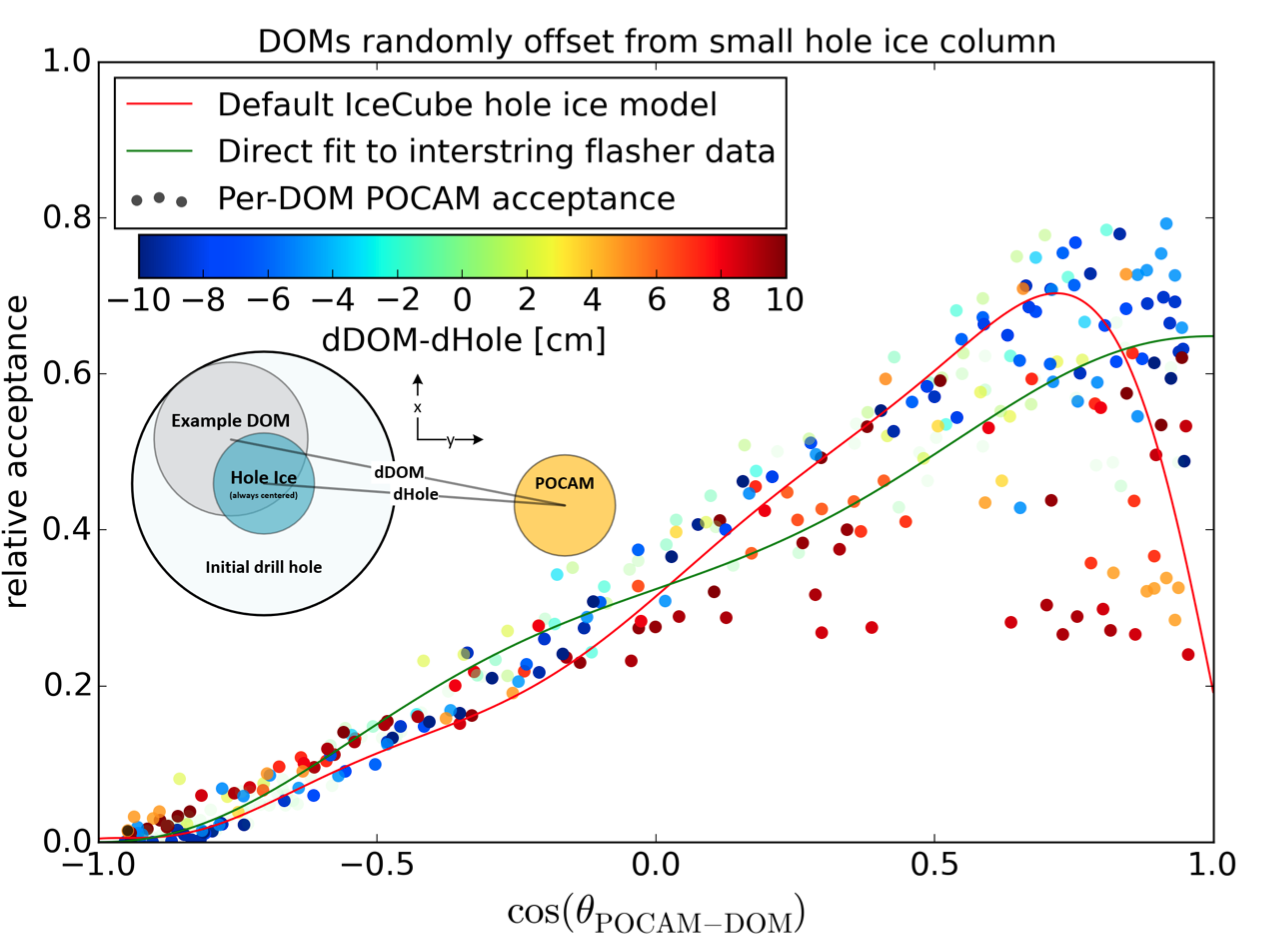}
  \caption{Results from a series of simulations where the position of the optical module is shifted randomly relative to the hole-ice column for each data point. The displacement of the module especially widens the spread of the angular-acceptance curve for all lower angles, $\cos \eta > 0$. The \enquote{Default IceCube hole ice model} refers to the a priori angular-acceptance curve with H2 parameters from \cite{icepaper}. The \enquote{Direct fit to interstring flasher data} refers to the model described in section \ref{sec:dimas_model}. Image source: \cite{icrc17pocam}}
  \label{fig:zao5Mah0}
\end{figure}

\rongen \cite{icrc17pocam} has performed a series of simulations where
the optical module is shifted randomly relative to the hole-ice column
for each data point. The results shown in figure \ref{fig:zao5Mah0}
indicate a wide spread of the effective acceptance curve for angles all
lower angles, \(\cos \eta > 0\).

  \subsection{Simulating Nested Hole-Ice Columns}
\label{sec:nested_cylinders}

Current descriptions of the hole ice characterize it as a relatively
clear outer region with a radius of the entire drill hole, and a central
column of about \(8\cm\) radius filled with air bubbles, resulting in a
small scattering length in this region.
\cite{instrumentation,icrc17pocam,rongenswedishcamera,martinspicehddard}

The new medium-propagation algorithm (section \ref{sec:algorithm_b})
allows to simulate both ice columns as independent cylinders. How to
arrange, shift, and size those cylinders, and how to configure their
scattering and absorption properties in production simulations, needs to
be determined in follow-up studies.\followup

As an example, the following simulation models two nested cylinders, an
outer cylinder of \(30\cm\) radius with a moderate geometric scattering
length of \(50\cm\) that resembles the drill hole, and an inner cylinder
of \(8\cm\) radius and a small geometric scattering length of \(1\cm\)
that represents the bubble column. With this configuration, the
simulation scans the effective angular acceptance of the optical module,
which is embedded in the center of the cylinders.

\docpar{The implementation of this simulation is documented in \issue{7}.}

\begin{figure}[htbp]
  \subcaptionbox{\steamshovel visualization of the simulation scenario. The outer column represents the entire drill hole. The inner column is filled with small air bubbles and is therefore called bubble column.}{\halfimage{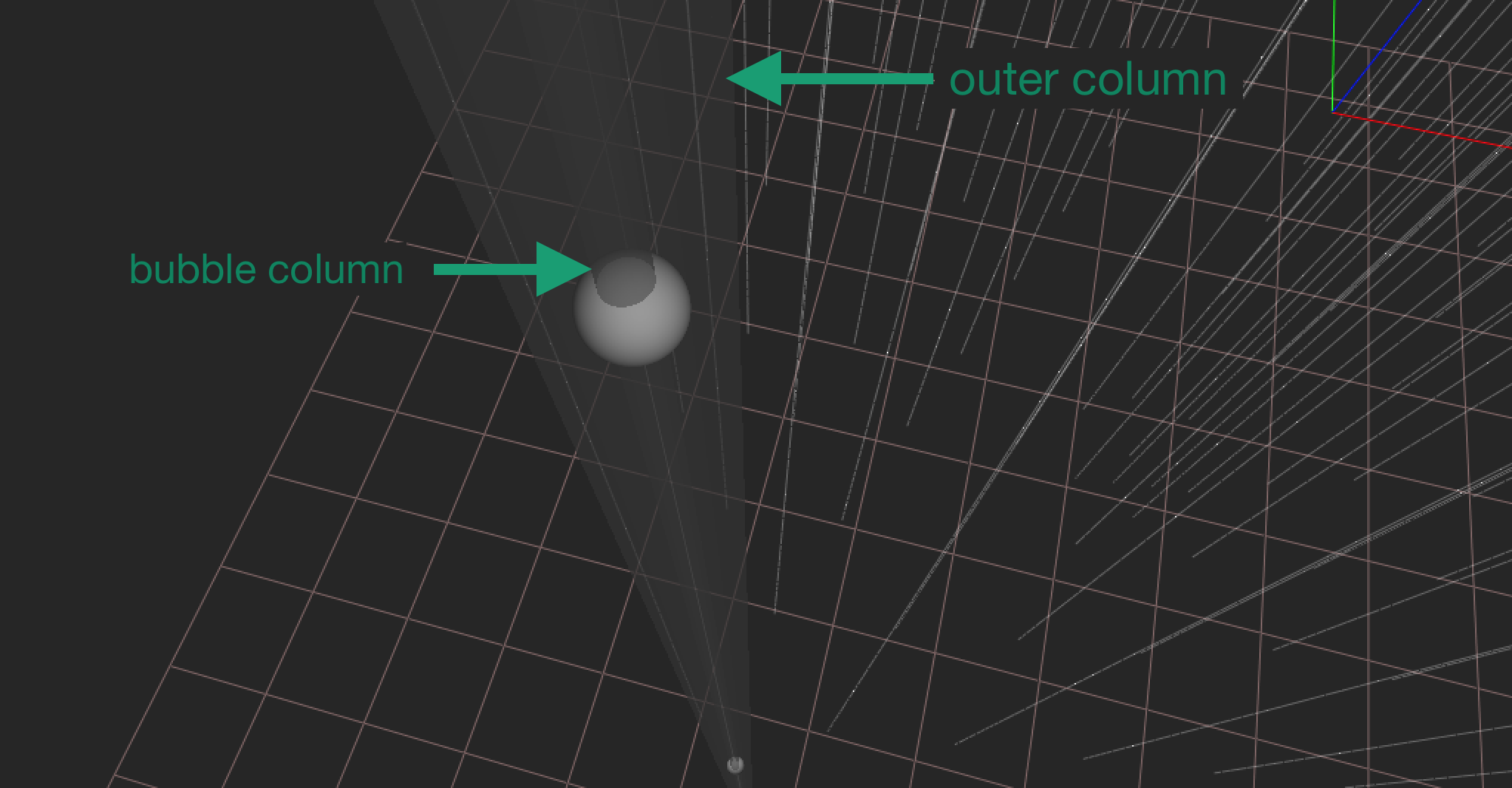}\vspace*{3mm}}\hfill
  \subcaptionbox{Effective angular acceptance of the optical module for different cylinder configurations, in the order of increasing hole-ice effect: Without hole-ice cylinders, with only a drill hole, with only a bubble column, and with drill hole and bubble column together.}{\halfimage{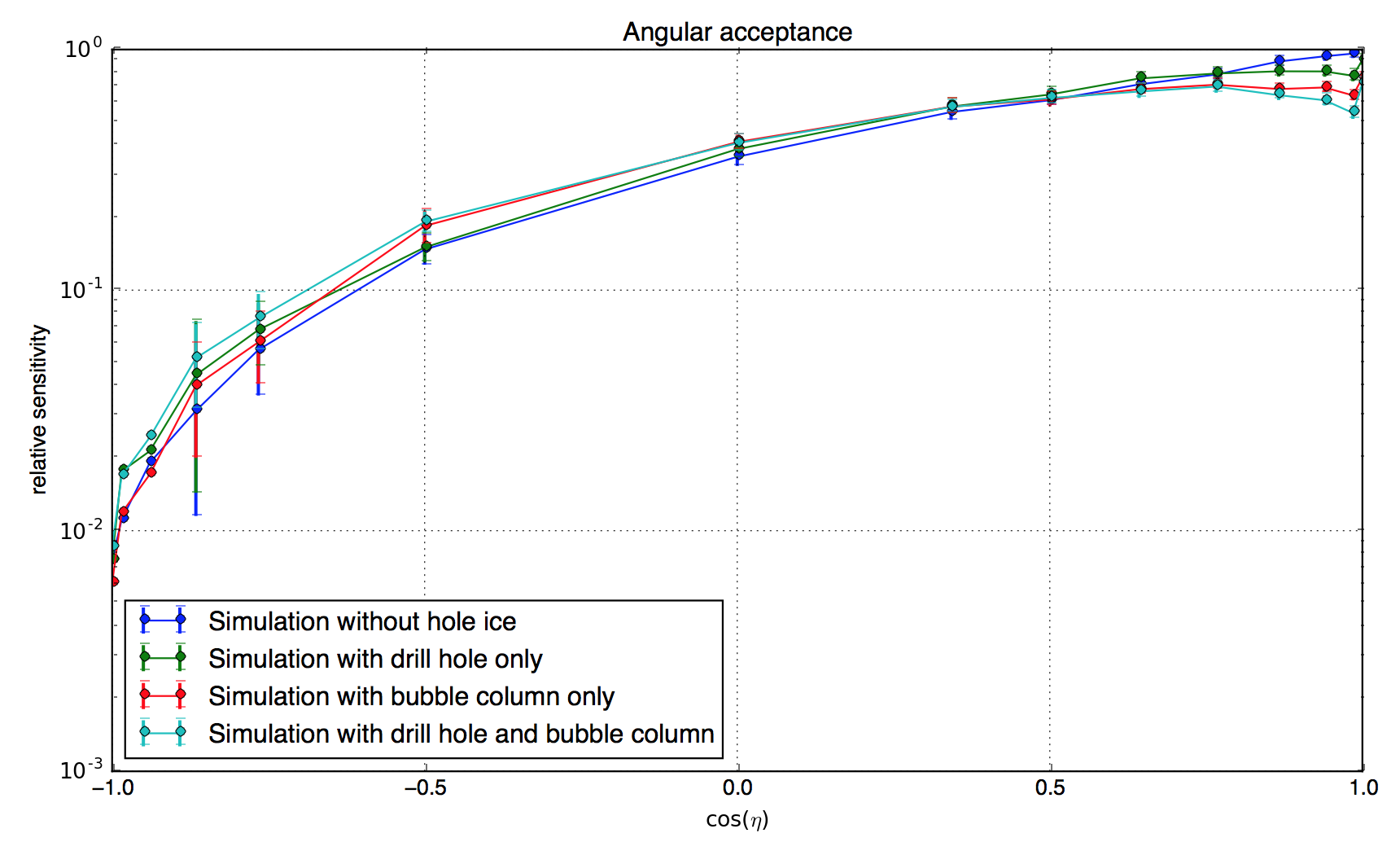}}
  \caption{Simulation of photon propagation through nested hole-ice cylinders.}
  \label{fig:haiv2IGi}
\end{figure}

The same simulation then is repeated, once with only the bubble column,
and once with only the drill hole. Figure \ref{fig:haiv2IGi} shows the
resulting effective angular-acceptance curves.

The simulation that implements both, drill hole and bubble column, shows
the strongest hole-ice effect. The next strongest hole-ice effect shows
the simulation that only implements the bubble column. The least
strongest effect shows the simulation that implements only the drill
hole. As reference, a simulation without any hole ice is shown.

Related studies favor different hole-ice models: A camera deployed in
the drill hole shows a diffuse ice volume, which only covers a part of
the module sphere, indicating the presence of a bubble column that is
not around the center of the optical modules but asymmetrically
arranged. \cite{instrumentation,rongenswedishcamera} LED measurements,
however, that measure the azimuthal dependencies of the ice properties
in the ice surrounding the optical modules, do not support the model of
a displaced bubble column that would effect the light from some of the
LEDs of the optical module but not all of them.
\cite{rongenswedishcamera}

The tool set provided by this study to recreate the different hole-ice
models with independent, displaced ice cylinders, allows follow-up
studies to compare the different models to flasher data, helping to
better understand the characteristics of the hole ice.\followup

  \subsection{Simulating Shadowing Cables as Opaque Cylinders}
\label{sec:cables}

\begin{figure}[htbp]
  \subcaptionbox{Optical module deployment schematics. Image source:  \cite{instrumentation}}{\includegraphics[width=0.25\textwidth]{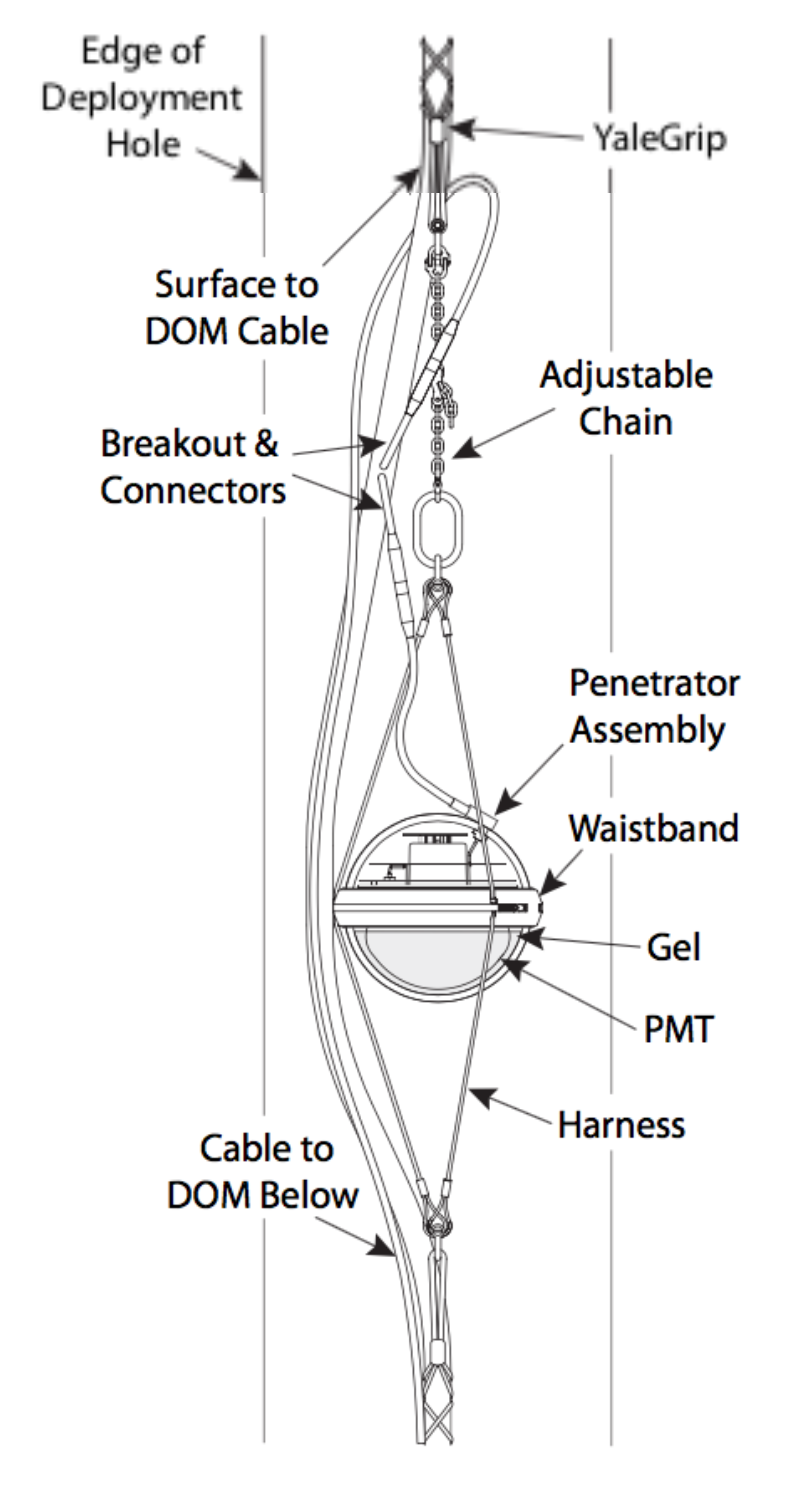}}\hfill
  \subcaptionbox{Rendered image of the optical module and the main cable. Image source: \cite{gallerydomcloseup}}{\includegraphics[width=0.35\textwidth]{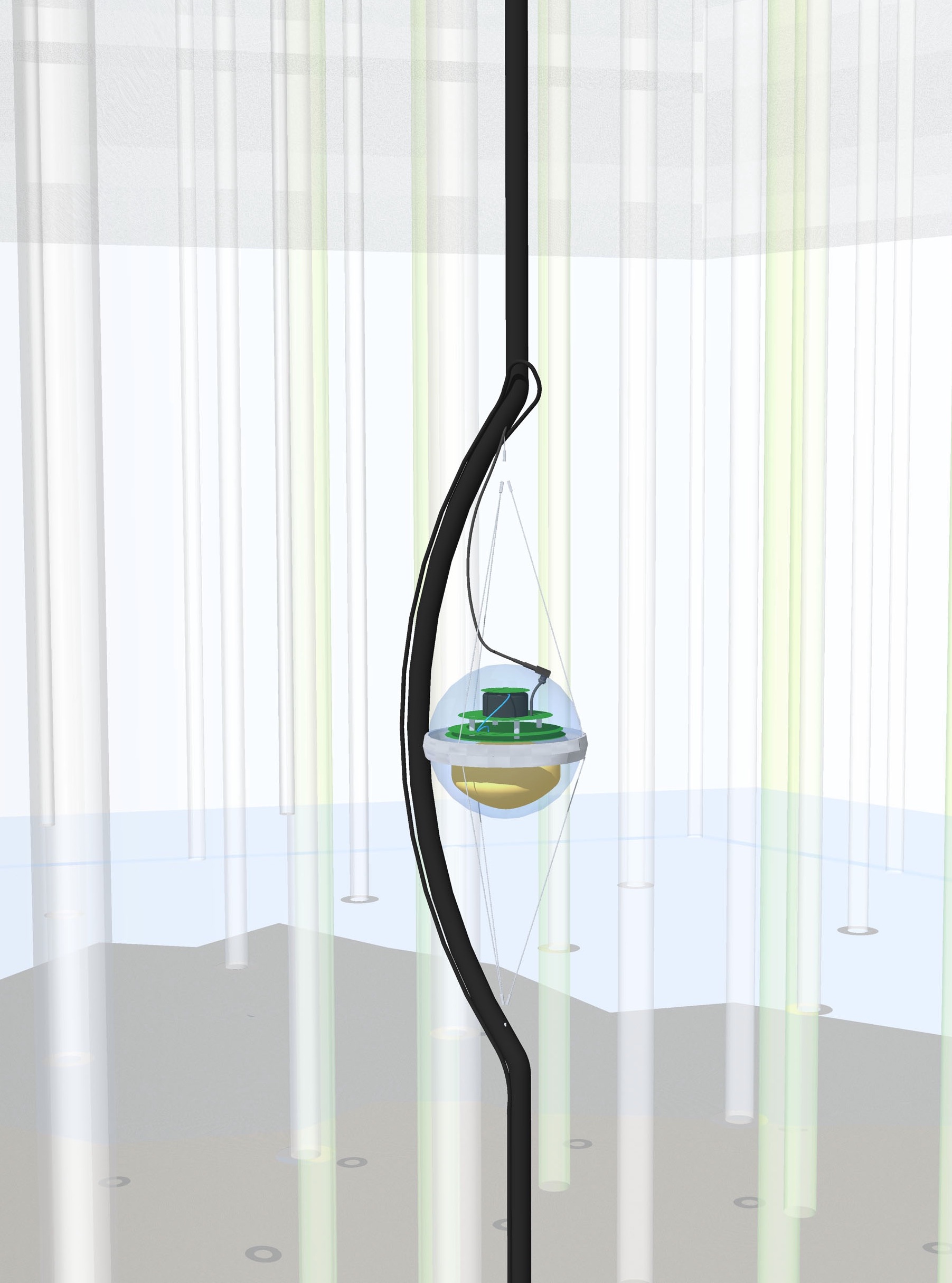}}\hfill
  \subcaptionbox{Photo: Optical module deployment. Image source: \cite{publicgallerydomdeployment}}{\includegraphics[width=0.35\textwidth]{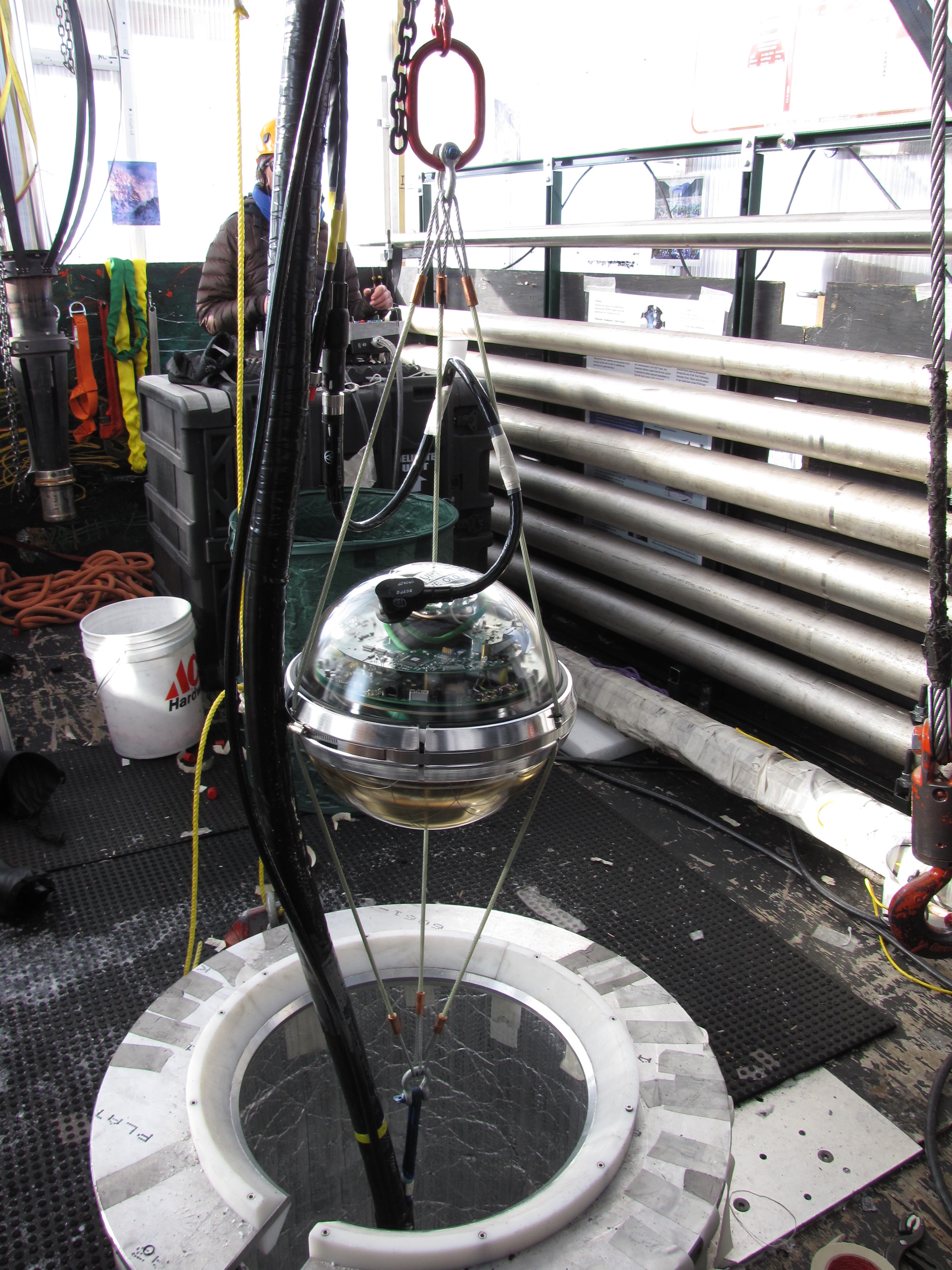}}
  \caption{\icecube's optical module in relation to the main cable.}
  \label{fig:ahyoi7Ma}
\end{figure}

The main cable of each detector string (figure \ref{fig:ahyoi7Ma})
allows to power the detector modules and to transfer data from the
modules to the surface. The cable has a diameter of \(46\mm\)
\cite{instrumentation} and, as it is located in close proximity to the
optical modules, shields the optical modules from a non-negligible
amount of incoming photons. The cable's mantle is black such that it
will absorb photons rather than reflect them.

\subsubsection{Asymmetric Shadowing Effect Caused by the Cable}

Using the new medium-propagation algorithm, the cable can be simulated
as one or several cylinders with limited \(z\)-range, which are
configured for instant absorption.

\begin{figure}[htbp]
  \subcaptionbox{\steamshovel visualization of the simulation scenario. Photons approaching from different angles are either shielded by the cable or can reach the optical module unhindered.}{\halfimage{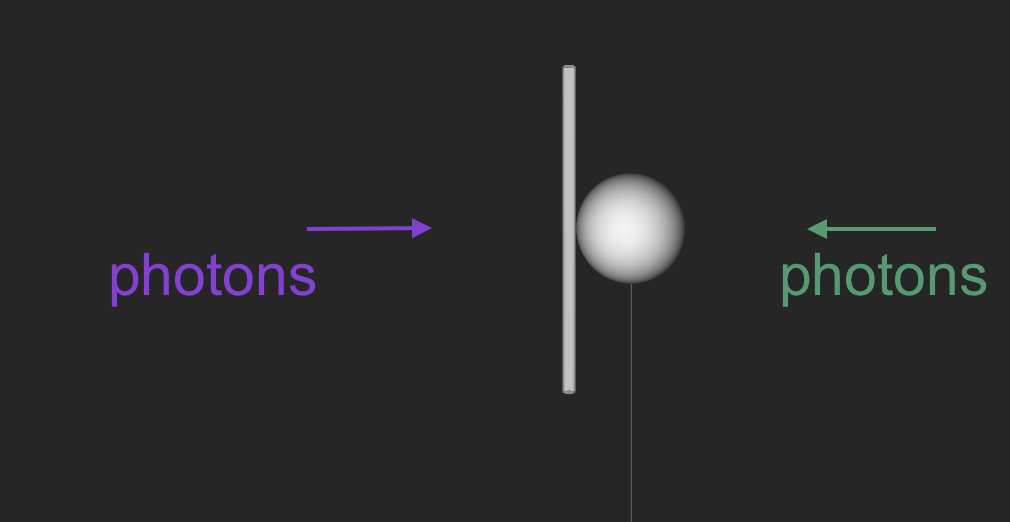}\vspace*{7mm}}\hfill
  \subcaptionbox{Effective angular acceptance of the optical module measured in the simulation. Photons approaching from the side of the cable are less likely to reach the optical module and be registered as a hit.}{\halfimage{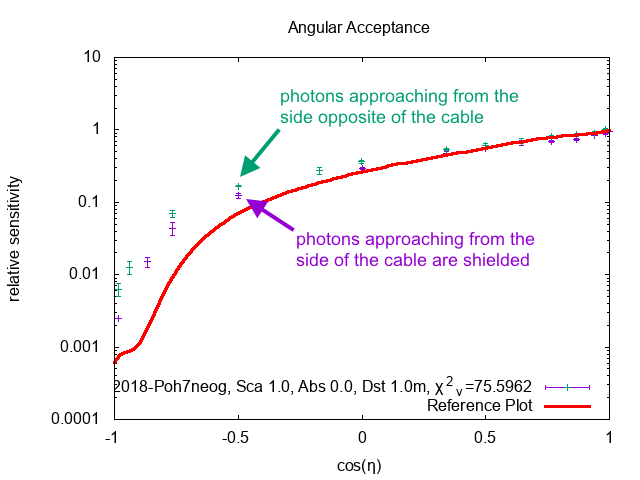}}
  \caption{Simulation with the main cable modeled as opaque cylinder placed besides the optical module.}
  \label{fig:ochoCh7o}
\end{figure}

As a first test, a simulation is performed to verify that the cable
causes an asymmetric shadowing effect, shielding photons approaching
from one direction while not shielding photons approaching from the
opposite direction. Figure \ref{fig:ochoCh7o} shows the effective
angular sensitivity of the optical module measured in this simulation
and confirms that the asymmetric shadowing can be observed in
simulations.

\docpar{This simulation, placing an opaque cylinder part besides the optical module, propagating the photons with the hole-ice-correction algorithm (section \ref{sec:algorithm_a}) and scanning the effective angular acceptance, is documented in \issue{35}.}

\subsubsection{Cable Inside the Bubble Column}

In one class of hole-ice models that has not been studied with
simulations, yet, the optical module is shifted such that the cable
resides fully or partially within the bubble column.
\cite{martinspicehddard}

\begin{figure}[htbp]
  \subcaptionbox{\steamshovel visualization of the simulation scenario. An opaque cable cylinder of $1\m$ height is placed besides the optical module, fully within the drill hole ice, and partially inside the bubble column.}{\halfimage{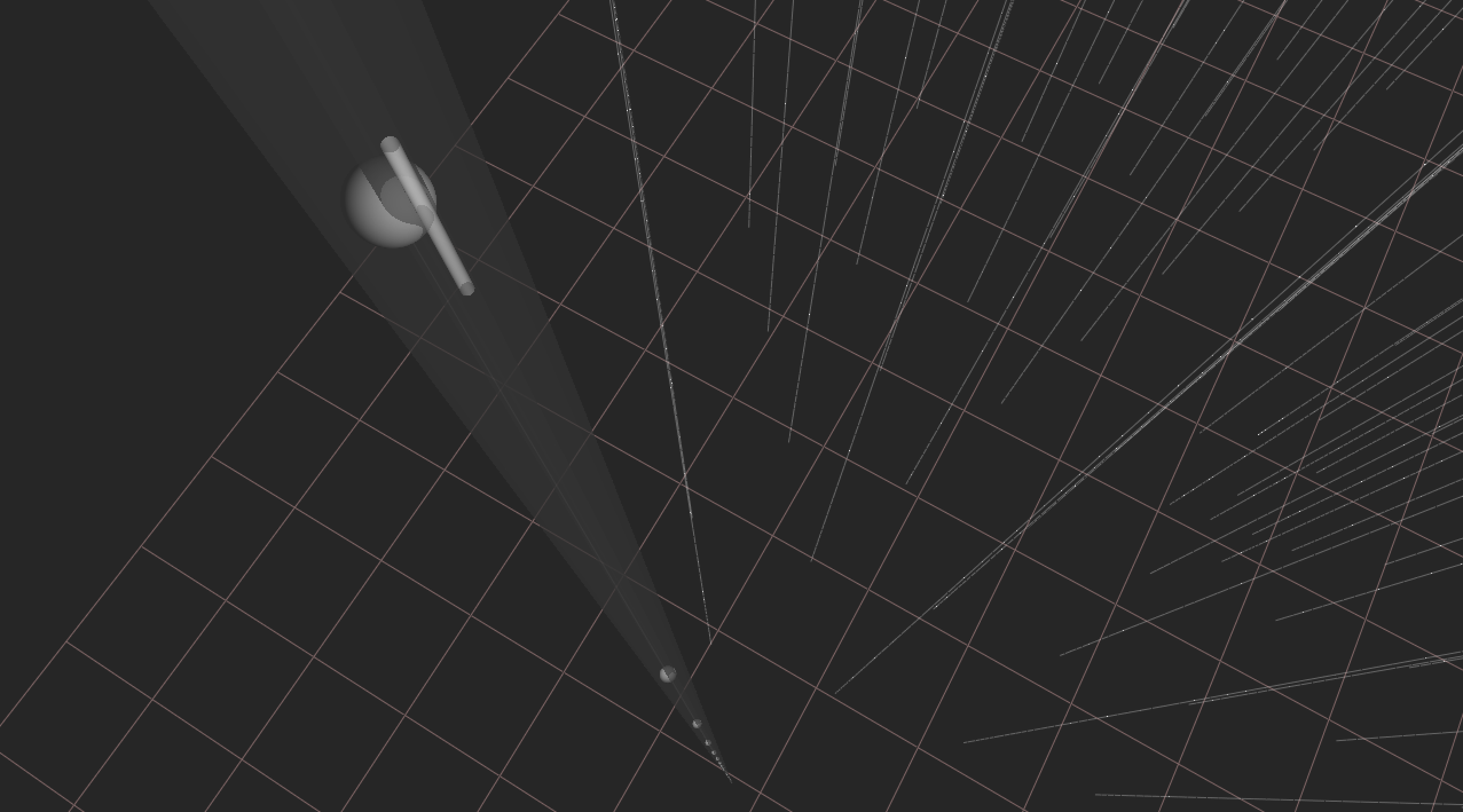}\vspace*{4mm}}\hfill
  \subcaptionbox{Effective angular acceptance of the optical module measured in simulations with and without cable.}{\halfimage{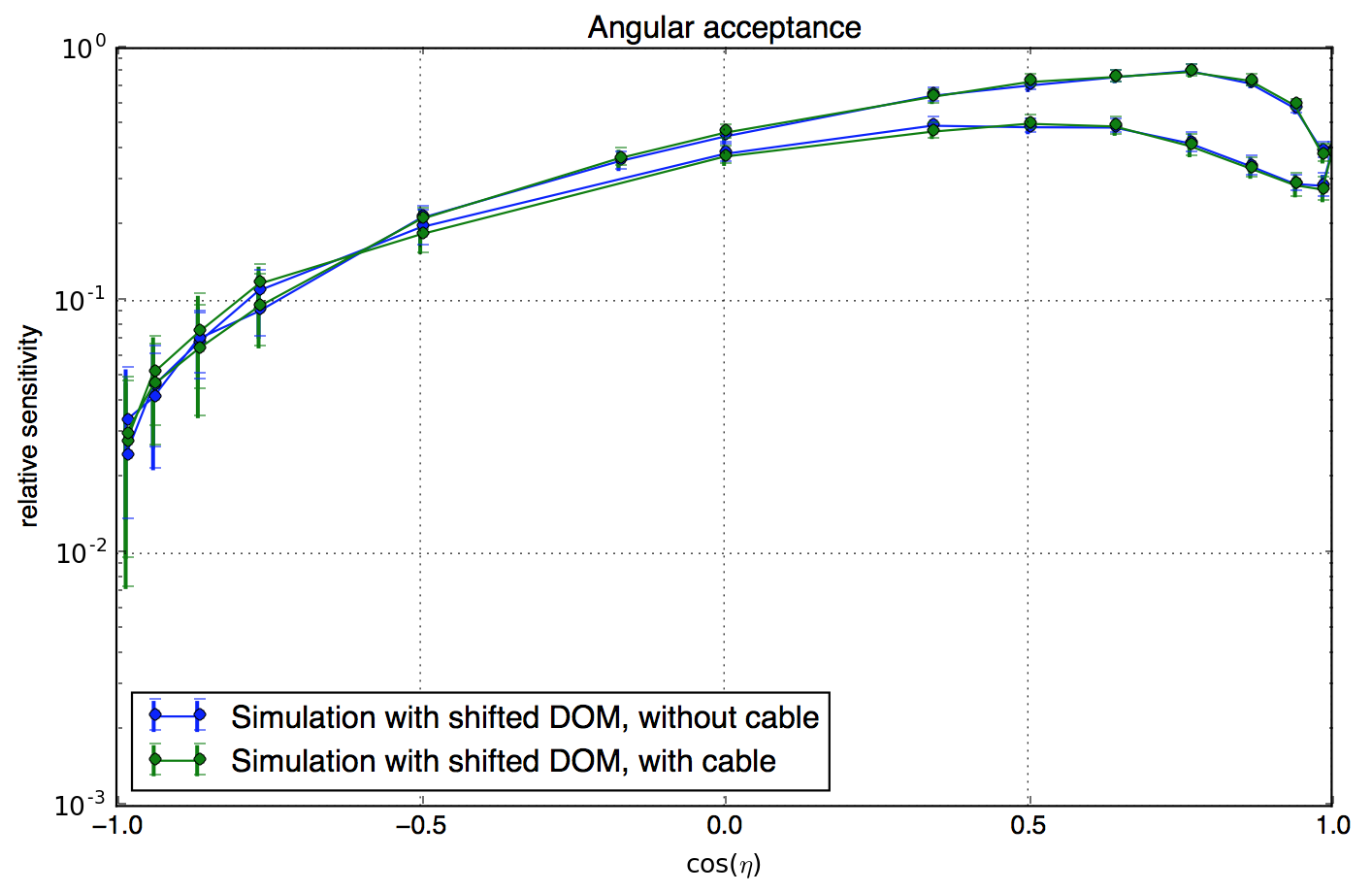}}\hfill
  \caption{Simulation with opaque cable partially inside bubble column.}
  \label{fig:caweNg6o}
\end{figure}

Placing a minimal opaque cable cylinder besides the optical module
partially inside the bubble column in a simulation, using plane waves as
photon sources, direct detection as hit acceptance criterion, and the
new medium-propagation algorithm (section \ref{sec:algorithm_b}), the
cable shadowing effect appears to be subdominant when measuring the
effective angular acceptance (figure \ref{fig:caweNg6o}) compared to the
effect of the bubble column and the drill hole ice. For photons
approaching the optical module from the side where the cable is located,
the number of photon hits is decreased compared to a simulation without
cable. But the effect is smaller than the statistical error.

\docpar{This simulation is documented in \issue{110}.}

\subsubsection{Can Cable Effects Account for the Observed Hole-Ice Effects?}

An open question raised by the \icecube Calibration Group is whether the
shielding by the main cable is sufficient to account for the effects
observed by other studies that are typically attributed to the hole
ice.\footnote{Internal correspondence. See for example \url{https://icecube-spno.slack.com/archives/C4FV72473/p1522086838000178}.}

\begin{figure}[htbp]
  \subcaptionbox{\steamshovel visualization of the simulation scenario with cables: The cable is modeled as three opaque cylinder parts. See also figure \ref{fig:ahyoi7Ma} for comparison.}{\halfimage{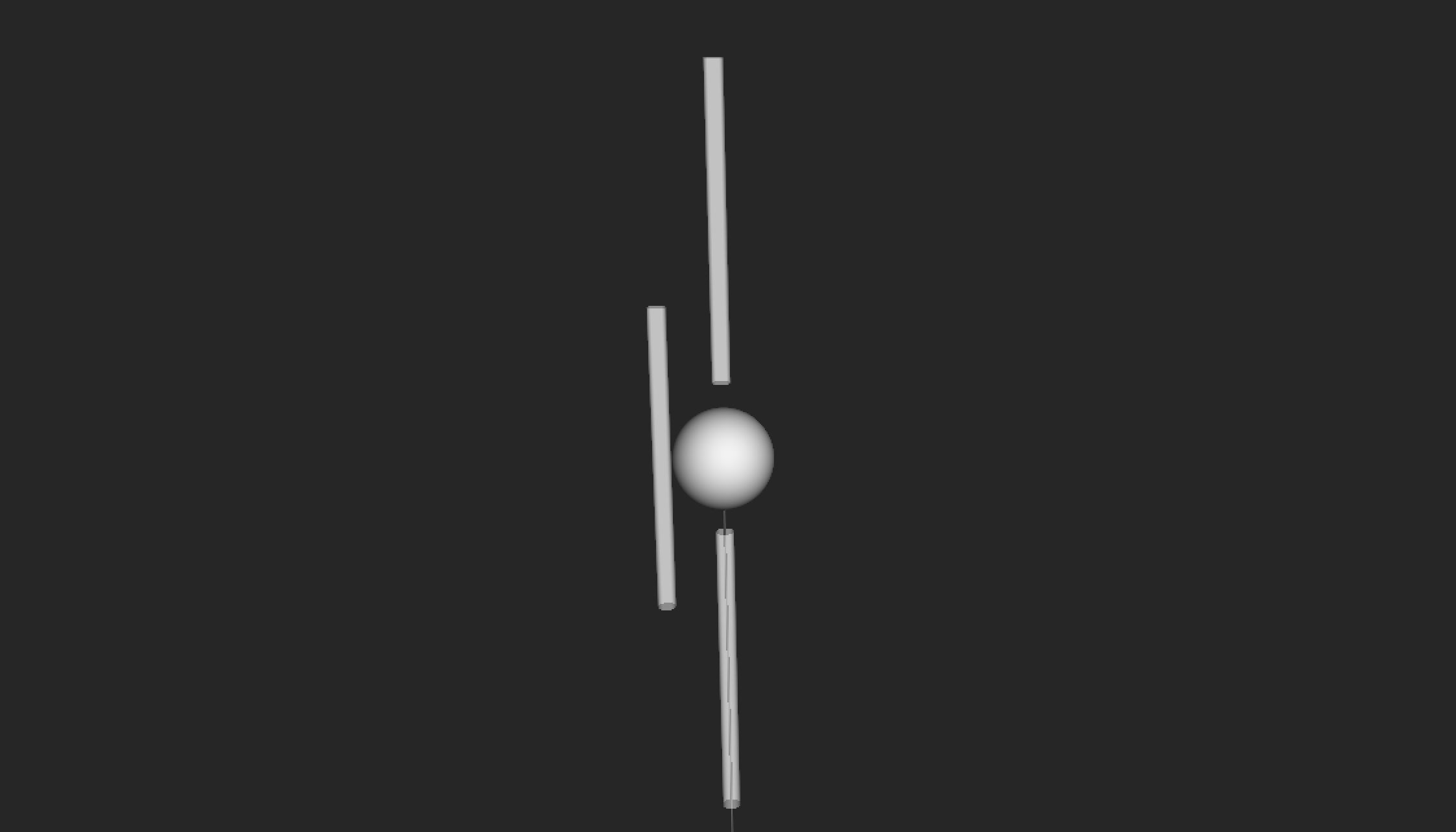}\vspace*{3mm}}\hfill
  \subcaptionbox{Effective angular acceptance curves for the optical module measured in simulations without hole ice, with only a cable, and with a hole-ice-cylinder of $30\cm$ radius and a geometric hole-ice scattering length of $50\cm$ (H2 model parameters).}{\halfimage{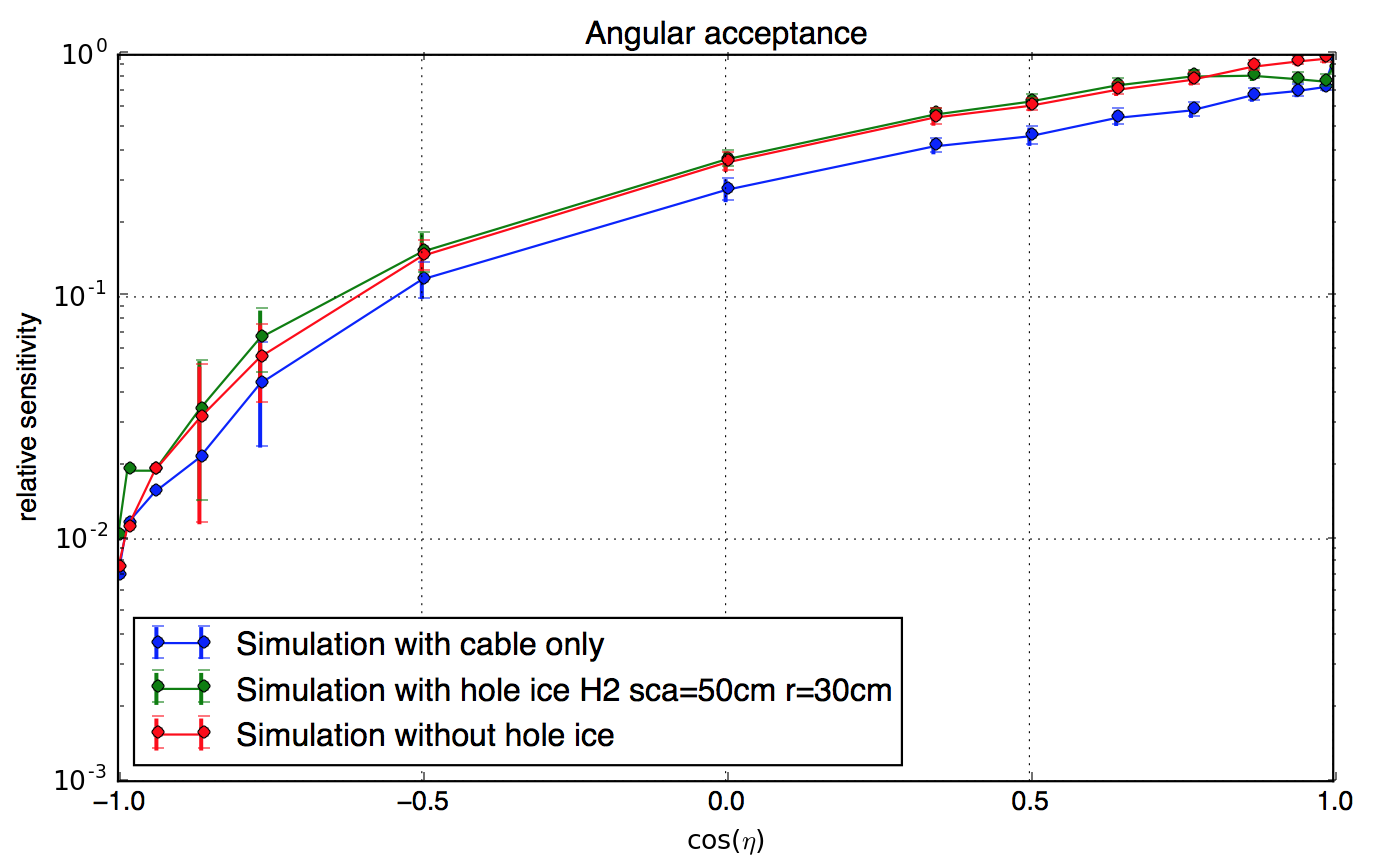}}\hfill
  \caption{Measuring the effective angular acceptance of an optical module in simulations: (1) without hole ice, (2) with only a drill hole, (3) with only a cable.}
  \label{fig:Ohw1aibu}
\end{figure}

The following series of simulations compares the effective angular
acceptance for a scenario where only a shadowing cable is considered,
and a typical drill-hole-ice scenario. This simulation uses the new
medium-propagation algorithm (section \ref{sec:algorithm_b}), plane
waves as photon sources, and direct detection as hit acceptance
criterion. The cable is modeled as three opaque cylinder parts (see
figure \ref{fig:Ohw1aibu} (a) in comparison to the illustrations in
figure \ref{fig:ahyoi7Ma}). The drill-hole ice is modeled with
properties suggested by the so-called \textit{H2 model}, a
hole-ice-cylinder radius of \(30\cm\) and a geometric scattering length
of \(50\cm\) \cite{holeicestudieswithyag}.

\docpar{This series of simulations is documented in \issue{101}.}

As shown in figure \ref{fig:Ohw1aibu} (b), the cable can account for a
total reduction in photon hits. For the decreased number of hits for
photons approaching from below, and an increased number of hits for
photons approaching from above, however, which are expected from earlier
hole-ice studies (see section \ref{sec:hole_ice_approximation}), the
simulated cable cannot account for.

  \subsection{Scanning Hole-Ice Parameters for Optimal Agreement with Flasher-Calibration Data}
\label{sec:flasher}

\begin{figure}[htbp]
  \subcaptionbox{Schematic diagram of an optical module. In the upper hemisphere, the flasher board is located, sustaining six horizontally emitting LEDs and six upward pointing LEDs. Image source: \cite{rongenswedishcamera}}{\halfimage{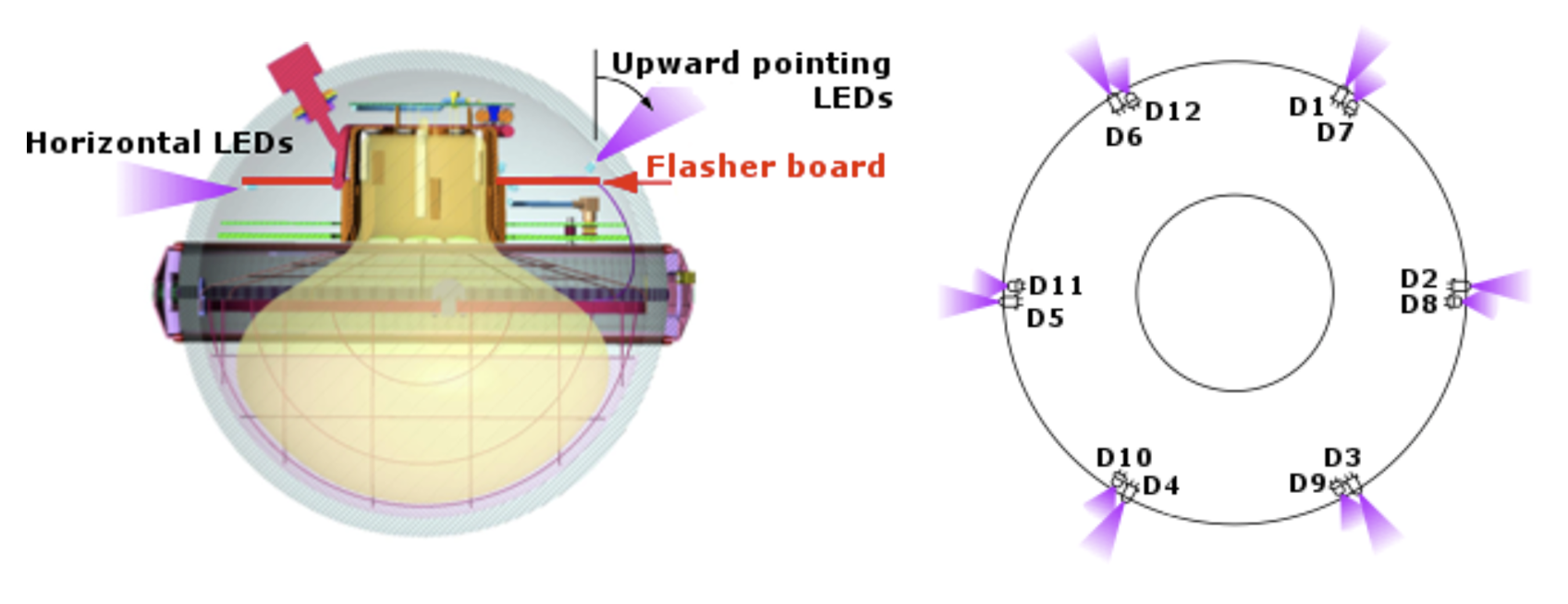}\vspace*{7mm}}\hfill
  \subcaptionbox{Principle of operation of the LED flasher calibration system: Photons emitted by the LED of one optical module propagate through the ice and are detected by another optical module. The amount of light observed at the receiving optical modules depends on the properties of the ice the photons propagate through on the path from the sending to the receiving optical modules. Image source: \cite{icepaper}}{\halfimage{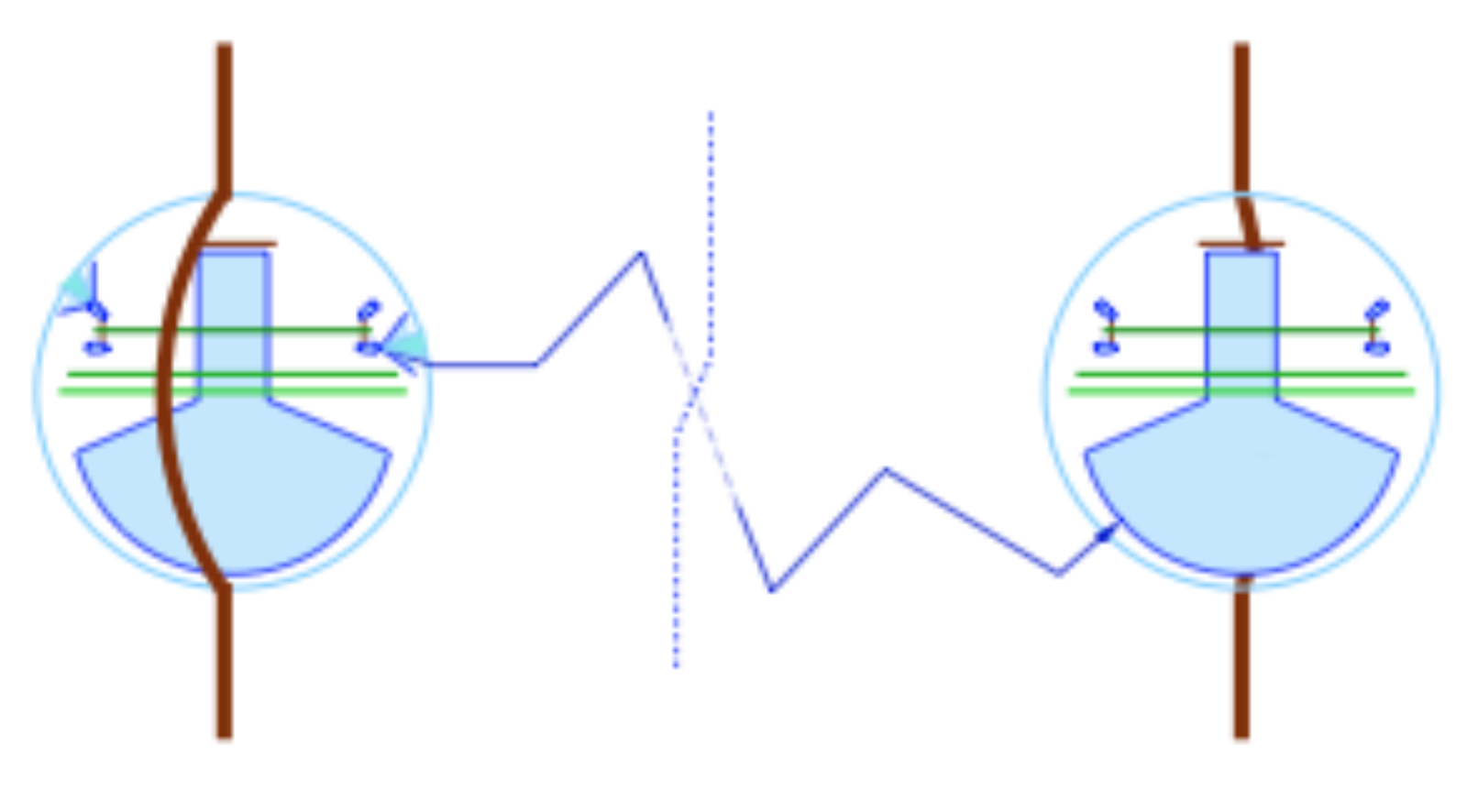}}
  \caption{\icecube's light-emitting diode (LED) flasher calibration system.}
  \label{fig:Quee3yui}
\end{figure}

For calibration purposes, each optical module in the \icecube detector
is equipped with a set of flasher light-emitting diodes (LEDs) as
illustrated in figure \ref{fig:Quee3yui}. Using these flasher LEDs, a
known amount of light can be produced at a given position and time
within the detector that can be measured by the optical modules and can
thereby be used to calibrate the detector. \cite{icepaper}

Light that is emitted at one optical module and detected by another
module has to travel through the ice in between those modules. The
amount of light arriving at the target module depends on the amount of
light absorbed or scattered away on the way from the sending to the
receiving optical module. Both, the properties of the bulk ice of the
South Polar glacier, and the properties of the hole ice surrounding the
sending and the receiving optical modules, contribute to this effect.

\begin{figure}[htbp]
  \subcaptionbox{Top view of the detector strings. Each green circle represents one detector string, containing 60 optical modules each. In this flasher study, the LEDs of the central optical module on string 63 have been activated, and the amount of light arriving at the optical modules of the surrounding strings 62, 54, 55, 64, 71, and 70 has been measured. Image based on \cite{geometry}}{\halfimage{flasher-scenario}}\hfill
  \subcaptionbox{Amount of light measured at the receiving optical modules on strings 62, 54, 55, 64, 71, and 70. Each column of this plot represents one receiving string. Within each column, the left most value represents the top most optical module of the string, the right most value represents the bottom most module of the string. Only optical modules are shown that receive at least one photon hit.}{\halfimage{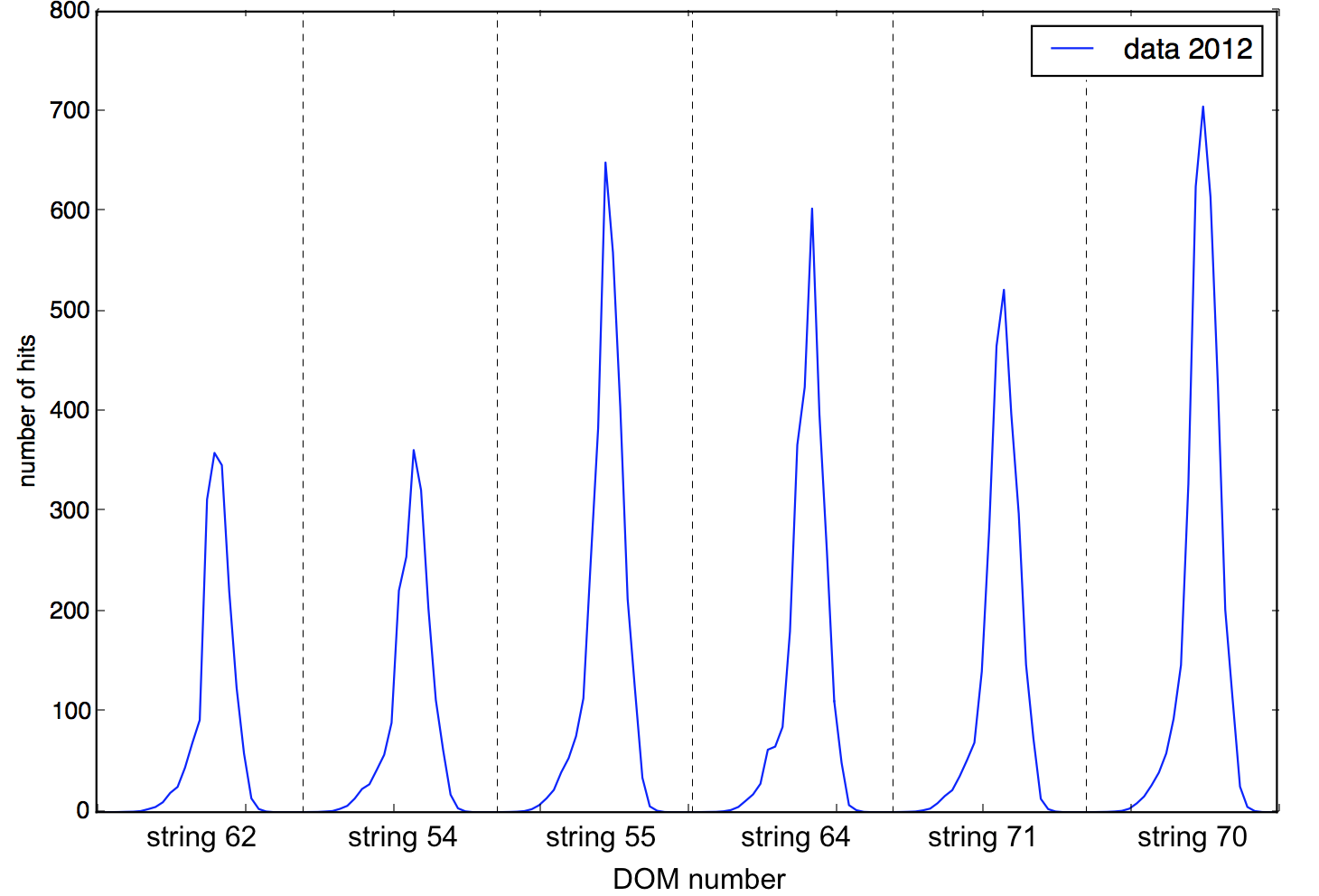}}
  \caption{7-string flasher data: A central optical module is emitting light using the mounted flasher LEDs. The surrounding optical modules are measuring the amount of light arriving there.}
  \label{fig:ea9Zieh0}
\end{figure}

Figure \ref{fig:ea9Zieh0} shows 7-string calibration
data.\footnote{Data source: \cite{flasherdata}} To produce this data
set, all LEDs on \texttt{DOM 60\_30}, which is the middle optical module
on string number 30, have been activated at once. The plot shows the
amount of light received at the optical modules of the surrounding
strings 62, 54, 55, 64, 71, and 70.

The aim of the following series of simulations is to adjust the
simulated hole-ice parameters until the the simulation is able to
reproduce the calibration data in order to find the hole-ice parameters
that best match the calibration data.

\docframe{
\docparwithoutframe{This preliminary 7-string flasher study is documented in \issue{59}.}\medskip

\sourceparwithoutframe{A script to configure and perform these kinds of simulations is provided in \script{FlasherSimulation}.}\medskip

\sourceparwithoutframe{A script for performing flasher simulations as grid scan is provided in \script{FlasherParameterScan}.}
}

As shown in figure \ref{fig:ea9Zieh0} (b), the amount of light received
at strings 62 and 54 is significantly reduced compared to the other
strings. The distances of the sending optical module to the receiving
strings range from \(122.0\m\) to \(125.4\m\) and average out at
\(124.5\m\), such that the different distances can only cause a peak
reduction of about \(2\,\%\) rather than the observed reduction of about
\(30\,\%\). The properties of the bulk ice cannot account for the
asymmetric reduction either. The reduction may be caused by some kind of
asymmetric shadowing effect, either by shadowing
cables\footnote{Flasher simulation with cable: See section \ref{sec:flasher_with_cable}.},
or by asymmetric hole ice in proximity of the sending optical module,
such that the emitted light is suppressed in the direction to strings 62
and 54. The reduction may also be caused by the hole ice of strings 62
and 54 having other optical properties than the hole ice of the other
strings, or by the position of the hole-ice cylinders of strings 62 and
54, shielding those strings less from the direction of the sending
string 63. A systematic simulation study to investigate those scenarios
is out of scope of this study, but is suggested as a follow-up
study.\followup

\begin{figure}[htbp]
  \centering
  \subcaptionbox{Total view of the detector. The flasher LEDs have been activated with full brightness. To see separate photon tracks in the image, only a fraction of $10^{-8}$ of the photons is shown in this image. The detector strings are about 125\m apart. The optical modules can detect light emitted up to 500\m away. \cite{instrumentation}\vspace*{3mm}}{\halfimage{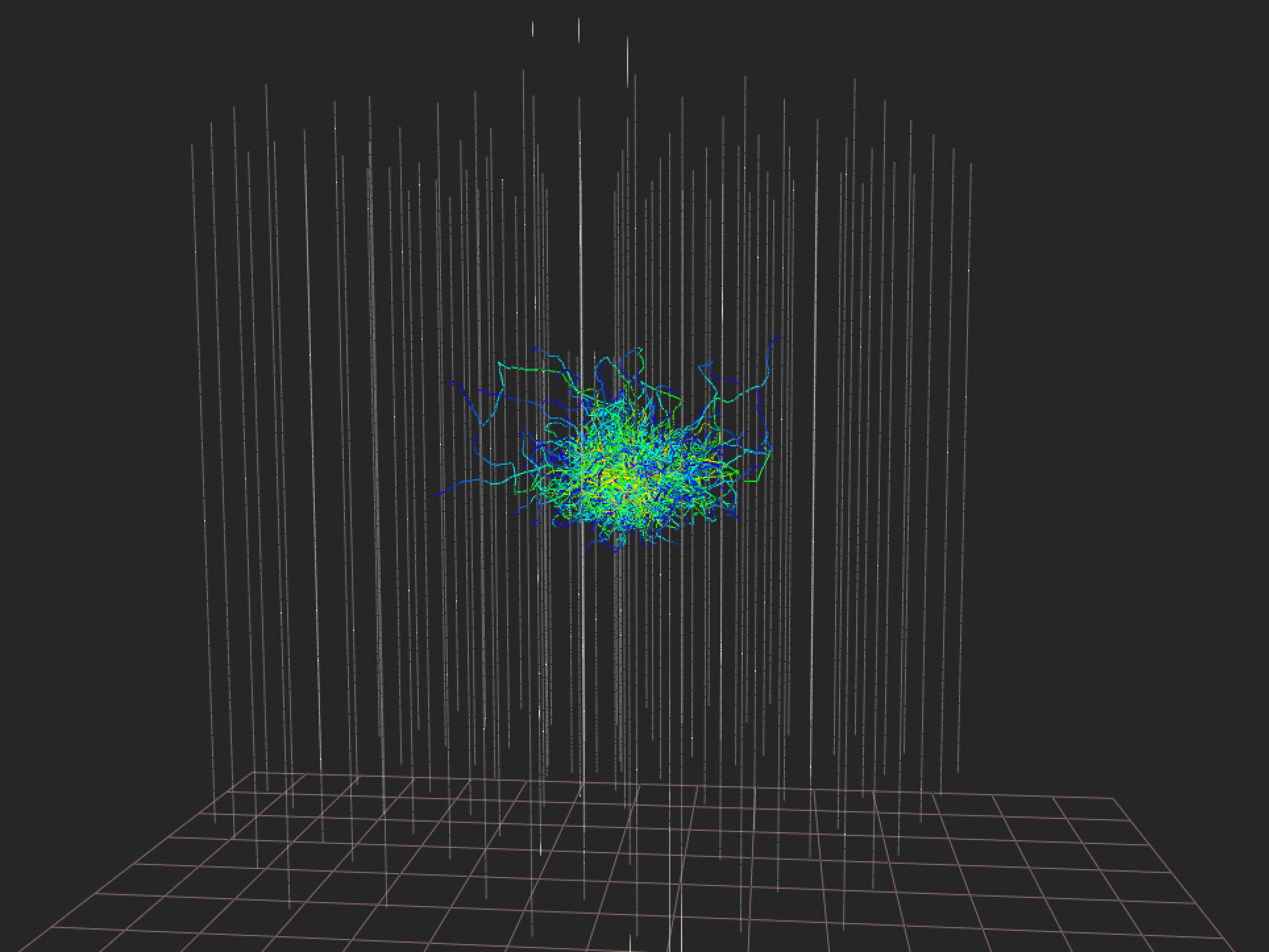}}\hfill
  \subcaptionbox{Light emitted by the LEDs at the sending optical module. The modules is positioned centered in a bubble column with a diameter of $60\,\%$ of the diameter of the optical module. Only a fraction of $10^{-8}$ of the photons is shown in this image.\vspace*{3mm}}{\halfimage{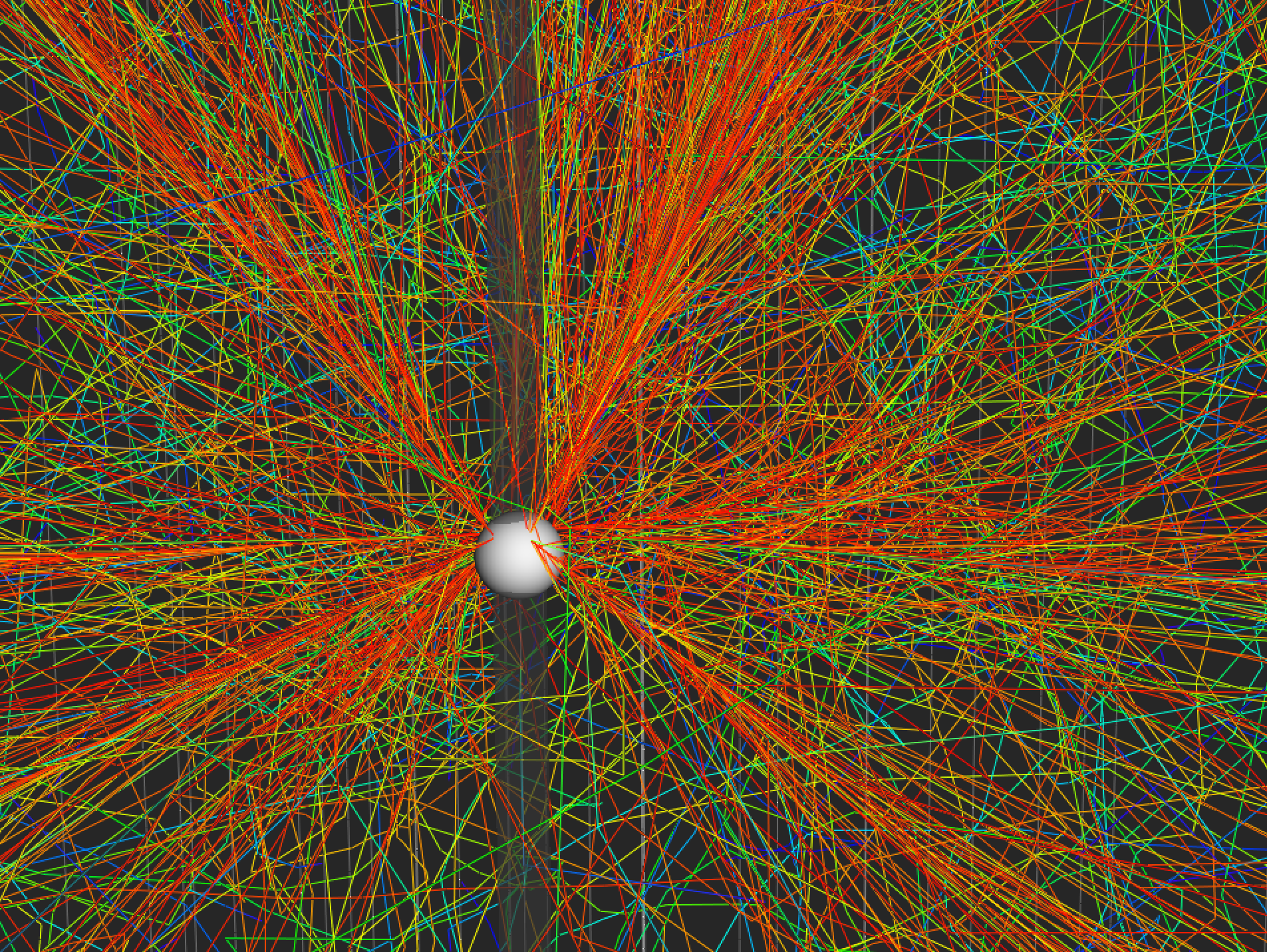}}\hfill
  \halfimage{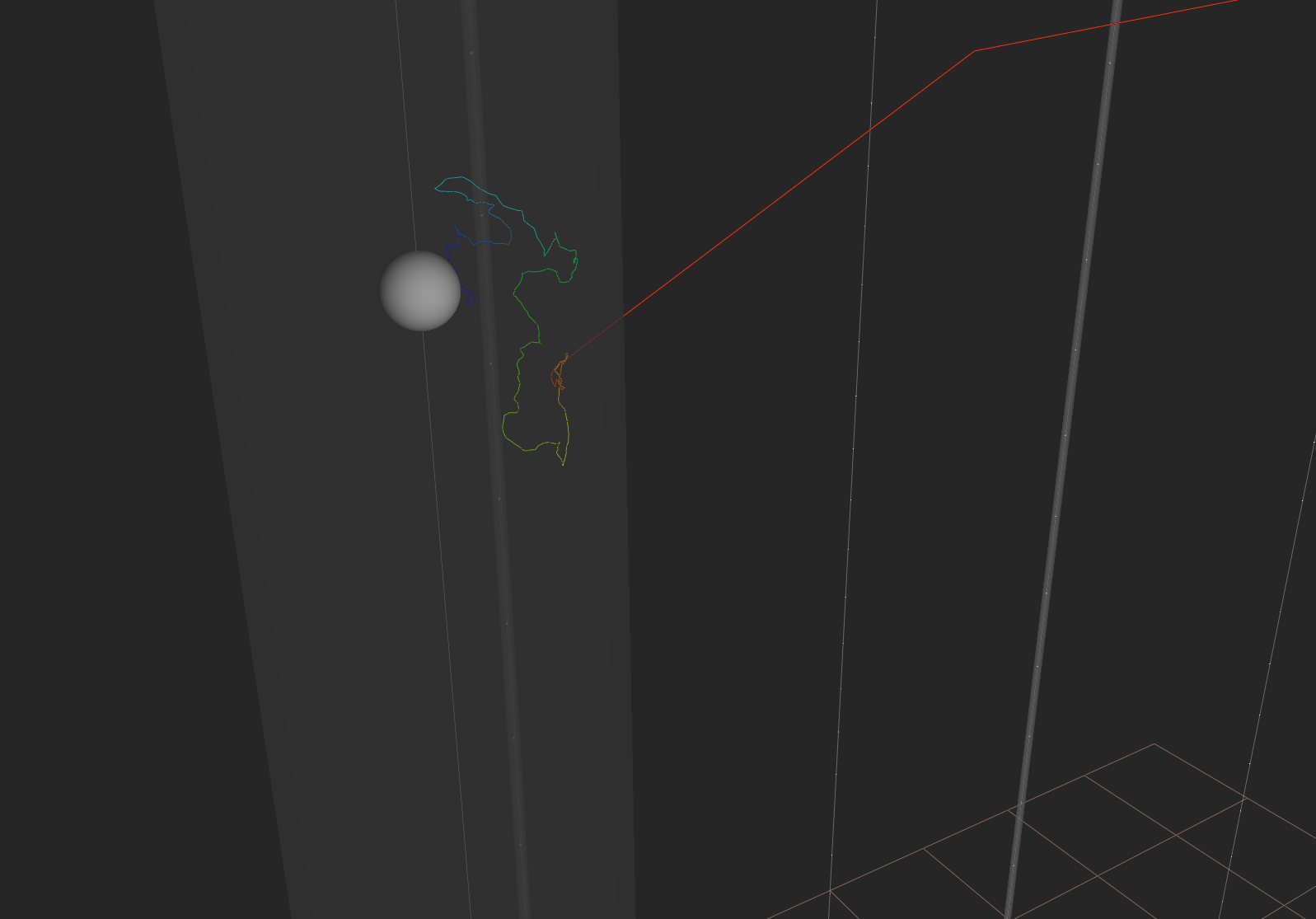}\hfill
  \begin{minipage}[b]{0.48\textwidth}
    \subcaption{To emphasize the effect of the hole ice, a single photon is isolated in this image, entering a hole-ice cylinder with a diameter of $500\,\%$ of the diameter of the optical module (larger than in the real simulation). The scattering probability increases significantly within the cylinder.}
  \end{minipage}
  \caption{\steamshovel visualization of the flasher scenario. LEDs at a central optical module emit light that propagates through the ice and is then received at surrounding optical modules.}
  \label{fig:Cahy7eej}
\end{figure}

In this flasher study, all optical modules are assumed to be positioned
central in their respective hole-ice cylinders. In the simulation, hits
are recorded in all optical modules of the detector. For a first
attempt, however, only the receiving strings 55, 64, 71, and 70 are used
to fit the properties of the hole
ice.\footnote{If the dominant effect causing the asymmetry is the displacement of the sending optical module within the hole ice, excluding strings 62 and 54 leads to a systematic error. If the dominant effect causing the asymmetry is the shadowing cable, including those strings leads to a systematic error. Further studies are required to account for both possibilities.}
Figure \ref{fig:Cahy7eej} shows \steamshovel visualizations of the
simulation scenario.

\paragraph{Likelihood}

As a measure for comparing the calibration data to the simulation, a
Poisson likelihood \(L\) is used as a basis.

\begin{equation}
  L = \prod_{i} \text{P}_{\lambda_i}(k_i) = \prod_{i} \frac{\lambda_i^{k_i}\,\e^{-\lambda_i}}{k_i !}
\end{equation}

The index \(i\) refers to the individual receiving optical modules. Each
factor \(\text{P}_{\lambda_i}(k_i)\) represents the probability that
module \(i\) registers \(k_i\) hits, given by the calibration data,
while \(\lambda_i\), which is the mean of the Poisson distribution,
represents the expected number of hits from the simulation.

To account for the finite statistics of the simulation, this likelihood
can be generalized. Rather than using just the number \(\lambda_i\) of
expected hits, which assumes infinite statistics for the simulation, one
uses the number \(k\imc\) of Monte-Carlo hits, which is the number of
hits seen in the simulation, and a weight \(w\), which quantifies the
ratio of calibration-data statistics and simulation statistics, to
account for the results from the simulations.
\cite[equation 21]{Gluesenkamp2018}

\begin{equation}
  L = \prod_{i} \frac%
    {\left(\frac{1}{w}\right)^{k\imc} \cdot \left(k_i + k\imc - 1\right)!}%
    {(k\imc - 1)! \cdot k_i! \cdot \left( 1 + \frac{1}{w} \right)^{k_i + k\imc}}
\end{equation}

If, in order to improve accuracy, the simulation has 10 times the number
of photons as compared to the calibration data, the weight would be
\(w=\sfrac{1}{10}\) as each simulated photon would correspond only to
\(\sfrac{1}{10}\) real photons. If, in order to save computation time,
only \(\sfrac{1}{10}\) of the number of photons from the calibration
data are simulated (section \ref{sec:thinning}), the weight is \(w=10\)
as each simulated photon corresponds to 10 photons from the calibration
data.

\paragraph{Results}

\begin{figure}[htbp]
  \smallerimage{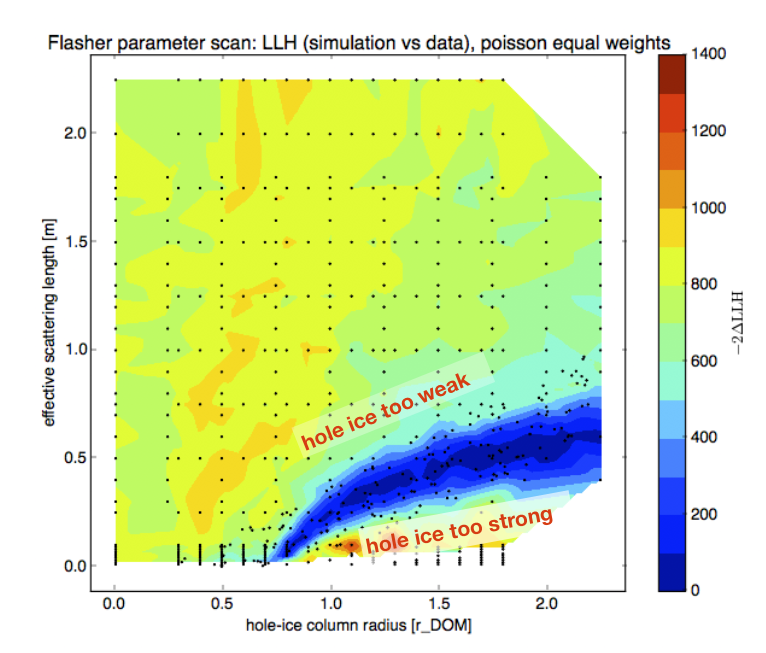}
  \caption{Agreement of flasher calibration data with flasher simulations for different hole-ice parameters. The axes show the radius of the simulated hole-ice column and the effective scattering length of the hole ice. The blue ``valley'' shows the region of best agreement of data and simulation. Above the valley, the simulated hole ice is too weak, below the valley it is too strong to account for the amount of photon hits in the calibration data.}
  \label{fig:ut4nao7X}
\end{figure}

Figure \ref{fig:ut4nao7X} shows a contour plot of the conducted
parameter scan. The likelihood \(L_\HH\) of agreement of calibration
data and the simulation using the hole-ice parameters \(\HH\) given by
the coordinate axes, or rather
\(-2\Delta\llh:= -2(\ln L_\HH - \ln L_\HHbest) = -2 \ln \left(\sfrac{L_\HH}{L_\HHbest}\right)\),
is plotted against the hole-ice parameters. \(L_\HHbest\) is the maximum
likelihood that corresponds to simulation with the hole-ice parameters
\(\HHbest\) that best describe the calibration data, such that
\(\Delta\llh = 0\) corresponds to the simulation that best agrees with
the calibration data.

The contour plot shows a valley (blue area) where the simulated hole ice
leads to good agreement of simulation and calibration data. If the
radius of the simulated hole-ice cylinder is too small, or the
scattering length of the hole ice is too large, the hole-ice effect is
too weak and the amount of light arriving at the receiving optical
modules is too high compared to the flasher calibration data. If, on the
other hand, the hole-ice cylinder is too large, or the scattering length
of the hole ice is too small, the hole-ice effect is too strong and the
amount of light arriving at the receiving optical modules is too low
compared to the flasher calibration data.

The optimal agreement of simulation and calibration data is achieved
with hole-ice parameters \(\HHbest\):
\(r = 0.83\,r\dom, \lambda\esca\hi = 0.10\m\), with a log likelihood
\(\llh = \ln L_\HHbest = -278.66\). Note, however, that this example
study is only to be considered a proof of concept for this kind of
flasher parameter scan. Due to a number of systematics, in particular
due to the unrealistic assumption that all optical modules are
symmetrically centered within their hole-ice columns, it is expected
that the study does not fit the correct hole-ice properties.

\paragraph{Systematics}

First, an outdated configuration for the positions of the optical module
has been used in these simulations. Preliminary investigations show that
based on newer calibration data, the positions of the optical modules
have been updated on the order of \(1\m\). Thus, newer position
calibration information should be used for follow-up studies.

Next, an outdated set of ice-model parameters has been used in this set
of simulations. In this study, the \texttt{spice\_mie} ice model from
2010 has been used. For a proper flasher study, however, current ice
models with improved fits for the scattering and absorption lengths,
which depend on the ice layer, in particular the dust concentration in
these layers, the ice temperature, photon wavelength, and photon
direction, should be used. (See
\cite{icepaper,flasherdataderivedicemodels}.) The tilt of the ice layers
and the absorption anisotropy (section \ref{sec:ice}) have not been
included in these simulations, because those features are not yet
re-implemented for the new medium-propagation
algorithm.\footnote{For the current status of re-implementing ice layers and ice anisotropy, see \url{https://github.com/fiedl/hole-ice-study/issues/48}.}
These ice properties are negligible for short propagation distances, but
become important for distances larger than about \(10\m\), which is
given for flasher studies where the spacing of sending and receiving
optical module is on the order of \(100\m\).

Most importantly, the displacement of the optical modules relative to
their hole-ice columns has not been considered in this study and should
be included in follow-up simulations. In this study, two receiving
strings have been deliberately excluded due to the observed asymmetry
(figure \ref{fig:ea9Zieh0}). After determining the effect of the
shadowing cable in simulations, however, the displacement of the optical
module along the direction suggested by the observed asymmetry should be
included as fit parameter. Preliminary investigations show that
neglecting the displacement of the optical module causes systematic
uncertainties for the fitted hole-ice scattering length on the order of
\(30\,\%\).

\subsection{Comparing Flasher-Calibration Data to a Flasher Simulation With Cable}
\label{sec:flasher_with_cable}

As shown in section \ref{sec:flasher}, evaluating calibration data
indicates a strongly asymmetric shadowing effect when flashing the LEDs
of a central optical module and measuring the amount of light detected
by the surrounding optical modules. This effect may either be caused by
the LEDs of the sending optical module being asymmetrically shielded, or
by the receiving optical modules in two of the receiving strings being
more shielded than in the other receiving strings, or by a combination
of the two.

\begin{figure}[htbp]
  \subcaptionbox{Top view of the detector strings with position of the shadowing cable. Image based on \cite{geometry}.}{\halfimage{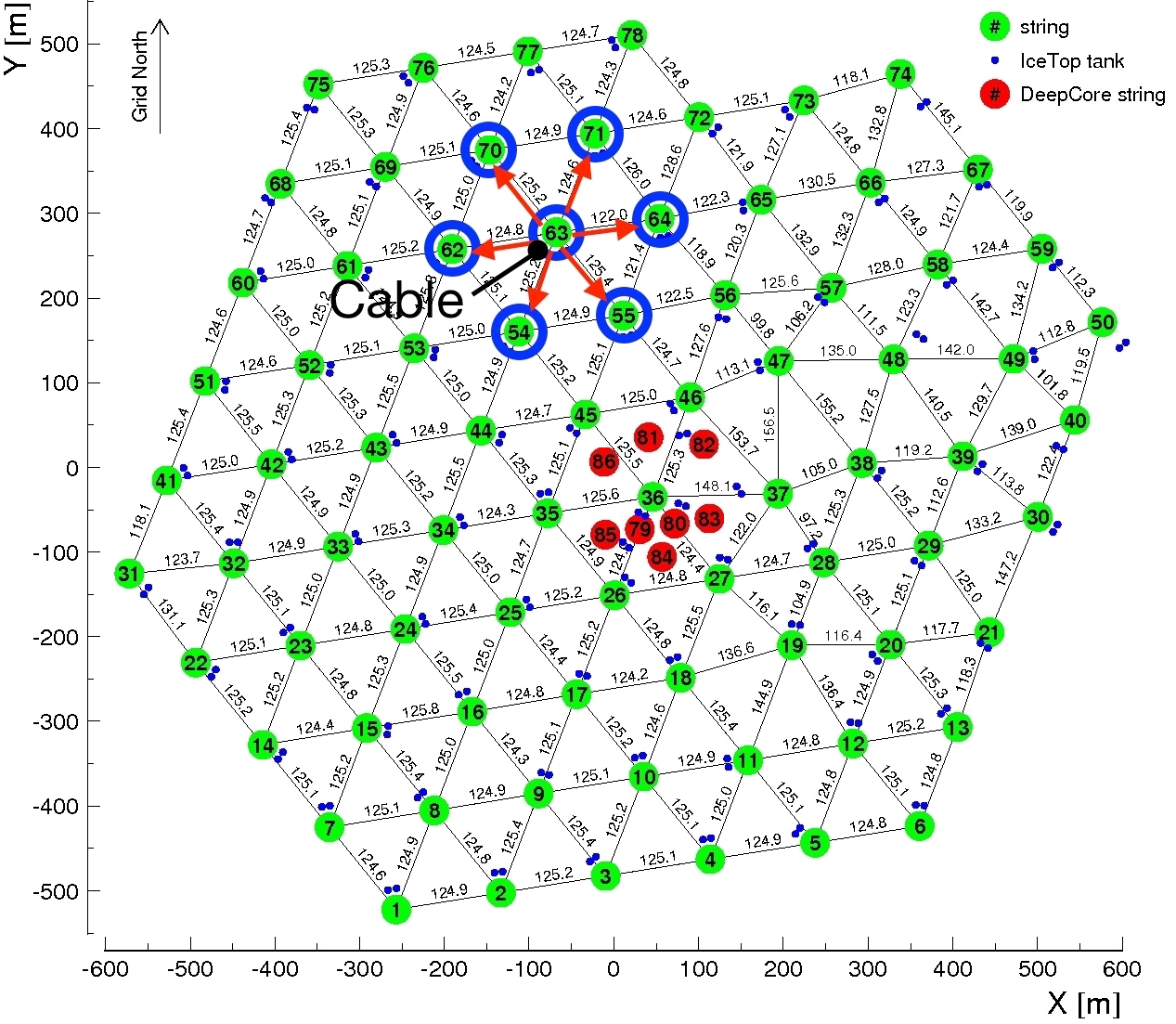}}\hfill
  \subcaptionbox{Amount of light seen by the receiving optical modules. From left to right each peak corresponds to one of the strings 62, 54, 55, 64, 71, and 70.}{\halfimage{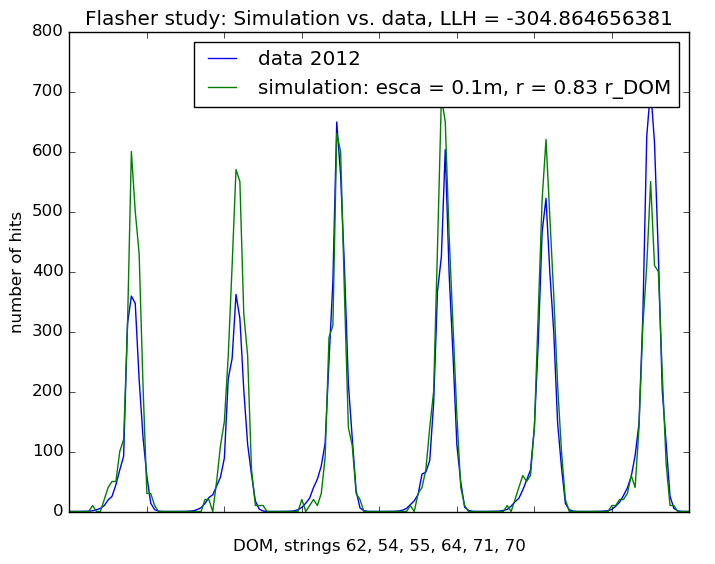}}
  \caption{Flasher-simulation scenario with a shadowing cable besides the sending optical module.}
  \label{fig:neen7Noo}
\end{figure}

As the main cable at the side of the sending optical module causes
asymmetric shielding as shown in section \ref{sec:cables}, the question
arises whether the cable shadow is sufficient to account for the
asymmetries observed in the flasher calibration data. Figure
\ref{fig:neen7Noo} (a) shows a flasher-simulation scenario where an
opaque cable with \(4\cm\) diameter is placed besides the sending
optical module such that it would reduce the emitted light in the
direction of the two strings that show a reduced amount of detected
light in the calibration data.

\docpar{This flasher simulation with shadowing cable is documented in \issue{97}.}

In this flasher simulation, the sending and the receiving optical
modules are embedded in a bubble column of a radius of \(83\,\%\) of the
optical module and an effective bubble-column scattering length of
\(10\cm\), which are the best-fit parameters of the flasher scan in
section \ref{sec:flasher} that only considers the supposedly unshadowed
receiving strings 55, 64, 71, and 70. All bubble columns are positioned
symmetrically relative to the embedded optical modules, such that the
bubble columns only cause an overall reduction, but no azimuthal
asymmetry. The distances of the sending optical module to the receiving
strings range from \(122.0\m\) to \(125.4\m\) and their differences can
only cause a peak reduction of about \(2\,\%\). Therefore, in this
scenario, the shadowing cable is the only remaining candidate to cause
the observed peak reduction of about \(32\,\%\) observed in the
calibration data. However, comparing results of the flasher simulation
with shadowing cable to the flasher-calibration data (figure
\ref{fig:neen7Noo} b) makes it implausible that the shadowing cable
alone could be the cause of the azimuthal asymmetry.

Hole ice, on the other hand, is a suitable candidate to account for the
observed asymmetric peak reduction. Assuming the entire drill hole being
filled with hole ice, varying the effective hole-ice scattering from
\(60\cm\) to \(40\cm\) can cause a peak reduction in the observer
magnitude of \(30\,\%\). As section \ref{sec:cylinder_shift} has shown
that the displacement of the optical modules relative to the hole ice
may cause strong asymmetric effects, it is plausible that a displacement
of the optical modules relative to a hole ice with suitable optical
properties is required to explain the asymmetric effects observed in
calibration data.

  \section{Discussion}
\label{sec:discussion}

\subsection{Comparison to Other Hole-Ice-Simulation Methods}
\label{sec:comparison_methods}

\subsubsection{A Priori Modification of Angular-Acceptance Curves}
\label{sec:a_priori_modification_of_angular_acception_curves}

\label{sec:a_priori_curve}

Using a modified angular-acceptance curve \(a(\eta)\) as acceptance
criterion of the optical modules in simulations to effectively account
for the effect of hole ice on the detection of photons, is the default
approach in \clsim (see sections \ref{sec:a_priori_angular_acceptance}
and \ref{sec:hole_ice_approximation}). The respective angular-acceptance
curves have been obtained using lab measurements and simulations using
the \photonics photon-propagation software.
\cite{icepaper, lundberg, photonics}

\docpar{Information on the nominal and the hole-ice-approximating angular-acceptance curves are provided or linked in \issue{10}.}

\begin{figure}[htbp]
  \smallerimage{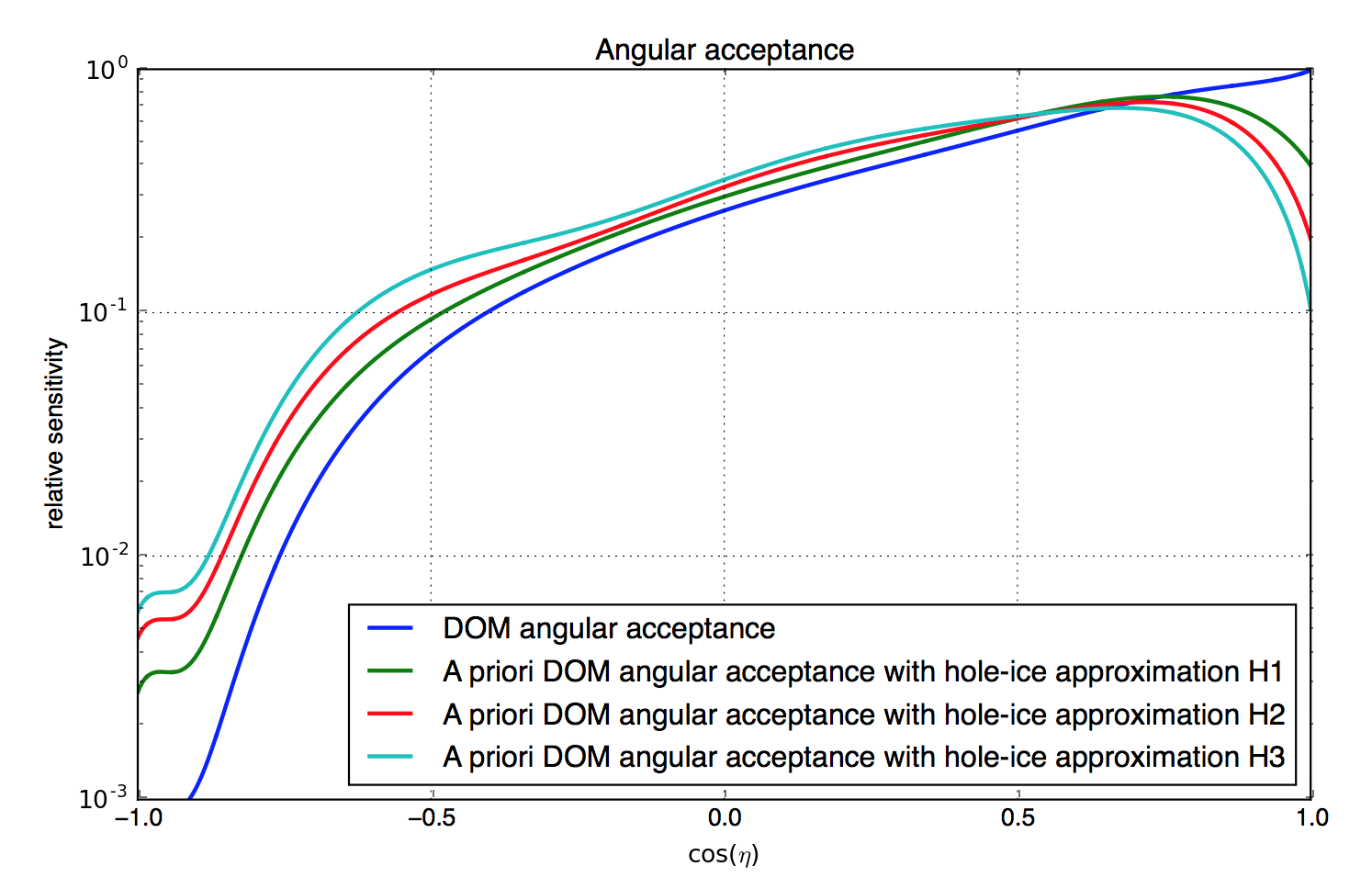}
  \caption{A priori angular-acceptance curves: \underline{blue}: without hole ice, \underline{H1}: assuming a hole-ice-cylinder radius if $30\cm$ and a geometric hole-ice scattering length of $\lambda\sca\hi = 1.00\m$, corresponding to an effective hole-ice scattering length of $\lambda\esca\hi = 16.67\m$, \underline{H2}: $\lambda\sca\hi = 0.50\m, \lambda\esca\hi = 8.33\m$, \underline{H3}: $\lambda\sca\hi = 0.30\m, \lambda\esca\hi = 5.00\m$. \cite{icepaper, yag, icemodelsdata}}
  \label{fig:Wee4ahYa}
\end{figure}

Figure \ref{fig:Wee4ahYa} shows three modified angular-acceptance curves
that approximate the hole-ice effect in comparison to the optical
module's angular acceptance based on the lab measurement. The modified
curves H1, H2, and H3 assume that the entire drill hole is filled with
hole ice, corresponding to a hole-ice-cylinder radius of \(30\cm\), and
assume geometric hole-ice scattering lengths of
\(\lambda\sca\hi = 100\cm\) (H1), \(\lambda\sca\hi = 50\cm\) (H2), and
\(\lambda\sca\hi = 30\cm\) (H3) respectively.

Using modified angular-acceptance curves for the optical modules to
approximate the hole-ice effect has the advantage that new parameters
gained from calibration studies can be inserted into the existing
simulations just by replacing the polynomial parameters used for the
angular acceptance of the optical modules. Also, using modified
acceptance curves is faster than simulating the hole ice directly as the
many scattering steps that correspond to the photons scattering within
the hole ice are not actually simulated and just effectively accounted
for using when the algorithm decides whether a photon hit is accepted by
an optical module.

This method, however, relies on strong assumptions: The lab measurements
the acceptance curves are based on, have been performed using
manufactured ice rather than South-Polar ice. The hole-ice modifications
are based purely on simulations and do not use any \icecube calibration
information. \cite{icepaper} Furthermore, this method assumes all
optical modules having exactly the same properties regarding positioning
and relative orientation. The optical modules are assumed to be
perfectly centered within the drill hole as well as in the bubble
column.

\begin{figure}[htbp]
  \subcaptionbox{\clsim simulation with H2 hole-ice parameters:\\ $r = 30\cm, \lambda\sca\hi = 50\cm, \lambda\sca\hi = 8.33\m$}{\includegraphics[height=0.27\textheight]{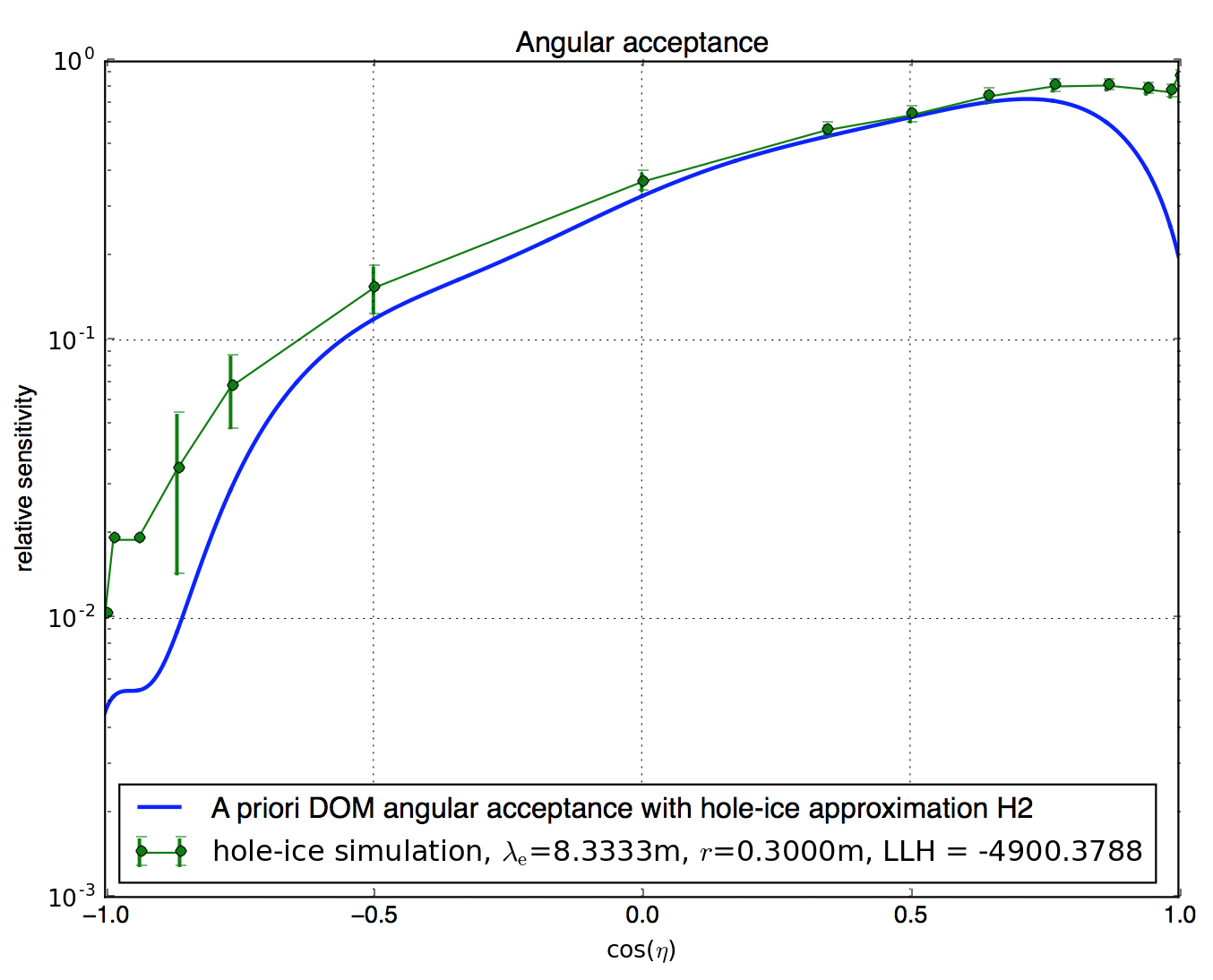}}\hfill
  \subcaptionbox{\clsim simulation with parameters\\ $r = r\dom = 0.1651\m, \lambda\sca\hi = 6\cm, \lambda\esca\hi = 1.0\m$}{\includegraphics[height=0.27\textheight]{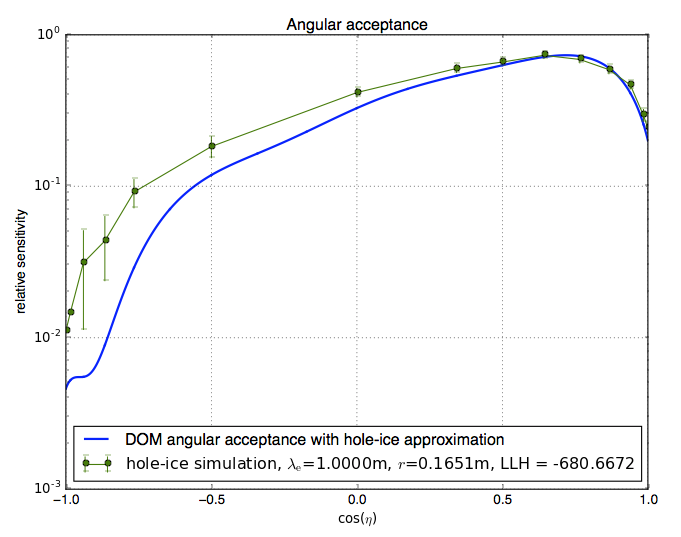}}\hfill
  \caption{Comparing the a priori angular-acceptance curve assuming H2 hole-ice parameters from \cite{icepaper,yag} to \clsim simulations.}
  \label{fig:iQu2aBuz}
\end{figure}

The a priori H2 acceptance curve, which is used in
standard-\clsim simulations by default, is used as reference curve for
other angular-acceptance plots in this study. However, comparing this
reference curve, which itself assumes a hole-ice-cylinder radius of
\(30\cm\) and a geometric scattering length of \(50\cm\) (H2 parameters)
\cite{icemodelsdata}, to a \clsim simulation using the same H2
parameters (figure \ref{fig:iQu2aBuz}), both angular-acceptance curves
do not match. Instead, the reference curve best matches a
\clsim simulation, which assumes a hole-ice radius of the same size as
the optical module, \(r = 0.17\m\), and an effective hole-ice scattering
length of \(\lambda\esca\hi \approx 1.0\m\) (section
\ref{sec:parameter_scan}).

\subsubsection{Angular-Acceptance Fitting Using Flasher Data}
\label{sec:dimas_model}

\chirkin \cite{flasherdataderivedicemodels} suggests a hole-ice model,
here referred to as \textit{Dima's model}, resulting in an
angular-acceptance curve, \(a\domdima(\eta;p)\), different from the a
priori curves for the H1, H2, H3 parameters. This angular-acceptance
curve can be used as acceptance criterion of the optical modules in
simulations instead of the a priori curves described in section
\ref{sec:a_priori_curve}.

The angular-acceptance curve of Dima's model is parameterized using a
single parameter \(p\in\reals\) that has been determined as
\(p \in [0.2;0.4]\) using flasher studies. \cite{msuforwardholeice} In
this model, the hole-ice-cylinder radius \(r\) and the scattering length
\(\lambda\sca\hi\) within the hole ice are
unknown.\footnote{As the hole-ice parameters are not given by Dima's model, one cannot directly compare the model to \clsim simulations with direct hole-ice propagation. A parameter scan to find the best hole-ice parameters for Dima's model is suggested as follow-up study. See: \url{https://github.com/fiedl/hole-ice-study/issues/114} \followup}

\begin{align}
  a\domdima(\eta;p) = & \ 0.34 \cdot (1 + 1.5\,\cos(\eta) - \cos(\eta)^{\sfrac{3}{2}}) \nonumber \\
      & + p \cdot \cos(\eta) \cdot  (\cos(\eta)^2 - 1)^3 \\
      & \ p \in [0.2;0.4] \nonumber
  \label{eq:adima}
\end{align}

\begin{figure}[htbp]
  \smallerimage{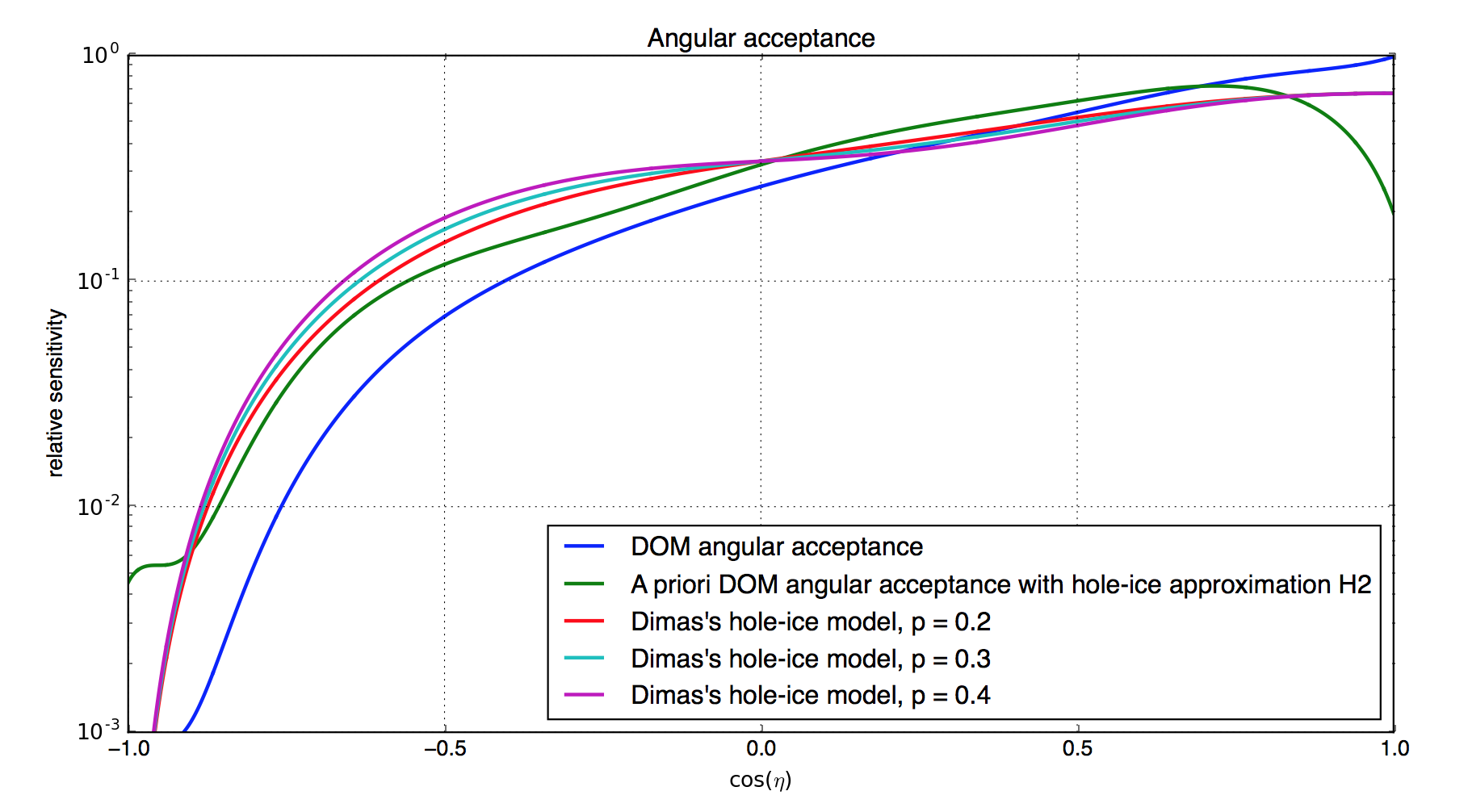}
  \caption{Comparing the a priori angular-acceptance curve H2 from \cite{icepaper} to Dimas's model \cite{flasherdataderivedicemodels}. Also, the angular acceptance of the optical module (DOM) without any hole ice is shown.}
  \label{fig:Vohn9Oov}
\end{figure}

Figure \ref{fig:Vohn9Oov} shows this angular-acceptance curve,
\(a\domdima(\eta;p)\), for the parameters \(p=0.2\), \(p=0.3\), and
\(p=0.4\), compared to the a priori angular-acceptance curve
\(a\domhi(\eta)\) of the H2 model and the angular-acceptance curve of an
optical module \(a\dom(\eta)\) without hole ice. Notably, the effect of
the hole ice on photons approaching the optical module directly from
above or below is weaker in this model compared to the H2 approximation
curve. However, the overall acceptance in the lower half sphere is
higher, and the overall acceptance in the upper half sphere is lower as
in the H2 approximation curve.

This approach does not rely on specific assumptions about the properties
of the hole ice, but uses flasher calibration data to measure the mean
angular-acceptance behavior of \icecube's optical modules. Nevertheless,
this method relies on the same assumptions regarding the equality of all
optical modules: All optical modules are assumed to have the same
relative position and orientation with respect to the hole ice.

\subsubsection{Direct Photon Propagation With \ppc}
\label{sec:direct_photon_propagation_with_ppc}

\label{sec:pocam}

The \noun{IceCube Simulation Framework} includes two independent
photon-propagation-simulation tools, \clsim and
\ppc (\noun{Photon Propagation Code}).
\cite{ppcpaper, ppcsource, ppcforhumans} Similar to this study, which
uses \clsim, simulations with direct propagation through hole ice have
been performed using \ppc.
\cite{martinspicehddard, martindardupdate, pocam, icrc17pocam, ppcpaper}

\sourcepar{The source code of \ppc can be found at \url{http://code.icecube.wisc.edu/projects/icecube/browser/IceCube/projects/ppc}.}

Simulating the propagation of photons through the hole ice by simulating
the scattering directly drops the assumption that all optical modules
need to be equally positioned and oriented with respect to the hole ice.
All optical modules can be individually calibrated using flasher data.
\cite{martinspicehddard}

The basic propagation algorithm is the same in \ppc and \clsim. The
algorithm for propagating the photons through different media, relies on
the same principle of converting randomized numbers of interaction
lengths to a geometric distances, but is implemented differently: Ice
layers and hole-ice cylinders are hard-coded in \ppc whereas in \clsim's
new medium-propagation algorithm, they are treated as generic medium
changes (see section \ref{sec:algorithm_b}). Both, \ppc and \clsim, can
be run on clusters of graphics processing units.
\cite{ppcpaper, ppcsource, ppcforhumans, clsimsource} Hardware
requirements and performance of \ppc and \clsim are comparable. Due to
micro optimizations, \ppc tends to run slightly
faster.\footnote{Internal correspondence suggests a performance factor of 1.5 to 2.0. See also section \ref{sec:performance} for performance considerations.}
\ppc does only support one hole-ice cylinder per string. The main cable
of the detector string is not implemented as absorbing object but rather
accounted for by using a direction-dependent shadowing factor for
incoming photons. \cite{ppcsource, ppcforhumans} In particular, the
cable cannot be modeled to reside inside the bubble-column cylinder.
\cite{martinspicehddard}

\begin{figure}[htbp]
  \subcaptionbox{No hole ice in both simulations.}{\halfimage{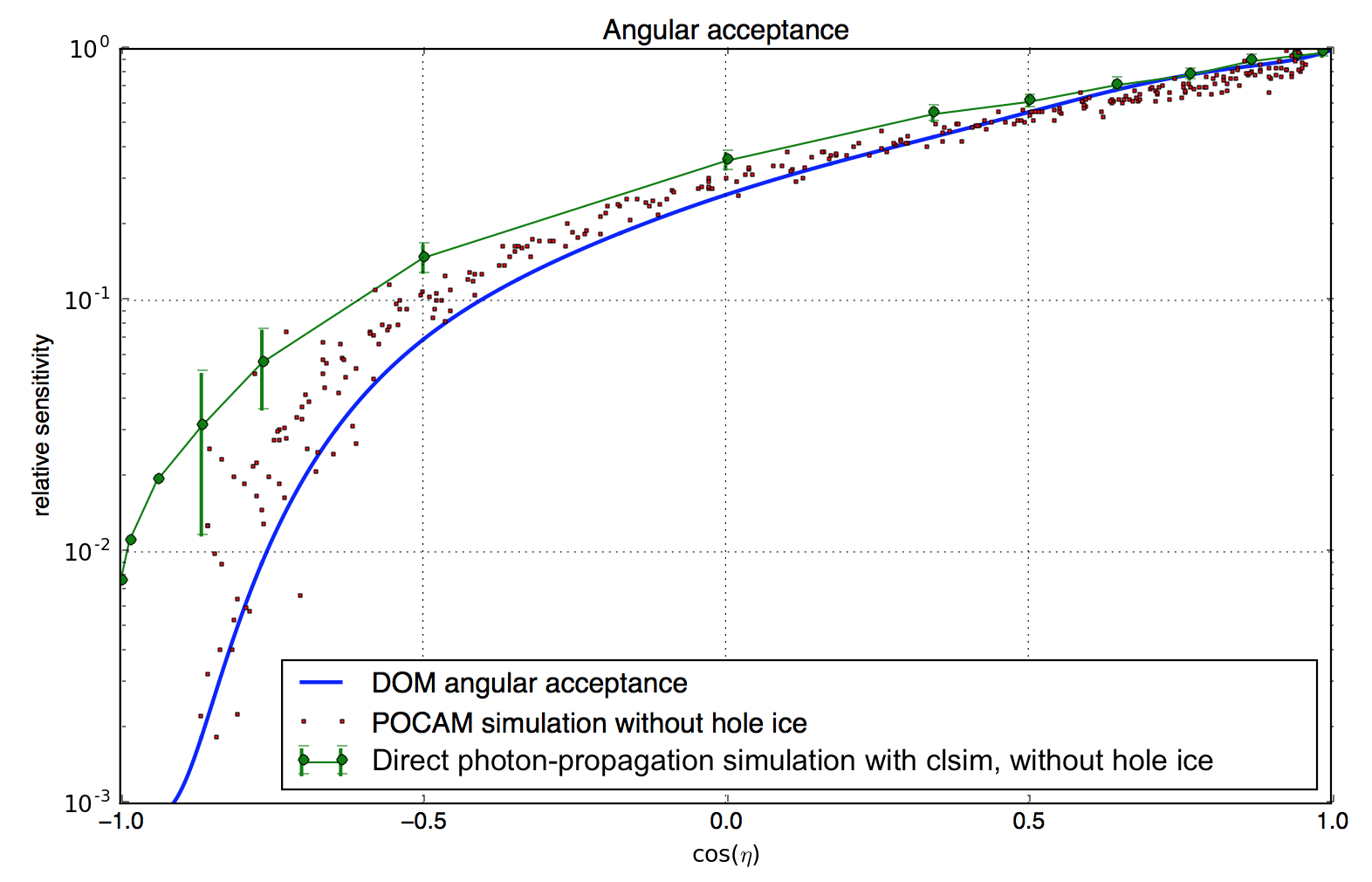}}\hfill
  \subcaptionbox{Hole-ice-cylinder radius $r = 0.6\,r\dom = 10\cm$, hole-ice effective scattering length $\lambda\esca\hi = 14\cm$, corresponding to a geometric scattering length $\lambda\sca\hi = 0.8\cm$ in both simulations.}{\halfimage{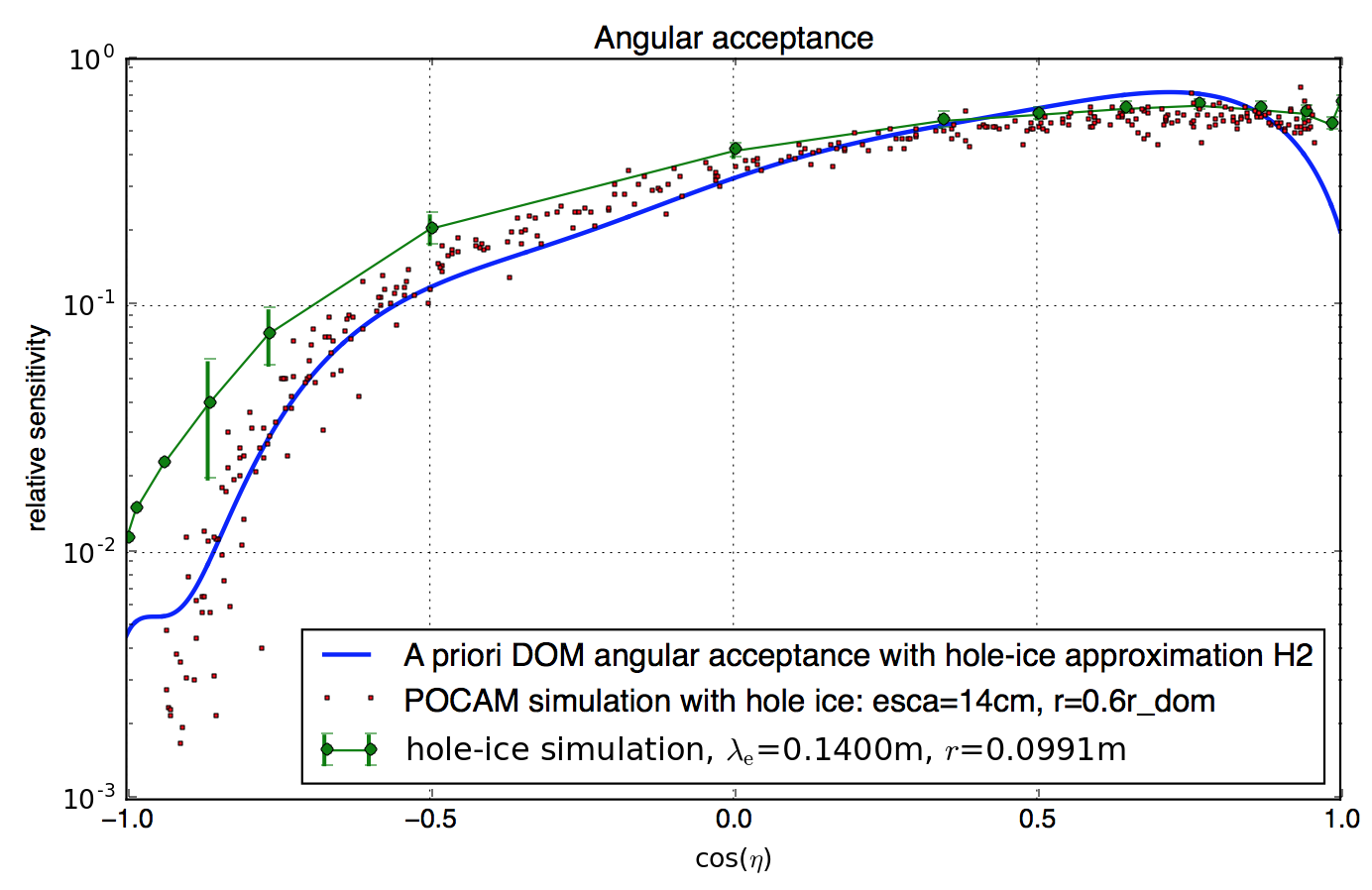}}\hfill
  \subcaptionbox{Hole-ice-cylinder radius $r = 1.8\,r\dom = 30\cm$, hole-ice effective scattering length $\lambda\esca\hi = 125\cm$, corresponding to a geometric scattering length $\lambda\sca\hi = 7.5\cm$ in both simulations.}{\halfimage{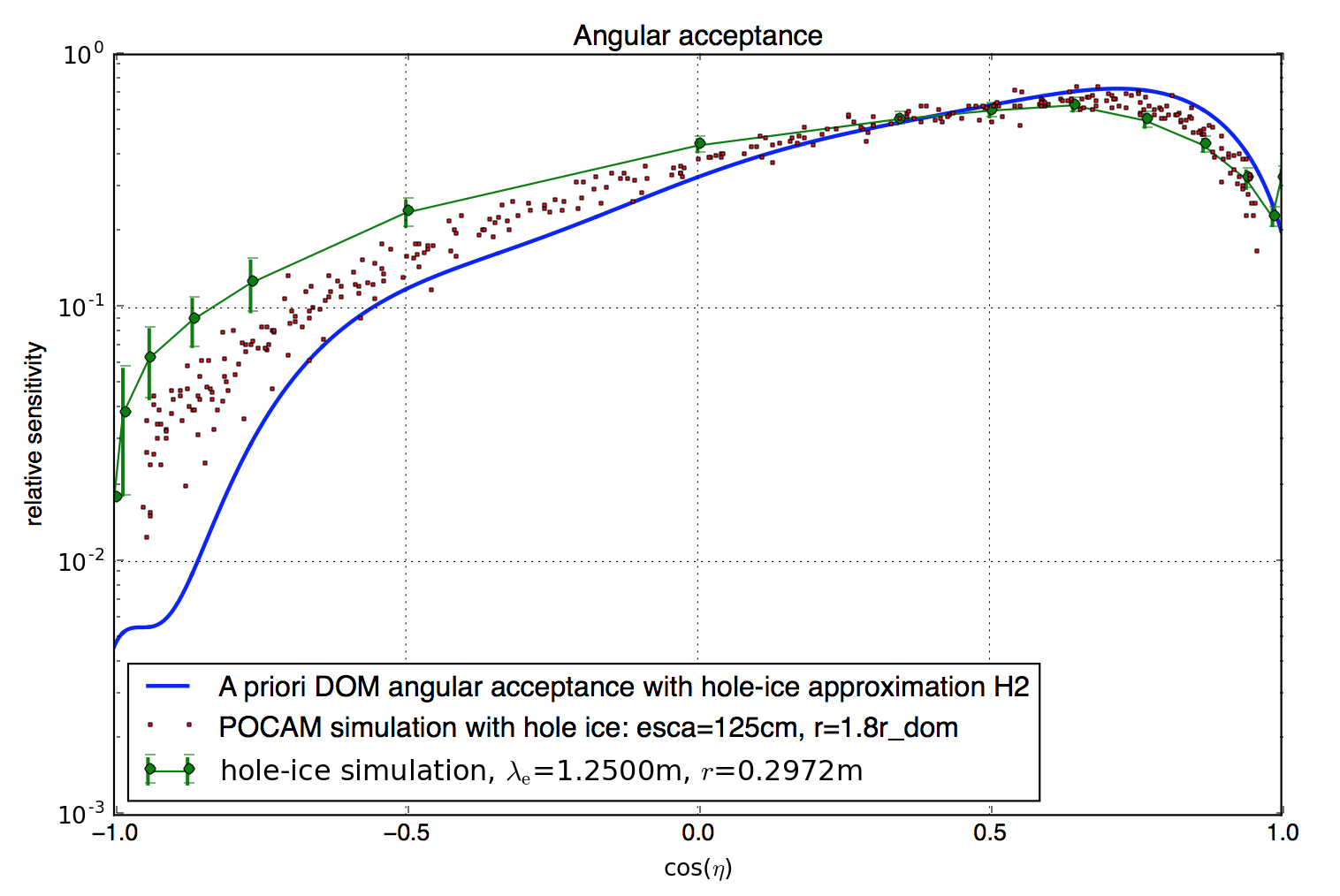}}\hfill
  \subcaptionbox{Hole-ice-cylinder radius $r = 1.8\,r\dom = 30\cm$, hole-ice effective scattering length $\lambda\esca\hi = 170\cm$, corresponding to a geometric scattering length $\lambda\sca\hi = 10.2\cm$ in both simulations.}{\halfimage{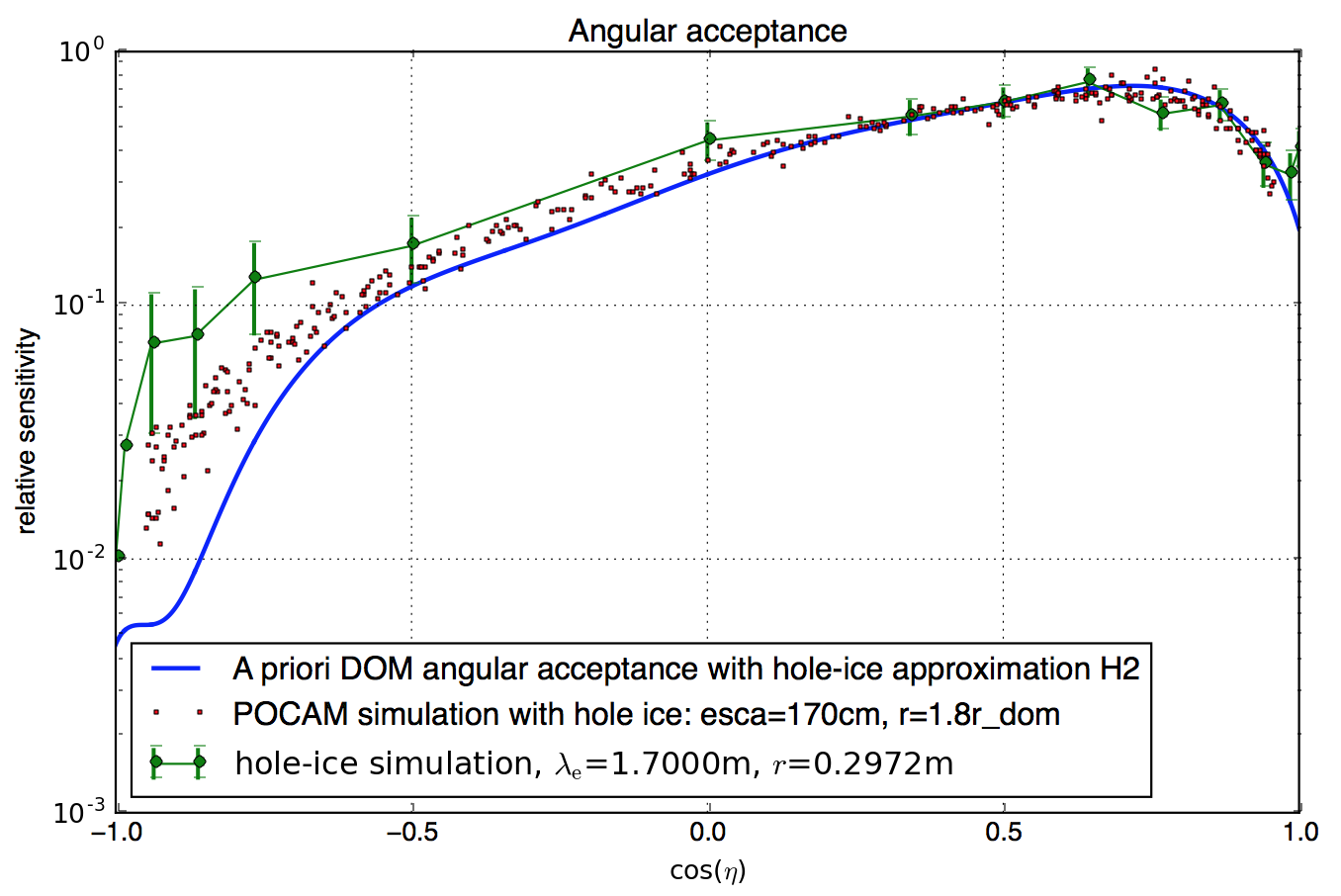}}\hfill
  \caption{Comparison of effective angular-acceptance curves, which include hole-ice effects for varying hole-ice parameters, created with two independent propagation tools, \ppc and \clsim. As reference, the a priori effective angular-acceptance curve for H2 parameters from \cite{icepaper} is shown. Data source of the \ppc simulations is a series of \noun{POCAM} simulations \cite{icrc17pocam}. The radius of the optical module is $r\dom = 16.510\cm$.}
  \label{fig:Ou7fux1o}
\end{figure}

Since both propagation softwares, \ppc and \clsim with the new
medium-propagation mechanism, are performing the same task, it is
important to know whether both simulations produce the same results
regarding the effect of the hole ice. A detailed comparison of \ppc and
\clsim results is out of scope of this study. But a first comparison can
be achieved using existing \ppc simulation results. \followup
In feasibility studies for the proposed
\noun{Precision Optical CAlibration Module} (\noun{POCAM}), direct
photon propagation simulations with \ppc have been performed, assuming
different hole-ice parameters, and producing effective
angular-acceptance curves corresponding to those hole-ice parameters.
\cite{pocam, icrc17pocam} Assuming the same hole-ice parameters
respectively, angular-acceptance simulations can be performed in \clsim,
allowing for a direct comparison of the \ppc and the \clsim results
(figure \ref{fig:Ou7fux1o}).

\docpar{The \clsim angular-acceptance simulations with hole-ice parameters from the \noun{POCAM} simulations are documented in \issue{103}.}

At a rough view, both simulations produce similar effective
angular-acceptance results for different hole-ice parameters. For upper
angles, however, this study's simulation systematically sees more photon
hits than \ppc. As this has already been observed in section
\ref{sec:angular_acceptance_simulations_without_hole_ice} (see figure
\ref{fig:Shai8yah}) when introducing plane waves as photon sources, this
effect could be a systematic problem of the implementation of the
angular-acceptance simulations rather than of the propagation algorithm.
Further investigations on this matter are required, but are out of scope
of this study.\followup

\subsection{Compatibility of Hole-Ice Parameters From Other Studies}
\label{sec:angular_acceptance_comparison}

\label{sec:parameter_comparison}

\newcommand\ok{\ding{51}} %
\newcommand\same{\cellcolor{black!25}}
\newcommand\greyedout{\cellcolor{black!25}}
\newcommand\bad{\ding{55}}

\newcommand\clsimppc{\noun{clsim+ppc}}

Simulations with direct photon propagation through hole ice of different
parameter configurations \(\HH\) allow to rate the compatibility of
different hole-ice models. For example, generating simulation-based
angular-acceptance curves allows to calculate the likelihood of two
models being compatible. This, however, would require a more involved
study of systematics of the specific simulation design, which is out of
scope of this study, but should be considered by follow-up
studies.\followup

As a first step, this section shows angular-acceptance curves from
simulations with hole-ice parameters from different studies. Table
\ref{tab:angular_acceptance_compatibility} at the end of this section
(page \pageref{tab:angular_acceptance_compatibility}) summarizes the
findings.

\subsubsection{YAG H2 Parameters}
\label{sec:yag_h2_parameters}

Based on laser measurements, using a Yttrium-Aluminium-Granat (YAG)
laser, \cite{holeicestudieswithyag} suggested a number of hole-ice
models, from which the so called \textit{H2 model} has become the
dominant hole-ice model in \icecube. This model suggests a geometric
scattering length of \(50\cm\) for the hole ice, and a hole-ice-cylinder
radius of \(30\cm\), and is the basis for the modified
angular-acceptance curve with hole-ice approximation described in
section \ref{sec:a_priori_modification_of_angular_acception_curves}.

\begin{figure}[htbp]
  \smallerimage{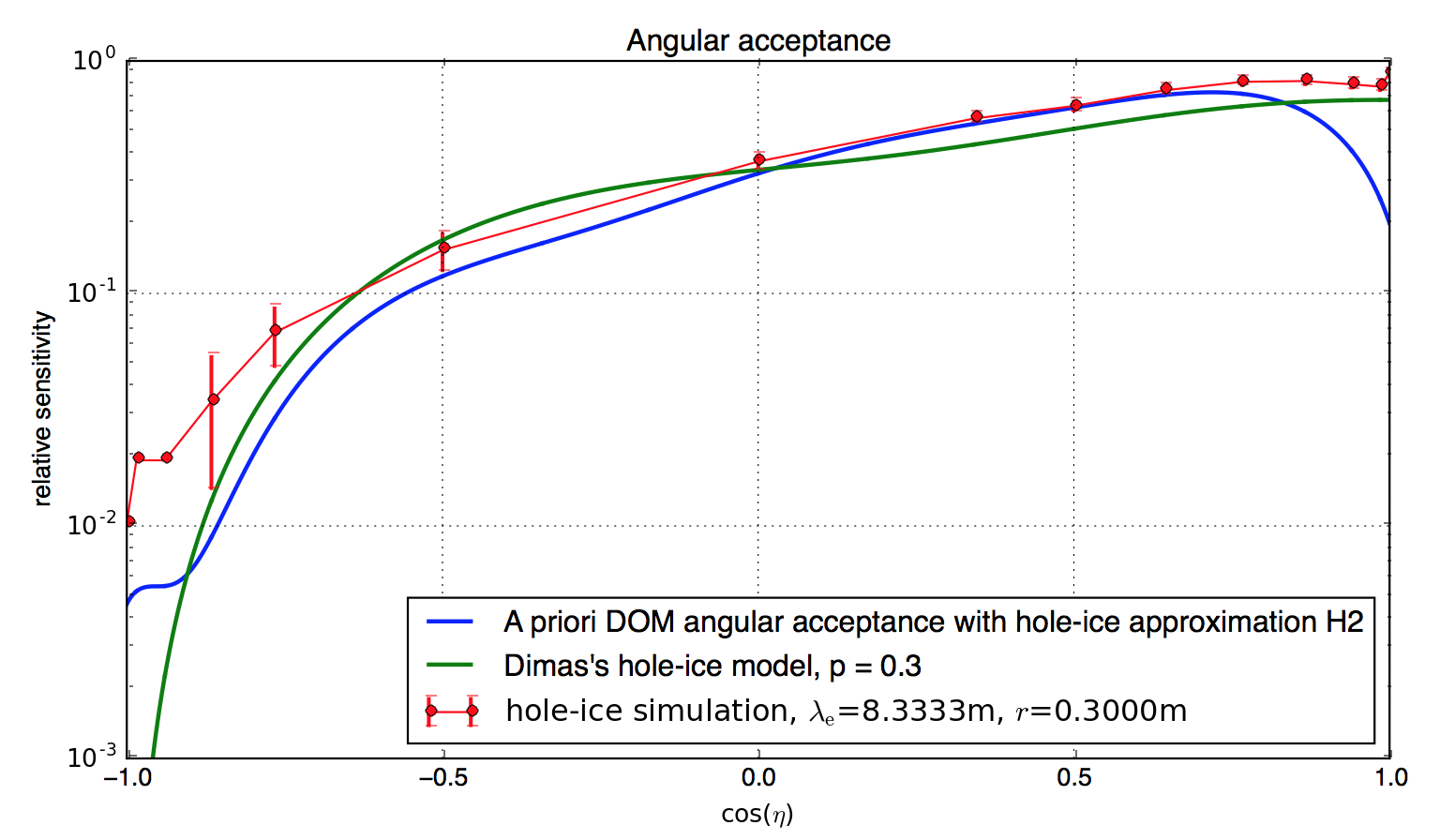}
  \caption{Angular-acceptance simulation with hole-ice parameters from the so called \textit{H2 model} \cite{holeicestudieswithyag}, which describes the hole ice as cylinder of $30\cm$ radius filling the entire drill hole, with a geometric scattering length of $50\cm$, corresponding to an effective scattering length of $\lambda\hi\esca = 8.33\m$, using the new medium-propagation algorithm (section \ref{sec:algorithm_b}) with direct detection as acceptance criterion and plane waves as photon sources. For comparison, the a priori angular-acceptance curve from \cite{icepaper} and the angular-acceptance curve of Dima's model \cite{flasherdataderivedicemodels} are shown.}
  \label{fig:xaeg2Mee}
\end{figure}

Using the new direct photon propagation though hole ice with \clsim, one
can simulate the propagation through hole ice with parameters suggested
by the H2 model, scan the angular acceptance of an optical module within
the simulation, and compare the result to the existing a priori
angular-acceptance curve (section \ref{sec:a_priori_curve})
\cite{icepaper} as well as the angular-acceptance curve suggested by
Dima's model (section \ref{sec:dimas_model})
\cite{flasherdataderivedicemodels}. Figure \ref{fig:xaeg2Mee} shows the
simulation results.

\docpar{This angular-acceptance simulation using the H2 hole-ice parameters is documented in \issue{80}.}

For photons approaching the optical module from below (right-hand side
of the plot in figure \ref{fig:xaeg2Mee}), the simulation rather follows
Dima's model, but counts more detected photons in total. For photons
approaching from above (left-hand side of the plot), the form of the
simulation curve is similar to the a priori curve, in particular with
respect to the plateau on the left hand side, but, again, the simulation
counts more photons in total.

\subsubsection{\noun{DARD} Parameters}
\label{sec:dard_parameters}

In the \noun{DARD} study (\noun{Data Acquisition for a flasheR DOM}),
LED measurements are used to determine the hole-ice properties. The
different flasher LEDs of an optical module are turned on in sequence,
and the light detected by the same DOM is measured. The process is then
compared to detailed \noun{Geant4}
simulations\footnote{Geant4 toolkit for the simulation of the passage of particles through matter, \url{https://geant4.web.cern.ch}}
for different hole-ice parameters in order to find the optimal hole-ice
parameters to explain the data.
\cite{martindardupdate, martinspicehddard}

The \noun{DARD} study assumes a fixed hole-ice radius of \(r=30\cm\).
The best fit value for the geometric scattering length of the hole ice
has been determined as \(\lambda\hi\sca = 10\cm\), which is considerably
lower than other estimates. \cite{martindardupdate}

\begin{figure}[htbp]
  \smallerimage{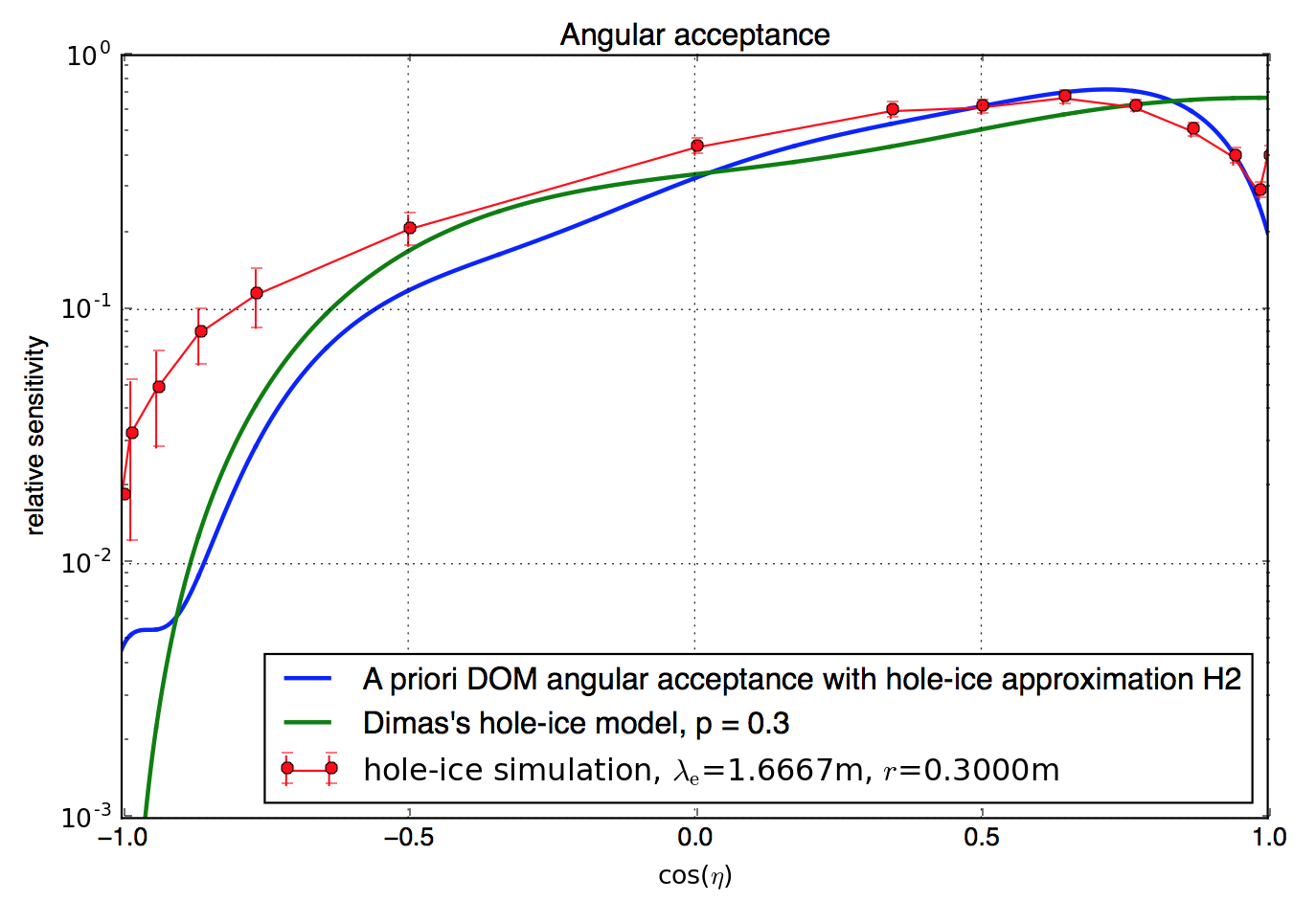}
  \caption{Angular-acceptance simulation with hole-ice parameters from the \noun{DARD} study \cite{martindardupdate}, which suggest a hole-ice cylinder of $30\cm$ radius and a geometric scattering length of $\lambda\sca\hi = 10\cm$, corresponding to an effective scattering length of $\lambda\hi\esca = 1.67\m$, using the new medium-propagation algorithm (section \ref{sec:algorithm_b}) with direct detection as acceptance criterion and plane waves as photon sources. For comparison, the a priori angular-acceptance curve from \cite{icepaper} and the angular-acceptance curve of Dima's model \cite{flasherdataderivedicemodels} are shown.}
  \label{fig:eePai1sh}
\end{figure}

Using the new direct photon propagation though hole ice with \clsim, one
can simulate the propagation through hole ice with parameters suggested
by \noun{DARD}, scan the angular acceptance of an optical module within
the simulation, and compare the result to the existing a priori
angular-acceptance curve \cite{icepaper} as well as the
angular-acceptance curve suggested by Dima's model
\cite{flasherdataderivedicemodels}. Simulation results are shown in
figure \ref{fig:eePai1sh}.

\docpar{This angular-acceptance simulation for the hole-ice parameters suggested by the \noun{DARD} study, is documented in \issue{105}.}

For photons approaching the optical module from below (right-hand side
of the plot in figure \ref{fig:eePai1sh}), the simulation curve rather
follows the a priori acceptance curve, but shows a less sharp hole-ice
drop-off effect. For higher angles, \(\cos\eta < 0.5\), the simulation
registers relatively more photons, both in comparison to the a priori
curve and to Dima's model.

\subsubsection{\noun{SpiceHD} Parameters}
\label{sec:spicehd_parameters}

The \noun{SpiceHD} study
(\noun{South Pole ICE model with Hole ice fitting and Direct detection})
uses flasher data and direct photon propagation simulations with \ppc to
determine the parameters of the hole ice. In this 7-string flasher
study, LED flashes are sent from a central optical module and the
numbers and time distributions of photon hits are observed at the
optical modules of the six surrounding
strings.\footnote{\noun{SpiceHD} uses the same methods as \cite{icepaper,dimaslikelihood}}

Comparing the flasher data to simulations with different hole-ice
parameters, scanning over the hole-ice-cylinder radius and the hole-ice
scattering length, a range of hole-ice parameters is found to be
suitable (section \ref{sec:spicehd_flasher_scan_contours}). When
assuming a compatible hole-ice column radius as observed by camera
images \cite{rongenswedishcamera}, the best fit hole-ice parameters have
been determined as a hole-ice-cylinder radius of \(60\,\%\) of the
radius of an optical module, \(r = 0.6\,r\dom = 10\cm\), and the
hole-ice effective scattering length \(\lambda\esca\hi = 14\cm\), which
corresponds to a geometric scattering length of
\(\lambda\sca\hi = 0.84\cm\), which is even smaller than the geometric
scattering length of \(10\cm\) suggested by \noun{DARD}.
\cite{martinspicehddard}

\begin{figure}[htbp]
  \smallerimage{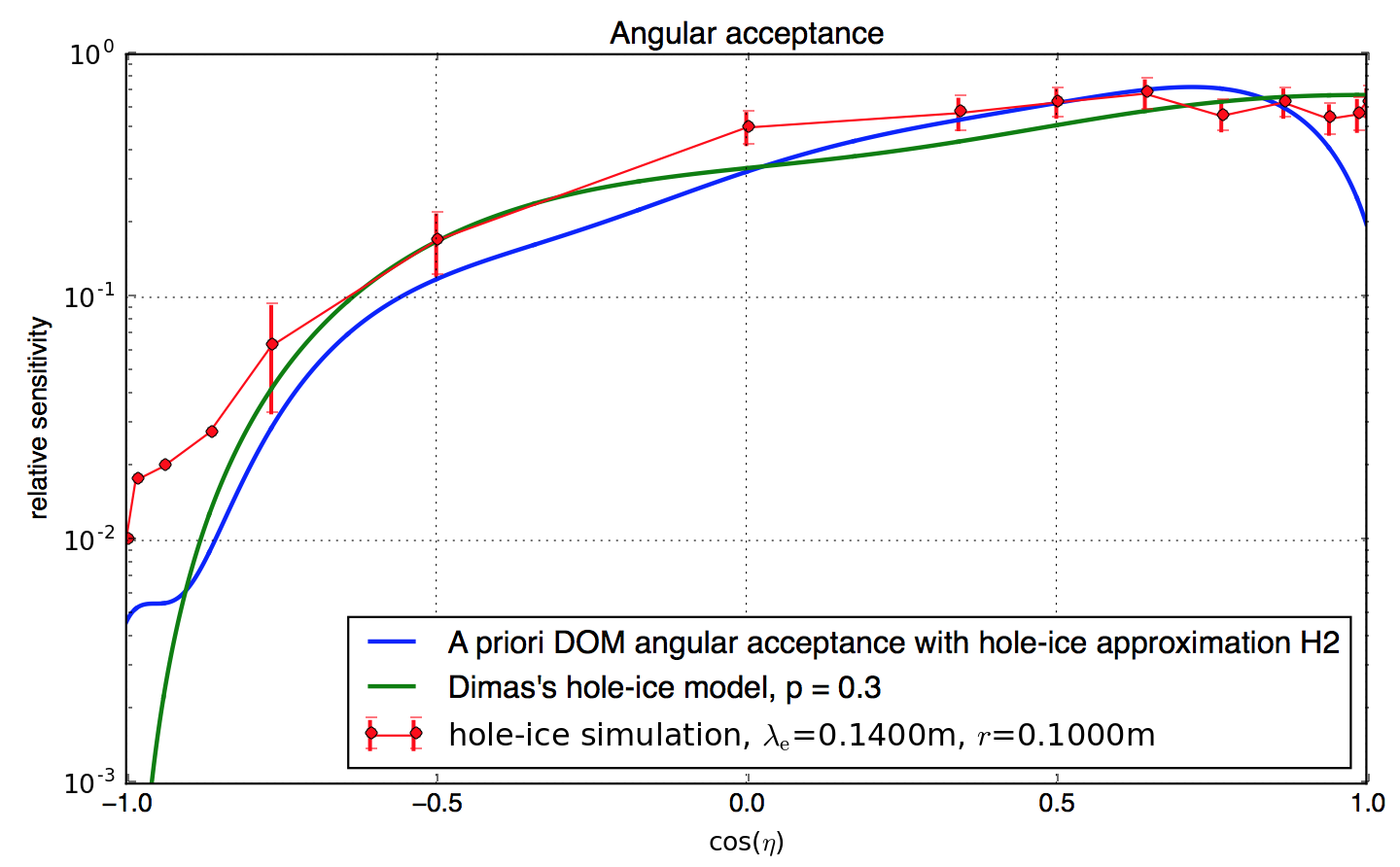}
  \caption{Angular-acceptance simulation with hole-ice parameters suggested by the \noun{SpiceHD} study \cite{martinspicehddard}: The dominant hole ice is assumed as a bubble column with radius $r = 0.6\,r\dom$ where $r\dom$ is the radius of the optical module. The bubble column is thinner than the optical module. The hole-ice effective scattering length is $\lambda\esca\hi = 14\cm$. The simulation uses the new medium-propagation algorithm (section \ref{sec:algorithm_b}) with direct detection as acceptance criterion and plane waves as photon sources. The simulation results are shown in comparison to the a priori angular-acceptance curve from \cite{icepaper} and the angular-acceptance curve of Dima's model \cite{flasherdataderivedicemodels}.}
  \label{fig:ku3Zie8z}
\end{figure}

Using the new direct photon propagation though hole ice with \clsim, one
can simulate the propagation through hole ice with parameters suggested
by \noun{SpiceHD}, scan the effective angular acceptance of an optical
module within the simulation, and compare the result to the existing a
priori angular-acceptance curve \cite{icepaper} as well as the
angular-acceptance curve suggested by Dima's model
\cite{flasherdataderivedicemodels} (figure \ref{fig:ku3Zie8z}).

\docpar{This angular-acceptance simulation for the hole-ice parameters suggested by the \noun{SpiceHD} study, is documented in \issue{87}.}

For photons that approach the optical module from below (right-hand side
of the plot in figure \ref{fig:ku3Zie8z}), the simulation shows a
hole-ice effect between the a priori curve and Dima's model. For photons
approaching from above, the simulation shows a plateau. For
\(\cos\eta < 0.5\), the simulation shows more photon hits than both, the
a priori curve and Dima's model.

\subsubsection{\noun{SpiceHD} Flasher-Scan Contours}
\label{sec:spicehd_flasher_scan_contours}

The \noun{SpiceHD} study
(\noun{South Pole ICE model with Hole ice fitting and Direct detection},
see also section \ref{sec:spicehd_parameters}) performs a similar
flasher study as described in section \ref{sec:flasher}. Both studies
compare simulations using direct photon propagation through hole ice of
varying parameters to flasher calibration data. Whereas section
\ref{sec:flasher} uses the new medium-propagation algorithm for \clsim,
\noun{SpiceHD} uses \ppc to propagate the photons.
\cite{martinspicehddard}

\begin{figure}[htbp]
  \subcaptionbox{\noun{SpiceHD} study using direct hole-ice propagation with \ppc \cite{martinspicehddard}, fitting the size and scattering length of the hole ice and the position of the optical modules relative to the hole ice.}{\halfimage{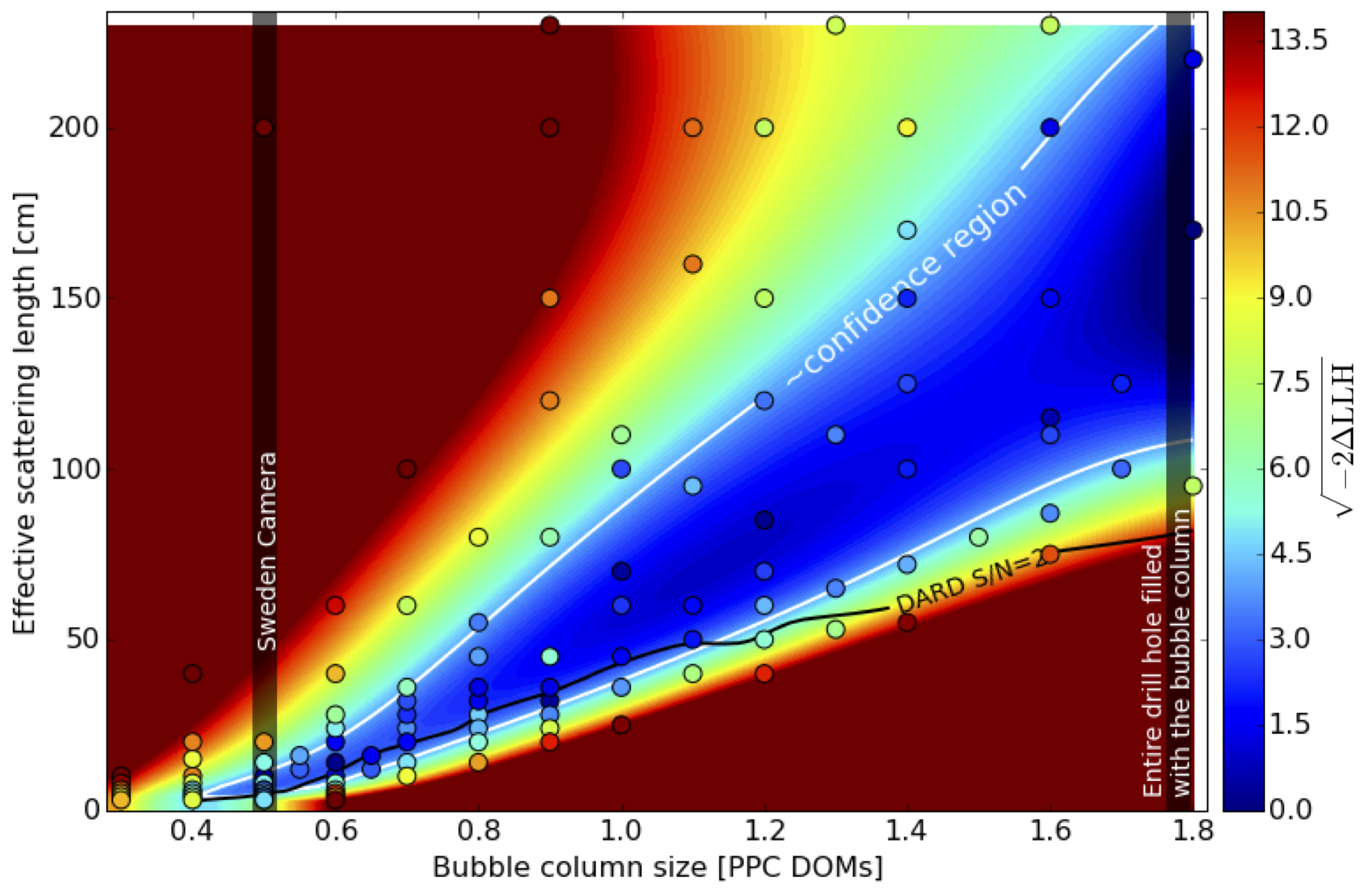}\vspace*{1mm}}\hfill
  \subcaptionbox{Flasher study with direct hole-ice propagation with \clsim (section \ref{sec:flasher}), fitting only the size and the scattering length of the hole ice.}{\halfimage{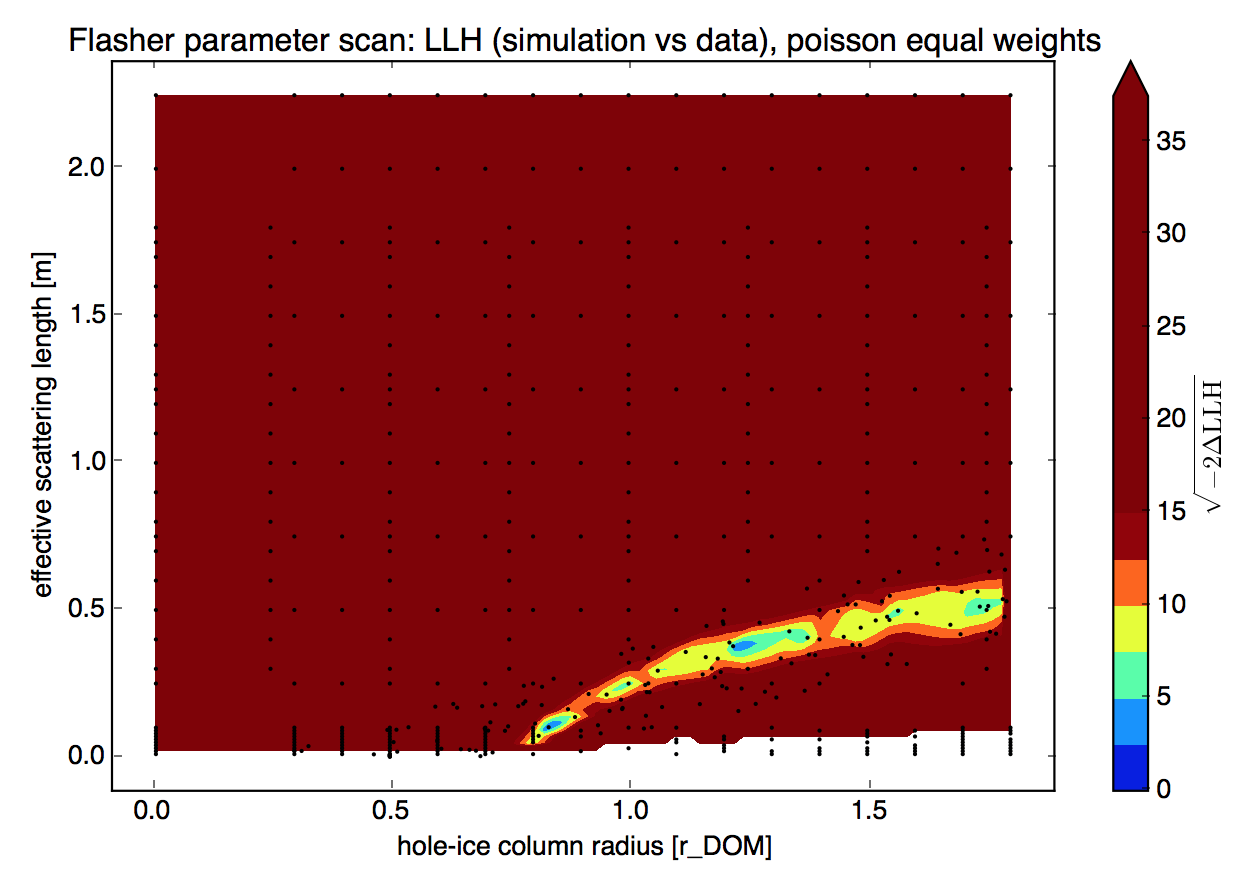}}
  \caption{Fitting flasher simulations for different hole-ice parameters to calibration data.}
  \label{fig:ahCoHee4}
\end{figure}

The most important difference between the two studies is that the
flasher scan of section \ref{sec:flasher} assumed all optical modules
being centered relative to their hole-ice column, but \noun{SpiceHD}
uses the positions of the optical modules relative to the hole ice as
additional fit parameters. These additional fit parameters make
\noun{SpiceHD}'s contour plot (figure \ref{fig:ahCoHee4} a) smoother and
the range of good fit parameters wider. This can best be seen when
adjusting the likelihood scale of the flasher scan of section
\ref{sec:flasher} to match the scale of \noun{SpiceHD}, as shown in
figure \ref{fig:ahCoHee4} (b).

Both studies show a ``valley''-like region of good agreement of data and
simulation. In the parameter region above the valley, where scattering
lengths are longer or hole-ice radii are smaller, the simulated hole ice
is too weak. In the parameter region below the valley, the simulated
hole ice is too strong. However, both, the parameter region of the
valley and the curvature are different.

As \clsim and \ppc show good agreement when comparing angular-acceptance
curves (section \ref{sec:pocam}), this difference is expected to be
caused by the systematics of the preliminary flasher study, discussed in
section \ref{sec:flasher}, and the different assumptions regarding the
positions of the optical modules relative to the hole ice. Preliminary
estimations show that this difference accounts roughly for a factor of
\(3\) for the fitted scattering length when assuming large hole-ice
radii. This agrees with the behavior seen in figure \ref{fig:ahCoHee4}.
For \(r=1.8\,r\dom\), figure \ref{fig:ahCoHee4} (b) shows an effective
scattering length of \(0.5\m\). Figure \ref{fig:ahCoHee4} (a) shows
about \(1.5\m\), which is three times as much.

The \noun{SpiceHD} contours are compatible to camera observations
(section \ref{sec:hole_ice}) that show a hole-ice radius of about
\(10\cm\). \cite{rongenswedishcamera} The matching \noun{SpiceHD}
hole-ice parameters are \(r = 0.6\,r\dom = 10\cm\),
\(\lambda\sca = 0.84\cm\), \(\lambda\esca = 14\cm\).
\cite{martinspicehddard}

\subsubsection{Compatibility of Hole-Ice Models}
\label{sec:compatibility_of_hole_ice_parameters}

Direct photon-propagation-simulation through hole ice allows to compare
different hole-ice models and hole-ice parameters. Both
photon-propagation-simulation tools, \ppc and \clsim, produce matching
effective angular-acceptance curves for direct photon propagation
through hole ice. For photons approaching the optical module from above,
the systematic uncertainties of the \clsim angular-acceptance scans
appear too large to consider this angular range. Therefore, this section
focuses on the angular range \(\cos \eta > 0\) with large statistics,
when assessing the compatibility of different hole-ice models.

The \noun{SpiceHD} flasher study is compatible to camera observations.
The resulting hole-ice parameters \(\HH_\text{SpiceHD}\) as well as the
parameters \(\HH_\text{YAG H2}\) suggested by earlier laser calibration
studies, both suggest a weaker hole-ice effect than approximated in the
modified a priori angular-acceptance curve used in standard
\clsim simulations. Dima's hole-ice model that only uses
flasher-calibration data and does not rely on direct hole-ice
simulations, also suggests a weaker hole ice. However, using the
hole-ice parameters \(\HH_\text{DARD}\) suggested by the single-LED
calibration study \noun{DARD}, angular-acceptance curves from direct
simulations look similar to the a priori angular-acceptance curve. Table
\ref{tab:angular_acceptance_compatibility} shows a summary (page
\pageref{tab:angular_acceptance_compatibility}).

This leaves no obvious conclusion on which current hole-ice model is to
be preferred. But as both direct hole-ice-simulation tools produce
compatible results, both tools can now be used to study the various
hole-ice scenarios in detail.

\begin{table}[p]
  \begin{center}
  \resizebox{\columnwidth}{!}{%
  \begin{tabular}{rr|ccccccc}
                            & \small{\textbf{Method}}     & \clsimppc           & \clsimppc         & \clsimppc            & A priori   & Dima \\
    \small{\textbf{Method}} & \small{\textbf{Parameters}} & $\HH_\text{YAG H2}$ & $\HH_\text{DARD}$ & $\HH_\text{SpiceHD}$ & $\HH_\text{YAG H2}$ & -- \\
    \hline
    \clsimppc               & $\HH_\text{YAG H2}$         & \same               & \greyedout        & \greyedout           & \greyedout & \greyedout \\
    \clsimppc               & $\HH_\text{DARD}$           & \bad                & \same             & \greyedout           & \greyedout & \greyedout \\
    \clsimppc               & $\HH_\text{SpiceHD}$        & \ok                 & \bad              & \same                & \greyedout & \greyedout \\
    A priori                & $\HH_\text{YAG H2}$         & \bad                & \ok               & \bad                 & \same      & \greyedout\\
    Dima's model            & --                          & \ok                 & \bad              & \ok                  & \bad       & \same
  \end{tabular}
  }
  \end{center}
  \caption{Compatibility of angular-acceptance curves $a(\eta)$. \textbf{Methods}: \underline{\clsim}: Angular-acceptance simulation with \clsim using the new medium-propagation algorithm with plane waves as photon sources and direct detection as acceptance criterion (section \ref{sec:angular_acceptance_scan}). \underline{\ppc}: POCAM-angular-acceptance simulation with \ppc using direct photon propagation through hole ice and direct detection (section \ref{sec:direct_photon_propagation_with_ppc}). \underline{\clsimppc}: Simulations performed with both, \clsim and \ppc, producing similar results. \underline{A priori}: A priori angular-acceptance curve $a\domhi(\eta)$ for $\HH_\text{YAG H2}$ parameters from \cite{icepaper} (section \ref{sec:hole_ice_approximation}). \underline{Dima}, \underline{Dima's model}: Angular-acceptance curve $a\domdima(\eta;p)$ from flasher unfolding studies \cite{flasherdataderivedicemodels} (section \ref{sec:dimas_model}). \textbf{Hole-Ice Parameters}: Each parameter set $\HH$ consists of a hole-ice-cylinder radius $r$ and a geometric hole-ice scattering length $\lambda\sca$, corresponding to an effective scattering length $\lambda\esca$. \underline{$\HH_\text{YAG H2}$}: from YAG laser measurements \cite{holeicestudieswithyag}, also called H2 model parameters, $r = 30\cm, \lambda\sca=50\cm, \lambda\esca = 8.33\m$ (section \ref{sec:yag_h2_parameters}). \underline{$\HH_\text{DARD}$}: from single-DOM LED measurements \cite{martindardupdate}, $r=30\cm$, $\lambda\sca = 10\cm, \lambda\esca = 1.67\m$ (section \ref{sec:dard_parameters}). \underline{$\HH_\text{SpiceHD}$}: from 7-string LED flasher studies \cite{martinspicehddard} and camera observations \cite{rongenswedishcamera}, $r = 0.6\,r\dom = 10\cm$, $\lambda\sca = 0.84\cm$, $\lambda\esca = 14\cm$ (section \ref{sec:spicehd_parameters}). \textbf{Rating}: \ok: Curves look roughly similar. \bad: Curves do not look similar.}
  \label{tab:angular_acceptance_compatibility}
\end{table}

  \subsection{Performance Considerations}
\label{sec:performance}

The performance of the different propagation algorithms, quantified by
the time it takes to simulate a certain number of propagation steps, is
about the same. The number of propagation steps, however, which is
determined by the number of scatterings, has a significant effect on the
total simulation time.

\begin{figure}[htbp]
  \includegraphics[width=\textwidth, trim = {1cm 0 10cm 0.7cm}, clip]{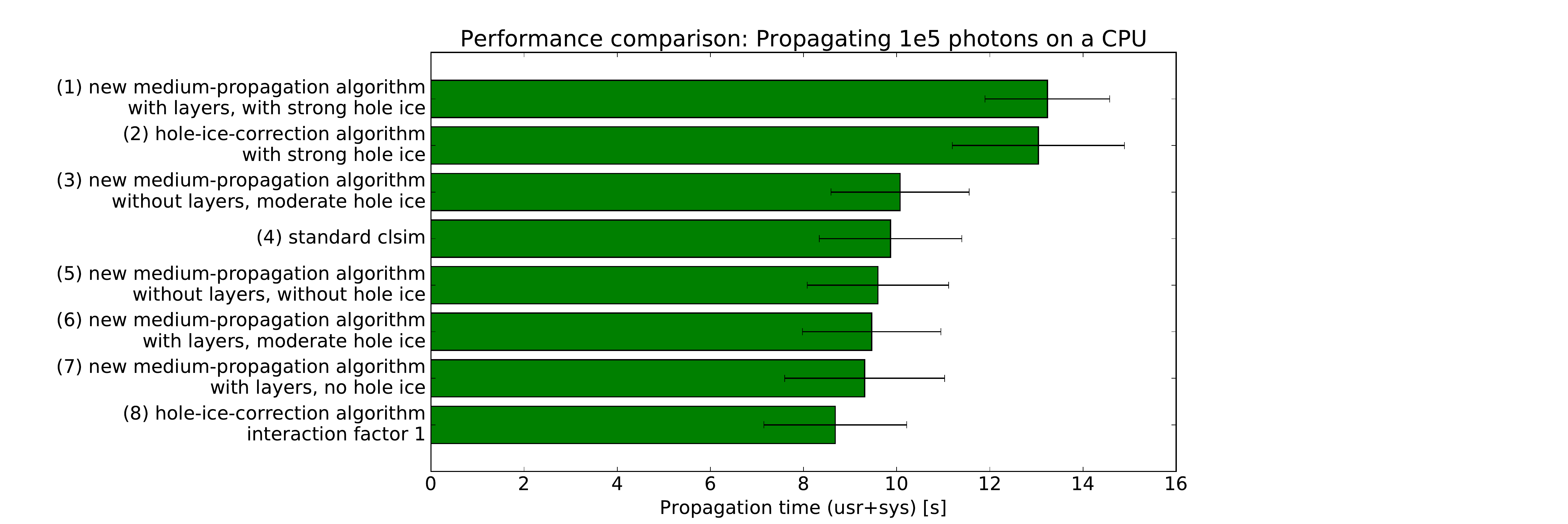}
  \caption{Performance comparison different hole-ice algorithms and different hole-ice scenarios.}
  \label{fig:Go7Maquo}
\end{figure}

Figure \ref{fig:Go7Maquo} shows a comparison of the
standard-\clsim algorithm (section \ref{sec:standard_clsim}), the
hole-ice-correction algorithm (section \ref{sec:algorithm_a}), and the
new medium-propagation algorithm with hole ice (section
\ref{sec:algorithm_b}), each running a simulation propagating \(10^5\)
photons on a
CPU.\footnote{Test system configuration for CPU simulation: OS: macOS Sierra 10.12.6 16G1212 x86\_64, Kernel: 16.7.0, CPU: Intel i7-4870HQ (8) @ 2.50GHz, GPU: Intel Iris Pro, NVIDIA GeForce GT 750M, RAM: 16384MiB.}

\docpar{The implementation of this performance measurement is documented in \issue{49}.}

The difference of the total simulation time for each algorithm is
smaller than the statistical uncertainty of the time measurement as long
as the scattering length of the hole-ice cylinder is kept moderate (rows
3 to 8 in figure \ref{fig:Go7Maquo}). When moving to smaller scattering
lengths within the hole-ice, however, the total simulation time
increases (rows 1 and 2 in figure \ref{fig:Go7Maquo}), as the total
number of simulation steps increases, because one simulation step
propagates the photon from one scattering point to the next scattering
point. For a fixed hole-ice-cylinder radius of \(30\cm\), going from a
geometric hole-ice scattering length of \(20\m\) to \(0.005\m\)
increases CPU propagation time per photon by about \(15\,\%\).

The same effect can be observed when comparing the run time of flasher
simulations on a
GPU.\footnote{Test system configuration for GPU simulation: Scientific Linux release 7.4 (Nitrogen) x86\_64, Kernel: 3.10.0-862.6.3.el7.x86\_64, CPU: Intel Xeon E5-2660 0 (32) @ 3.000GHz, GPU: NVIDIA Tesla K20m, RAM: 64215 MiB}
Running a flasher simulation with standard \clsim using hole-ice
approximation takes about 11 minutes. Running the same simulation with
the same number of photons, with the new medium-propagation algorithm,
but without any hole-ice cylinders takes about 10 minutes. Running the
same simulation, but adding hole-ice cylinders with \(36\cm\) radius and
a scattering length of \(\sfrac{1}{10}\) of the surrounding bulk ice
takes about 15 minutes, increasing the simulation time by about
\(50\,\%\).

As the number of simulation steps is the dominant factor for the
required simulation time, it is desirable to further optimize the
implementation of the simulation step (section
\ref{sec:technical_issues_and_optimizations}), because even small
improvements, multiplied by the number increased number of scattering
steps for hole-ice simulations, cause a considerable total performance
gain.\followup

These performance considerations also lead to the question when it is
most useful to use hole-ice propagation algorithms and when to use the
default hole-ice approximation with effective angular-sensitivity
curves. The approximation does not consider the effective reflection of
photons when entering the hole ice (section
\ref{sec:scattering_simulation}), but only removes a certain percentage
of photons for each incoming angle. Also, the approximation considers
each optical module perfectly centered within the hole ice rather than
accounting for the individual displacement of each module. In
high-energy events with large statistics where many photons and many
different optical modules are involved, the displacement effects are
expected to cancel out and the reflection effects are expected to be
negligible. For these scenarios, in particular for events with long
high-energy muon tracks, the hole-ice approximation is expected to be
sufficient. For lower-energy events and events where only a few optical
modules are involved, using a calibrated direct photon propagation
simulation is expected to reduce simulation systematics. Further
studies, however, are required to quantify this reduction of
systematics, or equivalently to quantify the gain in precision.\followup

  \subsection{Features Not Considered in This Study}
\label{sec:ice_features_not_considered}

This study provides the means for a
\textit{direct propagation of photons through hole ice} and, in general,
through cylinders of different ice properties (section
\ref{sec:algorithm_b}). \textit{Direct detection} is added as acceptance
criterion of the optical modules (section \ref{sec:direct_detection}).

For the propagation over large distances, the \textit{ice-layer tilt}
and the \textit{ice anisotropy} (section \ref{sec:ice}) are omitted in
this study and need to be re-implemented into the new medium-propagation
algorithm.\footnote{For the current status of the re-implementation of ice-layer tilt and ice anisotropy, check \url{https://github.com/fiedl/hole-ice-study/issues/48}.}

The hole ice is modeled as homogeneous cylinder with exact boundaries. A
gradient of the hole-ice properties in the radial direction, which could
model the radial gradient in the concentration of air bubbles in the
bubble column (section \ref{sec:hole_ice}) is not considered. Also, a
gradient in \(z\)-direction to model a pressure-caused gradient in the
ice properties of the hole ice (also section \ref{sec:hole_ice}) is not
included in this study.

The hole-ice cylinders are implemented perfectly aligned along the
\(z\)-direction. A tilt of the cylinders, or even deformations of the
cylinders which could be caused by oscillations of the drilling head are
not considered in this study.

For the direct-detection acceptance criterion (section
\ref{sec:direct_detection}), the optical modules are assumed perfectly
aligned along the \(z\)-direction. Tilted orientations of the optical
modules are not accounted for in this study, even though the orientation
of the optical modules could be fitted using calibration data. Also, the
impact angle is completely ignored when using direct detection, even if
transmission or quantum efficiency effects might be direction dependent.

This study also does not provide a ready-to-use geometry definition of
the whole detector with all optical-module positions, hole-ice cylinders
and cable
positions.\footnote{For the current status of the creation of a full geometry file, see \url{https://github.com/fiedl/hole-ice-study/issues/61}.}
Also, the performance impact when using such a full geometry file in
production simulations is not examined in this study and needs to be
evaluated in a follow-up study.\followup

  \section{Conclusion}
\label{sec:conclusion}

This study presented a new method to simulate the propagation of photons
through the hole ice around the detector strings of the
\icecube neutrino observatory as an extension of the standard
\clsim photon-propagation-simulation software.

Two algorithms were introduced to accomplish this task. As a first
approach, an algorithm was developed that accounts for the different
optical properties of the hole ice by adding subsequent corrections to
the calculations of the standard-\clsim algorithm. The changes to the
well-tested standard-\clsim codebase are kept minimal. The algorithm
requires the definition of the hole-ice properties to be relative to the
properties of the surrounding ice, which, however, does not allow to
implement more current hole-ice models. Therefore, a second algorithm
was devised that handles the transition from bulk ice to hole ice and
vice versa the same way it handles other medium transitions like the
propagation through ice layers, and allows to simulate hole ice with
absolute scattering and absorption lengths. Because this approach is
more generic, it also allows more complex hole-ice scenarios where a
photon crosses more than one hole-ice cylinder between two scatters.
This approach, however, required to rewrite the previous
\clsim media-propagation code. The layer tilt and ice anisotropy are
left out in the second algorithm for now, but can be added in the
future. The validity of the algorithms is supported by unit tests and a
series of statistical cross checks. \clsim hole-ice simulation results
were found to agree with simulation results of \ppc, which is a separate
\icecube simulation software that also supports direct hole-ice
propagation, but uses a different approach to implementing the medium
transitions. The performance difference per scattering step of the
standard-\clsim media propagation and the new hole-ice algorithms is
negligible within the statistical uncertainties. Simulating the direct
photon propagation through hole ice with small scattering lengths,
however, adds more scattering steps to the simulation, resulting in a
longer total simulation time. The performance of large-scale simulations
can still be improved by applying GPU-programming and memory
optimizations.

In general, the new method allows to simulate the propagation through
hole-ice cylinders with different absorption lengths, scattering
lengths, and radii. The optical modules can be positioned with
individual displacements relative to the hole ice. The method supports
the nesting of several hole-ice cylinders of different properties and
the simulation of light-absorbing cables. Optical modules support direct
detection and thereby can accept photon hits based on whether the photon
hit the sensitive area of the optical module rather than based on the
impact angle, as done in standard \clsim. These features allow to fit
additional calibration parameters such as the positions, sizes, and
scattering lengths of the individual hole-ice columns in the
\icecube detector.

Direct-photon-propagation simulations indicate that light propagation
through the \icecube detector on large scales is mostly unaffected by
the hole ice. Each photon, however, that is eventually detected by the
optical modules, and every photon that is emitted by the calibration
LEDs, needs to propagate through the hole ice and is affected by the
properties of the hole ice. For sufficiently sized hole-ice columns with
small scattering lengths, the optical modules are effectively shielded
by the hole ice. A fraction of the photons is absorbed during the random
walk through the hole ice, or effectively reflected by the hole ice.
Evaluating calibration data indicates a strongly asymmetric shielding of
the detector modules. Preliminary flasher simulations with direct photon
propagation hint that this cannot be accounted for by the shadow of
cables, but can be explained by simulating hole ice with a suitable
scattering length, size and position relative to the detector modules.

The new simulation method will be integrated into \icecube's simulation
framework. Low-statistics studies can use the new simulation method to
propagate all photons without hole-ice approximations in order to reduce
systematic uncertainties. As the direct hole-ice propagation requires
additional simulation time, high-statistics studies should continue to
use approximation techniques, which modify the angular-acceptance
behavior of the simulated optical modules to effectively account for the
average effect of the propagation through hole-ice columns. Comparing
the current standard hole-ice approximation to direct simulations, both
methods were found to disagree. Assuming the same hole-ice properties,
the new simulations indicate a less pronounced hole-ice effect than the
approximation curves. As recent calibration studies contradict each
other regarding the strength of the hole-ice properties, there is no
clear indication towards one of these models. With the new simulation
tools, however, more detailed hole-ice-simulation studies can be
performed aiming to find a hole-ice description that allows to reproduce
all calibration observations. After that, the approximation curves can
be updated accordingly to match the new hole-ice model.

  \appendix

\section{Appendix}

\nocite{*}
\hypersetup{urlcolor=redlink}
\printbibliography[heading=subbibnumbered]
\defaulthypersetup
  \cleardoublepage
\subsection{Contents of the CD-ROM}
\label{sec:cd_rom_contents}

\begin{tabelle}{lL}
  \texttt{algorithm\_a} & Source code of the hole-ice-correction algorithm (section \ref{sec:algorithm_a} and appendix \ref{sec:algorithm_a_source}) \\
  \texttt{algorithm\_b} & Source code of the new medium-propagation algorithm with hole ice (section \ref{sec:algorithm_b} and appendix \ref{sec:algorithm_b_source}) \\
  \texttt{clsim} & Source code of standard \clsim in version \texttt{V05-00-07} of the \icecube simulation framework \\
  \texttt{hole-ice-study} & scripts and data files used in this study, corresponding to the repository \url{https://github.com/fiedl/hole-ice-study} \\
  \texttt{issues} & Documentation of issues corresponding to \url{https://github.com/fiedl/hole-ice-study/issues} \\
  \texttt{latex} & \LaTeX\xspace version of this document, corresponding to the repository \url{https://github.com/fiedl/hole-ice-latex} \\
  \texttt{papers} & Papers from the bibliography, except for \icecube-internal material \\
  \texttt{pdf} & This document in portable document format (PDF) \\
  \texttt{unit\_tests} & Unit tests for the hole-ice-correction algorithm (section \ref{sec:unit_tests}) \\
  \texttt{video} & Animated visualization of the photon-propagation through a hole-ice cylinder (section \ref{sec:scattering_simulation}), corresponding to \url{https://youtu.be/BhJ6F3B-I1s} \\
\end{tabelle}

\subsection{List of Abbreviations}

\begin{tabelle}{lL}
  \clsim & OpenCL-based photon-tracking simulation using a (source-based) ray tracing algorithm modeling scattering and absorption of light in the deep glacial ice at the South Pole or Mediterranean sea water. See \cite{clsimreadme, clsimsource}. \\
  DOM & Digital Optical Module, primary instrumentation and detection unit in \icecube, see section \ref{sec:doms}. \\
  H0 & Hypothesis stating that no hole ice exists in the \icecube detector. See \cite{yag}. \\
  H1 & Hypothesis stating that hole ice exists in the \icecube detector with a hole-ice radius of $30\cm$, filling the entire drill hole, and a geometric hole-ice scattering length of $100\cm$. See \cite{yag}. \\
  H2 & Hypothesis stating that hole ice exists in the \icecube detector with a hole-ice radius of $30\cm$, filling the entire drill hole, and a geometric hole-ice scattering length of $50\cm$. See \cite{yag}. \\
  H3 & Hypothesis stating that hole ice exists in the \icecube detector with a hole-ice radius of $30\cm$, filling the entire drill hole, and a geometric hole-ice scattering length of $30\cm$. See \cite{yag}. \\
  H4 & Hypothesis stating that hole ice exists in the \icecube detector with a hole-ice radius of $30\cm$, filling the entire drill hole, and a geometric hole-ice scattering length of $10\cm$. See \cite{yag}. \\
  LED & Light-emitting diode, part of \icecube's flasher calibration system, see section \ref{sec:flasher}. \\
  \ppc & photon propagation code. A photon-propagation simulation software. See \cite{ppcpaper, ppcsource, ppcforhumans}. \\
  PMT & Photo Multiplier Tube, instrument to detect light, component of \icecube's optical modules, see section \ref{sec:doms}. \\
  YAG & Yttrium-Aluminium-Garnet, synthetic material, used as laser medium in ice studies, see section \ref{sec:yag_h2_parameters}. \\
\end{tabelle}

\newpage
\subsection{List of Quantities}

\begin{tabelle}{lL}
  $\lambda\sca$ & geometric scattering length, in general or within the bulk ice, see section \ref{sec:scattering}. \\
  $\lambda\sca\hi$ & geometric scattering length within the hole ice. \\
  $\lambda\esca$ & effective scattering length, in general or within the bulk ice, $\lambda\esca = \frac{\lambda\sca}{1 - \meancostheta}$, see section \ref{sec:scattering}. \\
  $\lambda\esca\hi$ & effective scattering length within the hole ice, $\lambda\esca\hi = \frac{\lambda\sca\hi}{1 - \meancostheta}$. \\
  $\lambda\abs$ & absorption length, in general or within the bulk ice, see section \ref{sec:scattering}. \\
  $\lambda\abs\hi$ & absorption length within the hole ice. \\
  $r$ & radius of the hole-ice cylinder, see section \ref{sec:hole_ice}. \\
  $r\dom$ & radius of \icecube's digital optical modules (DOMs), see section \ref{sec:doms}. \\
\end{tabelle}

\subsection{List of Units}

\paragraph{Units of Length} \mbox{}

\begin{tabelle}{lL}
  $\m$ & meter, base unit of length in the international system of units \\
  $\cm$ & centimeter, $10^{-2}\m$ \\
  $\mm$ & millimeter, $10^{-3}\m$ \\
  $\nm$ & nanometer, $10^{-9}\m$ \\
\end{tabelle}

\paragraph{Units of Time} \mbox{}

\begin{tabelle}{lL}
  $\unit{s}$ & second, base unit of time in the international system of units \\
  $\ns$ & nanosecond, $10^{-9}\unit{s}$ \\
\end{tabelle}

\newpage
\paragraph{Units of Energy} \mbox{}

\begin{tabelle}{lL}
  $\unit{eV}$ & electron volt, unit of energy, $1\unit{eV} = 1.6021766208 \cdot 10^{-19}$ joules, which is the basic unit of energy in the international system of units \\
  $\unit{MeV}$ & mega electron volt, $10^{6}\unit{eV}$ \\
  $\GeV$ & giga electron volt, $10^{9}\unit{eV}$ \\
  $\TeV$ & tera electron volt, $10^{12}\unit{eV}$ \\
  $\PeV$ & peta electron volt, $10^{15}\unit{eV}$ \\
\end{tabelle}
\vfill

\subsection{List of Mathematical Symbols}

\begin{tabelle}{lL}
  $:$ & property operator. $A:B$ means that object $A$ has the property $B$. The associativity is focused on $A$, such that \enquote{The radius of the cylinder is $r:B$} means \enquote{The radius of the cylinder is $r$ and $r$ has the property $B$.} \\
  $=$ & equality. $A=B$ means that the quantities $A$ and $B$ have the same value. \\
  $:=$ & short cut for property operator in conjunction with the equality operator. $A:=B$ means $A:A=B$. \\
  $\approx$ & approximate equality. $A \approx B$ means that the quantities $A$ and $B$ have approximately the same value, where the precision of the approximation is given by the context of the statement. \\
  $\equiv$ & definition equality operator. $A \equiv B$ means that the symbols $A$ and $B$ represent the same object. \\
  $\sum$ & sum. $\sum_{i=1}^n a_i$ means $a_1 + a_2 + \dots + a_n$ \\
  $\prod$ & product. $\prod_{i=1}^n a_i$ means $a_1 \cdot a_2 \cdot \dots \cdot a_n$ \\
  $\pi$ & circle constant, representing the ratio of a circle's circumference to its diameter, $\pi \approx 3.14159$ \\
  $\{ \}$ & set. $\{ A, B \}$ represents the set of the objects $A$ and $B$. \\
  $\naturals$ & set of all natural numbers \\
  $\reals$ & set of all real numbers \\
  $\reals^+$ & set of all positive real numbers \\
  $\reals^+_0$ & set of all positive real numbers and zero \\
  $\ket{A}$ & quantum mechanical bra-ket notation for the quantum state of $A$ as abstract column vector \\
\end{tabelle}

  \subsection{Additional Flow Charts}
\label{sec:flow_charts}

\subsubsection{\texorpdfstring{Standard \clsim's Media Propagation
Algorithm}{Standard 's Media Propagation Algorithm}}\label{standard-s-media-propagation-algorithm}

\begin{figure}[htbp]
  \image{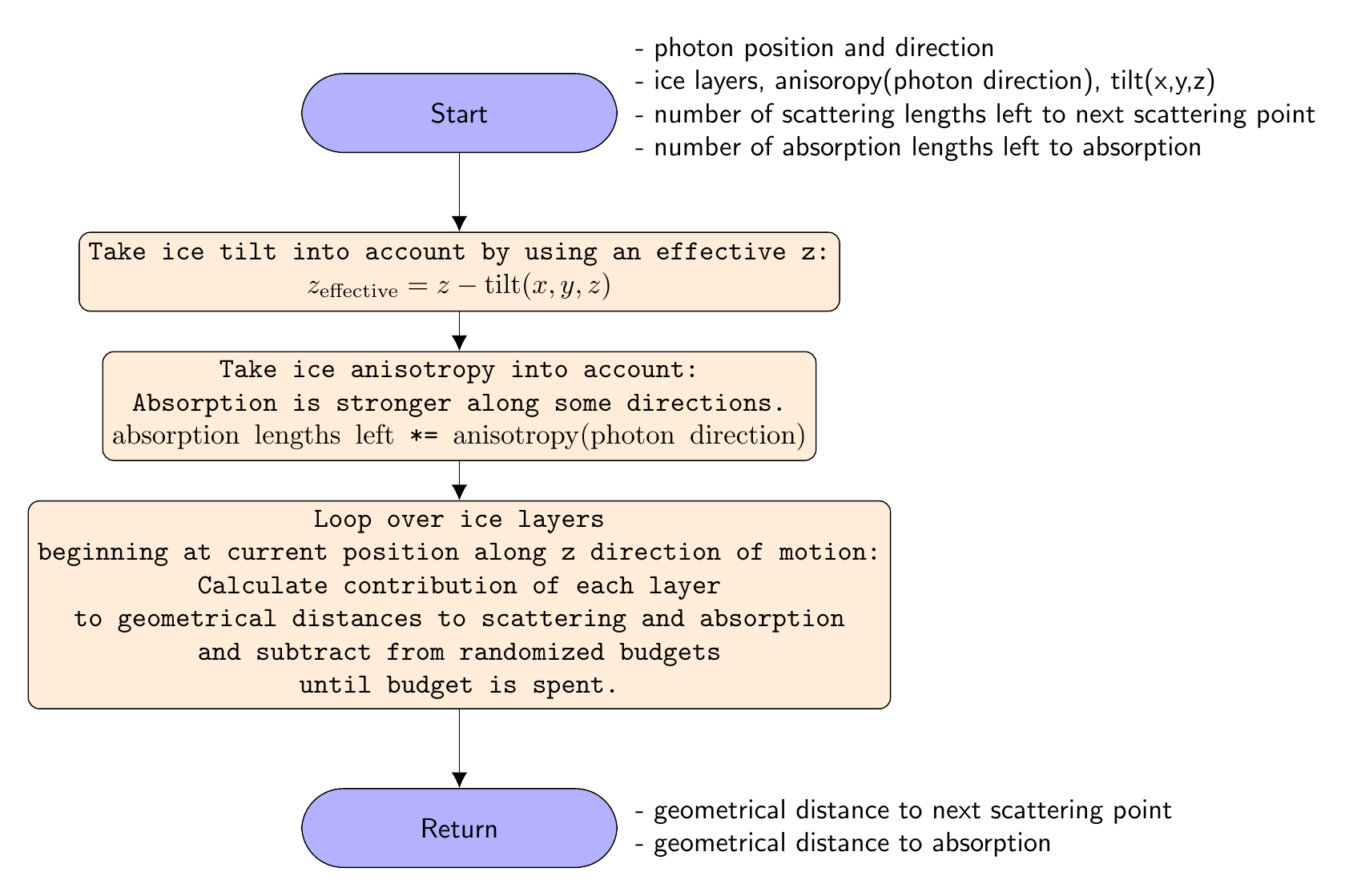}
  \caption{Flow chart of standard-\clsim's medium-propagation algorithm. Tilted ice layers and ice anisotropy are hard-coded. Media with other geometries such as hole-ice cylinders are not supported. This medium-propagation algorithm is replaced by the new algorithm described in section \ref{sec:algorithm_b}.}
  \label{fig:rieQu7sh}
\end{figure}

Figure \ref{fig:rieQu7sh} shows a flow chart of standard \clsim's
media-propagation algorithm (section \ref{sec:standard_clsim}) that does
not support cylinder-shaped media.

\newpage

\subsubsection{New Media-Propagation
Algorithm}\label{new-media-propagation-algorithm}

\begin{figure}[htbp]
  \image{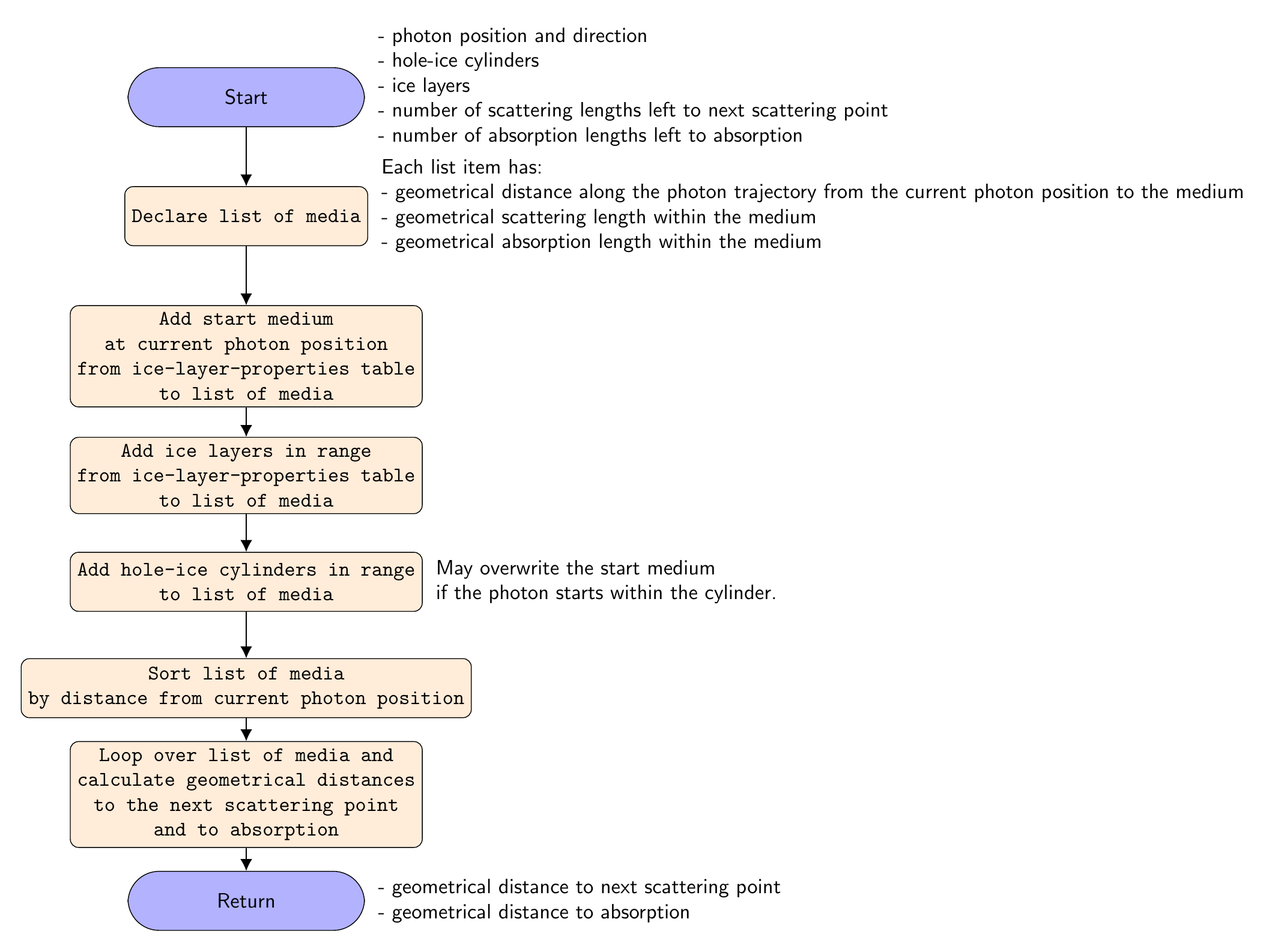}
  \caption{Flow chart of the new medium-propagation algorithm. The algorithm determines all medium changes (layers, hole ice) between two interaction points. Then the algorithm loops over all medium changes with different ice properties and calculates the geometrical distances to the next interaction points.}
  \label{fig:iez4Geih}
\end{figure}

A flow chart of the new medium-propagation algorithm introduced in
section \ref{sec:algorithm_b} that replaces standard-\clsim's
media-propagation algorithm is shown in figure \ref{fig:iez4Geih}.

\subsection{How to Install the Modified \clsim}
\label{sec:howtoclsim}

This study needs a modified version of \clsim that does support hole-ice simulations. Until this modified version has been merged into the main \icecube Simulation Framework source code, these patched version needs to be installed manually.

\begin{bash}
# Get clsim fork
git clone git@github.com:fiedl/clsim.git $SCRATCH/clsim
cd $SCRATCH/clsim

# Symlink it into the icesim source
cd $ICESIM_ROOT/src
rm -rf clsim
ln -s $SCRATCH/clsim clsim

# Compile it
cd $ICESIM_ROOT/build
make -j 6
\end{bash}

\docframe{
\docparwithoutframe{Detailed instructions on how to install the software on the Zeuthen computing center can be found at \url{https://github.com/fiedl/hole-ice-study/blob/master/notes/2018-01-23_Installing_IceSim_in_Zeuthen.md\#install-patched-clsim}.}\medskip

\docparwithoutframe{Instructions on how to install the software locally on macOS can be found at \url{https://github.com/fiedl/hole-ice-study/blob/master/notes/2016-11-15_Installing_IceSim_on_macOS_Sierra.md}.}
}

\subsection{Source Code of Algorithm A}
\label{sec:algorithm_a_source}

This is the source code of the first propagation algorithm through different media, described in section \ref{sec:algorithm_a}.

\sourcepar{The source code can also be found online at \url{https://github.com/fiedl/clsim/tree/sf/hole-ice-2017/resources/kernels/lib}.}

\begin{ccode}
// https://github.com/fiedl/clsim/blob/sf/hole-ice-2017/resources/kernels/lib/hole_ice/hole_ice.c

#ifndef HOLE_ICE_C
#define HOLE_ICE_C

#include "hole_ice.h"
#include "../intersection/intersection.c"

inline bool is_between_zero_and_one(floating_t a) {
  return ((!my_is_nan(a)) && (a > 0.0) && (a < 1.0));
}

inline bool not_between_zero_and_one(floating_t a) {
  return (! is_between_zero_and_one(a));
}

inline unsigned int number_of_medium_changes(HoleIceProblemParameters_t p)
{
  // These are ordered by their frequency of occurrence in order
  // to optimize for performance.
  if (not_between_zero_and_one(p.entry_point_ratio) && not_between_zero_and_one(p.termination_point_ratio)) return 0;
  if (not_between_zero_and_one(p.entry_point_ratio) || not_between_zero_and_one(p.termination_point_ratio)) return 1;
  if (p.entry_point_ratio == p.termination_point_ratio) return 0; // tangent
  return 2;
}

inline floating_t hole_ice_distance_correction(HoleIceProblemParameters_t p)
{
  // Depending on the fraction of the distance the photon is traveling
  // within the hole ice, there are six cases to consider.
  //
  // N  denotes the number of intersections.
  // H  means that the trajectory starts within hole ice.
  // !H means that the trajectory starts outside of the hole ice.
  //
  // Case 1: The trajectory is completely outside of the hole ice (!H, N=0).
  // Case 2: The trajectory is completely within the hole ice (H, N=0).
  // Case 3: The trajectory begins outside, but ends inside the hole ice (!H, N=1).
  // Case 4: The trajectory begins inside, but ends outside the hole ice (H, N=1).
  // Case 5: The trajectory starts and ends outside, but passes through the hole ice (!H, N=2).
  // Case 6: The trajectory begins within one hole-ice cylinder, passes through
  //           normal ice and ends in another hole-ice cylinder (H, N=2).
  //
  // For further information, please have look at:
  // https://github.com/fiedl/clsim/tree/sf/master/resources/kernels/lib/hole_ice

  const unsigned int num_of_medium_changes = number_of_medium_changes(p);

  // Case 1: The trajectory is completely outside of the hole ice.
  // Thus, needs no correction.
  if ((num_of_medium_changes == 0) && !p.starts_within_hole_ice) {
    return 0;
  }

  const floating_t ac = p.distance * p.termination_point_ratio;

  if (p.starts_within_hole_ice) {

    // Case 4: The trajectory begins inside, but ends outside the hole ice.
    if (p.interaction_length_factor * p.distance > ac) {
      return (1.0 - 1.0 / p.interaction_length_factor) * ac;

    // Case 2: The trajectory is completely within the hole ice.
    } else {
      return (p.interaction_length_factor - 1.0) * p.distance;
    }

  } else {

    const floating_t yb = p.distance * (1.0 - p.entry_point_ratio);
    floating_t yc = p.distance * (p.termination_point_ratio - p.entry_point_ratio);

    if (yc < 0) {
      printf("WARNING: YC SHOULD NOT BE NEGATIVE, BUT YC=%
      yc = 0;
    }

    // Case 5: The trajectory starts and ends outside, but passes through the hole ice.
    if (p.interaction_length_factor * yb > yc) {
      return (1.0 - 1.0 / p.interaction_length_factor) * yc;

    // Case 3
    } else {
      return (p.interaction_length_factor - 1.0) * p.distance * (1.0 - p.entry_point_ratio);
    }
  }

#ifdef PRINTF_ENABLED
  printf("WARNING: UNHANDLED INTERSECTION CASE. This point should not be reached.");
#endif

  return my_nan();
}

inline floating_t hole_ice_distance_correction_for_intersection_problem(floating_t distance, floating_t interaction_length_factor, IntersectionProblemParameters_t p)
{
  calculate_intersections(&p);
  HoleIceProblemParameters_t hip = {
    distance,
    interaction_length_factor,
    intersection_s1(p), // entry_point_ratio
    intersection_s2(p), // termination_point_ratio
    intersecting_trajectory_starts_inside(p) // starts_within_hole_ice
  };
  return hole_ice_distance_correction(hip);
}

#ifdef HOLE_ICE_TEST_H
inline floating_t apply_hole_ice_correction(floating4_t photonPosAndTime, floating4_t photonDirAndWlen, unsigned int numberOfCylinders, floating4_t *cylinderPositionsAndRadii, floating_t holeIceScatteringLengthFactor, floating_t holeIceAbsorptionLengthFactor, floating_t *distancePropagated, floating_t *distanceToAbsorption)
#endif
#ifndef HOLE_ICE_TEST_H
inline floating_t apply_hole_ice_correction(floating4_t photonPosAndTime, floating4_t photonDirAndWlen, unsigned int numberOfCylinders, __constant floating4_t *cylinderPositionsAndRadii, floating_t holeIceScatteringLengthFactor, floating_t holeIceAbsorptionLengthFactor, floating_t *distancePropagated, floating_t *distanceToAbsorption)
#endif
{

  if (!(my_is_nan(photonPosAndTime.x) || my_is_nan(*distancePropagated))) {

    // Find out which cylinders are in range in a separate loop
    // in order to improve parallelism and thereby performance.
    //
    // See: https://github.com/fiedl/hole-ice-study/issues/30
    //
    #ifdef NUMBER_OF_CYLINDERS
      // When running this on OpenCL, defining arrays using a constant
      // as array size is not possible. Therefore, we need to use a
      // pre-processor makro here.
      //
      // See: https://github.com/fiedl/hole-ice-study/issues/38
      //
      int indices_of_cylinders_in_range[NUMBER_OF_CYLINDERS];
    #else
      int indices_of_cylinders_in_range[numberOfCylinders];
    #endif
    {
      unsigned int j = 0;
      for (unsigned int i = 0; i < numberOfCylinders; i++) {
        indices_of_cylinders_in_range[i] = -1;
      }
      for (unsigned int i = 0; i < numberOfCylinders; i++) {
        if (sqr(photonPosAndTime.x - cylinderPositionsAndRadii[i].x) +
            sqr(photonPosAndTime.y - cylinderPositionsAndRadii[i].y) <=
            sqr(*distancePropagated + cylinderPositionsAndRadii[i].w /* radius */))
        {

          // If the cylinder has a z-range check if we consider that cylinder
          // to be in range. https://github.com/fiedl/hole-ice-study/issues/34
          //
          if ((cylinderPositionsAndRadii[i].z == 0) || ((cylinderPositionsAndRadii[i].z != 0) && !(((photonPosAndTime.z < cylinderPositionsAndRadii[i].z - 0.5) && (photonPosAndTime.z + *distancePropagated * photonDirAndWlen.z < cylinderPositionsAndRadii[i].z - 0.5)) || ((photonPosAndTime.z > cylinderPositionsAndRadii[i].z + 0.5) && (photonPosAndTime.z + *distancePropagated * photonDirAndWlen.z > cylinderPositionsAndRadii[i].z + 0.5)))))
          {
            indices_of_cylinders_in_range[j] = i;
            j += 1;
          }
        }
      }
    }

    // Now loop over all cylinders in range and calculate corrections
    // for `distancePropagated` and `distanceToAbsorption`.
    //
    for (unsigned int j = 0; j < numberOfCylinders; j++) {
      const int i = indices_of_cylinders_in_range[j];
      if (i == -1) {
        break;
      } else {

        IntersectionProblemParameters_t p = {

          // Input values
          photonPosAndTime.x,
          photonPosAndTime.y,
          cylinderPositionsAndRadii[i].x,
          cylinderPositionsAndRadii[i].y,
          cylinderPositionsAndRadii[i].w, // radius
          photonDirAndWlen,
          *distancePropagated,

          // Output values (will be calculated)
          0, // discriminant
          0, // s1
          0  // s2

        };

        calculate_intersections(&p);

        // Are intersection points possible?
        if (intersection_discriminant(p) > 0) {

          const floating_t scatteringEntryPointRatio = intersection_s1(p);
          const floating_t scatterintTerminationPointRatio = intersection_s2(p);

          HoleIceProblemParameters_t scatteringCorrectionParameters = {
            *distancePropagated,
            holeIceScatteringLengthFactor,
            scatteringEntryPointRatio,
            scatterintTerminationPointRatio,
            intersecting_trajectory_starts_inside(p)
          };

          const floating_t scaCorrection = hole_ice_distance_correction(scatteringCorrectionParameters);
          *distancePropagated += scaCorrection;

          // For the absorption, there are special cases where the photon is scattered before
          // reaching either the first or the second absorption intersection point.
          floating_t absorptionEntryPointRatio;
          floating_t absorptionTerminationPointRatio;
          floating_t absCorrection = 0.0;
          if (!(not_between_zero_and_one(scatteringCorrectionParameters.entry_point_ratio) && !scatteringCorrectionParameters.starts_within_hole_ice)) {
            // The photon reaches the hole ice, i.e. the absorption correction
            // needs to be calculated.
            p.distance = *distanceToAbsorption;
            calculate_intersections(&p);

            // If the photon is scattered away before reaching the far and of
            // the hole ice, the affected trajectory is limited by the
            // point where the photon is scattered away.
            absorptionEntryPointRatio = intersection_s1(p);
            absorptionTerminationPointRatio = min(
              *distancePropagated / *distanceToAbsorption,
              intersection_s2(p)
            );

            HoleIceProblemParameters_t absorptionCorrectionParameters = {
              *distanceToAbsorption,
              holeIceAbsorptionLengthFactor,
              absorptionEntryPointRatio,
              absorptionTerminationPointRatio,
              intersecting_trajectory_starts_inside(p),
            };

            absCorrection = hole_ice_distance_correction(absorptionCorrectionParameters);

          }
          *distanceToAbsorption += absCorrection;

        }

      }
    }
  }

}

#endif
\end{ccode}

\begin{ccode}
// https://github.com/fiedl/clsim/blob/sf/hole-ice-2017/resources/kernels/lib/hole_ice/hole_ice.h

#ifndef HOLE_ICE_H
#define HOLE_ICE_H

typedef struct HoleIceProblemParameters {
  floating_t distance;
  floating_t interaction_length_factor;
  floating_t entry_point_ratio;
  floating_t termination_point_ratio;
  bool starts_within_hole_ice;
} HoleIceProblemParameters_t;

#endif
\end{ccode}

\begin{ccode}
// https://github.com/fiedl/clsim/blob/sf/hole-ice-2017/resources/kernels/lib/intersection/intersection.c

#include "intersection.h"

inline void calculate_intersections(IntersectionProblemParameters_t *p)
{
  // Step 1
  const floating4_t vector_AM = {p->mx - p->ax, p->my - p->ay, 0.0, 0.0};
  const floating_t xy_projection_factor = my_sqrt(1 - sqr(p->direction.z));
  const floating_t length_AMprime = dot(vector_AM, p->direction) / xy_projection_factor;

  // Step 2
  p->discriminant = sqr(p->r) - dot(vector_AM, vector_AM) + sqr(length_AMprime);

  // Step 3
  const floating_t length_XMprime = my_sqrt(p->discriminant);

  // Step 4
  const floating_t length_AX1 = length_AMprime - length_XMprime;
  const floating_t length_AX2 = length_AMprime + length_XMprime;
  p->s1 = length_AX1 / p->distance / xy_projection_factor;
  p->s2 = length_AX2 / p->distance / xy_projection_factor;
}

inline floating_t intersection_s1(IntersectionProblemParameters_t p)
{
  return p.s1;
}

inline floating_t intersection_s2(IntersectionProblemParameters_t p)
{
  return p.s2;
}

inline floating_t intersection_discriminant(IntersectionProblemParameters_t p)
{
  return p.discriminant;
}

inline floating_t intersection_x1(IntersectionProblemParameters_t p)
{
  if ((p.s1 > 0) && (p.s1 < 1))
    return p.ax + p.direction.x * p.distance * p.s1;
  else
    return my_nan();
}

inline floating_t intersection_x2(IntersectionProblemParameters_t p)
{
  if ((p.s2 > 0) && (p.s2 < 1))
    return p.ax + p.direction.x * p.distance * p.s2;
  else
    return my_nan();
}

inline floating_t intersection_y1(IntersectionProblemParameters_t p)
{
  if ((p.s1 > 0) && (p.s1 < 1))
    return p.ay + p.direction.y * p.distance * p.s1;
  else
    return my_nan();
}

inline floating_t intersection_y2(IntersectionProblemParameters_t p)
{
  if ((p.s2 > 0) && (p.s2 < 1))
    return p.ay + p.direction.y * p.distance * p.s2;
  else
    return my_nan();
}

inline bool intersecting_trajectory_starts_inside(IntersectionProblemParameters_t p)
{
  return (intersection_s1(p) <= 0) &&
      (intersection_s2(p) > 0) &&
      (intersection_discriminant(p) > 0);
}

inline bool intersecting_trajectory_starts_outside(IntersectionProblemParameters_t p)
{
  return ( ! intersecting_trajectory_starts_inside(p));
}

inline bool intersecting_trajectory_ends_inside(IntersectionProblemParameters_t p)
{
  return (intersection_s1(p) < 1) &&
      (intersection_s2(p) >= 1) &&
      (intersection_discriminant(p) > 0);
}
\end{ccode}

\begin{ccode}
// https://github.com/fiedl/clsim/blob/sf/hole-ice-2017/resources/kernels/lib/intersection/intersection.h

#ifndef INTERSECTION_H
#define INTERSECTION_H

typedef struct IntersectionProblemParameters {

  // Input values
  //
  floating_t ax;
  floating_t ay;
  floating_t mx;
  floating_t my;
  floating_t r;
  floating4_t direction;
  floating_t distance;

  // Output values, which will be calculated in
  // `calculate_intersections()`.
  //
  floating_t discriminant;
  floating_t s1;
  floating_t s2;

} IntersectionProblemParameters_t;

#endif
\end{ccode}

\subsection{Source Code of Algorithm B}
\label{sec:algorithm_b_source}

This is the source code of the second propagation algorithm through different media, described in section \ref{sec:algorithm_b}.

\sourcepar{The source code can also be found online at \url{https://github.com/fiedl/clsim/tree/sf/hole-ice-2018/resources/kernels/lib}.}

\begin{ccode}
// https://github.com/fiedl/clsim/blob/sf/hole-ice-2018/resources/kernels/lib/propagation_through_media/propagation_through_media.c

#ifndef PROPAGATION_THROUGH_MEDIA_C
#define PROPAGATION_THROUGH_MEDIA_C

#include "propagation_through_media.h"
#include "../ice_layers/ice_layers.c"
#ifdef HOLE_ICE
  #include "../hole_ice/hole_ice.c"
#endif

// PROPAGATION THROUGH DIFFERENT MEDIA 2018: Layers, Cylinders
// -----------------------------------------------------------------------------

// We know how many scattering lengths (`sca_step_left`) and
// absorption lengths (`abs_lens_left`) the photon will
// travel in this step.
//
// Because the mean scattering and absorption lengths are local
// properties, i.e. depend on the ice layer or whether the photon
// is within a hole-ice cylinder, we need to convert `sca_step_left`
// and `abs_lens_left` to geometrical distances in order to determine
// where the next interaction point is, i.e. how far to propagate
// the photon in this step.

inline void apply_propagation_through_different_media(
  floating4_t photonPosAndTime, floating4_t photonDirAndWlen,
  #ifdef HOLE_ICE
    unsigned int numberOfCylinders, __constant floating4_t *cylinderPositionsAndRadii,
    __constant floating_t *cylinderScatteringLengths, __constant floating_t *cylinderAbsorptionLengths,
  #endif
  floating_t *distances_to_medium_changes, floating_t *local_scattering_lengths, floating_t *local_absorption_lengths,
  floating_t *sca_step_left, floating_t *abs_lens_left,
  floating_t *distancePropagated, floating_t *distanceToAbsorption)
{

  int number_of_medium_changes = 0;
  distances_to_medium_changes[0] = 0.0;
  int currentPhotonLayer = min(max(findLayerForGivenZPos(photonPosAndTime.z), 0), MEDIUM_LAYERS-1);
  local_scattering_lengths[0] = getScatteringLength(currentPhotonLayer, photonDirAndWlen.w);
  local_absorption_lengths[0] = getAbsorptionLength(currentPhotonLayer, photonDirAndWlen.w);

  // To check which medium boundaries are in range, we need to estimate
  // how far the photon can travel in this step.
  //
  const floating_t photonRange = *sca_step_left * local_scattering_lengths[0];

  add_ice_layers_on_photon_path_to_medium_changes(
    photonPosAndTime,
    photonDirAndWlen,
    photonRange,

    // These values will be updates within this function:
    &number_of_medium_changes,
    distances_to_medium_changes,
    local_scattering_lengths,
    local_absorption_lengths
  );

  #ifdef HOLE_ICE
    add_hole_ice_cylinders_on_photon_path_to_medium_changes(
      photonPosAndTime,
      photonDirAndWlen,
      photonRange,
      numberOfCylinders,
      cylinderPositionsAndRadii,

      // These values will be updates within this function:
      &number_of_medium_changes,
      distances_to_medium_changes,
      local_scattering_lengths,
      local_absorption_lengths
    );
  #endif

  sort_medium_changes_by_ascending_distance(
    number_of_medium_changes,

    // These values will be updates within this function:
    distances_to_medium_changes,
    local_scattering_lengths,
    local_absorption_lengths
  );

  loop_over_media_and_calculate_geometrical_distances_up_to_the_next_scattering_point(
    number_of_medium_changes,
    distances_to_medium_changes,
    local_scattering_lengths,
    local_absorption_lengths,

    // These values will be updates within this function:
    sca_step_left,
    abs_lens_left,
    distancePropagated,
    distanceToAbsorption
  );
}

inline void sort_medium_changes_by_ascending_distance(int number_of_medium_changes, floating_t *distances_to_medium_changes, floating_t *local_scattering_lengths, floating_t *local_absorption_lengths)
{
  // Sort the arrays `distances_to_medium_changes`, `local_scattering_lengths` and
  // `local_absorption_lengths` by ascending distance to have the medium changes
  // in the right order.
  //
  // https://en.wikiversity.org/wiki/C_Source_Code/Sorting_array_in_ascending_and_descending_order
  //
  for (int k = 0; k <= number_of_medium_changes; k++) {
    for (int l = 0; l <= number_of_medium_changes; l++) {
      if (distances_to_medium_changes[l] > distances_to_medium_changes[k]) {
        floating_t tmp_distance = distances_to_medium_changes[k];
        floating_t tmp_scattering = local_scattering_lengths[k];
        floating_t tmp_absorption = local_absorption_lengths[k];

        distances_to_medium_changes[k] = distances_to_medium_changes[l];
        local_scattering_lengths[k] = local_scattering_lengths[l];
        local_absorption_lengths[k] = local_absorption_lengths[l];

        distances_to_medium_changes[l] = tmp_distance;
        local_scattering_lengths[l] = tmp_scattering;
        local_absorption_lengths[l] = tmp_absorption;
      }
    }
  }
}

inline void loop_over_media_and_calculate_geometrical_distances_up_to_the_next_scattering_point(int number_of_medium_changes, floating_t *distances_to_medium_changes, floating_t *local_scattering_lengths, floating_t *local_absorption_lengths, floating_t *sca_step_left, floating_t *abs_lens_left, floating_t *distancePropagated, floating_t *distanceToAbsorption)
{
  // We know how many scattering lengths (`sca_step_left`) and how many
  // absorption lengths (`abs_lens_left`) we may spend when propagating
  // through the different media.
  //
  // Convert these into the geometrical distances `distancePropagated` (scattering)
  // and `distanceToAbsorption` (absorption) and decrease `sca_step_left` and
  // `abs_lens_left` accordingly.
  //
  // Abort when the next scattering point is reached, i.e. `sca_step_left == 0`.
  // At this point, `abs_lens_left` may still be greater than zero, because
  // the photon may be scattered several times until it is absorbed.
  //
  for (int j = 0; (j < number_of_medium_changes) && (*sca_step_left > 0); j++) {
    floating_t max_distance_in_current_medium = distances_to_medium_changes[j+1] - distances_to_medium_changes[j];

    if (*sca_step_left * local_scattering_lengths[j] > max_distance_in_current_medium) {
      // The photon scatters after leaving this medium.
      *sca_step_left -= my_divide(max_distance_in_current_medium, local_scattering_lengths[j]);
      *distancePropagated += max_distance_in_current_medium;
    } else {
      // The photon scatters within this medium.
      max_distance_in_current_medium = *sca_step_left * local_scattering_lengths[j];
      *distancePropagated += max_distance_in_current_medium;
      *sca_step_left = 0;
    }

    if (*abs_lens_left * local_absorption_lengths[j] > max_distance_in_current_medium) {
      // The photon is absorbed after leaving this medium.
      *abs_lens_left -= my_divide(max_distance_in_current_medium, local_absorption_lengths[j]);
      *distanceToAbsorption += max_distance_in_current_medium;
    } else {
      // The photon is absorbed within this medium.
      *distanceToAbsorption += *abs_lens_left * local_absorption_lengths[j];
      *abs_lens_left = 0;
    }
  }

  // Spend the rest of the budget with the last medium properties.
  if (*sca_step_left > 0) {
    *distancePropagated += *sca_step_left * local_scattering_lengths[number_of_medium_changes];
    *distanceToAbsorption += *abs_lens_left * local_absorption_lengths[number_of_medium_changes];
    *abs_lens_left -= my_divide(*distancePropagated, local_absorption_lengths[number_of_medium_changes]);
  }

  // If the photon is absorbed, only propagate up to the absorption point.
  if (*distanceToAbsorption < *distancePropagated) {
    *distancePropagated = *distanceToAbsorption;
    *distanceToAbsorption = ZERO;
    *abs_lens_left = ZERO;
  }
}

#endif
\end{ccode}

\begin{ccode}
// https://github.com/fiedl/clsim/blob/sf/hole-ice-2018/resources/kernels/lib/propagation_through_media/propagation_through_media.h

#ifndef PROPAGATION_THROUGH_MEDIA_H
#define PROPAGATION_THROUGH_MEDIA_H

inline void apply_propagation_through_different_media(
  floating4_t photonPosAndTime, floating4_t photonDirAndWlen,
  #ifdef HOLE_ICE
    unsigned int numberOfCylinders, __constant floating4_t *cylinderPositionsAndRadii,
    __constant floating_t *cylinderScatteringLengths, __constant floating_t *cylinderAbsorptionLengths,
  #endif
  floating_t *distances_to_medium_changes, floating_t *local_scattering_lengths, floating_t *local_absorption_lengths,
  floating_t *sca_step_left, floating_t *abs_lens_left,
  floating_t *distancePropagated, floating_t *distanceToAbsorption);

inline void sort_medium_changes_by_ascending_distance(int number_of_medium_changes, floating_t *distances_to_medium_changes, floating_t *local_scattering_lengths, floating_t *local_absorption_lengths);

inline void loop_over_media_and_calculate_geometrical_distances_up_to_the_next_scattering_point(int number_of_medium_changes, floating_t *distances_to_medium_changes, floating_t *local_scattering_lengths, floating_t *local_absorption_lengths, floating_t *sca_step_left, floating_t *abs_lens_left, floating_t *distancePropagated, floating_t *distanceToAbsorption);

#endif
\end{ccode}

\begin{ccode}
// https://github.com/fiedl/clsim/blob/sf/hole-ice-2018/resources/kernels/lib/ice_layers/ice_layers.c

#ifndef ICE_LAYERS_C
#define ICE_LAYERS_C

#include "ice_layers.h"

inline void add_ice_layers_on_photon_path_to_medium_changes(floating4_t photonPosAndTime, floating4_t photonDirAndWlen, floating_t photonRange, int *number_of_medium_changes, floating_t *distances_to_medium_changes, floating_t *local_scattering_lengths, floating_t *local_absorption_lengths)
{

  // The closest ice layer is special, because we need to check how far
  // it is away from the photon. After that, all photon layers are equidistant.
  //
  floating_t z_of_closest_ice_layer_boundary =
      mediumLayerBoundary(photon_layer(photonPosAndTime.z));
  if (photonDirAndWlen.z > ZERO) z_of_closest_ice_layer_boundary +=
      (floating_t)MEDIUM_LAYER_THICKNESS;

  *number_of_medium_changes += 1;
  distances_to_medium_changes[*number_of_medium_changes] =
      my_divide(z_of_closest_ice_layer_boundary - photonPosAndTime.z, photonDirAndWlen.z);
  int next_photon_layer =
      photon_layer(z_of_closest_ice_layer_boundary + photonDirAndWlen.z);
  local_scattering_lengths[*number_of_medium_changes] =
      getScatteringLength(next_photon_layer, photonDirAndWlen.w);
  local_absorption_lengths[*number_of_medium_changes] =
      getAbsorptionLength(next_photon_layer, photonDirAndWlen.w);

  // Now loop through the equidistant layers in range.
  //
  const floating_t max_trajectory_length_between_two_layers =
      my_divide((floating_t)MEDIUM_LAYER_THICKNESS, my_fabs(photonDirAndWlen.z));
  while (distances_to_medium_changes[*number_of_medium_changes] + max_trajectory_length_between_two_layers < photonRange)
  {
    *number_of_medium_changes += 1;
    distances_to_medium_changes[*number_of_medium_changes] =
        distances_to_medium_changes[*number_of_medium_changes - 1]
        + max_trajectory_length_between_two_layers;
    next_photon_layer = photon_layer(photonPosAndTime.z
        + (distances_to_medium_changes[*number_of_medium_changes] + 0.01) * photonDirAndWlen.z);
    local_scattering_lengths[*number_of_medium_changes] =
        getScatteringLength(next_photon_layer, photonDirAndWlen.w);
    local_absorption_lengths[*number_of_medium_changes] =
        getAbsorptionLength(next_photon_layer, photonDirAndWlen.w);
  }

}

inline int photon_layer(floating_t z)
{
  return min(max(findLayerForGivenZPos(z), 0), MEDIUM_LAYERS-1);
}

#endif
\end{ccode}

\begin{ccode}
// https://github.com/fiedl/clsim/blob/sf/hole-ice-2018/resources/kernels/lib/ice_layers/ice_layers.h

#ifndef ICE_LAYERS_H
#define ICE_LAYERS_H

inline void add_ice_layers_on_photon_path_to_medium_changes(floating4_t photonPosAndTime, floating4_t photonDirAndWlen, floating_t photonRange, int *number_of_medium_changes, floating_t *distances_to_medium_changes, floating_t *local_scattering_lengths, floating_t *local_absorption_lengths);

inline int photon_layer(floating_t z);

#endif
\end{ccode}

\begin{ccode}
// https://github.com/fiedl/clsim/blob/sf/hole-ice-2018/resources/kernels/lib/hole_ice/hole_ice.c

#ifndef HOLE_ICE_C
#define HOLE_ICE_C

#include "hole_ice.h"
#include "../intersection/intersection.c"

inline void add_hole_ice_cylinders_on_photon_path_to_medium_changes(floating4_t photonPosAndTime, floating4_t photonDirAndWlen, floating_t photonRange, unsigned int numberOfCylinders, __constant floating4_t *cylinderPositionsAndRadii, int *number_of_medium_changes, floating_t *distances_to_medium_changes, floating_t *local_scattering_lengths, floating_t *local_absorption_lengths)
{
  // Find out which cylinders are in range in a separate loop
  // in order to improve parallelism and thereby performance.
  //
  // See: https://github.com/fiedl/hole-ice-study/issues/30
  //
  #ifdef NUMBER_OF_CYLINDERS
    // When running this on OpenCL, defining arrays using a constant
    // as array size is not possible. Therefore, we need to use a
    // pre-processor makro here.
    //
    // See: https://github.com/fiedl/hole-ice-study/issues/38
    //
    int indices_of_cylinders_in_range[NUMBER_OF_CYLINDERS];
  #else
    int indices_of_cylinders_in_range[numberOfCylinders];
  #endif
  {
    unsigned int j = 0;
    for (unsigned int i = 0; i < numberOfCylinders; i++) {
      indices_of_cylinders_in_range[i] = -1;
    }
    for (unsigned int i = 0; i < numberOfCylinders; i++) {
      if (sqr(photonPosAndTime.x - cylinderPositionsAndRadii[i].x) +
          sqr(photonPosAndTime.y - cylinderPositionsAndRadii[i].y) <=
          sqr(photonRange + cylinderPositionsAndRadii[i].w /* radius */))
      {

        // If the cylinder has a z-range check if we consider that cylinder
        // to be in range. https://github.com/fiedl/hole-ice-study/issues/34
        //
        if ((cylinderPositionsAndRadii[i].z == 0) || ((cylinderPositionsAndRadii[i].z != 0) && !(((photonPosAndTime.z < cylinderPositionsAndRadii[i].z - 0.5) && (photonPosAndTime.z + photonRange * photonDirAndWlen.z < cylinderPositionsAndRadii[i].z - 0.5)) || ((photonPosAndTime.z > cylinderPositionsAndRadii[i].z + 0.5) && (photonPosAndTime.z + photonRange * photonDirAndWlen.z > cylinderPositionsAndRadii[i].z + 0.5)))))
        {
          indices_of_cylinders_in_range[j] = i;
          j += 1;
        }
      }
    }
  }

  // Now loop over all cylinders in range and calculate corrections
  // for `*distancePropagated` and `*distanceToAbsorption`.
  //
  for (unsigned int j = 0; j < numberOfCylinders; j++) {
    const int i = indices_of_cylinders_in_range[j];
    if (i == -1) {
      break;
    } else {

      IntersectionProblemParameters_t p = {

        // Input values
        photonPosAndTime.x,
        photonPosAndTime.y,
        cylinderPositionsAndRadii[i].x,
        cylinderPositionsAndRadii[i].y,
        cylinderPositionsAndRadii[i].w, // radius
        photonDirAndWlen,
        1.0, // distance used to calculate s1 and s2 relative to

        // Output values (will be calculated)
        0, // discriminant
        0, // s1
        0  // s2

      };

      calculate_intersections(&p);

      if (intersection_discriminant(p) > 0) {
        if ((intersection_s1(p) <= 0) && (intersection_s2(p) >= 0)) {
          // The photon is already within the hole ice.
          local_scattering_lengths[0] = cylinderScatteringLengths[i];
          local_absorption_lengths[0] = cylinderAbsorptionLengths[i];
        } else if (intersection_s1(p) > 0) {
          // The photon enters the hole ice on its way.
          *number_of_medium_changes += 1;
          distances_to_medium_changes[*number_of_medium_changes] = intersection_s1(p);
          local_scattering_lengths[*number_of_medium_changes] = cylinderScatteringLengths[i];
          local_absorption_lengths[*number_of_medium_changes] = cylinderAbsorptionLengths[i];
        }
        if (intersection_s2(p) > 0) {
          // The photon leaves the hole ice on its way.
          *number_of_medium_changes += 1;
          distances_to_medium_changes[*number_of_medium_changes] = intersection_s2(p);
          if (i == 0) // there is no larger cylinder
          {
            const int photonLayerAtTheCylinderBorder =
                photon_layer(photonPosAndTime.z + photonDirAndWlen.z * intersection_s2(p));
            local_scattering_lengths[*number_of_medium_changes] =
                getScatteringLength(photonLayerAtTheCylinderBorder, photonDirAndWlen.w);
            local_absorption_lengths[*number_of_medium_changes] =
                getAbsorptionLength(photonLayerAtTheCylinderBorder, photonDirAndWlen.w);
          } else {
            // There is a larger cylinder outside this one, which is the one before in the array.
            // See: https://github.com/fiedl/hole-ice-study/issues/47
            //
            local_scattering_lengths[*number_of_medium_changes] = cylinderScatteringLengths[i - 1];
            local_absorption_lengths[*number_of_medium_changes] = cylinderAbsorptionLengths[i - 1];
          }
        }
      }

    }
  }

}

#endif
\end{ccode}

\begin{ccode}
// https://github.com/fiedl/clsim/blob/sf/hole-ice-2018/resources/kernels/lib/hole_ice/hole_ice.h

#ifndef HOLE_ICE_H
#define HOLE_ICE_H

inline void add_hole_ice_cylinders_on_photon_path_to_medium_changes(floating4_t photonPosAndTime, floating4_t photonDirAndWlen, floating_t photonRange, unsigned int numberOfCylinders, __constant floating4_t *cylinderPositionsAndRadii, int *number_of_medium_changes, floating_t *distances_to_medium_changes, floating_t *local_scattering_lengths, floating_t *local_absorption_lengths);

#endif
\end{ccode}

\begin{ccode}
// https://github.com/fiedl/clsim/blob/sf/hole-ice-2018/resources/kernels/lib/intersection/intersection.c

#include "intersection.h"

inline void calculate_intersections(IntersectionProblemParameters_t *p)
{
  // Step 1
  const floating4_t vector_AM = {p->mx - p->ax, p->my - p->ay, 0.0, 0.0};
  const floating_t xy_projection_factor = my_sqrt(1 - sqr(p->direction.z));
  const floating_t length_AMprime = dot(vector_AM, p->direction) / xy_projection_factor;

  // Step 2
  p->discriminant = sqr(p->r) - dot(vector_AM, vector_AM) + sqr(length_AMprime);

  // Step 3
  const floating_t length_XMprime = my_sqrt(p->discriminant);

  // Step 4
  const floating_t length_AX1 = length_AMprime - length_XMprime;
  const floating_t length_AX2 = length_AMprime + length_XMprime;
  p->s1 = length_AX1 / p->distance / xy_projection_factor;
  p->s2 = length_AX2 / p->distance / xy_projection_factor;
}

inline floating_t intersection_s1(IntersectionProblemParameters_t p)
{
  return p.s1;
}

inline floating_t intersection_s2(IntersectionProblemParameters_t p)
{
  return p.s2;
}

inline floating_t intersection_discriminant(IntersectionProblemParameters_t p)
{
  return p.discriminant;
}

inline floating_t intersection_x1(IntersectionProblemParameters_t p)
{
  if ((p.s1 > 0) && (p.s1 < 1))
    return p.ax + p.direction.x * p.distance * p.s1;
  else
    return my_nan();
}

inline floating_t intersection_x2(IntersectionProblemParameters_t p)
{
  if ((p.s2 > 0) && (p.s2 < 1))
    return p.ax + p.direction.x * p.distance * p.s2;
  else
    return my_nan();
}

inline floating_t intersection_y1(IntersectionProblemParameters_t p)
{
  if ((p.s1 > 0) && (p.s1 < 1))
    return p.ay + p.direction.y * p.distance * p.s1;
  else
    return my_nan();
}

inline floating_t intersection_y2(IntersectionProblemParameters_t p)
{
  if ((p.s2 > 0) && (p.s2 < 1))
    return p.ay + p.direction.y * p.distance * p.s2;
  else
    return my_nan();
}

inline bool intersecting_trajectory_starts_inside(IntersectionProblemParameters_t p)
{
  return (intersection_s1(p) <= 0) &&
      (intersection_s2(p) > 0) &&
      (intersection_discriminant(p) > 0);
}

inline bool intersecting_trajectory_starts_outside(IntersectionProblemParameters_t p)
{
  return ( ! intersecting_trajectory_starts_inside(p));
}

inline bool intersecting_trajectory_ends_inside(IntersectionProblemParameters_t p)
{
  return (intersection_s1(p) < 1) &&
      (intersection_s2(p) >= 1) &&
      (intersection_discriminant(p) > 0);
}
\end{ccode}

\begin{ccode}
// https://github.com/fiedl/clsim/blob/sf/hole-ice-2018/resources/kernels/lib/intersection/intersection.h

#ifndef INTERSECTION_H
#define INTERSECTION_H

typedef struct IntersectionProblemParameters {

  // Input values
  //
  floating_t ax;
  floating_t ay;
  floating_t mx;
  floating_t my;
  floating_t r;
  floating4_t direction;
  floating_t distance;

  // Output values, which will be calculated in
  // `calculate_intersections()`.
  //
  floating_t discriminant;
  floating_t s1;
  floating_t s2;

} IntersectionProblemParameters_t;

#endif
\end{ccode}
  \subsection{Calculating Intersections of Photon Trajectories With Hole-Ice Cylinders}
\label{sec:intersections}

\subsubsection{Analytic Approach}\label{analytic-approach}

\image{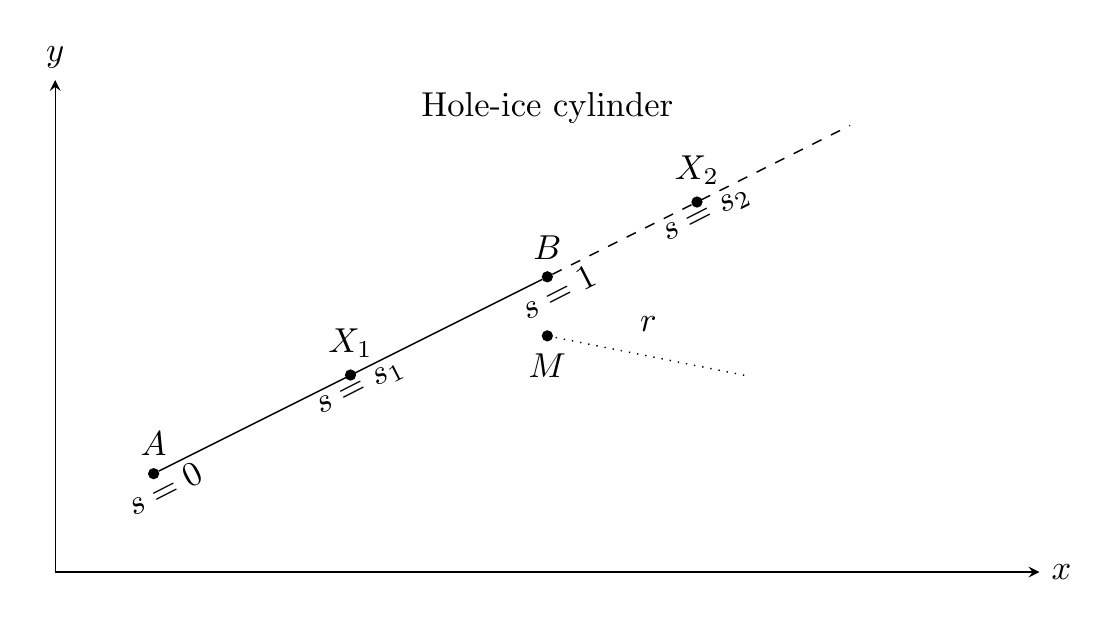}

Consider the starting point \(A := (A_x, A_y)\) and the ending point
\(B := (B_x, B_y)\) of the trajectory.

The length \(\len{AB}\) of the trajectory is given by the norm
\(\norm{.}\) of the vector difference \(\vec{AB}\) of starting and
ending point.

\begin{equation} \notag
  \len{AB} = \norm{\vec{AB}}, \ \ \vec{AB} \identical \vec{B} - \vec{A}, \ \ \norm{\vec{v}} \identical \sqrt{v_x^2 + v_y^2}
\end{equation}

In order to find the intersection points \(X_1\) and \(X_2\), solve the
vectorial system of equations

\begin{equation}
  \vec{A} + s \, \vec{AB} = \vec{M} + \vec{MX}
\end{equation}\begin{equation}
  \norm{\vec{MX}} = r
\end{equation}

for the scalar parameter \(s\). The equation system is quadratic in
\(s\) resulting in zero, one or two solutions.

\begin{equation}
  s_{1,2} = \frac{-\beta \mp \sqrt{\beta^2 - 4\alpha\gamma}}{2\alpha}
\end{equation}\begin{equation}
  \alpha = (B_x - A_x)^2 + (B_y - A_y)^2
\end{equation}\begin{equation}
  \beta = 2\,A_x(B_x-A_x) + 2\,A_y(B_y-Ay) - 2\,M_x(B_x-A_x) - 2\,M_y(B_y-A_y)
\end{equation}\begin{equation}
  \gamma = A_x^2 - 2\,A_x\,M_x^2 + M_x^2 + A_y^2 - 2\,A_y\,M_y + M_y^2 - r^2
\end{equation}

The term under the square root is also called \textbf{discriminant}
\(D\).

\begin{equation}
  D = \beta^2 - 4\alpha\gamma
\end{equation}

For \(D < 0\), the square root does not exist in \(\reals\) and
therefore, no intersection point exists. For \(D = 0\), there is only
one intersection point, which is a tangent point. For \(D > 0\), there
are two intersection points.

Note that \(s = 0\) at the starting point \(A\), \(s = 1\) at the ending
point \(B\), \(s = s_1\) at the first intersection point \(X_1\) and
\(s = s_2\) at the second intersection point \(X_2\).

Therefore, the intersection point coordinate vectors \(\vec{X_1}\) and
\(\vec{X_2}\) are given by:

\begin{equation}
  \vec{X_1} = \vec{A} + s_1 \, \vec{AB}
\end{equation}\begin{equation}
  \vec{X_2} = \vec{A} + s_2 \, \vec{AB}
\end{equation}

This result can be verified using a symbolic mathematics library like
\noun{SymPy}\footnote{\textbf{SymPy} is a Python library for symbolic mathematics. \url{http://sympy.org}}.

\begin{python}
# python

from sympy import *
init_printing(use_unicode=True)

# Variables
# ---------

Ax, Ay = symbols("Ax Ay")
Bx, By = symbols("Bx By")
Mx, My = symbols("Mx My")
Px, Py = symbols("Px Py")
r = symbols("r")
s = symbols("s")

vx = Bx - Ax
vy = By - Ay

# First ansatz
# ------------

lhs = (Ax + s * vx - Mx)**2 + (Ay + s * vy - My)**2 - r**2
expanded_lhs = expand(lhs)

intersection_equation = Eq(lhs, 0)

solution = solve(intersection_equation, s)
s1 = solution[0]
s2 = solution[1]

# Second ansatz
# -------------

alpha = (By - Ay)**2 + (Bx - Ax)**2
beta = 2 * Ay * (By - Ay) + 2 * Ax * (Bx - Ax) - 2 * My * (By - Ay) - 2 * Mx * (Bx - Ax)
gamma = Ay**2 - 2 * Ay * My + My**2 - r**2 + Ax**2 - 2 * Ax * Mx + Mx**2
discriminant = (beta**2 - 4 * alpha * gamma)

ss1 = (- beta - sqrt(beta**2 - 4 * alpha * gamma)) / (2 * alpha)
ss2 = (- beta + sqrt(beta**2 - 4 * alpha * gamma)) / (2 * alpha)

# Compare ansatzes
# ----------------

# See also: http://docs.sympy.org/latest/tutorial/gotchas.html

# # Does not work, because `==` checks object (structural) equality
# s1 == ss1
# s1 == ss2
# s2 == ss1
# s2 == ss2

# Should be 0 for the matching pairs:
simplify(s1 - ss1)  # => 0
simplify(s1 - ss2)
simplify(s2 - ss1)
simplify(s2 - ss2)  # => 0

# Or numerically:
s1.equals(ss1)  # True
s2.equals(ss2)  # True
\end{python}

\subsubsection{Geometric Approach}\label{geometric-approach}

It turns out, the simulation on the GPU is faster and more precise if
one does not treat the coordinates, e.g. \(A_x, A_y\) as separate
quantities but keeps the calculations as much as possible vectorial and
uses the GPU-native vector operation functions like \texttt{dot} for the
vector scalar product (``dot product'').

\image{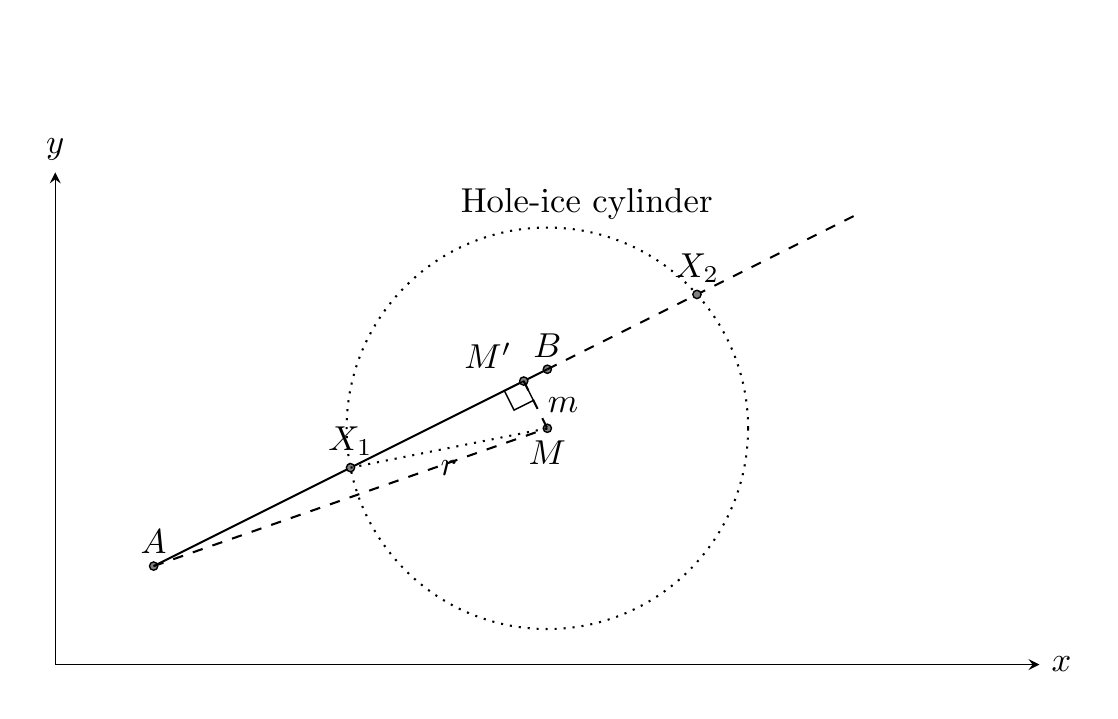}

\paragraph{Goal}

The goal is to find the distances \(\len{AX_1}, \len{AX_2}\) from the
starting point \(A\) to the intersection points \(X_1, X_2\).

\paragraph{Relations}

Consider a projection of \(M\) onto \(AB\). The projected point is
called \(M'\).

Given \(\len{A M'}\) and \(\len{X_1 M'}\), the distances to the
intersection points can be calculated as:

\begin{equation} \len{A X_{1,2}} = \len{A M'} \mp \len{X_1 M'} \end{equation}

Given \(\vec{AM}\) and the unit vector \(\vec{v}\) in \(\vec{AB}\)
direction, \(\len{AM'}\) can be calculated using a vector projection:

\begin{equation} \len{AM'} = \vec{AM} \cdot \vec{v} \end{equation}

The operator \(\cdot\) is the scalar product (dot product). The unit
vector \(\vec{v}\) in \(\vec{AB}\) direction is
\(\vec{v} = \vec{AB} / \len{AB}\).

The distance \(m:=\len{MM'}\) on right hand side of the triangle
\(\triangle AM'M\) can be calculated using the pythagorean theorem:

\begin{equation} m: \len{AM'}^2 + m^2 = \len{AM}^2 \end{equation}

The distance \(\len{X_1 M}\), interpreted as part of the triangle
\(\triangle X_1 M' M\) can be calculated using the pythagorean theorem:

\begin{equation} \len{X_1 M'}: \len{X_1 M'}^2 + m^2 = r^2 \end{equation}

The length \(\len{AM}\) of the vector \(\vec{AM}\) can be calculated
using the scalar product, i.e.~projecting the vector on itself, and
taking the square root:

\begin{equation} \len{AM} = \sqrt{\vec{AM} \cdot \vec{AM}} \end{equation}

Using both pythagorean equations, eliminating \(m\), one finds:

\begin{equation}\begin{split}
  \len{X_1 M'} &= \sqrt{r^2 - m^2} \\
    &= \sqrt{r^2 - (\len{AM}^2 - \len{AM'}^2)} \\
    &= \sqrt{r^2 - \len{AM}^2 + \len{AM'}^2} \\
    &= \sqrt{r^2 - \vec{AM} \cdot \vec{AM} + \len{AM'}^2}
\end{split}\end{equation}

\paragraph{Algorithm}

Using these relations, the desired distances \(\len{AX_1}, \len{AX_2}\)
can be calculated with the following steps:

\begin{enumerate}
\def\labelenumi{\arabic{enumi}.}
\tightlist
\item
  Calculate the length \(\len{AM'}\) by projecting \(\vec{AM}\) onto the
  unit vector \(\vec{v}\) in \(\vec{AB}\)-direction:
  \[ \len{AM'} = \vec{AM} \cdot \vec{v} \]
\item
  Calculate \(\len{X_1M'}^2\):
  \[ \len{X_1 M'}^2 = r^2 - \vec{AM} \cdot \vec{AM} + \len{AM'}^2 \]
\item
  Calculate \(\len{X_1M'}\): \[ \len{X_1 M'} = \sqrt{\len{X_1 M'}^2} \]
  In this approach, \(\len{X_1M'}^2\) plays the role of the
  \textbf{discriminant}. If it is greater than zero, there are two
  intersection points. If it is zero, the intersection point \(X\) falls
  onto \(M'\), i.e.~the line is actually a tangent. If the discriminant
  is less than zero, there is no intersection, because the radius \(r\)
  is too small, i.e.~the line is out of reach, resulting in the
  discriminant becoming negative.
\item
  Calculate the desired distances \(\len{AX_1}\) and \(\len{AX_2}\):
  \[ \len{A X_{1,2}} = \len{A M'} \mp \len{X_1 M'} \]
\end{enumerate}

Or, as \noun{C} code using the \noun{OpenCL} native vector functions:

\begin{ccode}
  // Step 1
  const floating4_t vector_AM = (const floating4_t)(M.x - A.x, M.y - A.y, 0, 0);
  const floating_t length_AMprime = dot(vector_AM, v);

  // Step 2
  const floating_t discriminant = r * r - dot(vector_AM, vector_AM) + length_AMprime * length_AMprime;

  // Step 3
  const floating_t length_XMprime = native_sqrt(discriminant);

  // Step 4
  const floating_t length_AX1 = length_AMprime - length_XMprime;
  const floating_t length_AX2 = length_AMprime + length_XMprime;
\end{ccode}

  \subsection{How to Switch Back to Standard-\clsim's Media-Propagation Algorithm}
\label{sec:how_to_switch_media_propagation}

In order to switch back to the standard \clsim algorithm, even after the
new code has been merged into the \icecube simulation framework, the
standard \clsim code is provided as drop-in replacement.

\sourcepar{The standard \clsim code as drop-in replacement can be found at \url{https://github.com/fiedl/clsim/blob/sf/hole-ice-2018/resources/kernels/lib/propagation_through_media/standard_clsim.c}}.

To apply this code, switch in \texttt{propagation\_kernel.c.cl} \newline
from calling \texttt{apply\_propagation\_through\_different\_media()}
\newline
to
\texttt{apply\_propagation\_through\_different\_media\_with\_standard\_clsim()}.

Also check \url{https://github.com/fiedl/hole-ice-study/issues/115}
whether an \texttt{IceTray} switch has already been implemented.

  \subsection{Exponential Distribution of the Total Photon Path Length}
\label{sec:exponential_distribution}

This section shows why the total-path-length distribution of photons
propagating through a homogeneous medium described in section
\ref{sec:total_path_length_distribution} is expected to follow an
exponential curve.

Let \(N\) be the total number of started within the medium. Let
\(\lambda\abs\) be the photon absorption length within the medium.

The probability \(p(x):=f\,\dx\) for a photon to be absorbed within its
path length interval \([x; x + \dx[\) is the same as long as the photon
stays within the homogeneous medium.

Let \(n(x)\) be the number of photons with a path length of \(x\) or
more, which is the number of photons that still exist with a path length
greater or equal \(x\).

The number of photons \(m(x)\) that are absorbed within the interval
\([x; x + \dx[\) is determined by the change of the number of remaining
photons, which in the limit of many photons is proportional to the
absorption probability \(p(x)\).

\[ m(x) = -\dn(x) = p(x)\ n(x) = f\,\dx\,n(x) \]

\[ \frac{\d}{\dx}\ n(x) = -f\ n(x) \]

As the derivative of \(n\) is proportional to \(n\), \(n\) is an
exponential.

\[
  n(x) = a\,\e^{b\,x}, \ \ \
  \frac{\d}{\dx}\,n(x) = b\,a\,\e^{b\,x} = b\,n(x) = -f\,n(x)
\]

The absorption length \(\lambda\abs\) is defined as the distance after
that the number of photons has dropped to \(\sfrac{1}{\e}\) of the
original number.

\[
  n(x) = a\,\e^{b\,x}, \ \
  n(\lambda\abs) = \frac{1}{\e}\, n(0)
  \ \ \Rightarrow \ \
  b = \sfrac{-1}{\lambdaabs}
\]

\[
  n(x) = n(0) \cdot \e^{\sfrac{-x}{\lambda\abs}}, \ \ \ n(0) = N
\]

In a histogram of the total path lengths \(X\), the bin height is
proportional to the number \(m(x)\) of photons that are absorbed within
the interval \([x; x + \dx[\), not the number \(n(x)\) of remaining
photons.

\[
  m(x) = p(x)\,n(x) = p(x)\,N\,\e^{\sfrac{-x}{\lambda\abs}}
\]

In the case of a homogeneous medium where \(p(x)\) is constant for all
\(x\), \(p(x) = p_0\), the histogram should also follow an exponential
curve.

\[
  m(x) = p_0\,N\,\e^{\sfrac{-x}{\lambda\abs}}
\]

From the rate of the exponential decay, one can read the absorption
length \(\lambda\abs\).

If, on the other hand, there is a medium border at \(x_0\) such that the
absorption probability is piecewise defined,

\[
  p(x) = \begin{cases}
    p_0 & : x \leq x_0 \\
    p_1 & : x > x_0
  \end{cases}
\]

then the histogram follows a piecewise defined curve, consisting of two
exponential curves.

\[
  m(x) = \begin{cases}
    p_0\,N\,\e^{\sfrac{-x}{\lambda_0}} & : x \leq x_0 \\
    p_1\,N\,\e^{\sfrac{-x}{\lambda_1}} & : x > x_0
  \end{cases}
\]

From the rates of the exponential decays, one can read the absorption
lengths \(\lambda_0\) and \(\lambda_1\) of both media from the
histogram.

Note that as the histogram shows the number \(m(x)\) of decayed photons
within an interval, not the number \(n(x)\) of remaining photons, the
curve the histogram follows does not need to be continuous but may have
a jump discontinuity at the position \(x_0\) of the medium border.

The number \(n(x)\) of remaining photons, however, is continuous, also
at the position \(x_0\) of the medium border.

\[
  n(x) = N - \int_0^x m(x) = \begin{cases}
    N - f_0\,N\,\lambda_0\,\e^{\sfrac{-x}{\lambda_0}} & : x \leq x_0 \\
    n(x_0) - f_1\,N\,\lambda_1\,\e^{\sfrac{-x}{\lambda_1}} & : x > x_0
  \end{cases}
\]

  \subsection{Angular-Acceptance Simulation For H2 Parameters}
\label{sec:angular_acceptance_simulation_for_h2_parameters}

The so-called H2 hole-ice parameters assume a hole-ice radius of
\(30\cm\), corresponding to the entire drill hole being filled with hole
ice, and a geometric hole-ice scattering length of \(50\cm\),
corresponding to an effective scattering length of \(8.33\m\).
\cite{holeicestudieswithyag}

Using the new medium-propagation algorithm (section
\ref{sec:algorithm_b}) with direct detection and plane waves as photon
sources (section \ref{sec:angular_acceptance_scan}) to generate an
effective angular-acceptance curve for these H2 hole-ice parameters:

\hspace{1cm}

\begin{center}
  \includegraphics[width=0.75\textwidth]{img/angular-acceptance-karle-h2-vs-reference}
\end{center}

\hspace{1cm}

The blue a priori curve is based on previous \photonics simulations
assuming the same H2 hole-ice parameters. \cite{lundberg, icepaper}

\newpage

The same simulation assuming an effective hole-ice scattering length of
\(50\cm\), corresponding to a geometric scattering length of \(3\cm\):

\hspace{1cm}

\begin{center}
  \includegraphics[width=0.75\textwidth]{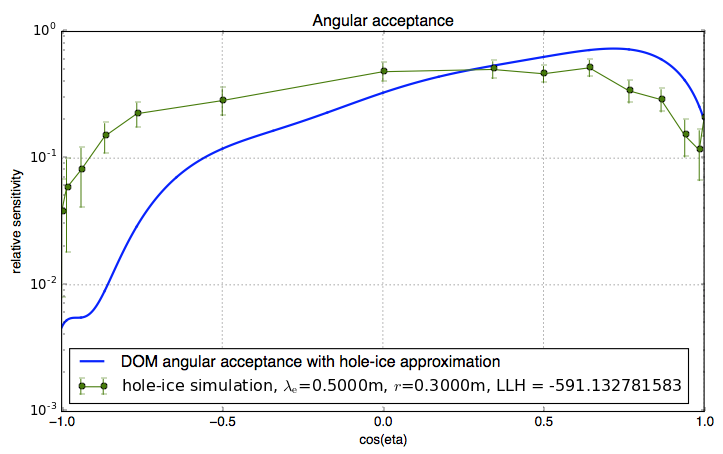}
\end{center}

\hspace{1cm}

\docpar{These simulations are documented in \issue{80}.}

  \subsection{Technical Issues Concerning Graphics Processing Units}
\label{sec:gpu_technical_issues}

\paragraph{Driver Issue Concerning \texttt{isnan}}

With some GPUs or GPU drivers, basic functions like
\texttt{bool\ isnan(float\ a)} may have bugs and may not always return
the expected results. These kind of issues are especially hard to track
down because they do not show up in unit tests when running the tests on
a different architecture, for example on the local CPU. They also don't
raise exceptions.

In this case, the workaround has been to use a custom implementation of
the \texttt{isnan} function that uses another basic operator to
determine the result and circumvents the native implementation of
\texttt{isnan}.

\begin{ccode}
// Fixed custom replacement of `isnan`:
bool my_is_nan(floating_t a) { return (a != a); }
\end{ccode}

\docpar{The issue is documented in \issue{14}. The workaround is documented in \issue{16}.}

\paragraph{Driver Issue Concerning Kernel Caching}

When changing the kernel code has no effect when running the simulation
again, this could be due to a caching issue: The cache for kernels that
are included using the \texttt{\#include} pre-processor statement is not
reset automatically when the included files change.

The workaround is to disable caching using an environment variable, or
to reset the cache by hand.

\begin{bash}
# Disable caching globally. Despite the name this also works with OpenCL.
export CUDA_CACHE_DISABLE=1

# Reset cache manually:
rm -r ~/.nv/ComputeCache
\end{bash}

\docpar{This problem and the workaround are documented in \issue{15}.}

\paragraph{Numerical Inconsistencies}

Computations that are mathematically equivalent are not guaranteed to be
numerically equivalent as well. Performing the same calculation using
different formula, or calculating the same quantity using two different,
but mathematically equivalent algorithms, may lead to different results.

If such two calculations are performed within one algorithm, this may
lead to meaningless results or simulation crashes.

\begin{figure}[htbp]
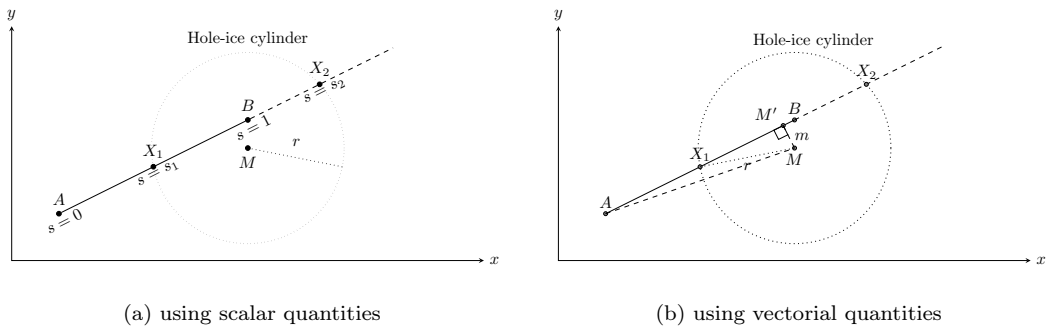

  \subcaptionbox{using scalar quantities}{\halfimage{intersection-Kahm4UeY}}
  \subcaptionbox{using vectorial quantities}{\halfimage{intersection-vectorial-auloLai2}}
  \caption{Calculating intersections of photon trajectory $AB$ and the hole-ice cylinder represented as circle.}
  \label{fig:aeQuae2U}
\end{figure}

For example, one part of the hole-ice-correction algorithm used the
\noun{Pythagorean Theorem} to calculate the distance of the photon from
the hole-ice center. Comparing this to the hole-ice radius, the
algorithm decided whether the photon is within or outside the cylinder.
Another part of the algorithm calculated intersections of the photon
trajectory with the hole-ice cylinder. Evaluating the scale parameters
\(s_1\) and \(s_2\) (see figure \ref{fig:aeQuae2U} a), the algorithm
decided again whether the photon is within or outside the hole ice.

If the photon is near the hole-ice radius, one mechanism may come to the
conclusion that the photon is within the hole-ice cylinder while the
other mechanism considers the photon outside the cylinder. If both
mechanisms are used in the same algorithm, this leads to
inconsistencies.

The solution to this problem is to only use one way to determine whether
the photon is within or outside the cylinder. The decision may still be
``wrong'' compared to a high-precision calculation, but it will be
consistent and not cause the program to crash.

Another, less destructive, but even more demonstrative example for
numerical inconsistencies is comparing two ways of calculating
intersections as shown in figure \ref{fig:aeQuae2U}.

\begin{figure}[htbp]
  \subcaptionbox{using scalar quantities to calculate the intersection points coordinate-wise}{\includegraphics[width=0.48\textwidth, clip, trim = {0 2cm 0 0}]{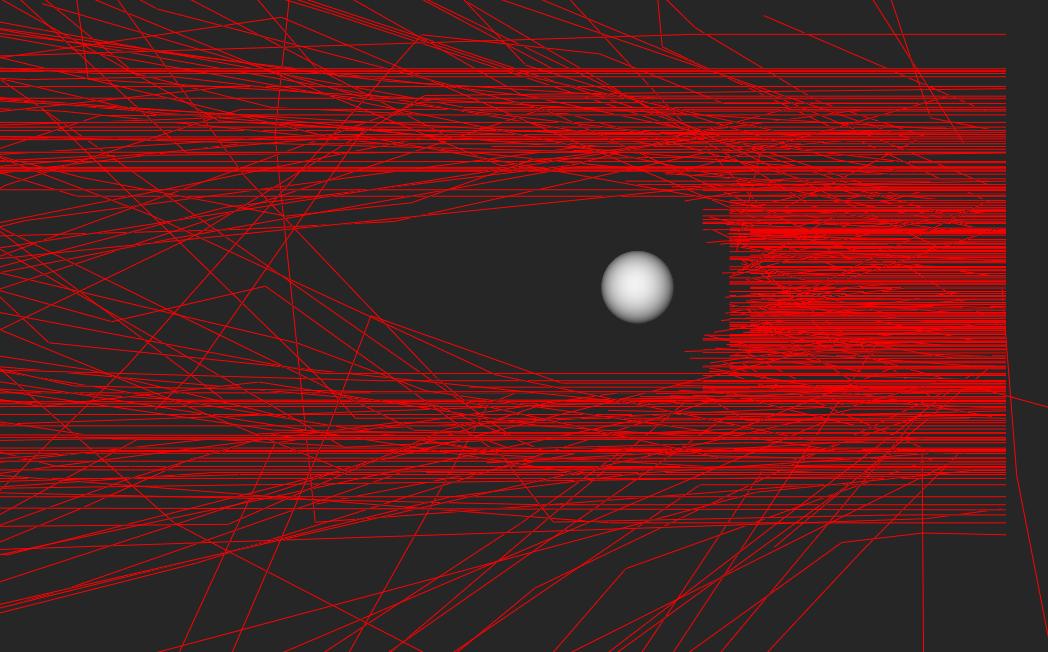}}\hfill
  \subcaptionbox{using native vector operations to calculate the intersection points}{\halfimage{instant-absorption-steamshovel-moo9Eiqu}}\hfill
  \caption{Simulation of photons propagating towards a cylinder configured for instant absorption. Both simulations are the same except for the method how intersections of photon rays and cylinders are calculated. Using native vector operations (b) leads to numerically more precise results.}
  \label{fig:usie5Ohj}
\end{figure}

Both methods are described in section \ref{sec:intersections}. The
important difference is that one method uses scalar quantities to
calculate the intersection points, the other method uses vectorial
quantities and native vector operations. While both methods are
mathematically equivalent, their results differ in their numerical
precisions, which is visualized in figure \ref{fig:usie5Ohj}: Both
simulations show photons hitting an instantly absorbing cylinder. While
native vector operations cut the photons cleanly at the cylinder border,
using only scalar quantities shows a scraggy border.

\docpar{This issue is documented in \issue{28}.}

\paragraph{Memory Issues: ``OpenCL worker thread died unexpectedly''}

While the user might wholeheartedly agree that the crash is rather
unexpected, the error message does not help to find the underlaying
issue, which is most probably a memory issue, for example allocating too
few or too much memory on the GPU.

To circumvent this issue, this study adds the propagation configuration
parameter \texttt{MaxNumOutputPhotonsCorrectionFactor} to \clsim, which
acts as a factor within the calculation that determines how much memory
will be allocated for storing photons.

As a rule of thumb, using values from
\texttt{MaxNumOutputPhotonsCorrectionFactor\ =\ 0.1} to
\texttt{MaxNumOutputPhotonsCorrectionFactor\ =\ 1e-3} may resolve the
issue. But the memory might then be insufficient to store all propagated
photons, especially when photon paths are to be saved as well as
compared to only the hits.

Another workaround when creating visualizations is to propagate less
photons and/or run the simulation on a CPU. The memory is much larger
when running on a CPU such that this issue might not occur on the CPU.
But the simulation time will be longer on the CPU.

\docpar{This issue is documented in \issue{23}.}

  \include{text/erklaerung}

\end{document}